# Phase equilibria of the Al–Ti–Nb–Zr–Ta system

Jiří Kozlík[1,*], František Lukáč[2], Mariano Casas-Luna[1], Jozef Veselý[1], Eliška Jača[1], Kateřina Ficková[1], Stanislav Šašek[1], Kristína Bartha[1], Adam Strnad[1], Tomáš Chráska[2], and Josef Stráský[1]

[1] *Department of Physics of Materials, Faculty of Mathematics and Physics, Charles University, Ke Karlovu 5, Prague 121 16, Czechia*
[2] *Institute of Plasma Physics of the Czech Academy of Sciences, U Slovanky 2525/1a, Prague 182 00, Czechia*

## ABSTRACT

Phase equilibria in the Al–Ti–Nb–Zr–Ta refractory complex concentrated alloy system were investigated using a high-throughput experimental approach. A pseudo-ternary section of the quinary compositional space was prepared by a honeycomb type powder metallurgy design, consolidated by spark plasma sintering and subsequently homogenized at 1400 °C for 168 h. Phase constitution and chemical partitioning were characterized by SEM/EDS, XRD, EBSD, and TEM, supported by a custom EDS phase clustering workflow. Equilibrium microstructures consisting primarily of BCC, B2, and secondary phases were identified across the sampled compositions, with nanoscale precipitates forming in Zr- and Ta-rich regions. Measured phase compositions were compared with CALPHAD predictions, revealing both agreements and systematic deviations linked to CALPHAD database limitations. The results provide new experimental insight into phase stability and microstructural trends in Al–Ti–Nb–Zr–Ta alloys and demonstrate the effectiveness of high-throughput combinatorial approaches for mapping complex multicomponent systems.

## Introduction

Refractory complex concentrated alloys (RCCAs) are a subset of complex concentrated alloys (CCAs), also known as high entropy alloys (HEAs), which are rich in high melting point transition metals. They have emerged as candidates for elevated temperature structural service. The RCCAs based on refractory elements (Ti, Nb, Mo, W, Ta, Zr,…) generally exhibit BCC structure as the major phase, although many other phases can be present, depending on the actual composition. The current research is focused on alloys which could potentially outperform the Ni-based superalloys in terms of high-temperature strength, while having at least a minimum level of room temperature ductility and fracture toughness [1].

It is in the very nature of the RCCAs, that the compositional space is large and thus not that much explored so far. In an exhaustive review paper by Miracle, Senkov and co-workers [1], the lack of reliable data for calculating even the basic alloy properties was identified as one of the key challenges in the field.

Address correspondence to E-mail: jiri.kozlik@matfyz.cuni.cz

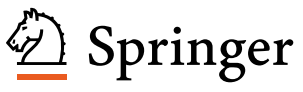

The thermodynamic information about the materials is generally calculated using the CALPHAD approach [2], which is fundamentally based on experimental data. Hence, solid, reliable and rich experimental data are needed which are, however, generally scarce for RCCAs. The aim of this paper is to provide such data for a subregion of the compositional space, namely of the Al–Ti–Nb–Zr–Ta subsystem, by applying efficient high-throughput approach.

The prominent phenomenon observed in this subsystem is a phase decomposition of the major BCC phase into two BCC or B2 phases, e.g. [3, 4], resembling the cuboidal microstructures typically observed in Ni superalloys [5, 6]. Prediction of whether these phases will form in the material and to what extent is commonly done via CALPHAD calculations before physically preparing the alloy, which calls for maximum accuracy of CALPHAD predictions.

The CALPHAD approach is based on assessing the data for the lower order subsystems first (i.e., binary or ternary) and then extrapolating the thermodynamic data for higher dimensions (more complex solutions or compounds). Recently, we have shown that assessing the higher-order data is equally important for accurate predictions, since there are element combinations not encountered in the binary or ternary systems, but obviously stable and confirmed in the material experimentally, such as $Al_3(Zr,Ti,Ta,Nb)_5$ intermetallics [7]. The newer versions of the available thermodynamic databases (e.g. TCHEA8 for Thermo-Calc) include some quaternary diagrams, but more detailed assessment is necessary for improved and more accurate predictions.

The number of alloying elements and therefore the number of possible alloy compositions calls for optimizing the fundamental acquisition of experimental data by incorporating high-throughput approaches into the workflow. The high-throughput methods for sample preparation and analysis typically incorporate miniaturization of the specimens, standardization of the geometry and automation of the analyses [8, 9], multiple compositions are often prepared at once within a single specimen (which is then called a material library), either separated, or in a form of a gradient.

This approach can be used for getting the information about the equilibrium chemical and phase even when the system is compositionally heterogeneous at the large scale, but *locally homogeneous* (at the scale applicable for the selected experimental method), then the results can be taken as representative for the bulk material.

The smallest specimens can be prepared in a form of 2D compositionally graded thin films [9, 10] via magnetron sputtering. Although being able to prepare a wide range of composition easily, the interaction with the substrate and possible presence of residual stresses can influence the phase stability. Also, mechanical properties of the bulk material cannot be assessed from such samples, and the undergoing phase transitions may not be relevant for bulk material.

Additive manufacturing is a promising candidate for high-throughput manufacturing of material libraries, as it provides the freedom of geometry and allows to adjust the composition during the run [11]. Directed energy deposition (DED) was used for high-throughput screening of RCCAs by compositional grading along build direction [12–14].

However, some alloys are difficult to prepare using melting-based techniques (DED or even arc-melting), mainly due to dissimilar melting points of the components, resulting in significant evaporation losses. Therefore, using purely solid-state methods is preferable for this type of alloys. Solid-state sintering is applied in this study.

Recently, we have shown that a heterogeneous RCCA sample can be prepared with a honeycomb geometry by a combination of field-assisted sintering (FAST), also known as spark plasma sintering (SPS) at moderate temperature of 1300 °C, followed by long homogenization treatment at 1400 °C for 168 h [15]. Such treatment was shown to be able to homogenize the Al–Ti–Nb–Zr–Ta equiatomic alloy sufficiently [7]. Independently, a similar method was proposed by Zhao et al. [16], who prepared a material library of the Cantor-type (FCC-based) CCA using pre-milled powder mixtures filled into honeycomb compartments and processed by hot isostatic pressing (HIP) at 1000 °C.

The scope of this paper is to investigate the equilibrium phase and chemical compositions of the Al–Ti–Nb–Zr–Ta system and to openly provide the experimental data, so it can be further used for optimization of the available thermodynamic and other models. Given the breadth of the dataset, the main text focuses on methodology and general trends, while detailed compartment-level microstructural analysis is provided in the supplementary materials. The most intriguing results are highlighted in the final parts of the Results section, followed by comparison to CALPHAD predictions for all prepared compositions.

## Methods

### Material preparation

Within a single specimen, we prepared a two-dimensional (pseudo-ternary) section of the four-dimensional compositional space of a quinary Al–Ti–Nb–Zr–Ta system and two quaternary (Al–Nb–Zr–Ta and Ti–Nb–Zr–Ta) subsystems using hexagonal compartments. The specimen was prepared by blended elemental powder metallurgy (BEPM), as described in detail in [15]. The sample consists of 19 compartments arranged in a honeycomb fashion (see Fig. 1), each with a different elemental powder blend to form the desired composition after sintering. The list of nominal compositions of each compartment is provided in Supplementary material A.

Table 1 summarizes properties of the initial elemental powders used for sintering. Only spherical powders were used, as their low specific surface area results in lower oxygen and nitrogen contents compared to irregular powders.

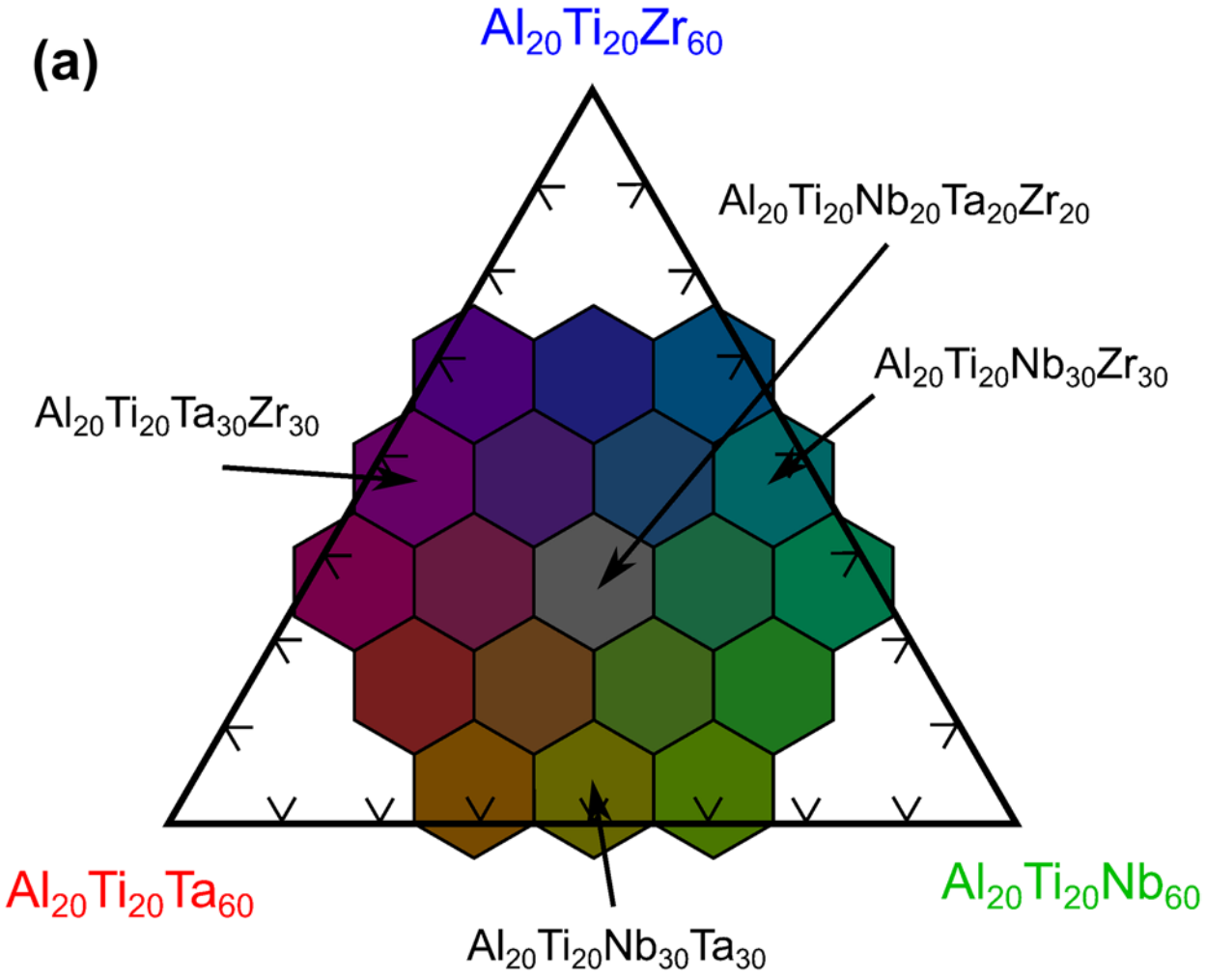

**Figure 1** A schematic of the mapping strategy—a two-dimensional ternary section of the full four-dimensional compositional space overlaid with the hexagonal compartments, each with a different nominal alloy composition. The concentration step between the compartments is 10 at.% (for the quinary Al–Ti–Nb–Zr–Ta system).

Samples were prepared using field-assisted sintering technology (FAST) / spark plasma sintering (SPS) using an in-house designed tool for separating compartments containing different powder compositions.

The sintering was performed in SPS 10–4 furnace (Thermal Technology LLC, USA) in a vacuum of the order of 1 Pa at 1300 °C for 30 min using thin tungsten foil separators inside a graphite crucible. As the sintering at 1300 °C for 30 min is not sufficient to homogenize the elemental powder blend [7], the sample was homogenized for 168 h at 1400 °C in an Ar-filled (99.9999%) tube furnace, wrapped in a Zr foil (acting as an additional oxygen getter) and water quenched directly from the annealing temperature. The samples were then cross-sectioned and grinded using SiC papers, followed by polishing at Vibromet polisher using 0.3 and 0.05 µm alumina suspension and 0.05 µm silica suspension (8 h each step).

More details about the sample preparation methodology can be found in a dedicated paper published earlier [15].

### Experimental characterization

The samples were characterized by carrier-gas hot extraction (CGHE, Bruker G8 Galileo) for oxygen and nitrogen content.

The microstructural observation was done by scanning electron microscopy (SEM) using ThermoFisher Apreo 2S and Zeiss Auriga Compact Crossbeam microscopes, both equipped with energy-dispersive X-ray spectroscopy (EDS, EDAX Octane Elite) and EBSD (EDAX Velocity). SEM imaging and EDS acquisition was partly automatized using newly developed predefined scripts.

The XRD measurements were carried out on vertical powder $\theta$-$\theta$ diffractometer D8 Discover (Bruker AXS, Germany) using Cu K$\alpha$ radiation and Rigaku

**Table 1** The results of a chemical analysis of powders by a carrier-gas hot extraction (CGHE) method. The numbers in parentheses are standard deviations at the last digit

| Powder | Manufacturer | Size and morphology | Oxygen (wt%) | Nitrogen (wt%) |
|---|---|---|---|---|
| Al | Eckart TLS | spherical (20–53 µm) | 0.083(4) | 0.0020(5) |
| Ti | Eckart TLS | spherical (20–53 µm) | 0.091(4) | 0.0098(8) |
| Nb | CAMEX | spherical (15–45 µm) | 0.221(5) | 0.051(4) |
| Zr | TLS Technik | spherical (20–80 µm) | 0.175(1) | 0.009(1) |
| Ta | CAMEX | spherical (15–45 µm) | 0.076(8) | 0.007(1) |

Rapid II using an Ag K$\alpha$ radiation, equipped with a graphite monochromator and a curved 2D detector for measurement in reflection mode. Rocking of the sample in $\varphi$ and $\chi$ axes was employed for better statistics. Phase identification and refinement was done using PDF-5 + database of crystalline phases in TOPAS V5 (Bruker AXS, Germany).

The lamella for TEM was prepared by Zeiss Auriga Compact Crossbeam using Ga ions and investigated using JEOL 2200FS transmission electron microscope operated at 200 kV.

Microhardness was measured with a Qness 10A automated testing device, using Vickers HV0.5 method (0.5 kgf load).

## EDS phase analysis

The ultimate aim of the phase and compositional analysis are phase identification and determination of their fractions, chemical composition of each phase and the average composition allowing us to place the studied equilibrium to the compositional space. EDS can be used to identify different present phases (based on different chemical composition). The goal is to assign each map point (one EDS spectrum) to a cluster of similar spectra (hence corresponding to a phase) and to compute an average spectrum for every cluster (i.e., for every phase). Due to the amount of the studied nominal overall compositions, the EDS phase analysis needs to be effective, reproducible and semi-automated. Because the vendor software provides only very limited control over the clustering algorithm and its parameters, a custom tool was developed to expose the full pipeline and allow parameter tuning.

The raw data file contains one EDS spectrum per each map point. First, the map is spatially binned to improve the count statistics. Technically, the spectrum is a vector of detector counts of a length equal to the number of energy bins (4096), which, when multiplied by the number of points in the map, makes the clustering problem large and difficult to compute. Therefore, the spectra are decomposed using non-negative matrix factorization (NMF) algorithm [17] to reduce the problem dimensionality (i.e., from the original dimension equal to the number of energy bins to the selected number of *components*–the major features of the spectra). Then, all spectra are expressed as the weighted sum of the components and a vector of weights is assigned to each map point.

The weight vectors are then clustered using a hierarchical density-based algorithm (HDBSCAN) [18], which identifies dense regions in feature space as clusters while labeling sparse points as noise. Optionally, SEM detector signal can be appended as an additional feature to help separate phases. A simple GUI visualizes component loadings, dendrograms, and cluster assignments in real time, allowing the user to adjust the number of NMF components and clustering sensitivity until a meaningful segmentation is achieved.

Spectra of the individual phases and the spectrum of all validly clustered points are then exported, along with the phase statistics and the processing parameters. The spectra are exported in the .msa format, which can be loaded into quantification software (NIST DTSA-II Neptune was used in this work).

The uncertainties of the chemical compositions are a standard output of the NIST DTSA-II Neptune. The uncertainties of the phase volumes (or pixels) are estimated using a simple model. Let $N_i$ be the number of pixels assigned to a phase $i$ and $N$ be the total number of pixels within the map. Then $\sigma_{N_i} = 0.01N_i + \sqrt{N}$. The first term comes from the noise within each phase and was estimated from number of noisy pixels in the single-phase maps. The second term corresponds to the average linear dimension and serves as a proxy for the number of misindexed pixels at the phase boundaries.

The workflow diagram is shown in Fig. 2. The Python-based software is available through a Zenodo repository, along with a more detailed documentation (see the Data Availability section).

## Naming convention

As there are several samples and many different compositions appearing in the paper, we feel that a consistent naming convention is of utmost importance. Following naming template will be used throughout the paper: Sx_Cyy_Pz, where “x” is the sample (S) number (S1—S3), “yy” is a two-digit identifier of the compartment (C, e.g. C22) within the sample (as shown in Fig. 3) and “z” is the (0-based) phase (P) index within each compartment. For example, the third identified phase in compartment C22 of the third sample is labeled as S3_C22_P2.

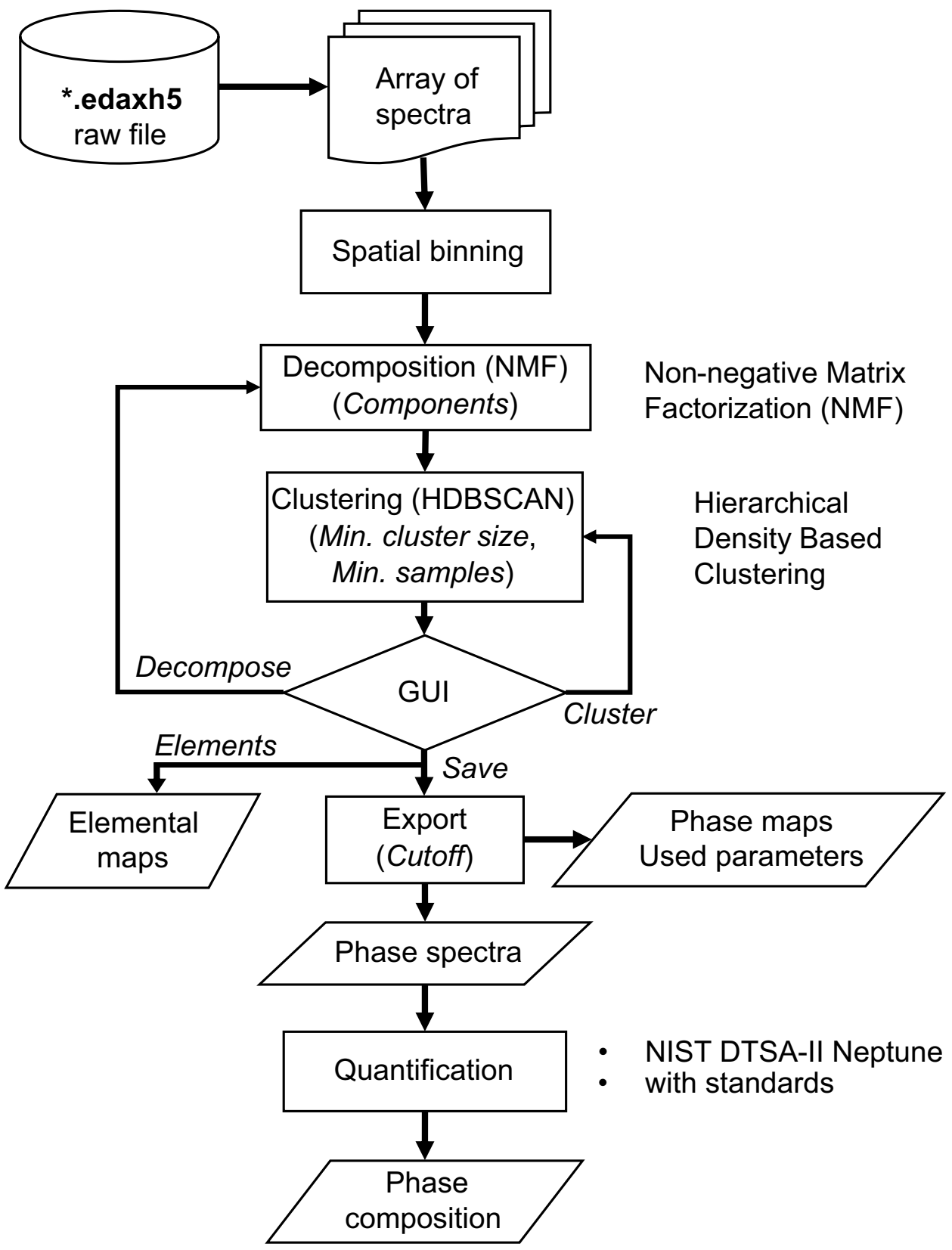

**Figure 2** The flow chart of the EDS clustering tool.

## Results

It is assumed that during annealing for 168 h at 1400 °C followed by immediate water quench, a thermodynamic equilibrium at 1400 °C was reached.

The Results section begins with a general overview of the samples, including integrity and compositional features, followed by microhardness measurements. The core part presents detailed phase and chemical analysis using combined SEM/EDS, EBSD, XRD, and TEM, highlighting recurring microstructural motifs. Finally, experimental phase equilibria are compared with CALPHAD predictions.

## Sample overview

In all homogenized samples, the concentration of oxygen and nitrogen were first measured on specimens randomly selected from the sample. The results are shown in Table 2.

The measured oxygen contents after homogenization exceed those expected from the initial powders (cf. Table 1). This increase is not attributed to the SPS step, which proceeds under high vacuum with short dwell and typically results in negligible oxygen uptake [19]. Instead, it most likely originates during the prolonged 1400 °C homogenization in Ar: even a very low residual oxygen partial pressure, combined with the long duration, can lead to measurable oxygen ingress despite Zr foil gettering. Nitrogen levels remained essentially unchanged.

The low-magnification micrographs and corresponding EDS maps are shown in Fig. 4. The structure of the compartment can be clearly identified, with gradients in Nb, Zr and Ta rotated by 120° with respect to each other as designed.

**Table 2** The results of a chemical analysis of the annealed samples by a carrier-gas hot extraction (CGHE) method. The numbers in parentheses are standard deviations at the last digits; three specimens were measured for each sample

| Sample | System | Oxygen (wt%) | Nitrogen (wt%) |
|---|---|---|---|
| S1 | Al–Ti–Nb–Zr–Ta | 0.22(10) | 0.011(2) |
| S2 | Al–Nb–Zr–Ta | 0.19(9) | 0.008(5) |
| S3 | Ti–Nb–Zr–Ta | 0.33(2) | 0.011(3) |

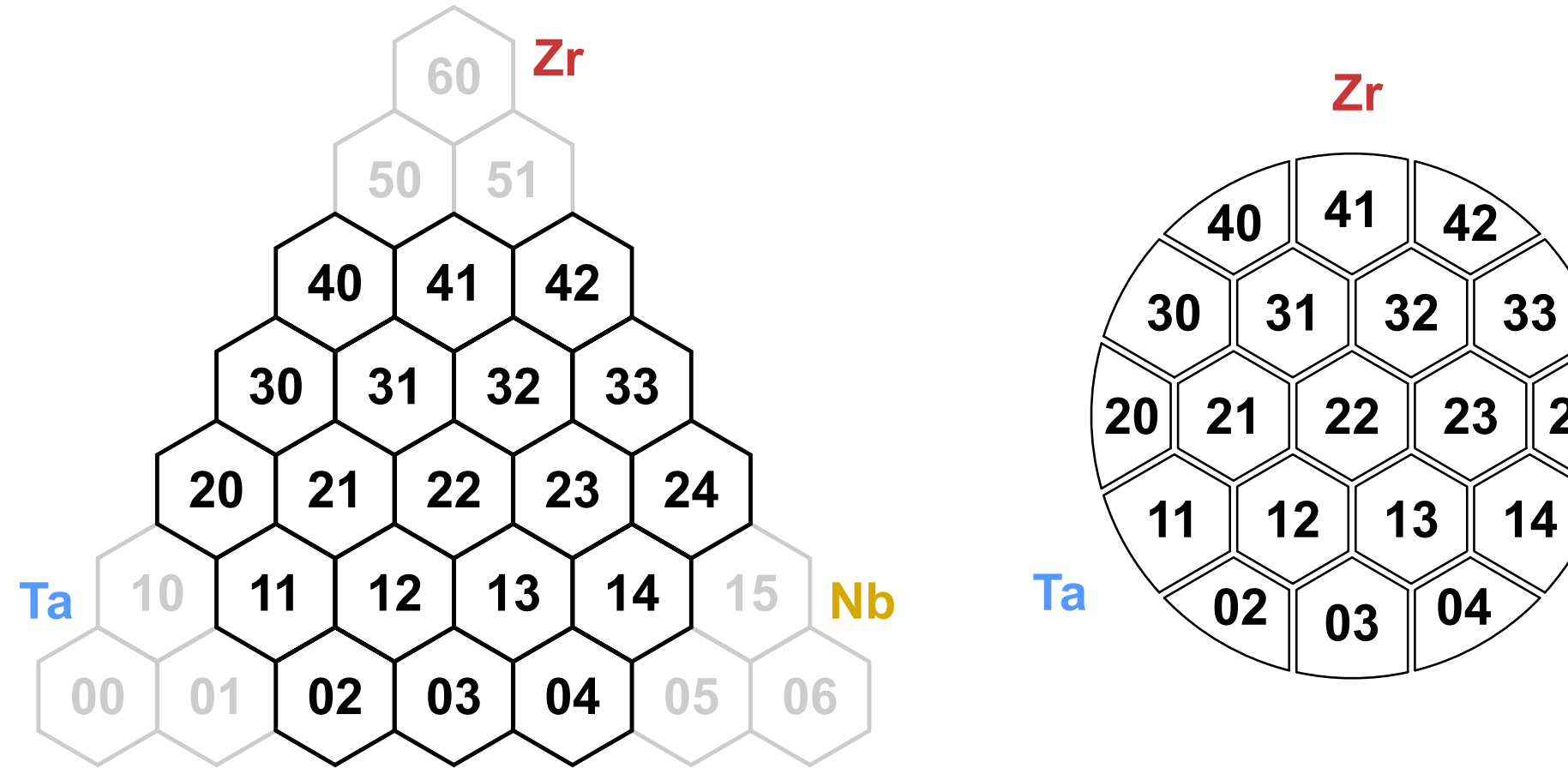

**Figure 3** Two-digit numbering of compartments within the sample. (**a**) The ternary section of compositional space with the sampled compositions highlighted, (**b**) the actual circular geometry of the sintered specimen. All micrographs presented in the paper have this orientation of the concentration gradients.

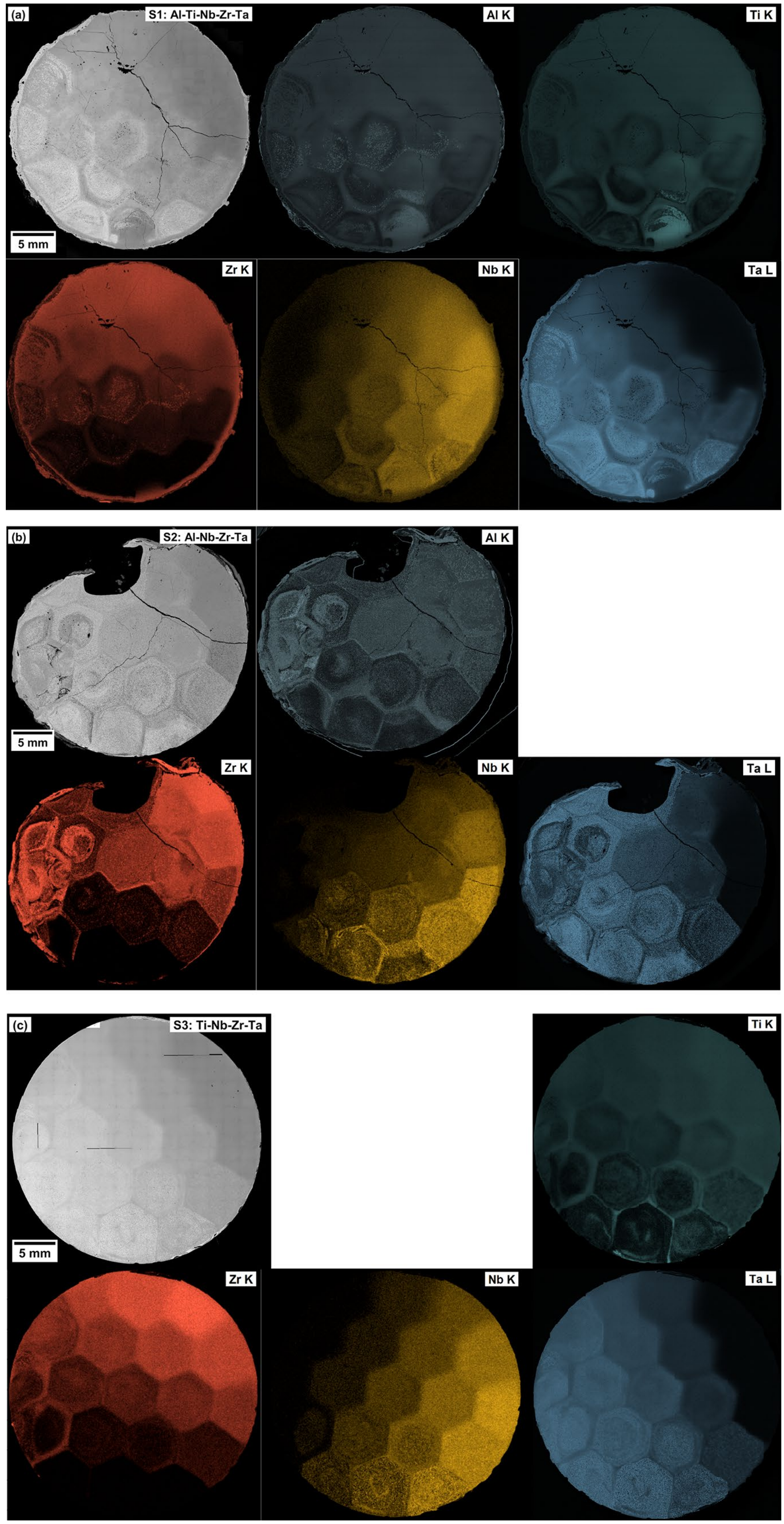

**Figure 4** Low magnification BSE SEM and EDS stitched micrographs, showing the preserved compartment structure. Full-resolution BSE micrographs are available through the linked Zenodo repository.

Occasional macroscopic cracks were observed; however, these do not affect the conclusions of this study, which relies on local phase equilibria rather than bulk mechanical response. The cracks are attributed to the intrinsic brittleness of certain regions, rich in intermetallic phases, in combination with thermal stresses generated during rapid quenching of a compositionally and hence mechanically heterogeneous body.

Compartments C30 and C40–C42 in S2 have melted due to the formation of Al–Zr-rich intermetallics with low melting points. The presence of liquid phase at the annealing temperature will be further demonstrated by eutectic microstructures (see the section on detailed phase analysis below).

Highly porous and/or inhomogeneous areas were found in some compartments (an example is shown in Fig. 5). There are two possible reasons why this happened. First, an uneven distribution of the uniaxial pressure over the heterogeneously filled sintering die could lead to insufficient sintering of the sample resulting in high porosity. Second, as this is predominantly observed in the Ta-rich compartments, its low diffusion rate could limit both the sintering rate and the subsequent homogenization as well.

As the presence of undissolved particles and concentration gradients is a clear mark that no equilibrium was reached, these compartments were excluded from further analysis, as the obtained results would be meaningless. Such heterogeneous compartments are marked as 'htg' within the plots below.

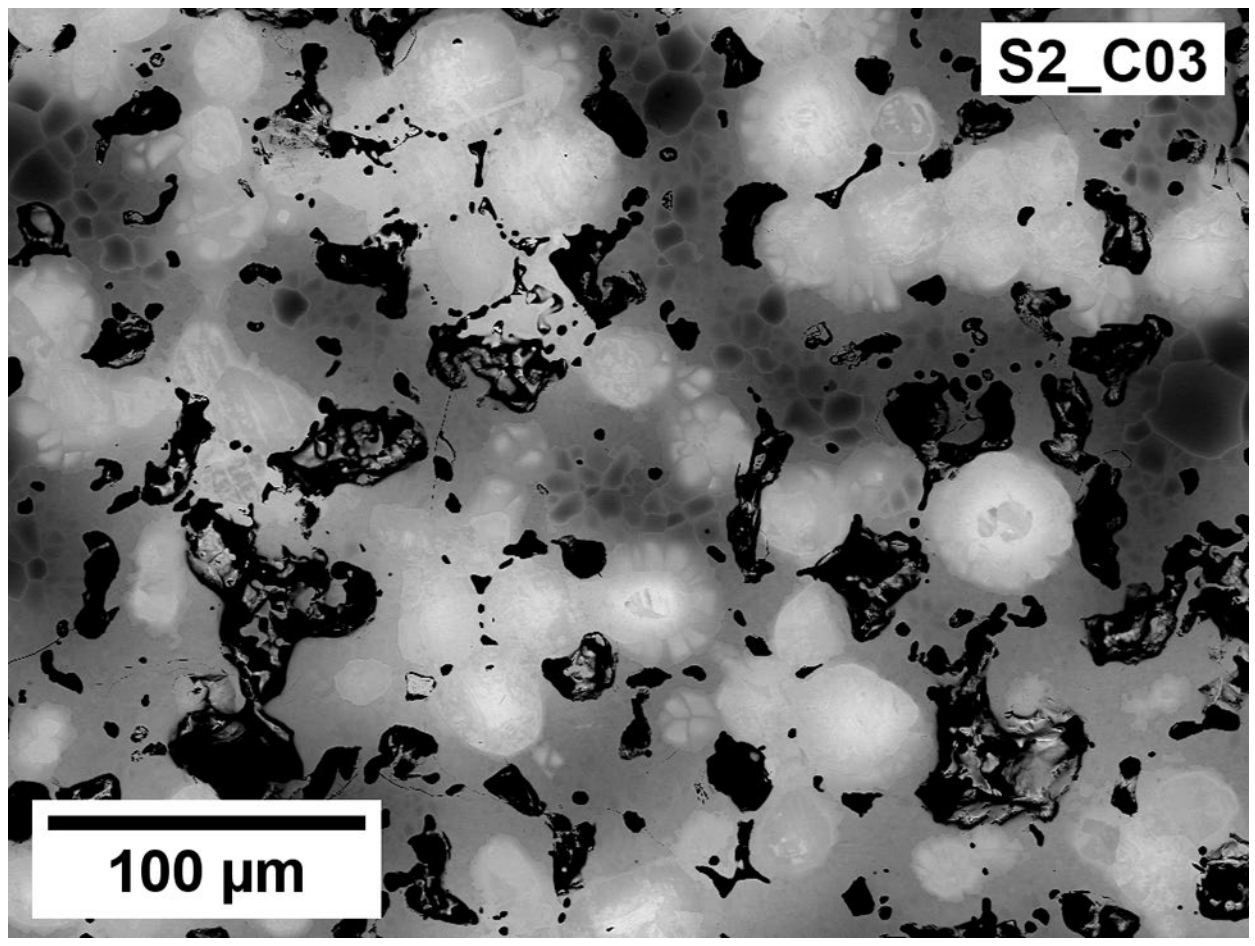

**Figure 5** An example of inhomogeneous S2_C03 compartment with very high porosity. The structure of powder particles is clearly visible, especially the brighter original Ta particles.

## Microhardness

Vickers microhardness was measured on the BCC matrix in all compartments deemed homogeneous and sufficiently dense using a HV0.5 method, the mean values and standard deviations are shown in Fig. 6. In S2, reliable isolation of the BCC phase under the on-board light microscope was not possible due to the intermixing of phases of similar morphology; hence, values for S2 are not reported. Markedly elevated hardness was recorded in Ta and Zr rich regions of S1 and S3, consistent with the presence of nanoscale strengthening precipitates (i.e., not visible during microhardness assessment), as described in detail below. The high standard deviation of the measured microhardness, observed in some compartments (e.g. C02—C12) can be attributed to the high residual porosity from sintering.

## Detailed analysis of phase and chemical composition

The phase analysis within each compartment is the core part of the paper. In each compartment, the EDS map was acquired, the data were clustered into phases, and these were quantified. XRD measurements were performed and the diffraction patterns were analyzed. The matching of structural XRD and chemical SEM results and final phase classification were expert-based and done using combined information from all available methods, including general trends provided by CALPHAD. The quantification is provided as Supplementary material A, along with the phase composition predicted by CALPHAD (using databases TCHEA4 and TCHEA8). Here, we will focus on particular aspects of this supplementary table (Supplementary material A) construction and phase identification; major physically based trends will then be commented on in the Discussion section.

### *XRD*

Conventional Bragg–Brentano geometry was used for the sample S2. For S1 and S3, very large grains necessitated a 2D detector to mitigate missing diffraction peaks, this improved sampling statistics at the expense of angular resolution. Owing to the coarse grain size, phase fraction estimates from XRD are unreliable for all samples and therefore not reported.

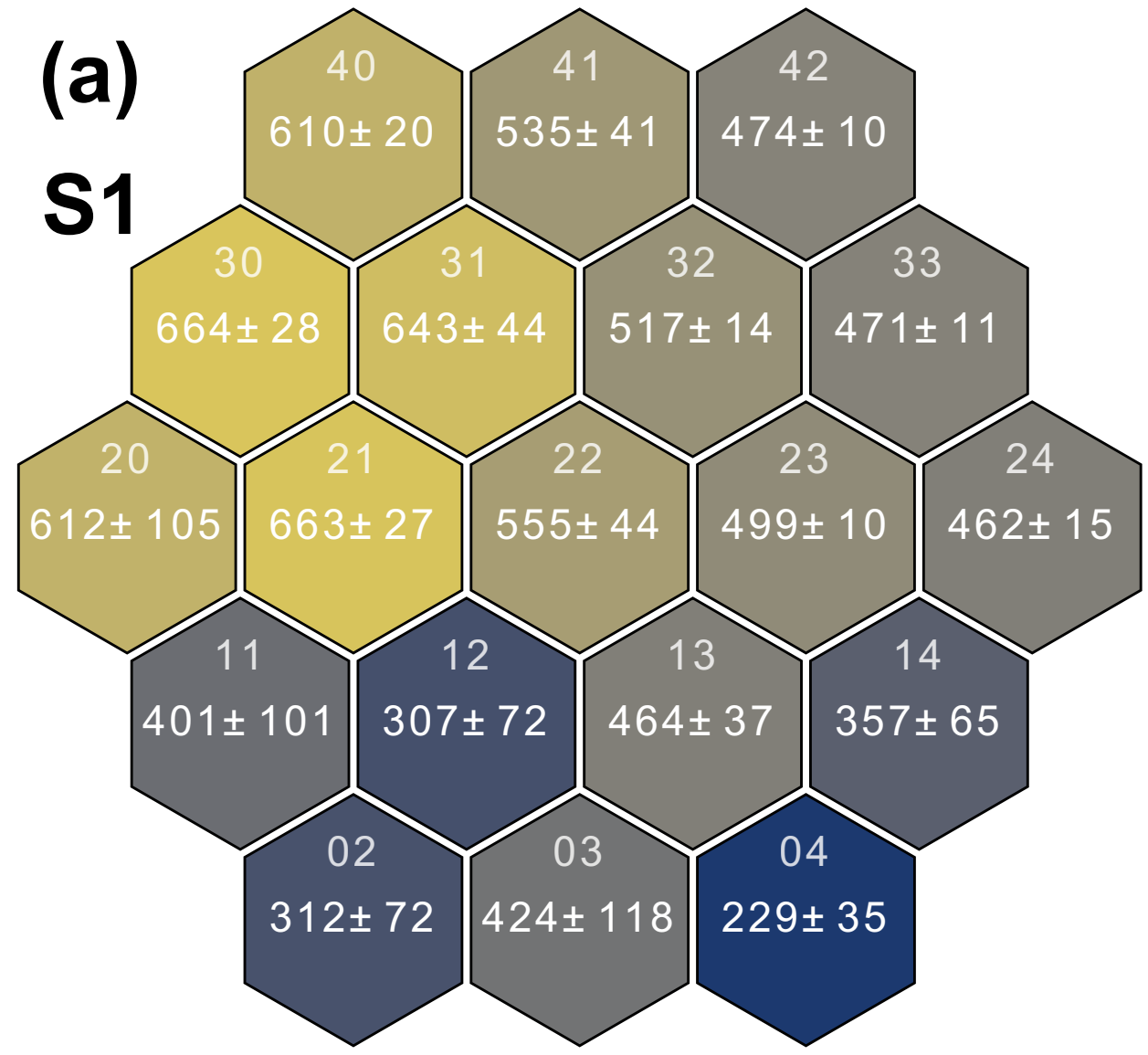

**Figure 6** Microhardness of samples S1 and S3. In S2 sample, the BCC phase cannot be unambiguously distinguished among others in the light microscope available at the device and therefore reliably measured. The underlying data can be found in Supplementary material A.

Complete diffraction patterns for each compartment are compiled in Supplementary materials B1–B3, and refined lattice parameters are summarized in Supplementary material A. Here we emphasize only overall trends in the BCC lattice parameter (Fig. 7), which is the sole phase present across vast majority of the studied compositions.

### *SEM + EDS + EBSD*

Representative SEM micrographs, elemental maps, and clustered EDS phase maps for each compartment are provided in Supplementary materials B1–B3, with quantitative phase compositions tabulated in

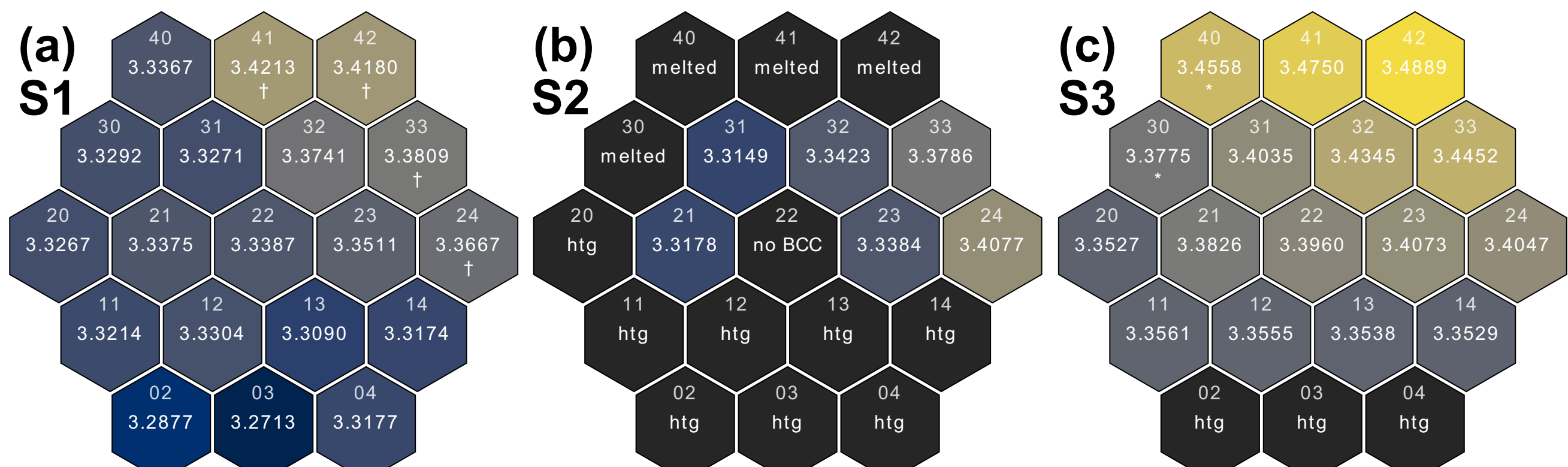

**Figure 7** Trends in BCC lattice parameters. Notes: (htg) Heterogeneous compartments, excluded from analysis as explained above. (melted) Melted compartments in S2, not available for analysis. (†) B2 phase lattice parameter plotted (as BCC is not present or overlaps with B2). (*) An average of the two measured BCC phases. The underlying data can be found in Supplementary material A.

Supplementary material A. Below, we highlight several microstructural motifs that recur across composition space and materially influence phase equilibria at 1400 °C.

Eutectic microstructure was observed in some S2 compartments resulting from partial melting at 1400 °C, see the highly refined microstructure marked in Fig. 8. A remark is also made about this in the supplementary quantification table. The eutectic region was chemically analyzed as if it was a single phase. It is assumed that this part of the material is liquid at the annealing temperature. However, the fraction of liquid was only minor, and the surrounding solid phases were able to maintain the material integrity in several compartments, while some of the neighboring compartments have melted away completely (e.g. S2_C30).

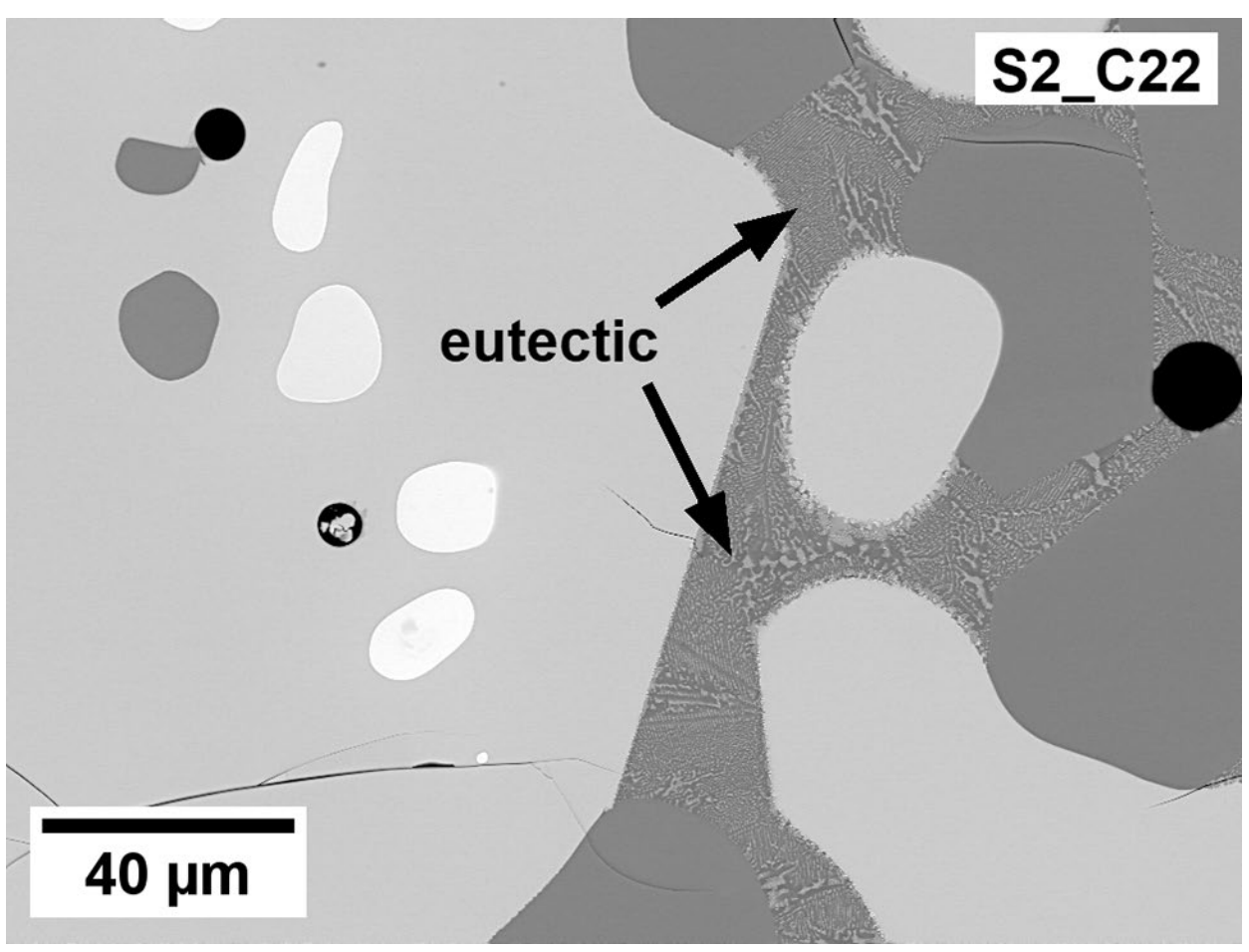

**Figure 8** Eutectic microstructures in S2_C22.

In several compartments in the sample S2 (specifically C22-24 and C33), the information for reliable phase identification cannot be determined even from combining XRD and EDS results. Hence, we have measured correlated EDS + EBSD maps; with EBSD indexing using the phases identified by XRD. One such analysis is shown in Fig. 9 for S2_C22. The phase which appears brightest is a Ta-rich BCC phase. The dark region consists of $Al_3Zr_5$-based intermetallic and a eutectic region of the $Al_3Zr_5$ and $Zr_4NbAl_3$-based intermetallic—both with an $Al_3M_5$ composition, but with slightly different structure. The rest is composed of the σ-$Nb_2Al$ phase.

Similar analysis was performed for compartments C23, C24 and C33 and is presented in Supplementary material C.

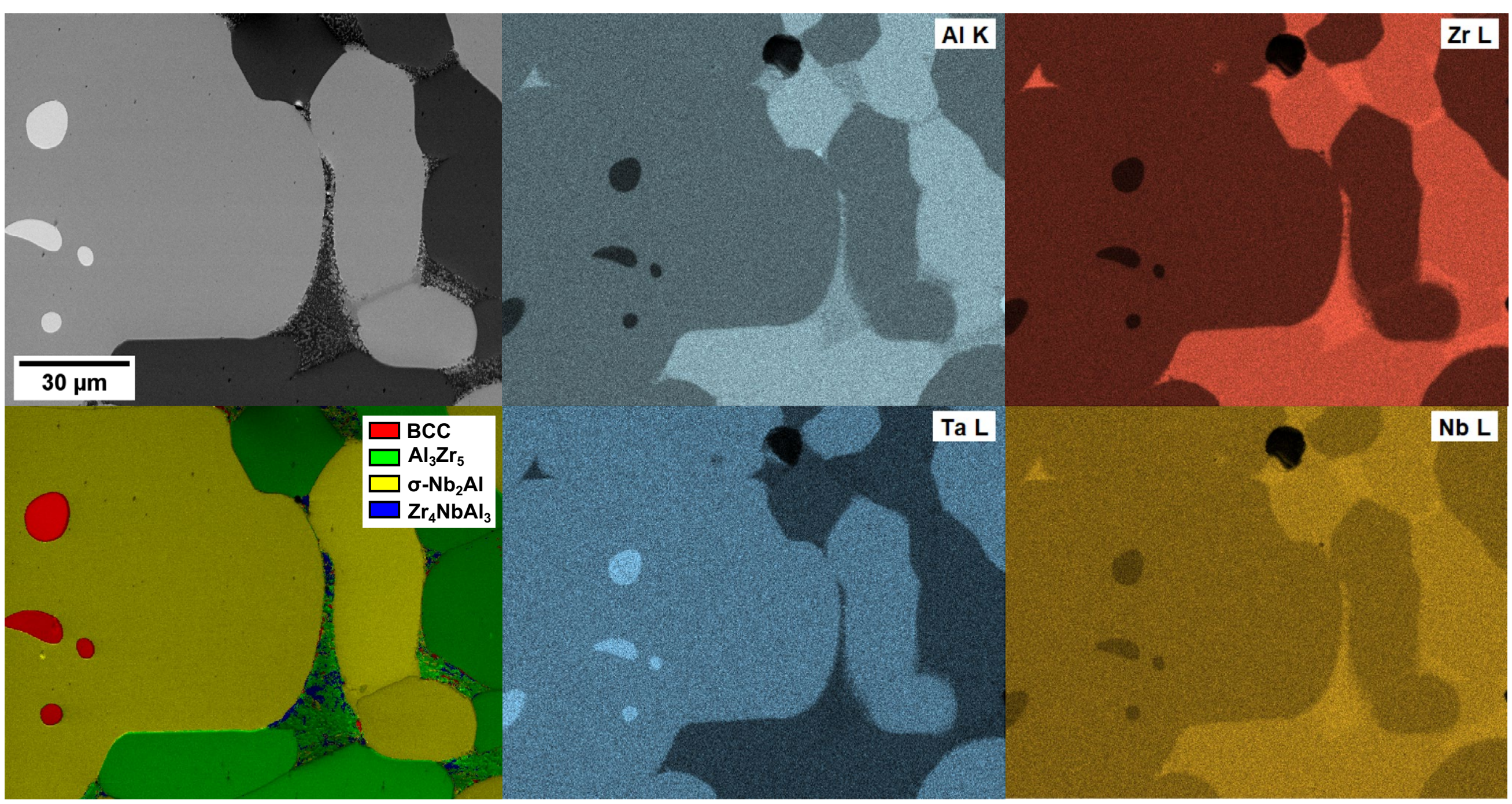

**Figure 9** Combined EDS and EBSD analysis in compartment S2_C22.

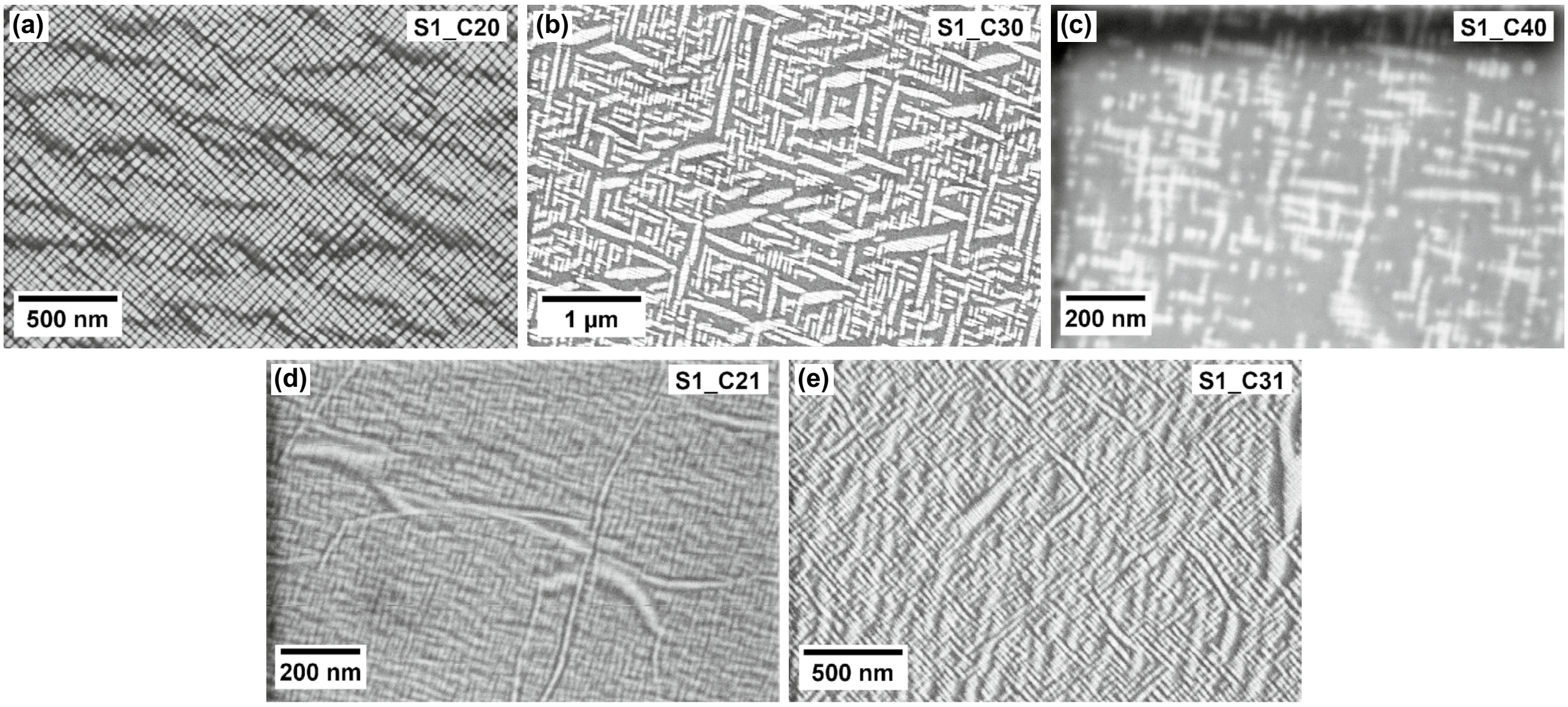

**Figure 10** Nanoprecipitates in S1 sample. Note that the characteristic size of the microstructure varies significantly.

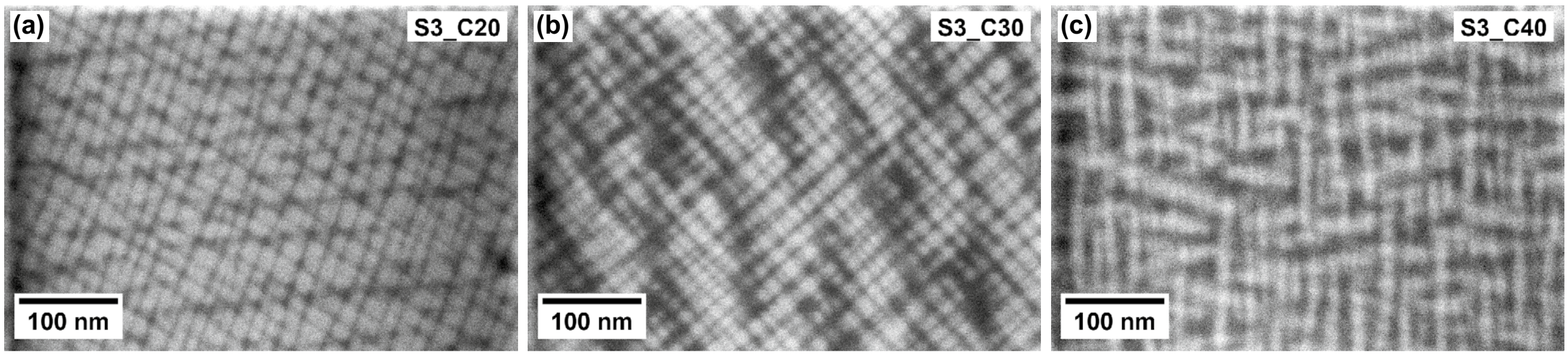

**Figure 11** Nanoprecipitates in S3 samples. The size of these precipitates is relatively uniform across different compositions when compared to those observed in the S1 sample (Fig. 10).

*Nanostructured precipitates*

A pervasive feature of Ta- and Zr-rich compositions is a dispersion of cuboidal nanoprecipitates with edge lengths from ~ 10 nm up to several hundred nanometers (shown in Figs. 10 and 11), depending on local chemistry and thermal history. TEM-EDS analysis, shown in Fig. 12, reveals Ta partitioning into the precipitates, whereas Al and Zr partition to the surrounding matrix; Ti remains comparatively uniform. Diffraction shows that the cubes are BCC. Together with their morphology, it implies an ordering/spinodal decomposition pathway akin to BCC/B2 partitioning reported in related RCCAs [20]. This type of dual-phase microstructure accounts for the pronounced hardness increase measured in these regions.

No nanoprecipitates were detected by SEM in S3_C21, S3_C31 and S3_C41. These compartments contain more Nb, so Nb is supposed to suppress the decomposition observed.

Given the precipitate size, SEM-EDS quantification of the chemical composition was not possible, as the interaction volume is significantly larger than the region of interest. As such, the phase fractions could not be quantitatively assessed and the chemical compositions reported in the Supplementary material A for BCC/B2 phase correspond to the phase mixture, i.e., as if the decomposition did not occur. A note about this is also added to the Supplementary material A in the respective compartments.

Phase analysis at the nanoscale requires caution because FIB-induced amorphization may have occurred, and damage from ion beam could obscure weak reflections from ordered or metastable phases. We therefore combine diffraction, dark field imaging, and real space EDS partitioning to cross-validate phase assignments.

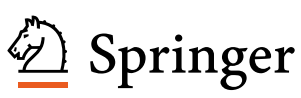

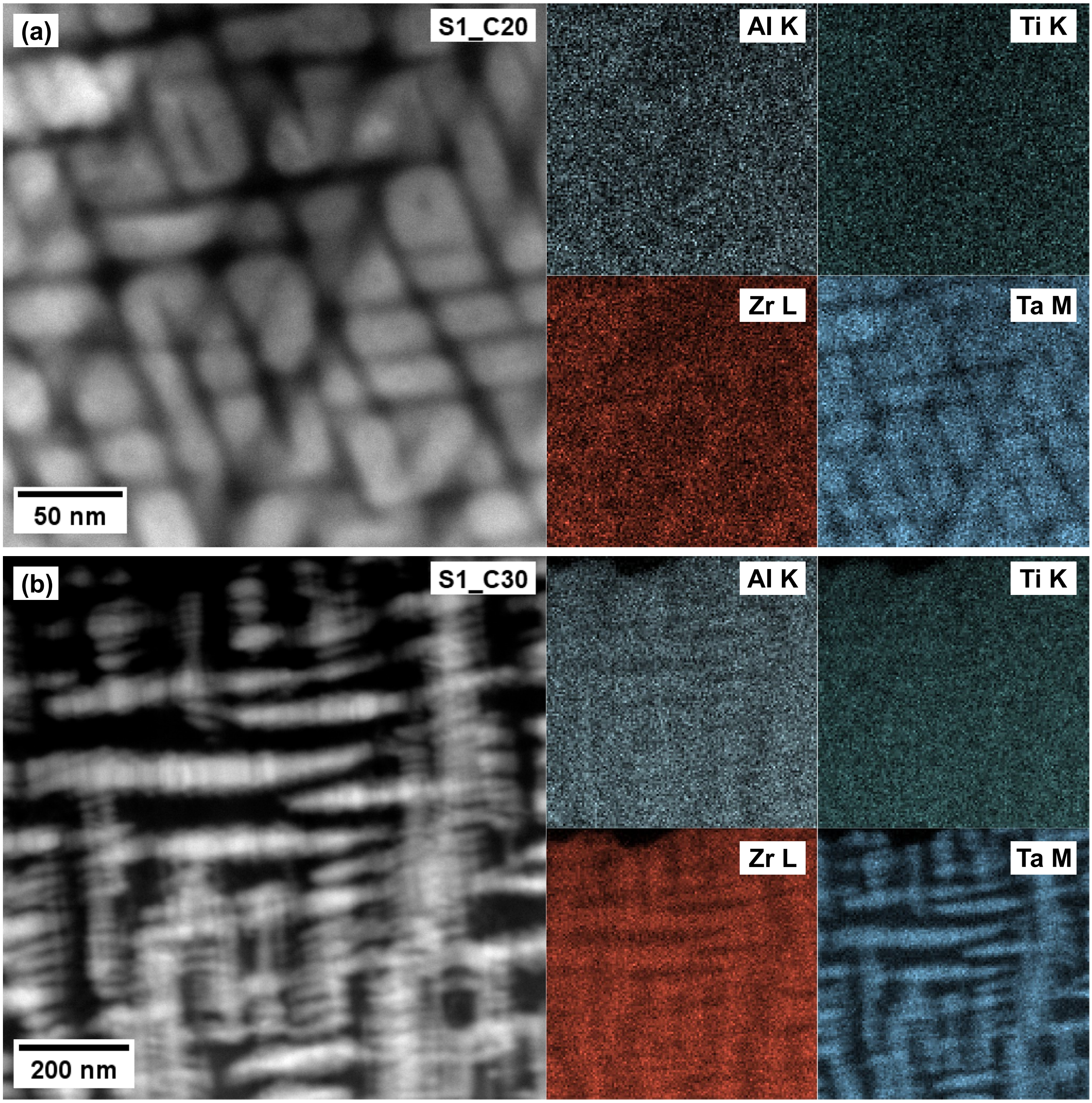

**Figure 12** Results of the TEM-EDS analysis of the (**a**) S1_C20 and (**b**) C30 samples. The cuboidal particles are rich in Ta, while the channels between them are enriched with Al and Zr. The distribution of Ti remains mostly homogeneous.

In S1_C20, as shown in Fig. 13, precipitates display BCC diffraction, whereas the inter-precipitate channels appear amorphous or extremely nanocrystalline. Given FIB finish, a thin amorphous rind cannot be excluded; nonetheless, the absence of superlattice reflections indicates that long range B2 order is weak or even absent in this composition.

For S1_C30, both BCC and B2 diffraction spots are detected in Fig. 14a, with B2 having slightly larger lattice parameters (the split of diffraction spots is visible in the diffraction pattern). B2 is shown in Fig. 14d in green. Amorphous phase is also present, shown in blue. Interestingly, additional diffraction spots are visible in Fig. 14c, marked in red—these correspond to hexagonal $B8_2$ structure, also belonging to the $AlZr_2$ phase. They visually resemble diffraction pattern of $\omega$ phase in many metastable $\beta$ titanium alloys [21], which is also consistent with [22].

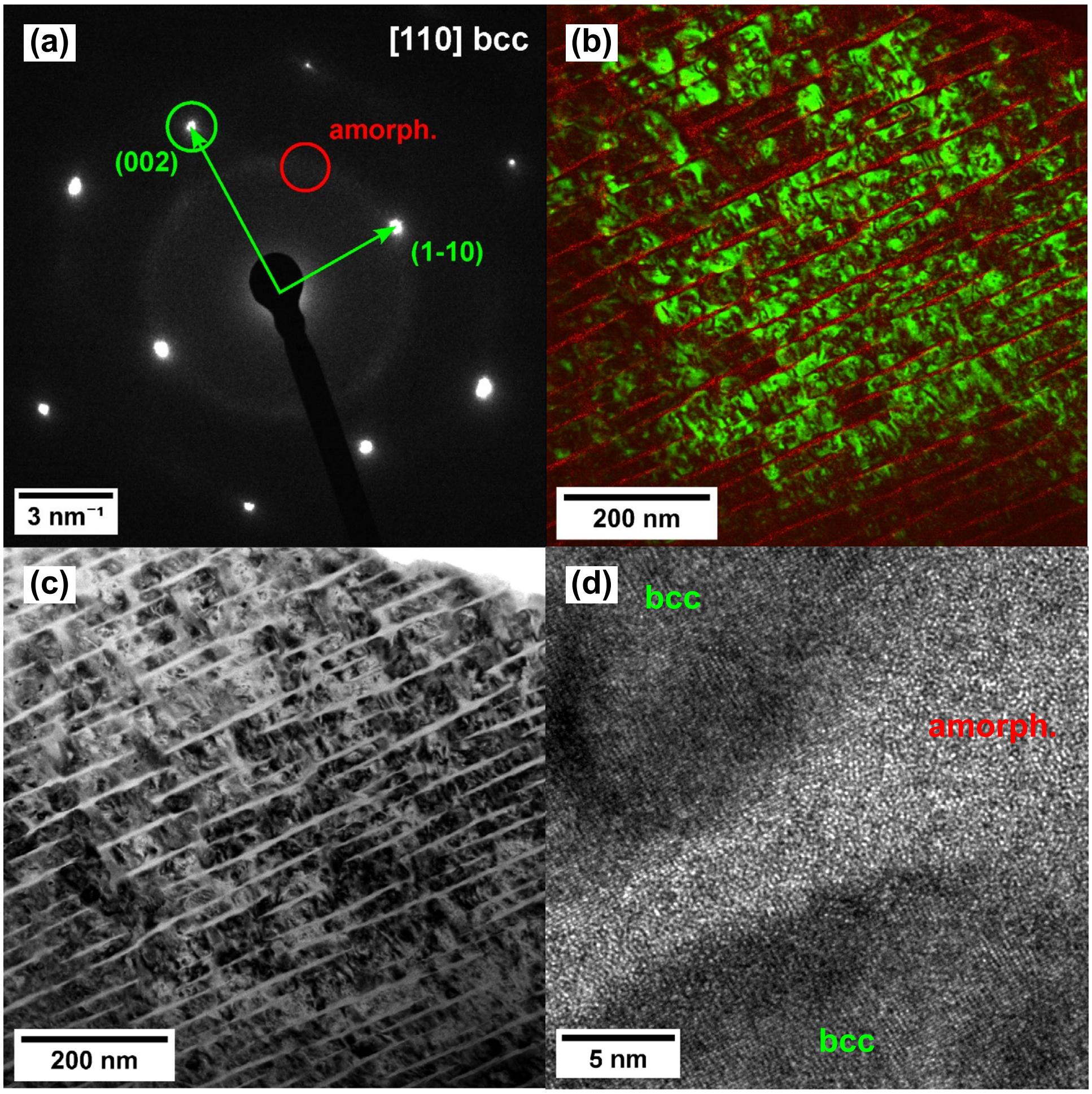

**Figure 13** TEM micrograph of the nanoscale microstructure in S1_C20. **a** diffraction, **b** composite DF micrograph from the regions marked **a**, **c** bright-field, **d** HR-TEM of the two regions with the amorphous / nanocrystalline region.

Comparison with CALPHAD predictions Finally, we have calculated a series of thermodynamic equilibria for the actual average compartment composition, as determined by EDS (referred to as 'valid' spectra in the Supplementary material A), using ThermoCalc software with TCHEA4 and TCHEA8 databases. As the new TCHEA8 database has the interstitial elements included, we have performed the same calculation with the oxygen and nitrogen concentrations from the CGHE (cf. Table 2) and without them.

Here we present the comparison of the experimental results with the calculated phase fractions for the three samples. The calculated chemical compositions of each phase can be found in Supplementary material A.

Note that there is a discrepancy between the definitions of the $Al_3Zr_5$ phase: the XRD prototype from the PDF5 database (entry 00-012-0674) has $P6_3/mcm$ symmetry, fits the diffraction data well and the chemical composition shows the corresponding stoichiometry (37.5 at.% of Al). On the other hand, the corresponding CALPHAD prototype AL3ZR5_D8M has, according to the documentation of the TCHEA8 database, the I4/mcm symmetry, while the $P6_3/mcm$ symmetry is defined for the AL4ZR5 phase. In [23], hexagonal $Al_4Zr_5$-type phase was also reported with a composition of 37.1 at.% Al, determined from TEM-EDS.

As neither of the phase definitions in TCHEA8 are complete (AL3ZR5_D8M has the (AL)3(TI, ZR)5

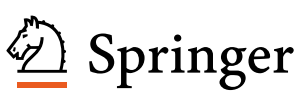

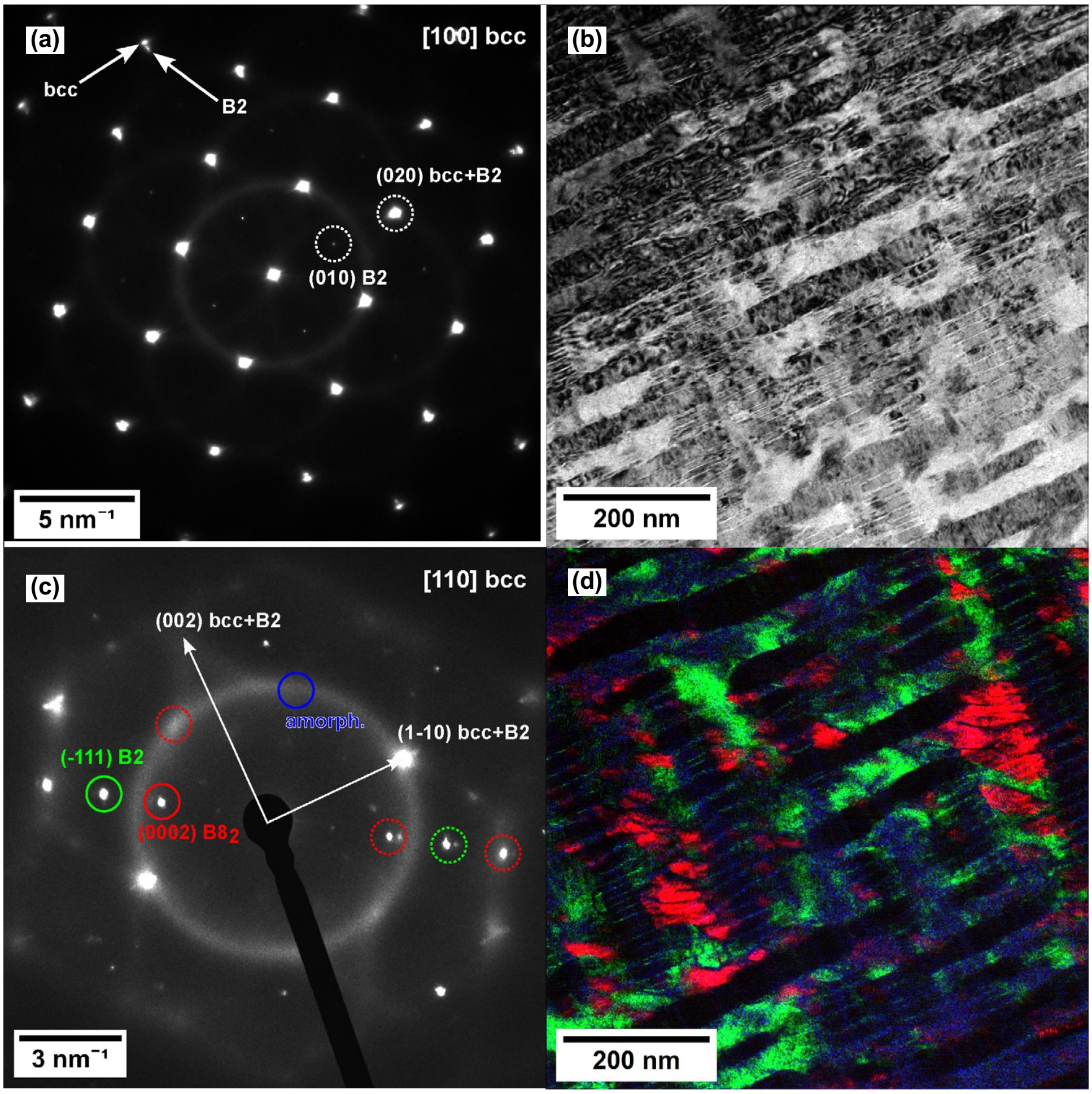

**Figure 14** TEM micrographs of the S1_C30 sample. **a** Diffraction pattern showing the presence of BCC and B2 phases (with B2 having larger lattice parameter. **b** BF micrograph of the sample in orientation corresponding to the diffraction pattern in **c**. **c** The diffraction pattern showing the presence of BCC, B2 and hexagonal $B8_2$ phase and amorphous phase. **d** the corresponding dark-field micrographs.

structure and AL4ZR5 is defined as (AL)4(ZR)5), i.e., both phases lack Ta and Nb (and the latter also Ti), the discrepancy can be caused by slight modifications of the phase stability with the alloying, not captured in the simplified ternary/binary definition.

For sample S1, as shown in Fig. 15, TCHEA8 improves the prediction of melting relative to TCHEA4, by not predicting formation of a liquid in S1_C30–C32 and C40–C42, which agrees with the experimental evidence.

In TCHEA8, the tendency to form dual-phase BCC/B2 microstructures in Ta- and Zr-rich region is correctly captured, as well as formation of the $\sigma$-phase in Zr-lean region. Including the interstitial elements does not change the phase composition significantly, it only adds minor fractions of $ZrO_2$ and HCP phases. However, the predicted oxygen levels in HCP reach 30 at.%, which is clearly a non-physical result.

For sample S2, the prediction accuracy is very low. The prediction tends to favor formation of intermetallic phases such as Al-Zr intermetallics, $\sigma$ phase or the hexagonal C14 Laves phase, while the experimental data also show disordered phases (BCC or liquid). Interestingly, adding interstitial elements improved the prediction, likely because the involved intermetallics accommodate only low amount of interstitial atoms in reality and no interstitial atoms by definition in the CALPHAD database. Hence, the presence of another phase is required to accommodate the prescribed oxygen and nitrogen content, as shown in the TCHEA8 with O and N plot in Fig. 16. Note that the experimentally identified Al-Zr intermetallic is different than the one predicted, but as the definition of the phase prototype is not entirely correct (some elements are omitted from the Zr sublattice), the prediction can be deemed to qualitatively capture the correct trends.

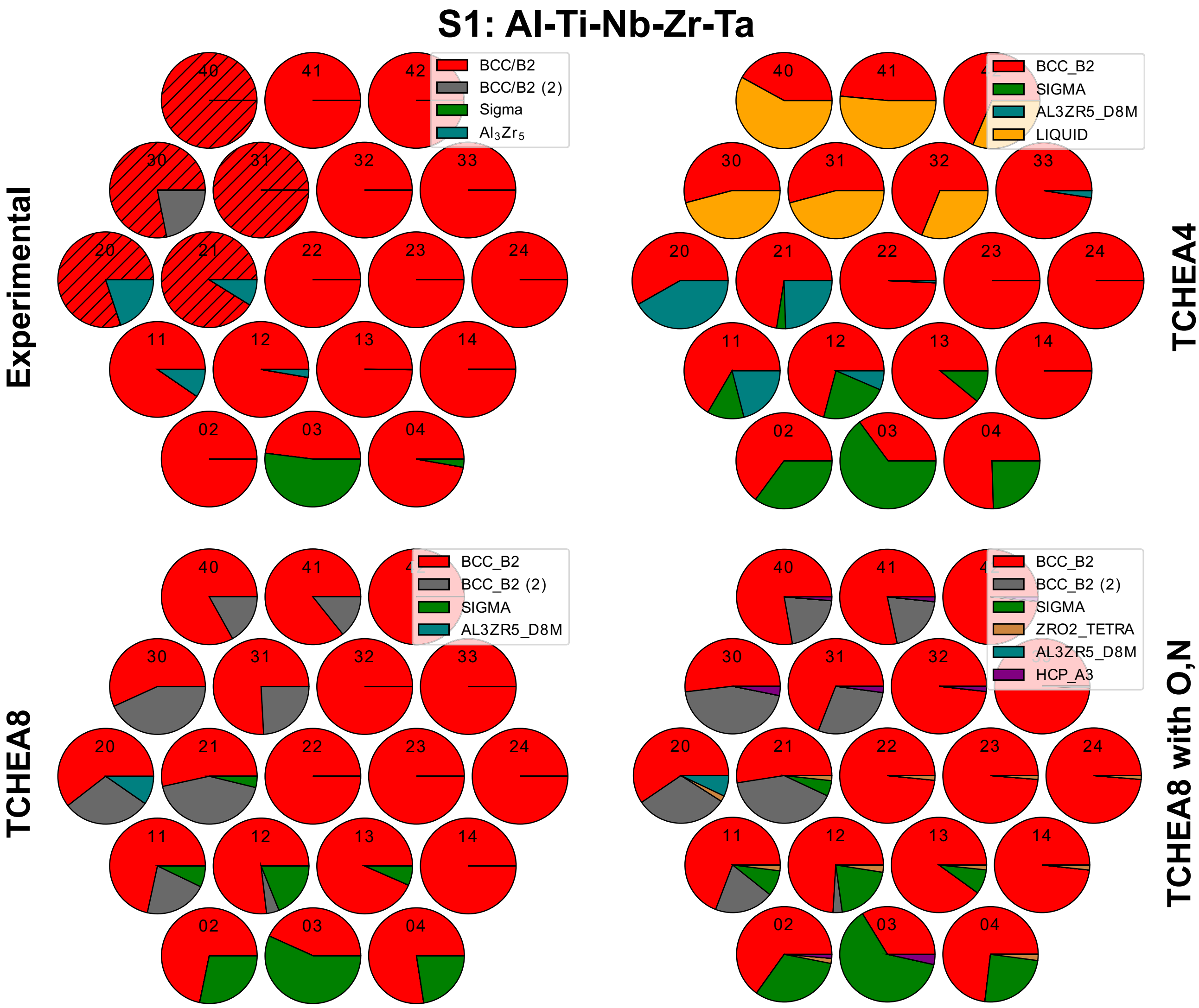

**Figure 15** Comparison of experimental phase composition and CALPHAD prediction using different database setups for sample S1. The more detailed results, including the chemical compositions of each phase can be found in Supplementary material A. The hatching of some BCC/B2 areas denotes the presence of cuboidal precipitates within the phase.

Finally, the prediction of the mostly-single-phase sample S3 is shown in Fig. 17. Without interstitials, both databases predict single-phase microstructure, while two or more phases are predicted with interstitials. Although neither prediction can be viewed as fully correct, the TCHEA with O and N predicts higher amount of the secondary phase in the Ta- and Zr-rich region (similarly to the S1 sample), capturing the experimentally observed trend.

## Discussion

### Confidence and limitations of the analysis

#### *Residual heterogeneity*

First, we will discuss some of the limiting factors of the presented analysis. At the beginning of the Results section, an assumption was made that after the annealing, thermodynamic equilibrium was reached. Conversely, presence of compositional gradients within a phase is a clear indication, as the equilibrium phases must have homogenous compositions.

**Figure 16** Comparison of experimental phase composition and CALPHAD prediction using different database setups for sample S2. The more detailed results, including the chemical compositions of each phase can be found in Supplementary material A.

Based on this, several compartments were outright excluded from the analysis because of the obvious residual heterogeneity (marked as 'htg' in the figures above), mostly due to the slow diffusion of Ta and Nb in the Ta-rich regions. This suggests that the selected annealing scheme is already the borderline for certain compositions to reach equilibrium. To check the residual heterogeneity, EDS line compositional profiles were extracted from three selected compartments from different samples and with different microstructures. The same EDS datasets were used as those for creating the phase maps. Therefore, the dwell time per pixel is rather low and the original noisy data must have been smoothed. For interpretation, recall that the initial powder particle sizes were in the range of 15–80 μm (cf. Table 1).

The compositional profile in Fig. 18a represents Ta-lean, single-phase compartment, showing good homogeneity over the measured range. The fluctuations are a result of the measurement noise only. The chemical homogeneity is also documented by the uniform brightness in the corresponding BSE micrograph.

The second profile in Fig. 18b comes from the multi-phase region of the sample S2. The different compositions of the three phases are clearly visible, but within each of them, no gradients are observed.

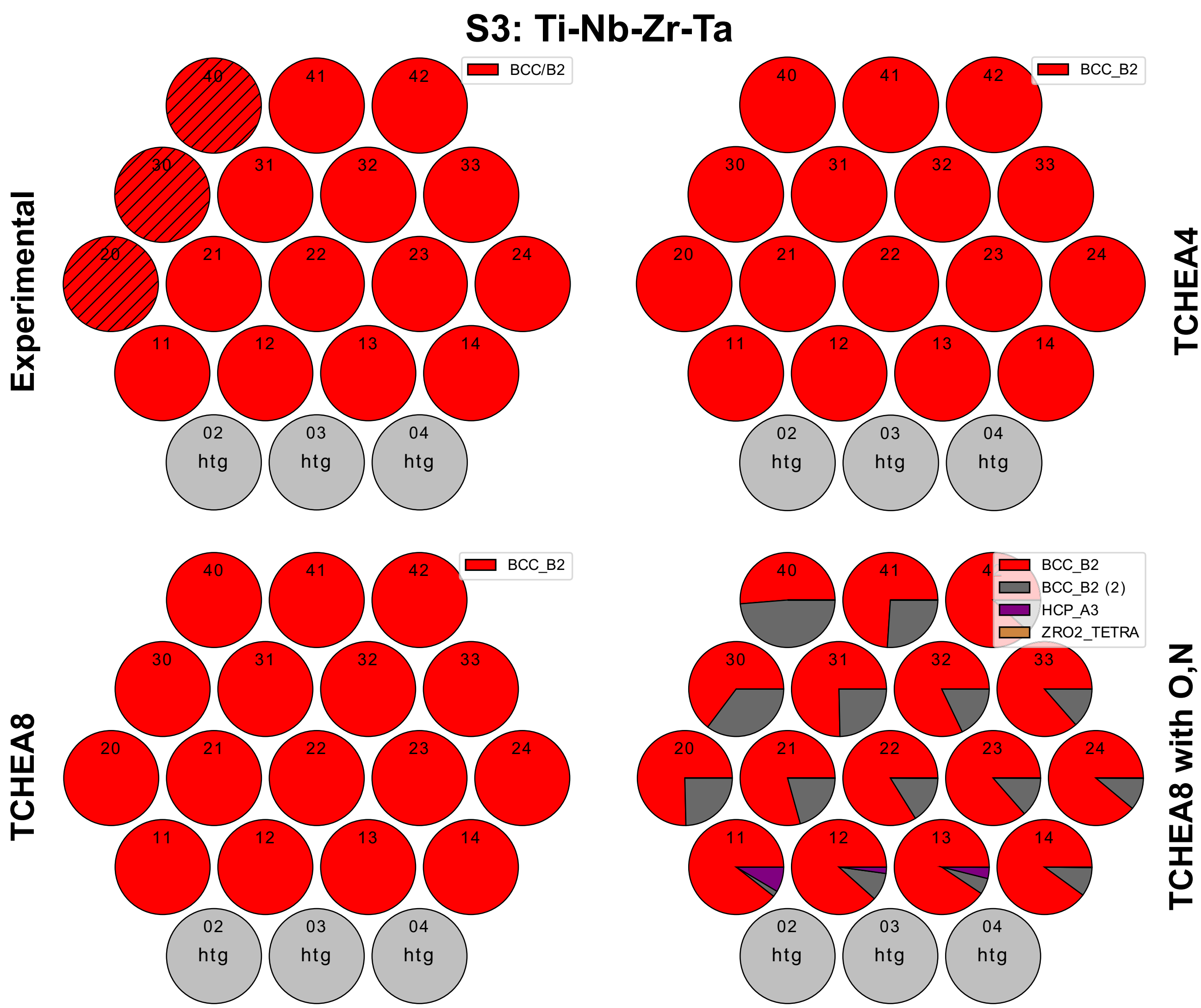

**Figure 17** Comparison of experimental phase composition and CALPHAD prediction using different database setups for sample S3. The more detailed results, including the chemical compositions of each phase can be found in Supplementary material A. The hatching of some BCC/B2 areas denotes the presence of cuboidal precipitates within the phase.

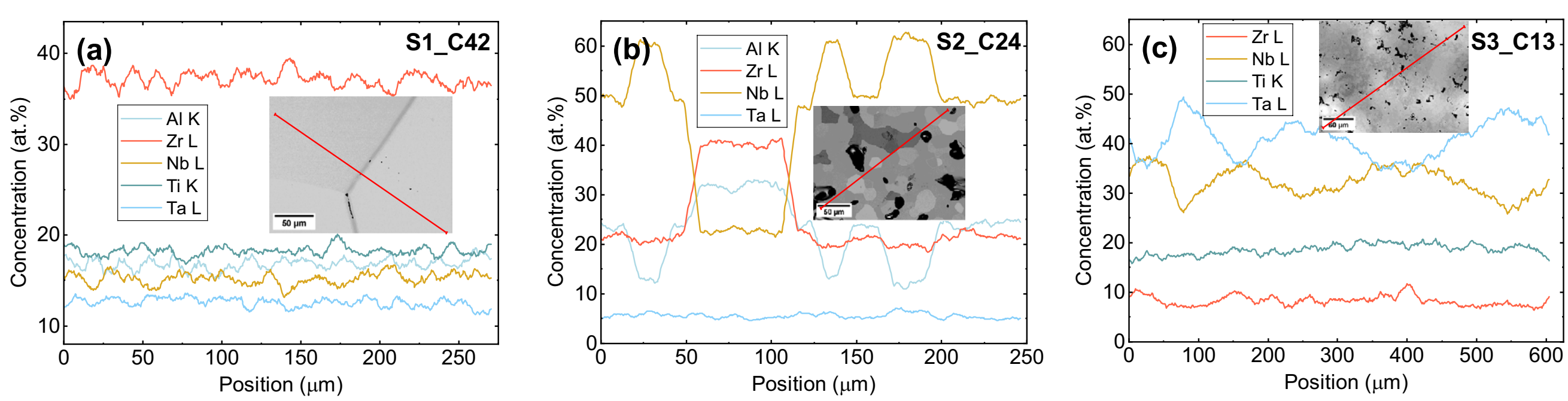

**Figure 18** EDS line compositional profiles extracted from the maps using EDAX APEX, after 20-point moving average smoothing. Three representative compartments were selected: **a** single-phase homogeneous compartment, **b** multi-phase compartment with homogeneous composition within phases and **c** single-phase Ta-rich compartment with residual heterogeneity.

Finally, the third figure (Fig. 18c) represents the compartment just next to the excluded region in Ta-rich corner of the sample S3. Although the sample was included in the analysis, residual heterogeneity in the range of 10 at.% is clearly visible in the Ta and Nb concentrations (this example represents the worst-case scenario of the non-excluded compartments). Luckily, since the S3 sample mostly consists of single-phase BCC, this residual heterogeneity does not compromise the validity of the results, although it leads to additional broadening of the XRD peaks due to compositional dependence of the lattice parameter.

The macroscopic heterogeneity (over the mm scale) originating from the powder demixing is present and visible in Fig. 4. However, for the sake of quantifying the phase equilibria, this does not play a significant role, provided that the heterogeneity length scale is significantly larger than the region of interest.

### *Quench-induced transformations*

Apart from residual heterogeneity, the thermodynamic equilibrium might be compromised due to microstructural processes occurring during quenching. During the quench, the sample is cooled to RT within several seconds, so it cannot induce any changes detectable by SEM-EDS. However, diffusionless phase transformations may occur. Despite no observed signs of martensitic transformation, the shuffle-dominated transformations (similar to formation of athermal $\omega$, of O′ phase in Ti alloys [21]) or short-range ordering of atoms may happen.

From the crystallographic standpoint, transformations during quenching can change the observed or measured phase composition (and it likely did, as can be seen from the presence of hexagonal diffraction spots in Fig. 14c). However, such phases are typically extremely small, hence their XRD peaks are broad and likely undetectable (which is confirmed by the absence of the $\omega$ or $B8_2$ peaks in the XRD data). Therefore, the overall analysis of the phase equilibria by SEM-EDS and XRD is not compromised by quenching.

## Decomposition and ordering pathways

A dominant phenomenon across the pseudo-ternary sections investigated is the strong chemical affinity between Al and Zr. These two elements are known for forming intermetallic phases of various stoichiometries. Their Gibbs energies are low enough to appear even in multicomponent alloys at low to medium temperatures (up to 1300–1500 °C [24]), but they melt at higher temperatures, as they do not mix well with the other refractory elements. This depresses the alloy liquidus, leading to the eutectic morphologies and local melting observed near Al–Zr rich corners in the S2 sample.

Even without direct presence of Al–Zr intermetallics, their interaction significantly influences the phase composition. Many compositions exhibit dual-phase BCC/B2 microstructure with pronounced partitioning: Ta segregates into the BCC phase while Al and Zr tend to form ordered B2 phase. The likely cause is that Al is the only non-transition element, having different properties than the others—e.g., its atomic radius is 125 pm, while Ti, Nb, Zr and Ta have larger radii of 140, 145, 155 and 145 pm, respectively [25]. On the other hand, sample S3, containing no Al, did not show any ordering and both phases were identified as BCC using XRD. This agrees well with the earlier results [26].

The lattice misfit between the BCC/B2 phases was shown to play a key role in the final morphology of phases. Even the misfit of 0.5% caused precipitation of cuboidal particles in Al–Ni–Co–Fe–Cr alloy [20]. The lattice misfit for the compartments S1_C30, C31 and C40 ranges from 1.7 to 2.7%, resulting in cuboidal or even regular weave-like morphology of the observed precipitates.

An interesting phenomenon is a different behavior of Nb and Ta with respect to the phase decomposition. In samples S1 and S3, decomposition is observed only in the Ta-rich region, while Nb-rich regions (with otherwise same composition) remain single phase (compare C30 vs C33 for both samples). Apart from XRD and SEM documentation, it is supported by the strong trend in microhardness, cf. Figure 6. Note that this concerns the *phase decomposition only*, but not the ordering.

In sample S1, the single-phase compartments show B2-exclusive peaks in the diffraction patterns. This observation is interesting, as Nb and Ta are chemically extremely similar—both having atomic radius of 145 pm, same group and thus valence electron structure, very similar elemental lattice parameters, etc. From the decomposition kinetic perspective, Ta is known as a slowly diffusing element and yet the decomposition is observed within Ta-rich region, so the kinetic origin can be excluded. On the other hand, there is a significant difference in atomic masses (92.9 for Nb vs. 180.9 for Ta), which affect the vibrational entropy. The vibrational entropy of mixing is typically smaller than the configurational entropy [27], but it can be important near the stability boundary. To our knowledge, no systematic study was done in this field. Limited data is available for binary systems [28, 29], but multidimensional calculations would be necessary to elucidate this phenomenon.

Finally, the hexagonal $B8_2$ phase visible in TEM SAED (Fig. 14) is sometimes misinterpreted as $\omega$ phase [22]. This confusion emerges from the fact that the $B8_2$ phase forms by the same collapse of {111} planes like the $\omega$ phase and the B2-$B8_2$ relationship is the same as the BCC-$\omega$ relationship. The difference between the phases lay in the prior ordering of the B2 parent phase. It was suggested that B2 is only an intermediate step between the original BCC and final $Al_4Zr_5$-type $\omega$ phase [23]—i.e., the intermetallic ordering arises gradually during annealing, and crystallographic transformation happens at some point. However, it is not clear whether the collapse of planes occurs already at the annealing temperature or during the quenching.

## Comparison with CALPHAD predictions

### *Intermetallic phases*

Comparison with CALPHAD (Thermo-Calc, TCHEA4 and TCHEA8 databases) indicates a broad qualitative agreement regarding the presence of BCC fields and the tendency toward BCC/B2 decomposition; however, some systematic deviations remain. The prediction of melting significantly improved over the past years of Thermo-Calc CALPHAD development. A key limitation is the restricted set of prototype elements allowed in sublattices for many intermetallics in the current databases. For example, AL3ZR5_D8M prototype excludes Nb and Ta substitution from the Zr sublattice, despite clear experimental evidence of their presence within the phase. Consequently, the excess Nb and Ta are artificially forced into other phases in the calculations, perturbing the predicted equilibria significantly.

### *Interstitial elements*

The calculations that include oxygen and nitrogen are somewhat disputable.

For sample S1, the phase fractions do not change significantly with the inclusion of interstitials and only minor amount of $ZrO_2$ and HCP phase appears, but the HCP phase has non-physical oxygen content reaching 30 at.%.

Considering sample S3, no parent phase decomposition is predicted without the interstitials in neither TCHEA4 nor TCHEA8 databases, while it is predicted for all compositions in TCHEA8 with O, N. As in the S1 sample, spuriously high oxygen is predicted in HCP_A3 (up to 27 at.%) and BCC_B2 (up to 23 at.%) for S3_C11–13.

These observations suggest that the interaction parameters of the interstitials are overestimated for certain elements (Zr and Ti, typically). Although the high affinity is qualitatively correct, a careful tuning of the parameters is required to improve the prediction accuracy. In sample S1, this effect is not that prominent in calculation, as the Al-Zr interactions are also strong there (while no Al is present in sample S3).

For sample S2, the calculations without the interstitials are clearly incorrect. The calculations with the interstitials can be seen as qualitatively correct (Fig. 16). Again, this is likely caused by a wide palette of intermetallics with $\Delta H_f \ll 0$, (although the numerical values of the interaction parameters cannot be accessed from the database), promoting formation of Al-Zr intermetallics, σ phase or the hexagonal C14 Laves phase. However, the interstitial elements are not allowed as solutes in either of them (which is a realistic assumption), so adding O and N forces presence of disordered phases (BCC/B2, FCC/L12 or liquid) to accommodate these interstitial atoms into the crystal lattices. The experimental data shows a combination of BCC, σ and $Al_3Zr_5$-type intermetallics while TCHEA8 with O and N predicts mostly $\sigma$, BCC_B2 and $Al_4Zr_5$ intermetallic phases. Note that the observed Al-Zr intermetallic is different from the predicted, but as the definition of the phase prototype is not entirely correct (some elements are omitted from the Zr sublattice), the prediction can be deemed to qualitatively capture the correct trends.

In the field of BCC Ti alloys, the content of intermetallic oxygen and nitrogen is an important parameter, having significant impact on the phase stability, not to mention immense effect on mechanical properties. The effect of interstitials is an order of magnitude larger than that of substitutional alloying elements. There is no reason to expect BCC/B2 RCCAs to behave differently. Accurate assessment of interstitial elements in CALPHAD is important for the future database development, but it requires systematic measurement of the interstitial concentration in all samples, whenever phase stability is investigated.

### *Recommendation for future development*

Finally, we would like to make few suggestions for the CALPHAD community for future development of thermodynamic databases. Some of the current challenges have roots in the historical design choices, which were justified at the time, but their limitations came up only recently with the advent of CCAs and combining thermodynamic data from different alloying systems.

The challenge starts at phase naming. In the current TCHEA8 database, it does not have any clear structure—some phases are named by its common name (BCC_B2), some by the Strukturbericht designation (C14_Laves), some by stoichiometry only (AL4ZR5, although other elements are allowed) and others by a field-specific Greek letter name (SIGMA). Standardization of the crystallography, sublattice structure and chemistry would be a great achievement.

Next, there are duplicate entries in the database, e.g., there are two $B8_2$ phases: B82_OMEGA defined as $(Al, Sn)_1(Co, Nb, Sn, Ta, Ti)_1(Ti)_1$ and ALZR2_B82 defined as $(Al)_1(Ti, Zr)_2$. Both phases share the crystallographic structure and, for example, a hypothetical $AlTi_2$ phase could be defined using both entries. The likely cause is using of Al–Ti and Al–Zr subsystems in a single database without resolving this conflict.

Some phases, like AL3ZR5_D8M, do not allow presence of other elements in the composition, even though their presence was clearly shown experimentally. As a first approximation, until more robust assessment is done, the elements can be included as ideal solid solution at the respective sublattice.

For future, merging related phases into a single phase with internal parameters can be also an option, like it was already done for BCC_B2 and FCC_L12 ordered phase (the ordering being the internal parameter). Similarly, BCC-$\omega$ phase with the {111} plane collapse as the internal parameter can be constructed, directly bridging CALPHAD with the phase-field modeling [30].

## Implications for alloy design

To produce next-generation high-temperature materials, the RCCA community focuses on preparation of dual phase superalloy-like BCC/B2 microstructures. In this perspective several recommendations can be made from the presented data:

- The volume fractions of the dual-phase system can be controlled by exchanging Ta (promoting decomposition) for Nb (suppressing decomposition).
- The B2 ordering largely relies on the interaction with Al—adding aluminum leads to stronger ordering.
- Attention should be paid to the actual phases formed, as excessive annealing and consequent ordering can cause further phase transformations to less symmetrical phases (with possible adverse implications for material ductility).

## Conclusions

Phase equilibria in the Al–Ti–Nb–Zr–Ta refractory complex concentrated alloy system were investigated using a high-throughput experimental approach. The phase content and chemical compositions were measured and summarized in the supplementary materials provided. The key findings are following:

- The equilibrium microstructures are dominated by BCC/B2 fields, $\sigma$ phase and Al–Zr intermetallics.
- Strong Al–Zr interactions drive eutectic formation and local melting.
- Nb was found to promote single-phase microstructure, while Ta promotes decomposition into two BCC/B2 phases. Al promotes BCC$\rightarrow$ B2 ordering.

Comparison of the experimental data and CALPHAD predictions showed partial agreement and revealed several challenges:

- TCHEA8 database shows significant improvement over the TCHEA4 thermodynamic database, especially in the prediction of melting of eutectics and BCC phase decomposition.
- The addition of interstitial elements into calculations strongly affects the resulting predicted phase composition. The correlation with the experiments was improved in some aspects, but chemical compositions of several phases are clearly incorrect, and the prediction cannot be relied upon.
- Some elements are absent from the phase prototypes (e.g., Nb and Ta in $Al_3Zr_5$-type intermetallics), despite confirmed to be present experimentally.

In addition, several suggestions for future database development were made, mostly concerning the phase systematization and database cleanup.

These findings, along with the published data, provide an experimental anchor for further database development and alloy design in the Al–Ti–Nb–Zr–Ta system.

## Acknowledgements

This work was financially supported by Czech Science Foundation under the Project No. 22-24563S. Financial support by the Operational Programme Johannes Amos Comenius of the MEYS of the Czech Republic, within the frame of the project Ferroic Multifunctionalities (FerrMion) [project No. CZ.02.01.01/00/22_008/0004591], co-funded by the European Union is also gratefully acknowledged.The annealing and (a part of the) diffraction experiments were performed in MGML (mgml.eu), which is supported within the program of Czech Research Infrastructures (project no. LM2023065).

## Author Contributions

J.K.: Conceptualization, Data Curation, Investigation, Methodology, Software, Visualization, Writing–original draft; F.L.: Investigation, Resources, Methodology, Data Curation, Visualization; M.C-L.: Investigation; J.V.: Investigation E.J.: Software, Visualization; K.F.: Investigation, Software, Visualization; S.Š.: Investigation; K.B.: Investigation, Resources; A.S.: Investigation, Resources; T.Ch.: Resources, Project administration, Funding acquisition; J.S.: Supervision, Project administration, Funding acquisition; All: Writing–review and editing.

## Funding

Open access publishing supported by the institutions participating in the CzechELib Transformative Agreement. Grantová Agentura České Republiky, 22-24563S, Dr. Tomáš Chráska. Ministerstvo Školství, Mládeže a Tělovýchovy, CZ.02.01.01/00/22_008/0004591. Ministerstvo Školství, Mládeže a Tělovýchovy, LM2023065.

## Data availability

Phase and chemical quantification results, the underlying data for plots and all processed XRD and EDS data are available in the respective supplementary materials. The EDS phase analysis tool is available through the Zenodo Software Archive at https://doi.org/10.5281/zenodo.19217166 under the CC-BY 4.0 license. The data that support the findings of this study are openly available in the Zenodo repository at https://doi.org/10.5281/zenodo.19182505 under the CC-BY 4.0 license.

**Supplementary Information** The online version contains supplementary material available at https://doi.org/10.1007/s10853-026-13083-2.

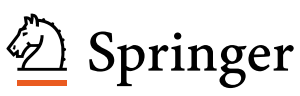

**Supplementary material A:**
Quantification data

**Supplementary of paper:**
Phase Equilibria of the Al-Ti-Nb-Zr-Ta System

**Authors:**
Jiří Kozlík [a] *, František Lukáč [b] , Mariano Casas-Luna [a] , Jozef Veselý [a] , Eliška Jača [a] , Kateřina Ficková [a], Stanislav Šašek [a] , Kristína Bartha [a] , Adam Strnad [a] , Tomáš Chráska [b] , Josef Stráský [a]
[a] *Charles University, Faculty of Mathematics and Physics, Department of Physics of Materials, Ke Karlovu 5, Prague 121 16, Czechia*
[b] *Institute of Plasma Physics of the Czech Academy of Sciences, U Slovanky 2525/1a, Prague 182 00, Czechia*
* Corresponding author jiri.kozlik@matfyz.cuni.cz

**Phase composition table legend:**

| | EDS chemical composition (at.%) | | | | | | | | | |
|---|---|---|---|---|---|---|---|---|---|---|
| **Phase ID** | **Al** | **U(Al)** | **Ti** | **U(Ti)** | **Zr** | **U(Zr)** | **Nb** | **U(Nb)** | **Ta** | **U(Ta)** |
| **S1_C03_valid** | 27.7 | 1.0 | 23.3 | 0.4 | 0.7 | 0.1 | 26.5 | 1.9 | 21.8 | 0.1 |
| **S1_C03_P0** | 30.0 | 1.1 | 18.2 | 0.3 | 0.7 | 0.1 | 27.9 | 2.0 | 23.2 | 0.1 |
| **S1_C03_P1** | 25.3 | 0.9 | 28.6 | 0.5 | 0.7 | 0.1 | 25.2 | 1.7 | 20.3 | 0.1 |

| | EDS phase composition | | | | | XRD | | | |
|---|---|---|---|---|---|---|---|---|---|
| **Phase ID** | **Pixels total** | **Pixels valid** | **Phase pixels** | **Volume fraction** | **Molar fraction** | **Phase label** | **Lattice parameter a (Å)** | **Lattice parameter c (Å)** | **Molar density (atoms/Å$^3$)** |
| **S1_C03_valid** | 51200 | 43150 | 43150 | 1.000 | | | | | *0.0571* |
| **S1_C03_P0** | 51200 | 43150 | 22448 | 0.520 | 0.520 | **Sigma** | 10.0417 | 5.2201 | 0.0570 |
| **S1_C03_P1** | 51200 | 43150 | 20702 | 0.480 | 0.480 | **BCC** | 3.2713 | | 0.0571 |
| | $V_{tot}$ | $V_i$ | $f_V = \frac{V_i}{V_{tot}}$ | $f_N = \frac{N_i}{N_{tot}} = \frac{V_i N_{V,i}}{\sum_i V_i N_{V,i}}$ | | | | | $N_{V,i} = \frac{N_{cell,i}}{V_{cell,i}}$ |

| | ThermoCalc - TCHEA4 ("valid" composition) | | | | | | |
|---|---|---|---|---|---|---|---|
| **Phase ID** | **Phase name** | **Phase fraction (mol.%)** | **Al (at.%)** | **Ti (at.%)** | **Zr (at.%)** | **Nb (at.%)** | **Ta (at.%)** |
| **S1_C03_valid** | | | | | | | |
| **S1_C03_P0** | **SIGMA** | **64.91** | 30.88 | 16.50 | | 29.17 | 23.44 |
| **S1_C03_P1** | **BCC_B2** | **35.09** | 21.69 | 35.96 | 2.02 | 21.67 | 18.66 |

## Nominal compositions

**Sample S1**

| Compartment | Al | Ti | Nb | Zr | Ta | Note |
|---|---|---|---|---|---|---|
| **C02** | 20.0 | 20.0 | 20.0 | 0.0 | 40.0 | |
| **C03** | 20.0 | 20.0 | 30.0 | 0.0 | 30.0 | |
| **C04** | 20.0 | 20.0 | 40.0 | 0.0 | 20.0 | |
| **C11** | 20.0 | 20.0 | 10.0 | 10.0 | 40.0 | |
| **C12** | 20.0 | 20.0 | 20.0 | 10.0 | 30.0 | |
| **C13** | 20.0 | 20.0 | 30.0 | 10.0 | 20.0 | |
| **C14** | 20.0 | 20.0 | 40.0 | 10.0 | 10.0 | |
| **C20** | 20.0 | 20.0 | 0.0 | 20.0 | 40.0 | |
| **C21** | 20.0 | 20.0 | 10.0 | 20.0 | 30.0 | |
| **C22** | 20.0 | 20.0 | 20.0 | 20.0 | 20.0 | equiatiomic |
| **C23** | 20.0 | 20.0 | 30.0 | 20.0 | 10.0 | |
| **C24** | 20.0 | 20.0 | 40.0 | 20.0 | 0.0 | |
| **C30** | 20.0 | 20.0 | 0.0 | 30.0 | 30.0 | |
| **C31** | 20.0 | 20.0 | 10.0 | 30.0 | 20.0 | |
| **C32** | 20.0 | 20.0 | 20.0 | 30.0 | 10.0 | |
| **C33** | 20.0 | 20.0 | 30.0 | 30.0 | 0.0 | |
| **C40** | 20.0 | 20.0 | 0.0 | 40.0 | 20.0 | |
| **C41** | 20.0 | 20.0 | 10.0 | 40.0 | 10.0 | |
| **C42** | 20.0 | 20.0 | 20.0 | 40.0 | 0.0 | |

**Sample S2**

| Compartment | Al | Nb | Zr | Ta | Note |
|---|---|---|---|---|---|
| **C02** | 25.0 | 25.0 | 0.0 | 50.0 | heterogeneous |
| **C03** | 25.0 | 37.5 | 0.0 | 37.5 | heterogeneous |
| **C04** | 25.0 | 50.0 | 0.0 | 25.0 | heterogeneous |
| **C11** | 25.0 | 12.5 | 12.5 | 50.0 | heterogeneous |
| **C12** | 25.0 | 25.0 | 12.5 | 37.5 | heterogeneous |
| **C13** | 25.0 | 37.5 | 12.5 | 25.0 | heterogeneous |
| **C14** | 25.0 | 50.0 | 12.5 | 12.5 | heterogeneous |
| **C20** | 25.0 | 0.0 | 25.0 | 50.0 | heterogeneous |
| **C21** | 25.0 | 12.5 | 25.0 | 37.5 | |
| **C22** | 25.0 | 25.0 | 25.0 | 25.0 | equiatiomic |
| **C23** | 25.0 | 37.5 | 25.0 | 12.5 | |
| **C24** | 25.0 | 50.0 | 25.0 | 0.0 | |
| **C30** | 25.0 | 0.0 | 37.5 | 37.5 | melted |
| **C31** | 25.0 | 12.5 | 37.5 | 25.0 | |
| **C32** | 25.0 | 25.0 | 37.5 | 12.5 | |
| **C33** | 25.0 | 37.5 | 37.5 | 0.0 | |
| **C40** | 25.0 | 0.0 | 50.0 | 25.0 | melted |
| **C41** | 25.0 | 12.5 | 50.0 | 12.5 | melted |
| **C42** | 25.0 | 25.0 | 50.0 | 0.0 | melted |

**Sample S3**

| Compartment | Ti | Nb | Zr | Ta | Note |
|---|---|---|---|---|---|
| **C02** | 25.0 | 25.0 | 0.0 | 50.0 | heterogeneous |
| **C03** | 25.0 | 37.5 | 0.0 | 37.5 | heterogeneous |
| **C04** | 25.0 | 50.0 | 0.0 | 25.0 | heterogeneous |
| **C11** | 25.0 | 12.5 | 12.5 | 50.0 | |
| **C12** | 25.0 | 25.0 | 12.5 | 37.5 | |
| **C13** | 25.0 | 37.5 | 12.5 | 25.0 | |
| **C14** | 25.0 | 50.0 | 12.5 | 12.5 | |
| **C20** | 25.0 | 0.0 | 25.0 | 50.0 | |
| **C21** | 25.0 | 12.5 | 25.0 | 37.5 | |
| **C22** | 25.0 | 25.0 | 25.0 | 25.0 | equiatiomic |
| **C23** | 25.0 | 37.5 | 25.0 | 12.5 | |
| **C24** | 25.0 | 50.0 | 25.0 | 0.0 | |
| **C30** | 25.0 | 0.0 | 37.5 | 37.5 | |
| **C31** | 25.0 | 12.5 | 37.5 | 25.0 | |
| **C32** | 25.0 | 25.0 | 37.5 | 12.5 | |
| **C33** | 25.0 | 37.5 | 37.5 | 0.0 | |
| **C40** | 25.0 | 0.0 | 50.0 | 25.0 | |
| **C41** | 25.0 | 12.5 | 50.0 | 12.5 | |
| **C42** | 25.0 | 25.0 | 50.0 | 0.0 | |

all values in at.%

**Schematics of the compartment numbering within the sample**

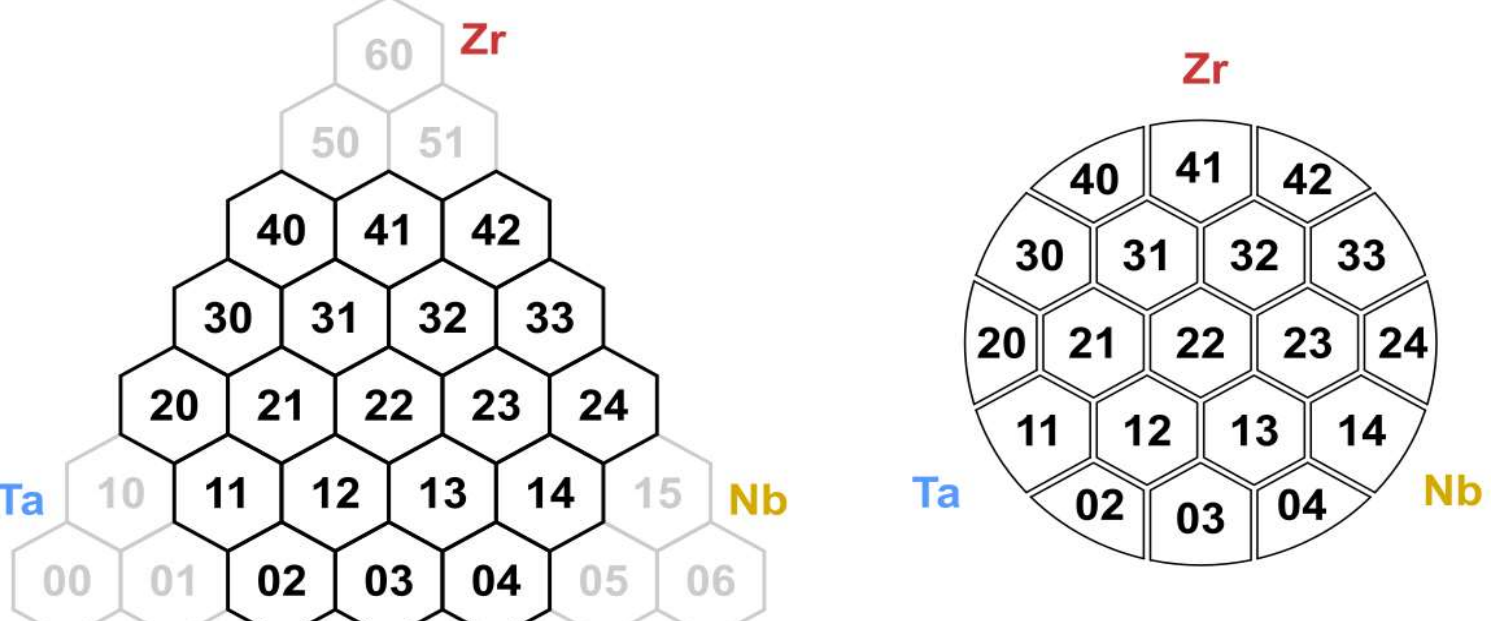

# Phase definitions

### XRD + EDS

| Label | Prototype used | Symmetry | PDF5 number | Atoms per unit cell | Unit cell coefficient | Per-atom coefficient |
|---|---|---|---|---|---|---|
| BCC | | Im-3m | | 2 | 1.000 | 2.000 |
| B2 | | Pm-3m | | 2 | 1.000 | 2.000 |
| Sigma | Nb2Al | P4_2/mnm | 04-003-6611 | 30 | 1.000 | 30.000 |
| Al3Zr5 | Al3Zr5 | P6_3/mcm | 00-012-0674 | 16 | 0.866 | 18.475 |
| | | | | $N_{cell,i}$ | $\alpha_i$ | $\beta_i$ |

**Calculation of molar density**

$$N_{V,i} = \frac{N_{cell,i}}{V_{cell,i}} = \frac{N_{cell,i}}{abc \cdot \alpha_i} = \frac{\beta_i}{abc}$$

the calculation is perfomed through the lookup table of phase-specific coefficients and the product of lattice parameters

### CALPHAD

| Label | Space group | Symmetry | Configuration (TCHEA8) |
|---|---|---|---|
| LIQUID | | - | (AL, AL1N1, AL2/3O1, AL2O4TI, AL4/3O2, B, BO3/2, C, CO, COO, COO3/2, CR, CRO, CRO3/2, CU, CU2O, CUO, FE, FEO, FEO3/2, HF, HF1/2O1, HF1N1, IR, MN, MNO, MNO3/2, MO, MO1/2O1, MOO3, N, NB, NB1O1, NBO5/2, NI, NIO, O1TI1, O1ZN1, O2TI1, RE, RE1/2O1, REO7/2, RH, RU, RU1/2O1, SI, SI1/2O1, SI2O4, SIO2, SN, SNO, SNO2, TA, TA2/5O1, TI, TIO3/2, V, V1O1, VO2, VO3/2, VO5/2, W, W1/3O1, Y, Y2/3O1, ZN, ZR, ZR1/2O1)1 |
| BCC_B2 | 221 | Pm-3m | (AL, CO, CR, CU, FE, HF, IR, MN, MO, NB, NI, RE, RH, RU, SI, SN, TA, TI, V, VA, W, Y, ZN, ZR)0.5(AL, CO, CR, CU, FE, HF, IR, MN, MO, NB, NI, RE, RH, RU, SI, SN, TA, TI, V, VA, W, Y, ZN, ZR)0.5(B, C, N, O, VA)3 |
| HCP_A3 | 194 | P6_3/mmc | (AL, CO, CR, CU, FE, HF, IR, MN, MO, NB, NI, RE, RH, RU, SI, SN, TA, TI, V, W, Y, ZN, ZR)1(B, C, N, O, VA)0.5 |
| SIGMA | 136 | P4_2/mnm | (AL, CO, CR, FE, IR, MN, MO, NB, NI, RE, RH, RU, SI, TA, TI, V, W)10(AL, CO, CR, FE, IR, MN, MO, NB, NI, RE, RH, RU, SI, TA, TI, V, W)4(AL, CO, CR, FE, IR, MN, MO, NB, NI, RE, RH, RU, SI, TA, TI, V, W)16 |
| C14_LAVES | 194 | P6_3/mmc | (AL, CO, CR, CU, FE, HF, MN, MO, NB, NI, RE, RU, SI, TA, TI, V, W, Y, ZN, ZR)2(AL, CO, CR, CU, FE, HF, MN, MO, NB, NI, RE, RU, SI, TA, TI, V, W, Y, ZN, ZR)1 |
| AL3ZR5_D8M | 140 | I4/mcm | (AL)3(TI, ZR)5 |
| AL3ZR2 | 43 | Fdd2 | (AL)3(HF, ZR)2 |
| AL4ZR5 | 193 | P6_3/mcm | (AL)4(ZR)5 |
| AL3TI_D022 | 139 | I4/mmm | (AL, CO, CR, FE, MO, NB, NI, SI, TI, V)3(AL, CO, CR, MO, NB, NI, SI, TA, TI, V, ZR)1 |
| FCC_L12 | 221 | Pm-3m | (AL, CO, CR, CU, FE, HF, IR, MN, MO, NB, NI, RE, RH, RU, SI, SN, TA, TI, V, W, Y, ZN, ZR)0.75(AL, CO, CR, CU, FE, HF, IR, MN, MO, NB, NI, RE, RH, RU, SI, SN, TA, TI, V, W, Y, ZN, ZR)0.25(B, C, N, O, VA)1 |
| TA1AL1 | 14 | P2_1/c | (TA)0.51515(AL)0.48485 |
| ZRO2_TETR | 137 | P4_2/nmc | (AL+3, CR+3, FE+2, HF+4, MN+2, MN+3, NB+5, NI+2, TI+4, VA, Y+3, ZR+4)2(O-2, VA)4 |

**CALPHAD subtable in the phase quantification tables**

Blank cell = element is not allowed by definition.

Zero cell = element is allowed, but not predicted to be in the phase.

| | EDS chemical composition (at.%) | | | | | | | | | | EDS phase composition | | | | | XRD | | | | ThermoCalc - TCHEA4 ("valid" composition) | | | | | | | ThermoCalc - TCHEA8 ("valid" composition) | | | | | | | ThermoCalc - TCHEA8 ("valid" composition) (with O and N) | | | | | | | | |
|---|---|---|---|---|---|---|---|---|---|---|---|---|---|---|---|---|---|---|---|---|---|---|---|---|---|---|---|---|---|---|---|---|---|---|---|---|---|---|---|---|---|---|
| Phase ID | Al | U(Al) | Ti | U(Ti) | Zr | U(Zr) | Nb | U(Nb) | Ta | U(Ta) | Pixels total | Pixels valid | Phase pixels | Volume fraction | Molar fraction | Phase label | Lattice parameter a (Å) | Lattice parameter c (Å) | Molar density (atoms/Å$^3$) | Phase name | Phase fraction (mol.%) | Al (at.%) | Ti (at.%) | Zr (at.%) | Nb (at.%) | Ta (at.%) | Phase name | Phase fraction (mol.%) | Al (at.%) | Ti (at.%) | Zr (at.%) | Nb (at.%) | Ta (at.%) | Phase name | Phase fraction (mol.%) | Al (at.%) | Ti (at.%) | Zr (at.%) | Nb (at.%) | Ta (at.%) | O (at.%) | N (at.%) |
| S1_C02_valid | 14.9 | 0.7 | 14.6 | 0.3 | 0.6 | 0.1 | 19.9 | 2.1 | 49.9 | 0.1 | 51200 | 49338 | 49338 | 1.000 | | | | | 0.0563 | | | | | | | | | | | | | | | | | | | | | | | |
| S1_C02_P0 | 14.9 | 0.7 | 14.6 | 0.3 | 0.6 | 0.1 | 19.9 | 2.1 | 49.9 | 0.1 | 51200 | 49338 | 49338 | 1.000 | 1.000 | BCC | 3.2877 | | 0.0563 | BCC_B2 | 64.91 | 9.05 | 19.50 | 0.95 | 19.55 | 50.95 | BCC_B2 | 71.75 | 11.08 | 18.35 | 0.86 | 18.90 | 50.82 | BCC_B2 | 65.14 | 10.53 | 18.13 | 0.06 | 19.09 | 51.99 | 0.06 | 0.14 |
| | | | | | | | | | | | | | | | | Sigma | 10.0351 | 5.2057 | 0.0572 | SIGMA | 35.09 | 25.84 | 5.61 | | 20.46 | 48.09 | SIGMA | 28.25 | 24.75 | 5.17 | | 22.34 | 47.74 | SIGMA | 31.78 | 24.58 | 5.20 | | 22.32 | 47.90 | | |
| | | | | | | | | | | | | | | | | | | | | | | | | | | | | | | | | | | ZRO2_TETR | 1.71 | 0.49 | 0.01 | 32.92 | 0.00 | | 66.58 | |
| | | | | | | | | | | | | | | | | | | | | | | | | | | | | | | | | | | HCP_A3 | 1.37 | 0.18 | 66.54 | 0.11 | 0.02 | 0.00 | 32.92 | 0.22 |
| S1_C03_valid | 27.7 | 1.0 | 23.3 | 0.4 | 0.7 | 0.1 | 26.5 | 1.9 | 21.8 | 0.1 | 51200 | 43150 | 43150 | 1.000 | | | | | 0.0571 | | | | | | | | | | | | | | | | | | | | | | | |
| S1_C03_P0 | 30.0 | 1.1 | 18.2 | 0.3 | 0.7 | 0.1 | 27.9 | 2.0 | 23.2 | 0.1 | 51200 | 43150 | 22448 | 0.520 | 0.520 | Sigma | 10.0417 | 5.2201 | 0.0570 | SIGMA | 64.91 | 30.88 | 16.50 | | 29.17 | 23.44 | SIGMA | 56.73 | 30.29 | 15.42 | | 31.13 | 23.16 | SIGMA | 62.76 | 30.39 | 15.00 | | 30.99 | 23.61 | | |
| S1_C03_P1 | 25.3 | 0.9 | 28.6 | 0.5 | 0.7 | 0.1 | 25.2 | 1.7 | 20.3 | 0.1 | 51200 | 43150 | 20702 | 0.480 | 0.480 | BCC | 3.2713 | | 0.0571 | BCC_B2 | 35.09 | 21.69 | 35.96 | 2.02 | 21.67 | 18.66 | BCC_B2 | 43.27 | 24.20 | 33.71 | 1.64 | 20.51 | 19.94 | BCC_B2 | 33.76 | 24.37 | 33.53 | 2.04 | 20.05 | 19.80 | 0.06 | 0.16 |
| | | | | | | | | | | | | | | | | | | | | | | | | | | | | | | | | | | HCP_A3 | 3.47 | 0.50 | 66.74 | 0.34 | 0.01 | 0.00 | 32.06 | 0.34 |
| S1_C04_valid | 13.2 | 0.5 | 11.5 | 0.2 | 1.0 | 0.1 | 45.5 | 3.4 | 28.8 | 0.1 | 51200 | 46044 | 46044 | 1.000 | | | | | 0.0548 | | | | | | | | | | | | | | | | | | | | | | | |
| S1_C04_P0 | 13.0 | 0.5 | 11.6 | 0.2 | 1.0 | 0.1 | 45.0 | 3.4 | 29.4 | 0.1 | 51200 | 46044 | 44827 | 0.974 | 0.972 | BCC | 3.3177 | | 0.0548 | BCC_B2 | 75.49 | 8.99 | 13.27 | 1.31 | 48.35 | 28.08 | BCC_B2 | 77.44 | 9.71 | 13.13 | 1.28 | 47.57 | 28.31 | BCC_B2 | 73.11 | 9.28 | 13.38 | 0.38 | 48.21 | 28.62 | 0.02 | 0.11 |
| S1_C04_P1 | 20.5 | 0.9 | 9.7 | 0.2 | 0.7 | 0.1 | 58.0 | 2.4 | 11.2 | 0.1 | 51200 | 46044 | 1217 | 0.026 | 0.028 | Sigma | not visible | | 0.0571 | SIGMA | 24.51 | 26.18 | 6.05 | | 36.74 | 31.04 | SIGMA | 22.56 | 25.19 | 5.90 | | 38.41 | 30.50 | SIGMA | 24.79 | 25.07 | 6.22 | | 38.64 | 30.07 | | |
| | | | | | | | | | | | | | | | | | | | | | | | | | | | | | | | | | | ZRO2_TETR | 2.11 | 0.27 | 0.00 | 33.10 | 0.00 | | 66.62 | |
| S1_C11_valid | 20.2 | 0.8 | 17.6 | 0.4 | 16.0 | 2.7 | 9.7 | 0.9 | 36.6 | 0.1 | 51200 | 49473 | 49473 | 1.000 | | | | | 0.0543 | | | | | | | | | | | | | | | | | | | | | | | |
| S1_C11_P0 | 17.7 | 0.8 | 19.2 | 0.4 | 10.6 | 1.9 | 10.7 | 1.1 | 41.8 | 0.1 | 51200 | 49473 | 44538 | 0.900 | 0.904 | BCC | 3.3214 | | 0.0546 | BCC_B2 | 66.61 | 13.51 | 21.86 | 7.90 | 11.60 | 45.14 | BCC_B2 | 71.59 | 16.75 | 19.98 | 9.87 | 10.08 | 43.33 | BCC_B2 | 69.21 | 16.88 | 20.30 | 9.66 | 10.00 | 43.10 | 0.02 | 0.04 |
| | | | | | | | | | | | | | | | | | | | | | | | | | | | BCC_B2 | 21.28 | 29.70 | 13.90 | 41.85 | 5.96 | 8.59 | BCC_B2 | 20.09 | 29.36 | 14.30 | 41.73 | 5.72 | 8.32 | 0.31 | 0.26 |
| S1_C11_P1 | 37.0 | 1.3 | 9.9 | 0.2 | 38.5 | 2.9 | 4.9 | 0.3 | 9.7 | 0.1 | 51200 | 49473 | 4935 | 0.100 | 0.096 | Al3Zr5 | 8.1002 | 5.4104 | 0.0520 | AL3ZR5_D8M | 21.00 | 37.50 | 11.50 | 51.00 | | | | | | | | | | | | | | | | | | |
| | | | | | | | | | | | | | | | | | | | | SIGMA | 12.39 | 26.87 | 4.91 | | 15.68 | 52.54 | SIGMA | 7.13 | 26.57 | 4.49 | | 16.59 | 52.35 | SIGMA | 8.71 | 26.66 | 4.59 | | 16.66 | 52.09 | | |
| | | | | | | | | | | | | | | | | | | | | | | | | | | | | | | | | | | ZRO2_TETR | 1.99 | 0.07 | 0.00 | 33.27 | 0.00 | | 66.65 | |
| S1_C12_valid | 13.6 | 0.6 | 10.2 | 0.2 | 7.9 | 1.5 | 18.1 | 1.9 | 50.2 | 0.1 | 51200 | 47623 | 47623 | 1.000 | | | | | 0.0541 | | | | | | | | | | | | | | | | | | | | | | | |
| S1_C12_P0 | 12.5 | 0.6 | 10.5 | 0.2 | 5.7 | 1.1 | 18.7 | 2.0 | 52.5 | 0.1 | 51200 | 47623 | 46262 | 0.971 | 0.973 | BCC | 3.3304 | | 0.0541 | BCC_B2 | 70.89 | 7.51 | 12.48 | 6.31 | 19.52 | 54.18 | BCC_B2 | 76.84 | 10.01 | 12.01 | 7.79 | 18.32 | 51.87 | BCC_B2 | 74.07 | 10.04 | 12.28 | 7.47 | 18.32 | 51.80 | 0.01 | 0.08 |
| | | | | | | | | | | | | | | | | | | | | | | | | | | | BCC_B2 | 4.27 | 27.76 | 10.79 | 45.25 | 7.72 | 8.48 | BCC_B2 | 3.11 | 26.91 | 11.35 | 45.62 | 6.85 | 7.80 | 0.32 | 1.17 |
| S1_C12_P1 | 36.4 | 1.3 | 6.6 | 0.1 | 38.1 | 3.0 | 8.7 | 0.5 | 10.2 | 0.1 | 51200 | 47623 | 1361 | 0.029 | 0.027 | Al3Zr5 | 8.1672 | 5.4397 | 0.0509 | AL3ZR5_D8M | 6.47 | 37.50 | 9.20 | 53.30 | | | | | | | | | | | | | | | | | | |
| | | | | | | | | | | | | | | | | | | | | SIGMA | 22.64 | 25.81 | 3.43 | | 18.75 | 52.01 | SIGMA | 18.88 | 24.98 | 2.81 | | 19.46 | 52.75 | SIGMA | 20.33 | 24.97 | 2.88 | | 19.60 | 52.55 | | |
| | | | | | | | | | | | | | | | | | | | | | | | | | | | | | | | | | | ZRO2_TETR | 2.48 | 0.05 | 0.00 | 33.29 | 0.00 | | 66.66 | |
| S1_C13_valid | 21.3 | 0.7 | 21.1 | 0.3 | 9.4 | 1.0 | 30.9 | 1.8 | 17.3 | 0.1 | 51200 | 50851 | 50851 | 1.000 | | | | | 0.0552 | | | | | | | | | | | | | | | | | | | | | | | |
| S1_C13_P0 | 21.3 | 0.7 | 21.1 | 0.3 | 9.4 | 1.0 | 30.9 | 1.8 | 17.3 | 0.1 | 51200 | 50851 | 50579 | 0.999 | 0.999 | BCC | 3.3090 | | 0.0552 | BCC_B2 | 89.07 | 20.37 | 22.51 | 10.61 | 30.51 | 15.99 | BCC_B2 | 93.44 | 20.89 | 22.00 | 10.11 | 30.40 | 16.60 | BCC_B2 | 90.05 | 20.85 | 22.38 | 9.75 | 30.39 | 16.51 | 0.04 | 0.07 |
| | | | | | | | | | | | 51200 | 50851 | 209 | porosity | | B2 | 3.2945 | | 0.0559 | SIGMA | 10.93 | 29.23 | 9.21 | | 33.86 | 27.69 | SIGMA | 6.56 | 27.79 | 7.69 | | 37.66 | 26.86 | SIGMA | 8.27 | 27.83 | 7.87 | | 37.85 | 26.45 | | |
| S1_C13_P1 | 36.0 | 1.3 | 9.7 | 0.2 | 35.4 | 1.8 | 13.4 | 0.8 | 5.4 | 0.3 | 51200 | 50851 | 63 | 0.001 | 0.001 | Al3Zr5 | not visible | | 0.0512 | | | | | | | | | | | | | | | ZRO2_TETR | 1.67 | 0.15 | 0.00 | 33.21 | 0.00 | | 66.64 | |
| S1_C14_valid | 17.7 | 0.7 | 16.5 | 0.3 | 8.8 | 0.8 | 43.2 | 2.1 | 13.8 | 0.1 | 51200 | 49703 | 49703 | 1.000 | | | | | 0.0548 | | | | | | | | | | | | | | | | | | | | | | | |
| S1_C14_P0 | 17.7 | 0.7 | 16.5 | 0.3 | 8.8 | 0.8 | 43.2 | 2.1 | 13.8 | 0.1 | 51200 | 49703 | 49703 | 1.000 | 1.000 | BCC | 3.3174 | | 0.0548 | BCC_B2 | 100.00 | 17.72 | 16.51 | 8.76 | 43.21 | 13.80 | BCC_B2 | 100.00 | 17.72 | 16.51 | 8.76 | 43.21 | 13.80 | BCC_B2 | 98.27 | 17.81 | 16.60 | 8.22 | 43.43 | 13.87 | 0.02 | 0.07 |
| | | | | | | | | | | | | | | | | | | | | | | | | | | | | | | | | | | ZRO2_TETR | 1.73 | 0.13 | 0.00 | 33.22 | 0.00 | | 66.64 | |
| S1_C20_valid | 20.2 | 0.9 | 14.0 | 0.3 | 25.1 | 4.3 | 0.3 | 0.0 | 40.5 | 0.1 | 51200 | 47995 | 47995 | 1.000 | | | | | 0.0536 | | | | | | | | | | | | | | | | | | | | | | | |
| S1_C20_P0 | 14.0 | 0.7 | 17.0 | 0.4 | 13.6 | 2.7 | 0.3 | 0.0 | 55.3 | 0.1 | 51200 | 47995 | 37925 | 0.790 | 0.801 | BCC | 3.3267 | | 0.0543 | BCC_B2 | 58.22 | 7.76 | 17.12 | 5.12 | 0.48 | 69.51 | BCC_B2 | 60.51 | 12.58 | 16.92 | 7.07 | 0.38 | 63.06 | BCC_B2 | 59.51 | 12.96 | 16.88 | 6.63 | 0.38 | 63.12 | 0.01 | 0.02 |
| + nanoscale cuboidal phase | | | | | | | | | | | | | | | | | | | | | | | | | | | BCC_B2 | 29.86 | 30.02 | 10.54 | 51.52 | 0.18 | 7.74 | BCC_B2 | 31.47 | 30.24 | 10.39 | 51.25 | 0.17 | 7.24 | 0.48 | 0.24 |
| S1_C20_P1 | 36.2 | 1.4 | 9.0 | 0.1 | 42.6 | 3.7 | 0.3 | 0.0 | 11.9 | 0.1 | 51200 | 47995 | 10070 | 0.210 | 0.199 | Al3Zr5 | 8.1220 | 5.5179 | 0.0508 | AL3ZR5_D8M | 41.78 | 37.50 | 9.56 | 52.94 | | | AL3ZR5_D8M | 9.63 | 37.50 | 6.01 | 56.49 | | | AL3ZR5_D8M | 7.05 | 37.50 | 6.07 | 56.43 | | | | |
| | | | | | | | | | | | | | | | | | | | | | | | | | | | | | | | | | | ZRO2_TETR | 1.97 | 0.05 | 0.00 | 33.29 | 0.00 | | 66.66 | |
| S1_C21_valid | 21.5 | 0.7 | 16.1 | 0.2 | 21.6 | 2.5 | 20.6 | 1.4 | 20.3 | 0.1 | 51200 | 51156 | 51156 | 1.000 | | | | | 0.0536 | | | | | | | | | | | | | | | | | | | | | | | |
| S1_C21_P0 | 19.7 | 0.7 | 17.2 | 0.3 | 18.8 | 2.3 | 22.2 | 1.5 | 22.1 | 0.1 | 51200 | 51156 | 46423 | 0.907 | 0.912 | BCC | 3.3375 | | 0.0538 | BCC_B2 | 72.51 | 15.92 | 17.98 | 12.57 | 27.18 | 26.34 | BCC_B2 | 53.44 | 18.73 | 17.24 | 16.55 | 22.60 | 24.88 | BCC_B2 | 52.47 | 18.80 | 17.50 | 16.36 | 22.64 | 24.63 | 0.04 | 0.04 |
| + nanoscale cuboidal phase | | | | | | | | | | | | | | | | | | | | | | | | | | | BCC_B2 | 42.75 | 24.62 | 15.63 | 29.73 | 17.30 | 12.72 | BCC_B2 | 40.56 | 24.69 | 15.89 | 29.89 | 17.02 | 12.28 | 0.12 | 0.11 |
| S1_C21_P1 | 36.2 | 1.3 | 8.0 | 0.1 | 39.3 | 2.3 | 9.8 | 0.6 | 6.7 | 0.1 | 51200 | 51156 | 4733 | 0.093 | 0.088 | Al3Zr5 | 8.1012 | 5.5055 | 0.0511 | AL3ZR5_D8M | 24.47 | 37.50 | 11.67 | 50.83 | | | | | | | | | | | | | | | | | | |
| | | | | | | | | | | | | | | | | | | | | SIGMA | 3.02 | 27.19 | 5.28 | | 28.29 | 39.24 | SIGMA | 3.81 | 26.26 | 4.22 | | 28.71 | 40.81 | SIGMA | 5.26 | 26.31 | 4.31 | | 28.93 | 40.44 | | |
| | | | | | | | | | | | | | | | | | | | | | | | | | | | | | | | | | | ZRO2_TETR | 1.71 | 0.07 | 0.00 | 33.27 | 0.00 | | 66.65 | |
| S1_C22_valid | 19.2 | 0.6 | 17.7 | 0.2 | 20.2 | 1.7 | 31.1 | 1.5 | 11.8 | 0.1 | 51200 | 50884 | 50884 | 1.000 | | | | | 0.0537 | | | | | | | | | | | | | | | | | | | | | | | |
| S1_C22_P0 | 19.2 | 0.6 | 17.7 | 0.2 | 20.2 | 1.7 | 31.1 | 1.5 | 11.8 | 0.1 | 51200 | 50884 | 50884 | 1.000 | 1.000 | BCC | 3.3387 | | 0.0537 | BCC_B2 | 99.29 | 19.05 | 17.72 | 20.00 | 31.32 | 11.91 | BCC_B2 | 100.00 | 19.18 | 17.68 | 20.21 | 31.09 | 11.83 | BCC_B2 | 98.41 | 19.26 | 17.76 | 19.75 | 31.22 | 11.88 | 0.06 | 0.07 |
| | | | | | | | | | | | | | | | | Al3Zr5 | 8.1171 | 5.4795 | 0.0512 | AL3ZR5_D8M | 0.71 | 37.50 | 12.74 | 49.76 | | | | | | | | | | ZRO2_TETR | 1.59 | 0.06 | 0.00 | 33.28 | 0.00 | | 66.66 | |
| S1_C23_valid | 20.7 | 0.8 | 19.3 | 0.3 | 21.5 | 0.6 | 38.0 | 1.2 | 0.5 | 0.0 | 51200 | 50738 | 50738 | 1.000 | | | | | 0.0531 | | | | | | | | | | | | | | | | | | | | | | | |
| S1_C23_P0 | 20.7 | 0.8 | 19.3 | 0.3 | 21.5 | 0.6 | 38.0 | 1.2 | 0.5 | 0.0 | 51200 | 50738 | 50738 | 1.000 | 1.000 | BCC | 3.3511 | | 0.0531 | BCC_B2 | 100.00 | 20.69 | 19.29 | 21.54 | 37.95 | 0.53 | BCC_B2 | 100.00 | 20.69 | 19.29 | 21.54 | 37.95 | 0.53 | BCC_B2 | 98.69 | 20.75 | 19.35 | 21.16 | 38.07 | 0.53 | 0.09 | 0.06 |
| | | | | | | | | | | | | | | | | | | | | | | | | | | | | | | | | | | ZRO2_TETR | 1.31 | 0.07 | 0.00 | 33.28 | 0.00 | | 66.66 | |
| S1_C24_valid | 18.8 | 0.8 | 19.6 | 0.3 | 22.2 | 0.6 | 39.4 | 1.3 | 0.0 | 0.0 | 51200 | 50535 | 50535 | 1.000 | | | | | 0.0524 | | | | | | | | | | | | | | | | | | | | | | | |
| S1_C24_P0 | 18.8 | 0.8 | 19.6 | 0.3 | 22.2 | 0.6 | 39.4 | 1.3 | 0.0 | 0.0 | 51200 | 50535 | 50535 | 1.000 | 1.000 | B2 | 3.3667 | | 0.0524 | BCC_B2 | 100.00 | 18.80 | 19.61 | 22.19 | 39.40 | 0.00 | BCC_B2 | 100.00 | 18.80 | 19.61 | 22.19 | 39.40 | 0.00 | BCC_B2 | 98.71 | 18.85 | 19.66 | 21.81 | 39.50 | 0.00 | 0.11 | 0.06 |
| | | | | | | | | | | | | | | | | | | | | | | | | | | | | | | | | | | ZRO2_TETR | 1.29 | 0.05 | 0.00 | 33.29 | 0.00 | | 66.66 | |
| S1_C30_valid | 17.1 | 0.7 | 16.9 | 0.3 | 33.5 | 5.1 | 0.6 | 0.1 | 32.0 | 0.1 | 51200 | 51184 | 51184 | 1.000 | | | | | 0.0509 | | | | | | | | | | | | | | | | | | | | | | | |
| S1_C30_P0 | 18.7 | 0.7 | 17.7 | 0.3 | 36.2 | 5.1 | 0.6 | 0.1 | 26.8 | 0.1 | 51200 | 51184 | 40675 | 0.795 | 0.781 | B2 | 3.4202 | | 0.0500 | BCC_B2 | 54.09 | 9.12 | 23.17 | 15.55 | 0.83 | 51.33 | BCC_B2 | 56.80 | 22.14 | 15.07 | 50.81 | 0.40 | 11.58 | BCC_B2 | 51.90 | 22.77 | 14.99 | 50.95 | 0.38 | 9.93 | 0.84 | 0.14 |
| + nanoscale cuboidal phase | | | | | | | | | | | | | | | | | | | | | | | | | | | BCC_B2 | 43.20 | 10.37 | 19.27 | 10.69 | 0.79 | 58.88 | BCC_B2 | 44.95 | 11.09 | 19.61 | 9.75 | 0.81 | 58.72 | 0.01 | 0.01 |
| S1_C30_P1 | 9.3 | 0.5 | 13.8 | 0.4 | 13.5 | 2.8 | 0.4 | 0.1 | 62.9 | 0.1 | 51200 | 51184 | 10509 | 0.205 | 0.219 | BCC | 3.3292 | | 0.0542 | LIQUID | 45.91 | 26.40 | 9.48 | 54.60 | 0.25 | 9.27 | | | | | | | | HCP_A3 | 3.15 | 0.00 | 1.47 | 68.65 | 0.00 | 0.00 | 29.82 | 0.06 |
| S1_C31_valid | 18.7 | 0.7 | 18.9 | 0.3 | 35.9 | 4.3 | 6.1 | 0.4 | 20.5 | 0.1 | 51200 | 50877 | 50877 | 1.000 | | | | | 0.0543 | | | | | | | | | | | | | | | | | | | | | | | |
| S1_C31_P0 | 18.7 | 0.7 | 18.9 | 0.3 | 35.9 | 4.3 | 6.1 | 0.4 | 20.5 | 0.1 | 51200 | 50877 | 50877 | 1.000 | 1.000 | BCC | 3.3271 | | 0.0543 | BCC_B2 | 54.14 | 12.43 | 24.90 | 22.38 | 8.66 | 31.63 | BCC_B2 | 75.86 | 20.77 | 18.06 | 42.63 | 5.25 | 13.29 | BCC_B2 | 69.06 | 21.49 | 17.94 | 43.73 | 4.90 | 11.04 | 0.81 | 0.09 |
| + nanoscale cuboidal phase | | | | | | | | | | | | | | | | B2 | 3.3800 | | 0.0518 | | | | | | | | BCC_B2 | 24.14 | 12.34 | 21.35 | 14.68 | 8.57 | 43.06 | BCC_B2 | 28.81 | 12.67 | 21.44 | 13.17 | 9.00 | 43.67 | 0.04 | 0.01 |
| | | | | | | | | | | | | | | | | | | | | LIQUID | 45.86 | 26.18 | 11.71 | 51.83 | 2.98 | 7.30 | | | | | | | | HCP_A3 | 2.13 | 0.00 | 2.36 | 67.08 | 0.01 | 0.00 | 30.52 | 0.03 |
| S1_C32_valid | 19.4 | 0.7 | 18.4 | 0.3 | 38.6 | 2.6 | 14.7 | 0.8 | 8.9 | 0.1 | 51200 | 51126 | 51126 | 1.000 | | | | | 0.0521 | | | | | | | | | | | | | | | | | | | | | | | |
| S1_C32_P0 | 19.4 | 0.7 | 18.4 | 0.3 | 38.6 | 2.6 | 14.7 | 0.8 | 8.9 | 0.1 | 51200 | 51126 | 51126 | 1.000 | 1.000 | BCC | 3.3741 | | 0.0521 | BCC_B2 | 68.83 | 16.48 | 20.63 | 33.52 | 18.15 | 11.22 | BCC_B2 | 100.00 | 19.38 | 18.41 | 38.57 | 14.69 | 8.95 | BCC_B2 | 98.23 | 19.51 | 18.47 | 37.63 | 14.79 | 9.00 | 0.54 | 0.06 |
| | | | | | | | | | | | | | | | | B2 | 3.6968 | | 0.0396 | LIQUID | 31.17 | 25.79 | 13.49 | 49.73 | 7.05 | 3.94 | | | | | | | | HCP_A3 | 1.77 | 0.00 | 3.08 | 66.08 | 0.02 | 0.00 | 30.79 | 0.03 |
| S1_C33_valid | 19.6 | 0.8 | 18.3 | 0.3 | 39.4 | 0.8 | 22.2 | 1.1 | 0.4 | 0.0 | 51200 | 50839 | 50839 | 1.000 | | | | | 0.0518 | | | | | | | | | | | | | | | | | | | | | | | |
| S1_C33_P0 | 19.6 | 0.8 | 18.3 | 0.3 | 39.4 | 0.8 | 22.2 | 1.1 | 0.4 | 0.0 | 51200 | 50839 | 50839 | 1.000 | 1.000 | B2 | 3.3809 | | 0.0518 | BCC_B2 | 97.60 | 19.17 | 18.48 | 39.11 | 22.79 | 0.45 | BCC_B2 | 100.00 | 19.61 | 18.34 | 39.37 | 22.24 | 0.44 | BCC_B2 | 99.49 | 19.50 | 18.25 | 38.99 | 22.12 | 0.44 | 0.64 | 0.06 |
| | | | | | | | | | | | | | | | | overlap with BCC possible | | | | AL3ZR5_D8M | 2.40 | 37.50 | 12.74 | 49.76 | | | | | | | | | | ZRO2_TETR | 0.51 | 0.02 | 0.00 | 33.31 | 0.00 | | 66.66 | |
| S1_C40_valid | 19.0 | 0.7 | 18.9 | 0.3 | 41.2 | 4.8 | 1.5 | 0.1 | 19.4 | 0.1 | 51200 | 50975 | 50975 | 1.000 | | | | | 0.0538 | | | | | | | | | | | | | | | | | | | | | | | |
| S1_C40_P0 | 19.0 | 0.7 | 18.9 | 0.3 | 41.2 | 4.8 | 1.5 | 0.1 | 19.4 | 0.1 | 51200 | 50975 | 50975 | 1.000 | 1.000 | BCC | 3.3367 | | 0.0538 | BCC_B2 | 42.01 | 11.47 | 26.94 | 24.29 | 2.40 | 34.90 | BCC_B2 | 83.10 | 20.55 | 18.14 | 46.90 | 1.36 | 13.06 | BCC_B2 | 77.78 | 20.96 | 17.89 | 48.04 | 1.26 | 10.70 | 1.05 | 0.09 |
| + nanoscale cuboidal phase | | | | | | | | | | | | | | | | B2 | 3.4143 | | 0.0502 | | | | | | | | BCC_B2 | 16.90 | 11.55 | 22.39 | 13.17 | 2.32 | 50.57 | BCC_B2 | 20.96 | 11.83 | 22.32 | 11.68 | 2.48 | 51.66 | 0.03 | 0.01 |
| | | | | | | | | | | | | | | | | | | | | LIQUID | 57.99 | 24.50 | 13.00 | 53.44 | 0.88 | 8.17 | | | | | | | | HCP_A3 | 1.26 | 0.00 | 2.14 | 67.94 | 0.00 | 0.00 | 29.88 | 0.03 |
| S1_C41_valid | 18.8 | 0.7 | 18.9 | 0.3 | 39.4 | 4.4 | 4.7 | 0.3 | 18.2 | 0.1 | 51200 | 50767 | 50767 | 1.000 | | | | | 0.0499 | | | | | | | | | | | | | | | | | | | | | | | |
| S1_C41_P0 | 18.8 | 0.7 | 18.9 | 0.3 | 39.4 | 4.4 | 4.7 | 0.3 | 18.2 | 0.1 | 51200 | 50767 | 50767 | 1.000 | 1.000 | B2 | 3.4213 | | 0.0499 | BCC_B2 | 48.52 | 12.26 | 25.49 | 25.57 | 6.97 | 29.71 | BCC_B2 | 85.74 | 19.96 | 18.46 | 43.50 | 4.29 | 13.79 | BCC_B2 | 78.34 | 20.61 | 18.26 | 44.88 | 3.99 | 11.29 | 0.88 | 0.08 |
| | | | | | | | | | | | | | | | | overlap with BCC possible | | | | | | | | | | | BCC_B2 | 14.26 | 11.80 | 21.73 | 14.87 | 7.03 | 44.57 | BCC_B2 | 20.02 | 12.03 | 21.72 | 13.26 | 7.47 | 45.47 | 0.04 | 0.01 |
| | | | | | | | | | | | | | | | | | | | | LIQUID | 51.48 | 24.95 | 12.74 | 52.47 | 2.53 | 7.31 | | | | | | | | HCP_A3 | 1.64 | 0.00 | 2.33 | 67.45 | 0.01 | 0.00 | 30.17 | 0.03 |
| S1_C42_valid | 19.0 | 0.7 | 18.5 | 0.3 | 40.2 | 2.5 | 14.2 | 0.8 | 8.1 | 0.1 | 51200 | 50737 | 50737 | 1.000 | | | | | 0.0501 | | | | | | | | | | | | | | | | | | | | | | | |
| S1_C42_P0 | 19.0 | 0.7 | 18.5 | 0.3 | 40.2 | 2.5 | 14.2 | 0.8 | 8.1 | 0.1 | 51200 | 50737 | 50737 | 1.000 | 1.000 | B2 | 3.4180 | | 0.0501 | BCC_B2 | 68.54 | 16.25 | 20.55 | 35.66 | 17.43 | 10.10 | BCC_B2 | 100.00 | 19.00 | 18.53 | 40.23 | 14.15 | 8.10 | BCC_B2 | 98.48 | 19.07 | 18.56 | 39.36 | 14.21 | 8.13 | 0.61 | 0.06 |
| | | | | | | | | | | | | | | | | overlap with BCC possible | | | | LIQUID | 31.46 | 24.98 | 14.12 | 50.17 | 7.01 | 3.72 | | | | | | | | HCP_A3 | 1.52 | 0.00 | 3.05 | 66.38 | 0.02 | 0.00 | 30.52 | 0.03 |

| | EDS chemical composition (at.%) | | | | | | | | EDS phase composition | | | | | XRD | | | | ThermoCalc - TCHEA4 ("valid" composition) | | | | | | ThermoCalc - TCHEA8 ("valid" composition) | | | | | | ThermoCalc - TCHEA8 ("valid" composition) (with O and N) | | | | | | | |
|---|---|---|---|---|---|---|---|---|---|---|---|---|---|---|---|---|---|---|---|---|---|---|---|---|---|---|---|---|---|---|---|---|---|---|---|---|---|
| Name | Al | U(Al) | Zr | U(Zr) | Nb | U(Nb) | Ta | U(Ta) | Pixels total | Pixels valid | Phase pixels | Volume fraction | Molar fraction | Phase label | Lattice parameter a (Å) | Lattice parameter c (Å) | Molar density (atoms/Å$^3$) | Phase name | Phase fraction (mol.%) | Al | Zr | Nb | Ta | Phase name | Phase fraction (mol.%) | Al | Zr | Nb | Ta | Phase name | Phase fraction (mol.%) | Al | Zr | Nb | Ta | O | N |
| S2_C21_valid | 30.2 | 1.1 | 43.2 | 4.6 | 8.9 | 0.6 | 17.7 | 0.1 | 51200 | 41331 | 41331 | 1.000 | | | | | 0.0491 | | | | | | | | | | | | | | | | | | | | |
| S2_C21_P0 | 36.9 | 1.5 | 46.4 | 3.4 | 7.3 | 0.5 | 9.4 | 0.1 | 51200 | 41331 | 26107 | 0.632 | 0.655 | Al3Zr5 | 8.1049 | 5.5273 | 0.0509 | | | | | | | | | | | | | AL4ZR5 | 35.48 | 44.44 | 55.56 | | | | |
| S2_C21_P1 | 9.2 | 0.5 | 11.7 | 2.4 | 19.3 | 2.1 | 59.8 | 0.1 | 51200 | 41331 | 14258 | 0.345 | 0.331 | Sigma | 10.9036 | 5.3611 | 0.0471 | SIGMA | 12.12 | 37.86 | | 28.31 | 33.82 | SIGMA | 12.12 | 37.86 | | 28.31 | 33.82 | SIGMA | 59.16 | 26.60 | | 19.00 | 54.41 | | |
| | | | | | | | | | | | | | | BCC | 3.3178 | | 0.0548 | | | | | | | | | | | | | | | | | | | | |
| S2_C21_P2 | 33.8 | 1.4 | 51.1 | 3.3 | 7.1 | 0.4 | 8.1 | 0.1 | 51200 | 41331 | 563 | 0.014 | 0.014 | Eutectic / Liquid at 1400 °C | | | 0.0495 | | | | | | | | | | | | | LIQUID | 3.35 | 36.47 | 53.12 | 3.34 | 6.96 | 0.13 | 0.23 |
| | | | | | | | | | | | | | | | | | | C14_LAVES | 87.99 | 66.47 | 3.15 | 2.22 | 1.59 | C14_LAVES | 87.99 | 66.47 | 3.15 | 2.22 | 1.59 | HCP_A3 | 0.26 | 0.26 | 7.79 | 1.58 | 5.50 | 0.22 | 22.46 |
| | | | | | | | | | | | | | | | | | | | | | | | | | | | | | | ZRO2_TETR | 1.76 | 0.56 | 33.29 | 0.00 | | 66.66 | |
| S2_C22_valid | 28.4 | 1.0 | 36.3 | 3.9 | 17.8 | 1.2 | 17.5 | 0.1 | 51200 | 47973 | 47973 | 1.000 | | | | | 0.0531 | | | | | | | | | | | | | | | | | | | | |
| S2_C22_P0 | 25.9 | 1.0 | 28.8 | 3.6 | 22.5 | 1.6 | 22.7 | 0.1 | 51200 | 47973 | 32478 | 0.677 | 0.685 | Sigma | 10.2562 | 5.3064 | 0.0537 | SIGMA | 18.31 | 39.70 | | 37.01 | 23.29 | SIGMA | 18.31 | 39.70 | | 37.01 | 23.29 | SIGMA | 62.03 | 27.18 | | 36.23 | 36.59 | | |
| S2_C22_P1 | 37.1 | 1.5 | 45.9 | 2.8 | 10.0 | 0.6 | 7.0 | 0.1 | 51200 | 47973 | 8851 | 0.184 | 0.176 | Al3Zr5 | 8.1126 | 5.5313 | 0.0508 | | | | | | | | | | | | | AL4ZR5 | 32.51 | 44.44 | 55.56 | | | | |
| S2_C22_P2 | 32.1 | 1.4 | 51.8 | 2.7 | 9.9 | 0.6 | 6.2 | 0.1 | 51200 | 47973 | 5218 | 0.109 | 0.109 | Eutectic / Liquid at 1400 °C | | | 0.0531 | | | | | | | | | | | | | LIQUID | 3.73 | 36.97 | 50.78 | 8.15 | 3.99 | 0.11 | 0.00 |
| S2_C22_P3 | 10.0 | 0.5 | 13.5 | 2.4 | 29.2 | 2.8 | 47.2 | 0.2 | 51200 | 47973 | 1426 | 0.030 | 0.030 | BCC | not visible | | 0.0542 | | | | | | | | | | | | | | | | | | | | |
| | | | | | | | | | | | | | | | | | | C14_LAVES | 81.69 | 66.17 | 26.72 | 5.46 | 1.65 | C14_LAVES | 81.69 | 66.17 | 26.72 | 5.46 | 1.65 | ZRO2_TETR | 1.60 | 0.07 | 33.27 | 0.00 | | 66.65 | |
| | | | | | | | | | | | | | | | | | | | | | | | | | | | | | | FCC_L12 | 0.12 | 0.00 | 55.70 | 0.06 | 0.03 | 0.01 | 44.19 |
| S2_C23_valid | 28.0 | 1.1 | 37.6 | 2.3 | 26.6 | 1.4 | 7.8 | 0.1 | 51200 | 50941 | 50941 | 1.000 | | | | | 0.0532 | | | | | | | | | | | | | | | | | | | | |
| S2_C23_P0 | 27.5 | 1.0 | 31.4 | 2.2 | 31.5 | 1.6 | 9.5 | 0.1 | 51200 | 50941 | 23351 | 0.458 | 0.461 | Sigma | 10.2713 | 5.3178 | 0.0535 | SIGMA | 22.85 | 35.17 | | 55.12 | 9.71 | SIGMA | 22.85 | 35.17 | | 55.12 | 9.71 | SIGMA | 48.50 | 30.40 | | 48.73 | 20.87 | | |
| S2_C23_P1 | 24.8 | 1.0 | 41.2 | 2.4 | 26.5 | 1.4 | 7.6 | 0.1 | 51200 | 50941 | 19505 | 0.383 | 0.387 | BCC | 3.3384 | | 0.0538 | | | | | | | | | | | | | BCC_B2 | 35.21 | 35.74 | 35.19 | 27.77 | 1.15 | 0.01 | 0.13 |
| S2_C23_P2 | 36.2 | 1.5 | 44.9 | 1.7 | 15.1 | 0.8 | 3.7 | 0.0 | 51200 | 50941 | 8085 | 0.159 | 0.152 | Al3Zr5 | 8.1183 | 5.4886 | 0.0511 | AL3ZR2 | 0.40 | 60.00 | 40.00 | | | AL3ZR2 | 0.40 | 60.00 | 40.00 | | | AL4ZR5 | 14.90 | 44.44 | 55.56 | | | | |
| | | | | | | | | | | | | | | | | | | C14_LAVES | 76.76 | 65.20 | 29.97 | 4.56 | 0.27 | C14_LAVES | 76.76 | 65.20 | 29.97 | 4.56 | 0.27 | | | | | | | | |
| | | | | | | | | | | | | | | | | | | | | | | | | | | | | | | ZRO2_TETR | 1.40 | 0.18 | 33.18 | 0.00 | | 66.64 | |
| S2_C24_valid | 27.1 | 1.3 | 23.2 | 0.7 | 49.7 | 1.6 | 0.0 | 0.0 | 51200 | 46198 | 46198 | 1.000 | | | | | 0.0516 | | | | | | | | | | | | | | | | | | | | |
| S2_C24_P0 | 27.7 | 1.3 | 21.0 | 0.7 | 51.3 | 1.5 | 0.0 | 0.0 | 51200 | 46198 | 33041 | 0.715 | 0.721 | Sigma | not visible | | 0.0520 | SIGMA | 37.71 | 37.09 | | 62.83 | 0.08 | SIGMA | 37.71 | 37.09 | | 62.83 | 0.08 | | | | | | | | |
| S2_C24_P1 | 14.3 | 0.8 | 18.5 | 0.7 | 67.2 | 1.5 | 0.0 | 0.0 | 51200 | 46198 | 7236 | 0.157 | 0.154 | BCC | 3.4077 | | 0.0505 | | | | | | | | | | | | | BCC_B2 | 98.75 | 33.05 | 20.15 | 46.76 | 0.00 | 0.00 | 0.04 |
| S2_C24_P2 | 36.5 | 1.6 | 40.7 | 0.9 | 22.7 | 1.2 | 0.0 | 0.0 | 51200 | 46198 | 5921 | 0.128 | 0.125 | Al3Zr5 | 8.1167 | 5.5665 | 0.0504 | | | | | | | | | | | | | | | | | | | | |
| | | | | | | | | | | | | | | | | | | C14_LAVES | 62.29 | 65.57 | 25.34 | 9.09 | 0.00 | C14_LAVES | 62.29 | 65.57 | 25.34 | 9.09 | 0.00 | | | | | | | | |
| | | | | | | | | | | | | | | | | | | | | | | | | | | | | | | ZRO2_TETR | 1.25 | 0.33 | 33.06 | 0.00 | | 66.61 | |
| S2_C31_valid | 29.5 | 1.1 | 41.4 | 4.2 | 12.6 | 0.8 | 16.4 | 0.1 | 51200 | 51199 | 51199 | 1.000 | | | | | 0.0521 | | | | | | | | | | | | | | | | | | | | |
| S2_C31_P0 | 36.9 | 1.5 | 46.6 | 2.9 | 9.2 | 0.6 | 7.3 | 0.1 | 51200 | 51199 | 33580 | 0.656 | 0.638 | Al3Zr5 | 8.1114 | 5.5387 | 0.0507 | | | | | | | | | | | | | AL4ZR5 | 53.57 | 44.44 | 55.56 | | | | |
| S2_C31_P1 | 10.1 | 0.5 | 14.5 | 2.7 | 27.4 | 2.7 | 48.1 | 0.2 | 51200 | 51199 | 17619 | 0.344 | 0.362 | BCC | 3.3149 | | 0.0549 | | | | | | | | | | | | | | | | | | | | |
| | | | | | | | | | | | | | | | | | | C14_LAVES | 90.05 | 66.33 | 25.21 | 5.65 | 2.81 | C14_LAVES | 90.05 | 66.33 | 25.21 | 5.65 | 2.81 | | | | | | | | |
| | | | | | | | | | | | | | | | | | | SIGMA | 9.95 | 41.93 | | 29.93 | 28.14 | SIGMA | 9.95 | 41.93 | | 29.93 | 28.14 | SIGMA | 44.62 | 29.16 | | 1.04 | 69.81 | | |
| | | | | | | | | | | | | | | | | | | | | | | | | | | | | | | ZRO2_TETR | 1.69 | 0.11 | 33.24 | 0.00 | | 66.65 | |
| | | | | | | | | | | | | | | | | | | | | | | | | | | | | | | FCC_L12 | 0.12 | 0.00 | 54.91 | 0.00 | 0.08 | 0.01 | 45.00 |
| S2_C32_valid | 26.6 | 1.0 | 42.0 | 2.9 | 22.3 | 1.3 | 9.0 | 0.1 | 51200 | 50755 | 50755 | 1.000 | | | | | 0.0532 | | | | | | | | | | | | | | | | | | | | |
| S2_C32_P0 | 24.9 | 0.9 | 41.3 | 3.0 | 24.0 | 1.3 | 9.8 | 0.1 | 51200 | 50755 | 44063 | 0.868 | 0.875 | BCC | 3.3423 | | 0.0536 | | | | | | | | | | | | | | | | | | | | |
| S2_C32_P1 | 36.7 | 1.5 | 45.6 | 2.0 | 13.2 | 0.8 | 4.6 | 0.0 | 51200 | 50755 | 6692 | 0.132 | 0.125 | Al3Zr5 | 8.1350 | 5.5272 | 0.0505 | AL3ZR2 | 32.06 | 60.00 | 40.00 | | | AL3ZR2 | 32.06 | 60.00 | 40.00 | | | AL4ZR5 | 48.57 | 44.44 | 55.56 | | | | |
| | | | | | | | | | | | | | | | | | | SIGMA | 22.55 | 35.33 | | 52.61 | 12.06 | SIGMA | 22.55 | 35.33 | | 52.61 | 12.06 | SIGMA | 44.13 | 26.74 | | 23.71 | 49.54 | | |
| | | | | | | | | | | | | | | | | | | C14_LAVES | 45.39 | 65.29 | 30.06 | 4.30 | 0.35 | C14_LAVES | 45.39 | 65.29 | 30.06 | 4.30 | 0.35 | | | | | | | | |
| | | | | | | | | | | | | | | | | | | | | | | | | | | | | | | LIQUID | 5.53 | 36.51 | 52.68 | 4.59 | 6.09 | 0.13 | 0.00 |
| | | | | | | | | | | | | | | | | | | | | | | | | | | | | | | ZRO2_TETR | 1.55 | 0.06 | 33.28 | 0.00 | | 66.66 | |
| | | | | | | | | | | | | | | | | | | | | | | | | | | | | | | HCP_A3 | 0.22 | 0.27 | 69.95 | 2.20 | 4.78 | 0.23 | 22.56 |
| S2_C33_valid | 24.0 | 1.1 | 35.4 | 0.9 | 40.6 | 1.7 | 0.0 | 0.0 | 51200 | 51022 | 51022 | 1.000 | | | | | 0.0521 | | | | | | | | | | | | | | | | | | | | |
| S2_C33_P0 | 21.6 | 1.0 | 36.1 | 0.9 | 42.3 | 1.7 | 0.0 | 0.0 | 51200 | 51022 | 38592 | 0.756 | 0.753 | BCC | 3.3786 | | 0.0519 | BCC_B2 | 54.64 | 41.46 | 16.27 | 42.20 | 0.07 | BCC_B2 | 54.64 | 41.46 | 16.27 | 42.20 | 0.07 | BCC_B2 | 25.98 | 31.27 | 38.84 | 27.34 | 2.35 | 0.02 | 0.17 |
| S2_C33_P1 | 28.3 | 1.2 | 28.8 | 0.8 | 43.0 | 1.6 | 0.0 | 0.0 | 51200 | 51022 | 8409 | 0.165 | 0.170 | Sigma | 10.2627 | 5.3009 | 0.0537 | SIGMA | 1.01 | 34.69 | | 63.33 | 1.98 | SIGMA | 1.01 | 34.69 | | 63.33 | 1.98 | SIGMA | 36.22 | 28.03 | | 43.73 | 28.24 | | |
| S2_C33_P2 | 36.4 | 1.6 | 43.4 | 0.9 | 20.3 | 1.1 | 0.0 | 0.0 | 51200 | 51022 | 4021 | 0.079 | 0.077 | Al3Zr5 | 8.1127 | 5.4780 | 0.0512 | AL3ZR2 | 19.91 | 60.00 | 40.00 | | | AL3ZR2 | 19.91 | 60.00 | 40.00 | | | AL4ZR5 | 36.40 | 44.44 | 55.56 | | | | |
| | | | | | | | | | | | | | | | | | | C14_LAVES | 24.45 | 64.90 | 29.68 | 5.38 | 0.04 | C14_LAVES | 24.45 | 64.90 | 29.68 | 5.38 | 0.04 | | | | | | | | |
| | | | | | | | | | | | | | | | | | | | | | | | | | | | | | | ZRO2_TETR | 1.39 | 0.09 | 33.25 | 0.00 | | 66.65 | |

| | EDS chemical composition (at.%) | | | | | | | | EDS phase composition | | | | | XRD | | | | ThermoCalc - TCHEA4 ("valid" composition) | | | | | | ThermoCalc - TCHEA8 ("valid" composition) | | | | | | ThermoCalc - TCHEA8 ("valid" composition) | | | | | | | |
|---|---|---|---|---|---|---|---|---|---|---|---|---|---|---|---|---|---|---|---|---|---|---|---|---|---|---|---|---|---|---|---|---|---|---|---|---|---|
| Name | Ti | U(Ti) | Zr | U(Zr) | Nb | U(Nb) | Ta | U(Ta) | Pixels total | Pixels valid | Phase pixels | Volume fraction | Molar fraction | Phase label | Lattice parameter a (Å) | Lattice parameter c (Å) | Molar density (atoms/Å$^3$) | Phase name | Phase fraction (mol.%) | Ti (at.%) | Zr (at.%) | Nb (at.%) | Ta (at.%) | Phase name | Phase fraction (mol.%) | Ti (at.%) | Zr (at.%) | Nb (at.%) | Ta (at.%) | Phase name | Phase fraction (mol.%) | Ti (at.%) | Zr (at.%) | Nb (at.%) | Ta (at.%) | O (at.%) | N (at.%) |
| S3_C11_valid | 16.9 | 0.4 | 11.5 | 2.3 | 13.5 | 1.4 | 58.0 | 0.1 | 51200 | 49565 | 49565 | 1.000 | | | | | 0.0529 | | | | | | | | | | | | | | | | | | | | |
| S3_C11_P0 | 16.9 | 0.4 | 11.5 | 2.3 | 13.5 | 1.4 | 58.0 | 0.1 | 51200 | 49565 | 49565 | 1.000 | 1.000 | BCC | 3.3561 | | 0.0529 | BCC_B2 | 100.00 | 16.94 | 11.55 | 13.48 | 58.03 | BCC_B2 | 100.00 | 16.94 | 11.55 | 13.48 | 58.03 | BCC_B2 | 89.46 | 16.83 | 5.49 | 14.63 | 63.00 | 0.01 | 0.03 |
| | | | | | | | | | | | | | | | | | | | | | | | | | | | | | | BCC_B2 | 2.01 | 34.85 | 43.94 | 0.42 | 0.89 | 19.87 | 0.03 |
| | | | | | | | | | | | | | | | | | | | | | | | | | | | | | | HCP_A3 | 8.53 | 8.22 | 63.56 | 0.02 | 0.00 | 27.30 | 0.91 |
| S3_C12_valid | 18.7 | 0.4 | 10.9 | 1.9 | 26.1 | 2.4 | 44.4 | 0.1 | 51200 | 49683 | 49683 | 1.000 | | | | | 0.0529 | | | | | | | | | | | | | | | | | | | | |
| S3_C12_P0 | 18.7 | 0.4 | 10.9 | 1.9 | 26.1 | 2.4 | 44.4 | 0.1 | 51200 | 49683 | 49683 | 1.000 | 1.000 | BCC | 3.3555 | | 0.0529 | BCC_B2 | 100.00 | 18.65 | 10.88 | 26.07 | 44.40 | BCC_B2 | 100.00 | 18.65 | 10.88 | 26.07 | 44.40 | BCC_B2 | 88.39 | 16.41 | 6.01 | 28.65 | 48.87 | 0.01 | 0.04 |
| | | | | | | | | | | | | | | | | | | | | | | | | | | | | | | BCC_B2 | 9.43 | 36.06 | 41.99 | 0.78 | 0.67 | 20.46 | 0.05 |
| | | | | | | | | | | | | | | | | | | | | | | | | | | | | | | HCP_A3 | 2.19 | 12.00 | 60.45 | 0.05 | 0.00 | 25.12 | 2.38 |
| S3_C13_valid | 15.8 | 0.3 | 8.5 | 1.3 | 37.6 | 3.2 | 38.0 | 0.1 | 51200 | 50271 | 50271 | 1.000 | | | | | 0.0530 | | | | | | | | | | | | | | | | | | | | |
| S3_C13_P0 | 15.8 | 0.3 | 8.5 | 1.3 | 37.6 | 3.2 | 38.0 | 0.1 | 51200 | 50271 | 50271 | 1.000 | 1.000 | BCC | 3.3538 | | 0.0530 | BCC_B2 | 100.00 | 15.81 | 8.54 | 37.63 | 38.02 | BCC_B2 | 100.00 | 15.81 | 8.54 | 37.63 | 38.02 | BCC_B2 | 90.61 | 14.14 | 4.49 | 40.43 | 40.88 | 0.01 | 0.04 |
| | | | | | | | | | | | | | | | | | | | | | | | | | | | | | | BCC_B2 | 5.48 | 38.21 | 37.45 | 0.87 | 0.41 | 23.03 | 0.02 |
| | | | | | | | | | | | | | | | | | | | | | | | | | | | | | | HCP_A3 | 3.81 | 13.43 | 57.14 | 0.05 | 0.00 | 27.92 | 1.46 |
| | | | | | | | | | | | | | | | | | | | | | | | | | | | | | | ZRO2_TETR | 0.10 | 0.00 | 33.33 | 0.00 | | 66.67 | |
| S3_C14_valid | 21.9 | 0.4 | 12.2 | 1.0 | 52.7 | 2.4 | 13.3 | 0.1 | 51200 | 49461 | 49461 | 1.000 | | | | | 0.0531 | | | | | | | | | | | | | | | | | | | | |
| S3_C14_P0 | 21.9 | 0.4 | 12.2 | 1.0 | 52.7 | 2.4 | 13.3 | 0.1 | 51200 | 49461 | 49461 | 1.000 | 1.000 | BCC | 3.3529 | | 0.0531 | BCC_B2 | 100.00 | 21.88 | 12.17 | 52.67 | 13.27 | BCC_B2 | 100.00 | 21.88 | 12.17 | 52.67 | 13.27 | BCC_B2 | 89.98 | 19.42 | 8.94 | 57.15 | 14.43 | 0.02 | 0.04 |
| | | | | | | | | | | | | | | | | | | | | | | | | | | | | | | BCC_B2 | 9.82 | 39.95 | 38.54 | 1.92 | 0.27 | 19.25 | 0.07 |
| | | | | | | | | | | | | | | | | | | | | | | | | | | | | | | HCP_A3 | 0.20 | 26.08 | 50.88 | 2.07 | 0.32 | 4.11 | 16.54 |
| S3_C20_valid | 25.5 | 0.6 | 25.2 | 4.6 | 0.4 | 0.0 | 48.9 | 0.1 | 51200 | 51029 | 51029 | 1.000 | | | | | 0.0531 | | | | | | | | | | | | | | | | | | | | |
| S3_C20_P0 | 25.5 | 0.6 | 25.2 | 4.6 | 0.4 | 0.0 | 48.9 | 0.1 | 51200 | 51029 | 51029 | 1.000 | 1.000 | BCC | 3.3527 | | 0.0531 | BCC_B2 | 100.00 | 25.50 | 25.16 | 0.44 | 48.89 | BCC_B2 | 100.00 | 25.50 | 25.16 | 0.44 | 48.89 | BCC_B2 | 75.37 | 23.27 | 14.19 | 0.55 | 61.94 | 0.01 | 0.03 |
| | + nanoscale cuboidal phase | | | | | | | | | | | | | | | | | | | | | | | | | | | | | BCC_B2 | 24.63 | 29.65 | 56.09 | 0.06 | 3.81 | 10.11 | 0.28 |
| S3_C21_valid | 21.8 | 0.4 | 26.3 | 4.2 | 13.4 | 1.2 | 38.4 | 0.1 | 51200 | 50936 | 50936 | 1.000 | | | | | 0.0517 | | | | | | | | | | | | | | | | | | | | |
| S3_C21_P0 | 21.8 | 0.4 | 26.3 | 4.2 | 13.4 | 1.2 | 38.4 | 0.1 | 51200 | 50936 | 50936 | 1.000 | 1.000 | BCC | 3.3826 | | 0.0517 | BCC_B2 | 100.00 | 21.83 | 26.34 | 13.44 | 38.39 | BCC_B2 | 100.00 | 21.83 | 26.34 | 13.44 | 38.39 | BCC_B2 | 79.34 | 19.89 | 17.32 | 16.17 | 46.58 | 0.02 | 0.03 |
| | | | | | | | | | | | | | | | | | | | | | | | | | | | | | | BCC_B2 | 20.62 | 26.74 | 57.85 | 1.37 | 2.43 | 11.33 | 0.28 |
| | | | | | | | | | | | | | | | | | | | | | | | | | | | | | | HCP_A3 | 0.04 | 13.76 | 66.92 | 0.53 | 1.26 | 2.39 | 15.13 |
| S3_C22_valid | 21.8 | 0.4 | 24.3 | 3.2 | 26.9 | 1.9 | 27.0 | 0.1 | 51200 | 50578 | 50578 | 1.000 | | | | | 0.0511 | | | | | | | | | | | | | | | | | | | | |
| S3_C22_P0 | 21.8 | 0.4 | 24.3 | 3.2 | 26.9 | 1.9 | 27.0 | 0.1 | 51200 | 50578 | 50578 | 1.000 | 1.000 | BCC | 3.3960 | | 0.0511 | BCC_B2 | 100.00 | 21.78 | 24.30 | 26.92 | 27.00 | BCC_B2 | 100.00 | 21.78 | 24.30 | 26.92 | 27.00 | BCC_B2 | 83.86 | 19.84 | 17.89 | 30.98 | 31.23 | 0.03 | 0.04 |
| | | | | | | | | | | | | | | | | | | | | | | | | | | | | | | BCC_B2 | 16.01 | 28.95 | 54.13 | 2.11 | 1.31 | 13.30 | 0.19 |
| | | | | | | | | | | | | | | | | | | | | | | | | | | | | | | HCP_A3 | 0.14 | 15.26 | 64.70 | 1.01 | 0.74 | 3.13 | 15.16 |
| S3_C23_valid | 22.7 | 0.3 | 25.1 | 2.0 | 39.7 | 1.9 | 12.5 | 0.1 | 51200 | 51118 | 51118 | 1.000 | | | | | 0.0506 | | | | | | | | | | | | | | | | | | | | |
| S3_C23_P0 | 22.7 | 0.3 | 25.1 | 2.0 | 39.7 | 1.9 | 12.5 | 0.1 | 51200 | 51118 | 51118 | 1.000 | 1.000 | BCC | 3.4073 | | 0.0506 | BCC_B2 | 100.00 | 22.72 | 25.09 | 39.68 | 12.51 | BCC_B2 | 100.00 | 22.72 | 25.09 | 39.68 | 12.51 | BCC_B2 | 86.46 | 20.93 | 20.40 | 44.51 | 14.08 | 0.05 | 0.04 |
| | | | | | | | | | | | | | | | | | | | | | | | | | | | | | | BCC_B2 | 13.47 | 30.97 | 51.39 | 3.10 | 0.63 | 13.73 | 0.19 |
| | | | | | | | | | | | | | | | | | | | | | | | | | | | | | | HCP_A3 | 0.06 | 16.82 | 62.67 | 1.57 | 0.36 | 3.30 | 15.29 |
| S3_C24_valid | 21.7 | 0.4 | 23.9 | 0.8 | 54.4 | 1.6 | 0.0 | 0.0 | 51200 | 50922 | 50922 | 1.000 | | | | | 0.0507 | | | | | | | | | | | | | | | | | | | | |
| S3_C24_P0 | 21.7 | 0.4 | 23.9 | 0.8 | 54.4 | 1.6 | 0.0 | 0.0 | 51200 | 50922 | 50922 | 1.000 | 1.000 | BCC | 3.4047 | | 0.0507 | BCC_B2 | 100.00 | 21.67 | 23.93 | 54.40 | 0.00 | BCC_B2 | 100.00 | 21.67 | 23.93 | 54.40 | 0.00 | BCC_B2 | 88.97 | 20.00 | 20.28 | 59.63 | 0.00 | 0.05 | 0.04 |
| | | | | | | | | | | | | | | | | | | | | | | | | | | | | | | BCC_B2 | 10.92 | 31.83 | 49.43 | 3.64 | 0.00 | 14.96 | 0.15 |
| | | | | | | | | | | | | | | | | | | | | | | | | | | | | | | HCP_A3 | 0.11 | 17.61 | 61.21 | 2.04 | 0.00 | 4.05 | 15.08 |
| S3_C30_valid | 25.1 | 0.5 | 36.7 | 5.9 | 0.6 | 0.1 | 37.6 | 0.1 | 51200 | 50877 | 50877 | 1.000 | | | | | 0.0529 | | | | | | | | | | | | | | | | | | | | |
| S3_C30_P0 | 25.1 | 0.5 | 36.7 | 5.9 | 0.6 | 0.1 | 37.6 | 0.1 | 51200 | 50877 | 50877 | 1.000 | 1.000 | BCC | 3.3567 | | 0.0529 | BCC_B2 | 100.00 | 25.07 | 36.72 | 0.56 | 37.64 | BCC_B2 | 100.00 | 25.07 | 36.72 | 0.56 | 37.64 | BCC_B2 | 64.76 | 23.54 | 22.50 | 0.77 | 53.15 | 0.02 | 0.02 |
| | + nanoscale cuboidal phase | | | | | | | | | | | | | BCC | 3.3984 | | 0.0510 | | | | | | | | | | | | | BCC_B2 | 35.24 | 26.18 | 60.37 | 0.14 | 6.60 | 6.50 | 0.21 |
| S3_C31_valid | 23.9 | 0.4 | 36.6 | 4.8 | 12.8 | 1.0 | 26.6 | 0.1 | 51200 | 51067 | 51067 | 1.000 | | | | | 0.0507 | | | | | | | | | | | | | | | | | | | | |
| S3_C31_P0 | 23.9 | 0.4 | 36.6 | 4.8 | 12.8 | 1.0 | 26.6 | 0.1 | 51200 | 51067 | 51067 | 1.000 | 1.000 | BCC | 3.4035 | | 0.0507 | BCC_B2 | 100.00 | 23.91 | 36.63 | 12.85 | 26.61 | BCC_B2 | 100.00 | 23.91 | 36.63 | 12.85 | 26.61 | BCC_B2 | 75.33 | 22.55 | 28.06 | 15.90 | 33.39 | 0.07 | 0.04 |
| | | | | | | | | | | | | | | | | | | | | | | | | | | | | | | BCC_B2 | 24.67 | 25.92 | 59.52 | 2.40 | 3.54 | 8.41 | 0.21 |
| S3_C32_valid | 23.1 | 0.3 | 36.6 | 3.0 | 27.3 | 1.5 | 12.9 | 0.1 | 51200 | 50926 | 50926 | 1.000 | | | | | 0.0494 | | | | | | | | | | | | | | | | | | | | |
| S3_C32_P0 | 23.1 | 0.3 | 36.6 | 3.0 | 27.3 | 1.5 | 12.9 | 0.1 | 51200 | 50926 | 50926 | 1.000 | 1.000 | BCC | 3.4345 | | 0.0494 | BCC_B2 | 100.00 | 23.12 | 36.65 | 27.32 | 12.91 | BCC_B2 | 100.00 | 23.12 | 36.65 | 27.32 | 12.91 | BCC_B2 | 82.12 | 21.80 | 31.22 | 31.70 | 15.10 | 0.13 | 0.05 |
| | | | | | | | | | | | | | | | | | | | | | | | | | | | | | | BCC_B2 | 17.88 | 26.65 | 57.54 | 4.20 | 1.43 | 10.00 | 0.18 |
| S3_C33_valid | 22.9 | 0.4 | 36.2 | 0.9 | 40.9 | 1.6 | 0.0 | 0.0 | 51200 | 50932 | 50932 | 1.000 | | | | | 0.0489 | | | | | | | | | | | | | | | | | | | | |
| S3_C33_P0 | 22.9 | 0.4 | 36.2 | 0.9 | 40.9 | 1.6 | 0.0 | 0.0 | 51200 | 50932 | 50932 | 1.000 | 1.000 | BCC | 3.4452 | | 0.0489 | BCC_B2 | 100.00 | 22.90 | 36.21 | 40.89 | 0.00 | BCC_B2 | 100.00 | 22.90 | 36.21 | 40.89 | 0.00 | BCC_B2 | 86.46 | 21.65 | 32.51 | 45.60 | 0.00 | 0.19 | 0.05 |
| | | | | | | | | | | | | | | | | | | | | | | | | | | | | | | BCC_B2 | 13.54 | 27.94 | 55.19 | 5.59 | 0.00 | 11.13 | 0.16 |
| S3_C40_valid | 22.6 | 0.4 | 48.4 | 6.6 | 0.6 | 0.1 | 28.4 | 0.1 | 51200 | 50779 | 50779 | 1.000 | | | | | 0.0519 | | | | | | | | | | | | | | | | | | | | |
| S3_C40_P0 | 22.6 | 0.4 | 48.4 | 6.6 | 0.6 | 0.1 | 28.4 | 0.1 | 51200 | 50779 | 50779 | 1.000 | 1.000 | BCC | 3.3770 | | 0.0519 | BCC_B2 | 100.00 | 22.63 | 48.38 | 0.62 | 28.37 | BCC_B2 | 100.00 | 22.63 | 48.38 | 0.62 | 28.37 | BCC_B2 | 51.28 | 22.49 | 63.04 | 0.28 | 9.88 | 4.17 | 0.15 |
| | + nanoscale cuboidal phase | | | | | | | | | | | | | BCC | 3.5346 | | 0.0453 | | | | | | | | | | | | | BCC_B2 | 48.72 | 21.72 | 30.72 | 0.96 | 46.54 | 0.04 | 0.02 |
| S3_C41_valid | 23.1 | 0.4 | 49.0 | 4.2 | 14.2 | 0.9 | 13.7 | 0.1 | 51200 | 50517 | 50517 | 1.000 | | | | | 0.0477 | | | | | | | | | | | | | | | | | | | | |
| S3_C41_P0 | 23.1 | 0.4 | 49.0 | 4.2 | 14.2 | 0.9 | 13.7 | 0.1 | 51200 | 50517 | 50517 | 1.000 | 1.000 | BCC | 3.4750 | | 0.0477 | BCC_B2 | 100.00 | 23.05 | 48.98 | 14.22 | 13.75 | BCC_B2 | 100.00 | 23.05 | 48.98 | 14.22 | 13.75 | BCC_B2 | 73.97 | 22.25 | 43.37 | 17.09 | 16.82 | 0.42 | 0.05 |
| | | | | | | | | | | | | | | | | | | | | | | | | | | | | | | BCC_B2 | 26.03 | 23.58 | 61.20 | 4.98 | 3.98 | 6.12 | 0.13 |
| S3_C42_valid | 24.1 | 0.4 | 49.6 | 1.0 | 25.4 | 1.3 | 0.8 | 0.0 | 51200 | 50909 | 50909 | 1.000 | | | | | 0.0471 | | | | | | | | | | | | | | | | | | | | |
| S3_C42_P0 | 24.1 | 0.4 | 49.6 | 1.0 | 25.4 | 1.3 | 0.8 | 0.0 | 51200 | 50909 | 50909 | 1.000 | 1.000 | BCC | 3.4889 | | 0.0471 | BCC_B2 | 100.00 | 24.11 | 49.64 | 25.44 | 0.81 | BCC_B2 | 100.00 | 24.11 | 49.64 | 25.44 | 0.81 | BCC_B2 | 88.43 | 23.46 | 47.58 | 26.97 | 0.87 | 1.06 | 0.06 |
| | | | | | | | | | | | | | | | | | | | | | | | | | | | | | | BCC_B2 | 11.57 | 25.49 | 57.92 | 9.90 | 0.27 | 6.33 | 0.10 |

## Microhardness (HV0.5)

| Compartment | Sample S1 | | Sample S3 | |
|---|---|---|---|---|
| | mean | sigma | mean | sigma |
| **C02** | 312.44 | 72.16 | 113.64 | 19.69 |
| **C03** | 423.64 | 117.51 | 87.54 | 9.65 |
| **C04** | 228.76 | 34.73 | 127.08 | 30.64 |
| **C11** | 401.09 | 100.61 | 375.57 | 204.09 |
| **C12** | 306.70 | 71.74 | 214.85 | 32.89 |
| **C13** | 463.75 | 36.80 | 177.20 | 27.65 |
| **C14** | 357.43 | 64.59 | 216.59 | 25.57 |
| **C20** | 612.26 | 104.78 | 565.07 | 23.44 |
| **C21** | 662.95 | 26.89 | 466.20 | 42.74 |
| **C22** | 554.84 | 44.48 | 340.65 | 24.01 |
| **C23** | 498.81 | 10.49 | 320.05 | 12.83 |
| **C24** | 462.05 | 15.44 | 285.48 | 7.94 |
| **C30** | 663.62 | 28.43 | 551.70 | 18.28 |
| **C31** | 643.10 | 43.55 | 501.00 | 33.32 |
| **C32** | 517.10 | 13.53 | 337.79 | 4.02 |
| **C33** | 471.20 | 10.64 | 304.47 | 4.83 |
| **C40** | 609.71 | 19.55 | 530.50 | 21.53 |
| **C41** | 535.29 | 40.90 | 340.68 | 3.97 |
| **C42** | 474.17 | 9.92 | 303.19 | 3.33 |

# Supplementary material B1: Phase identification: Sample S1 **Al-Ti-Nb-Zr-Ta system**

Supplementary of paper:

## Phase Equilibria of the Al-Ti-Nb-Zr-Ta System

Jiří Kozlík [a*], František Lukáč [b], Mariano Casas-Luna [a], Jozef Veselý [a], Eliška Jača [a], Kateřina Ficková [a], Stanislav Šašek [a], Kristína Bartha [a], Adam Strnad [a], Tomáš Chráska [b], Josef Stráský [a]

[a] *Charles University, Faculty of Mathematics and Physics, Department of Physics of Materials, Ke Karlovu 5, Prague 121 16, Czechia*

[b] *Institute of Plasma Physics of the Czech Academy of Sciences, U Slovanky 2525/1a, Prague 182 00, Czechia*

* Corresponding author: jiri.kozlik@matfyz.cuni.cz

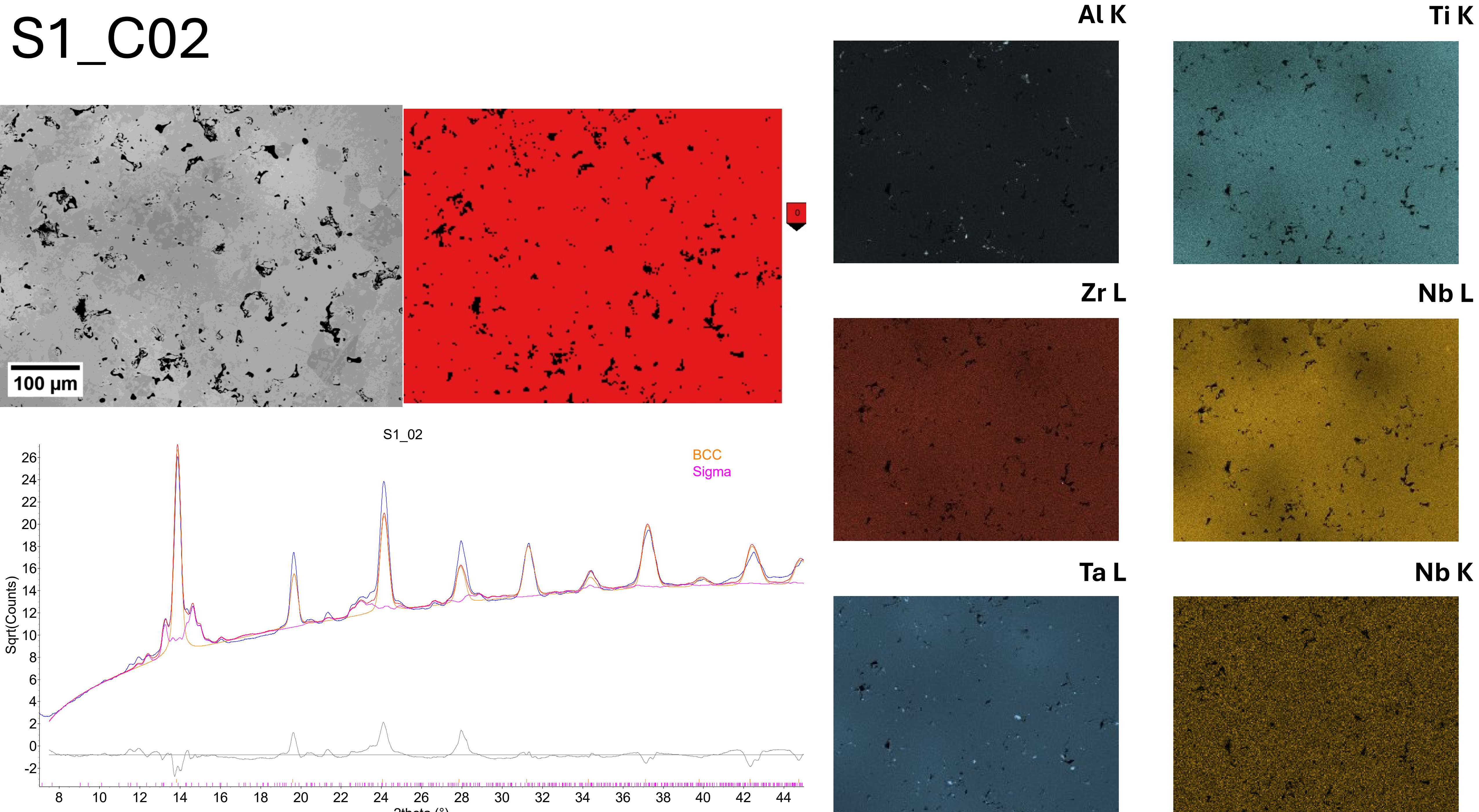

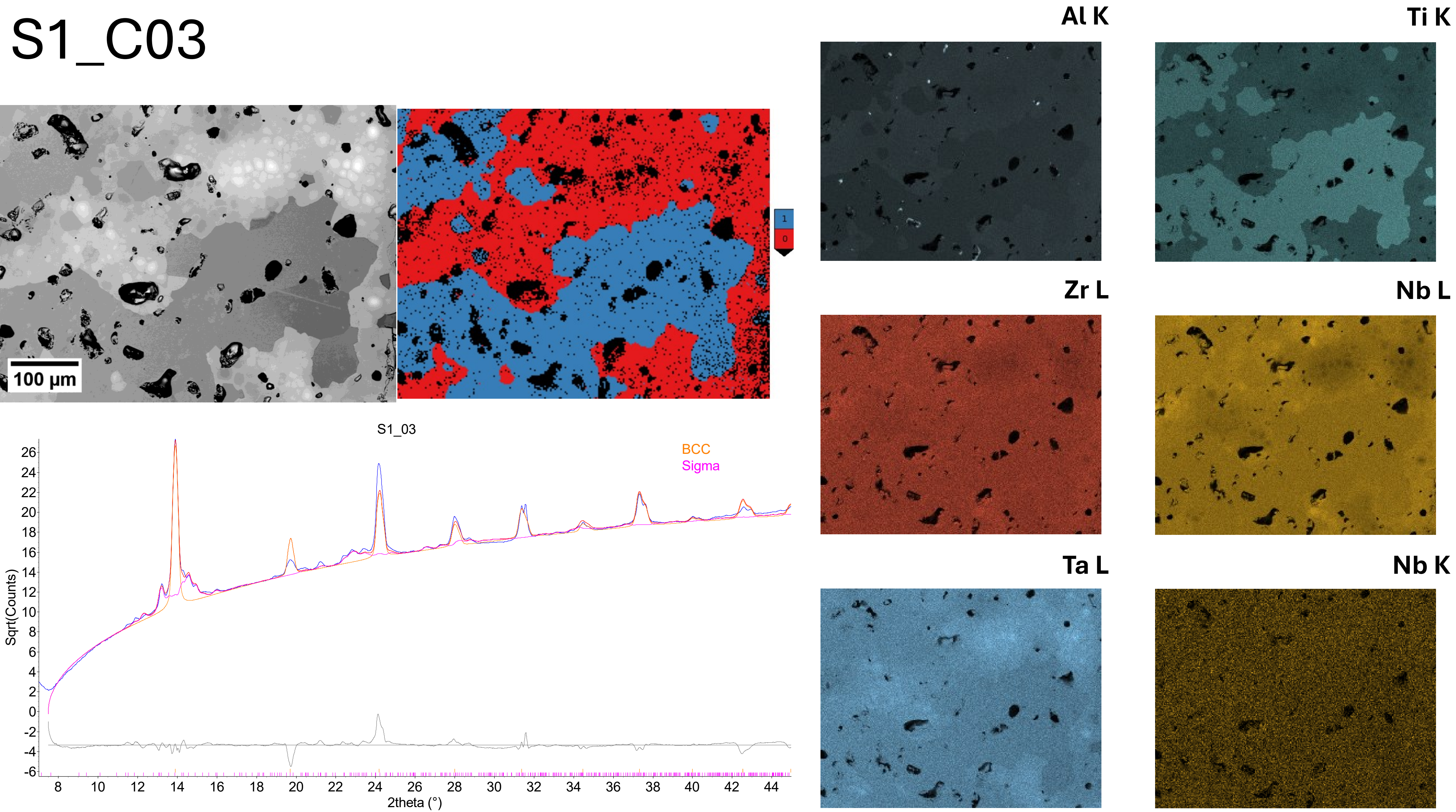

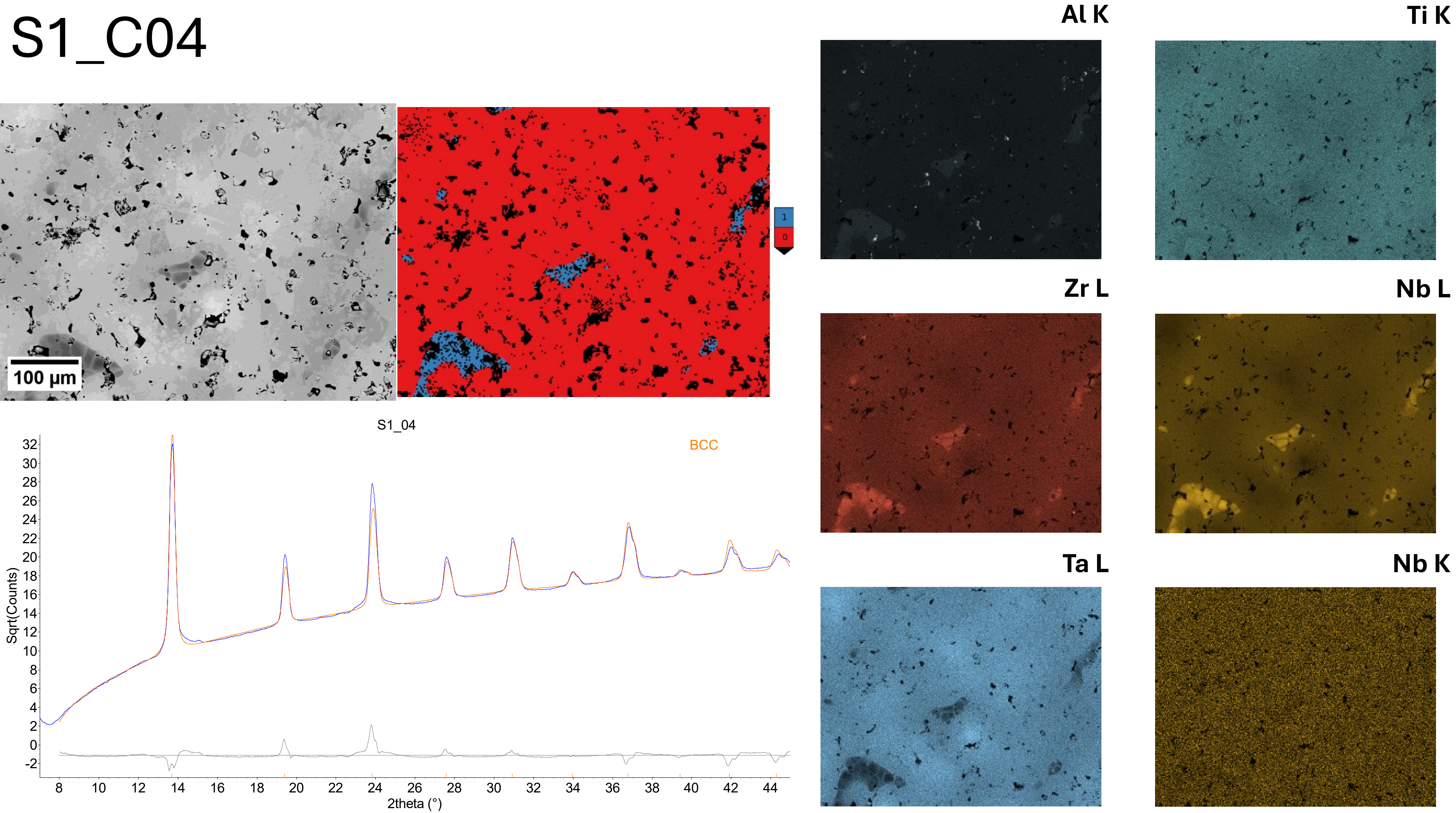

# S1_C11

80 µm

1
0

S1_11

BCC
Al3Zr5

Sqrt(Counts)
2theta (°)

Al K
Ti K
Zr L
Nb L
Ta L
Nb K

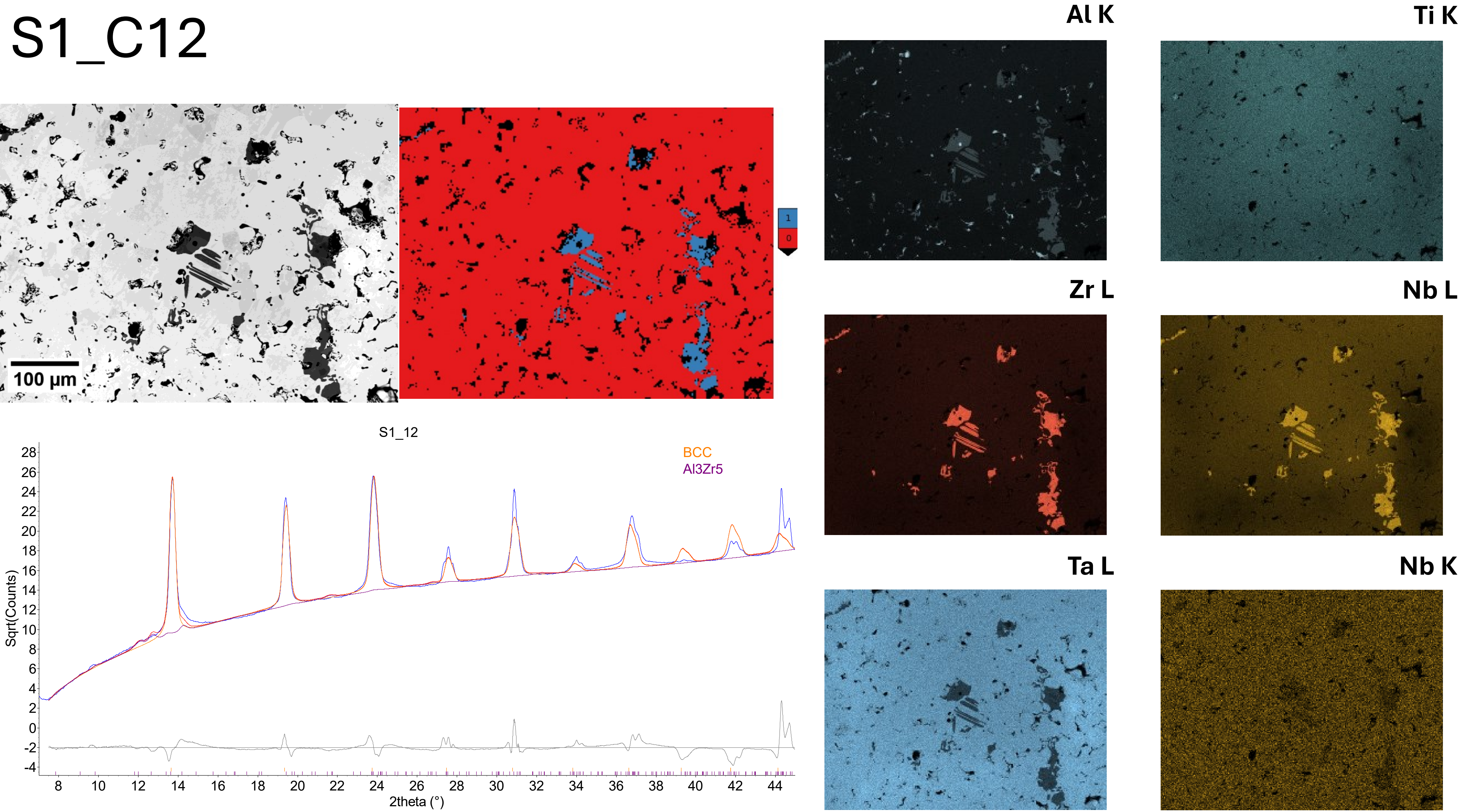

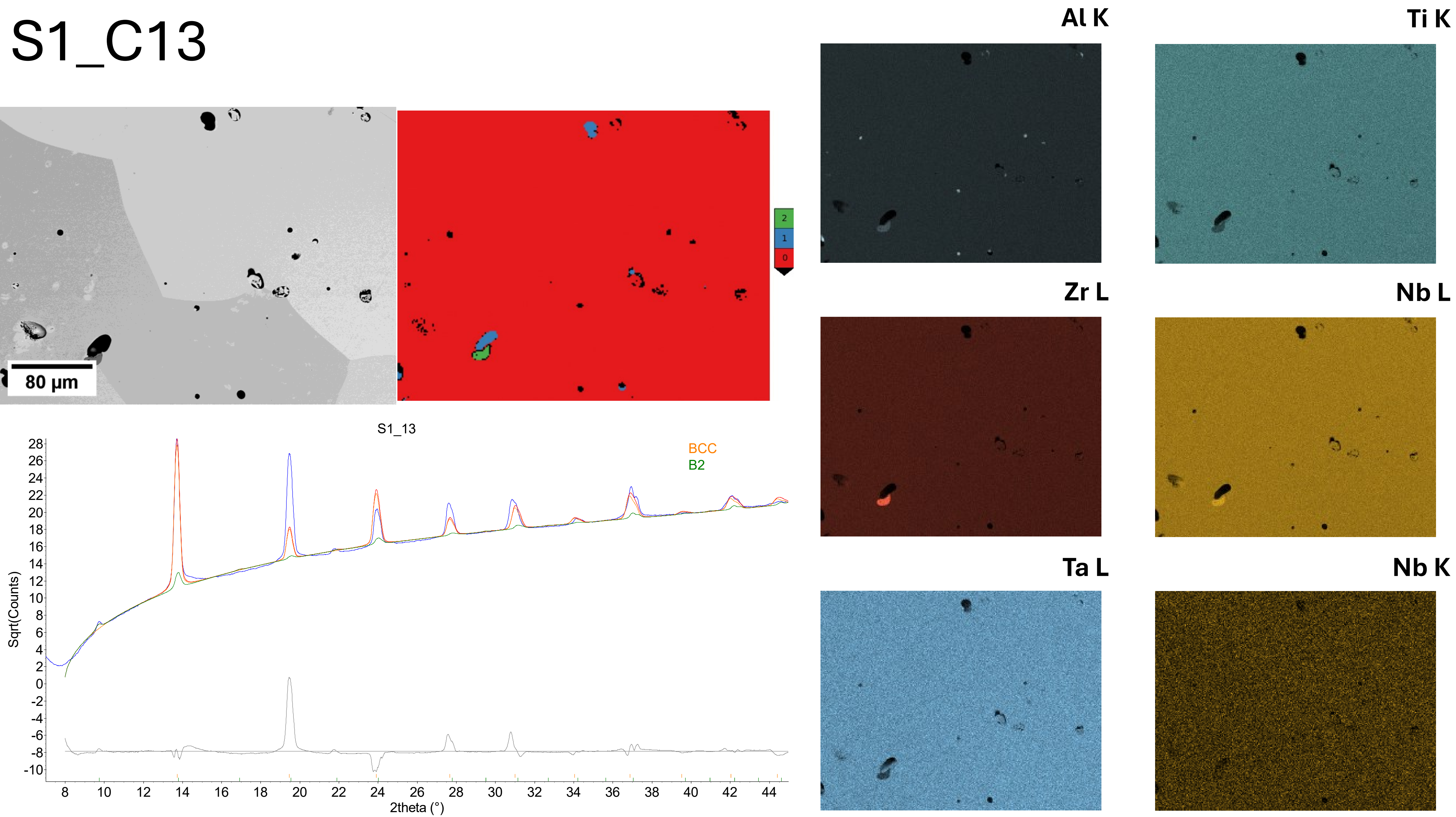

# S1_C14

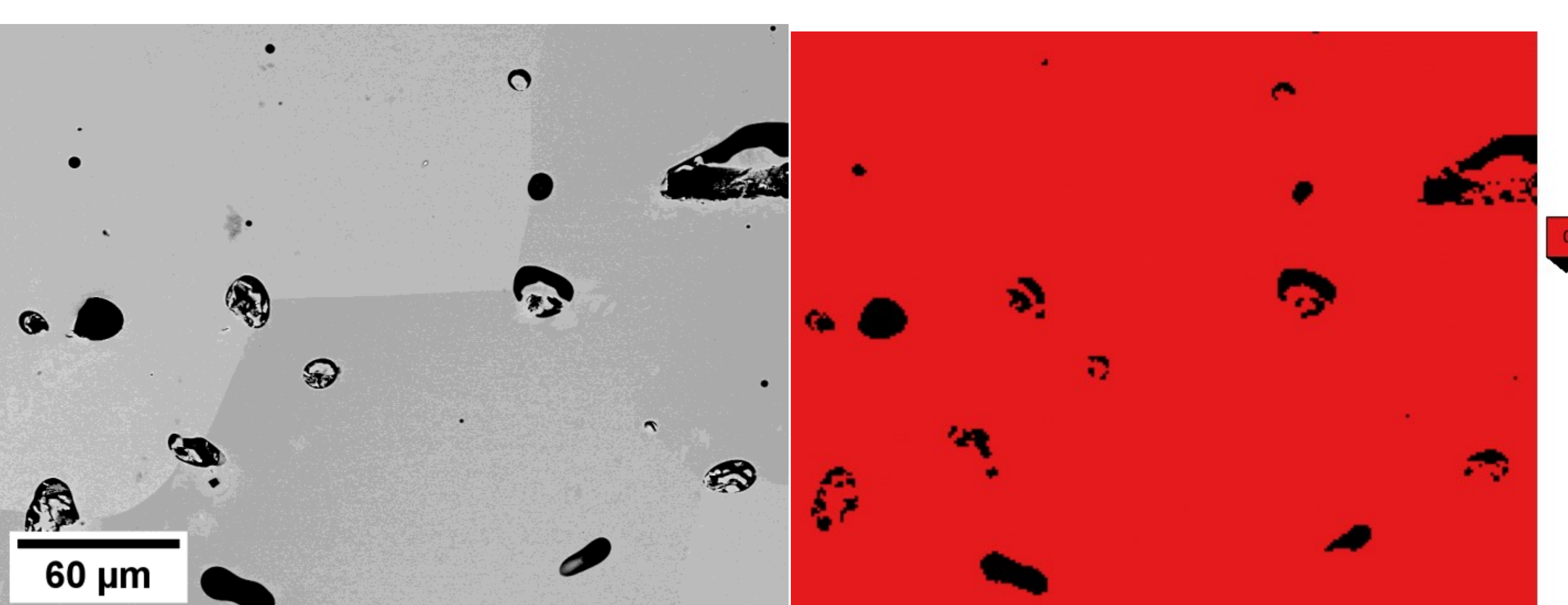

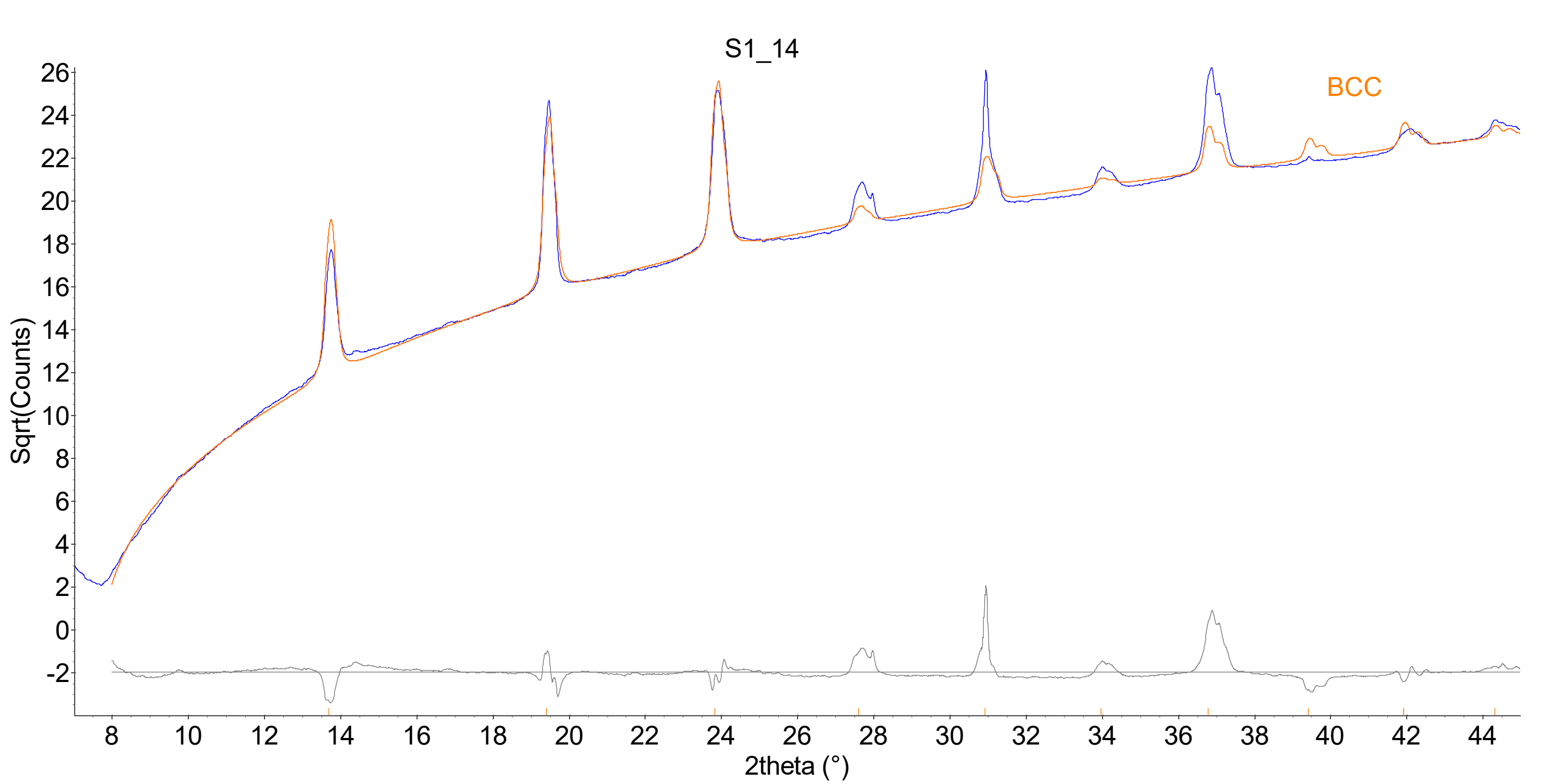

Al K

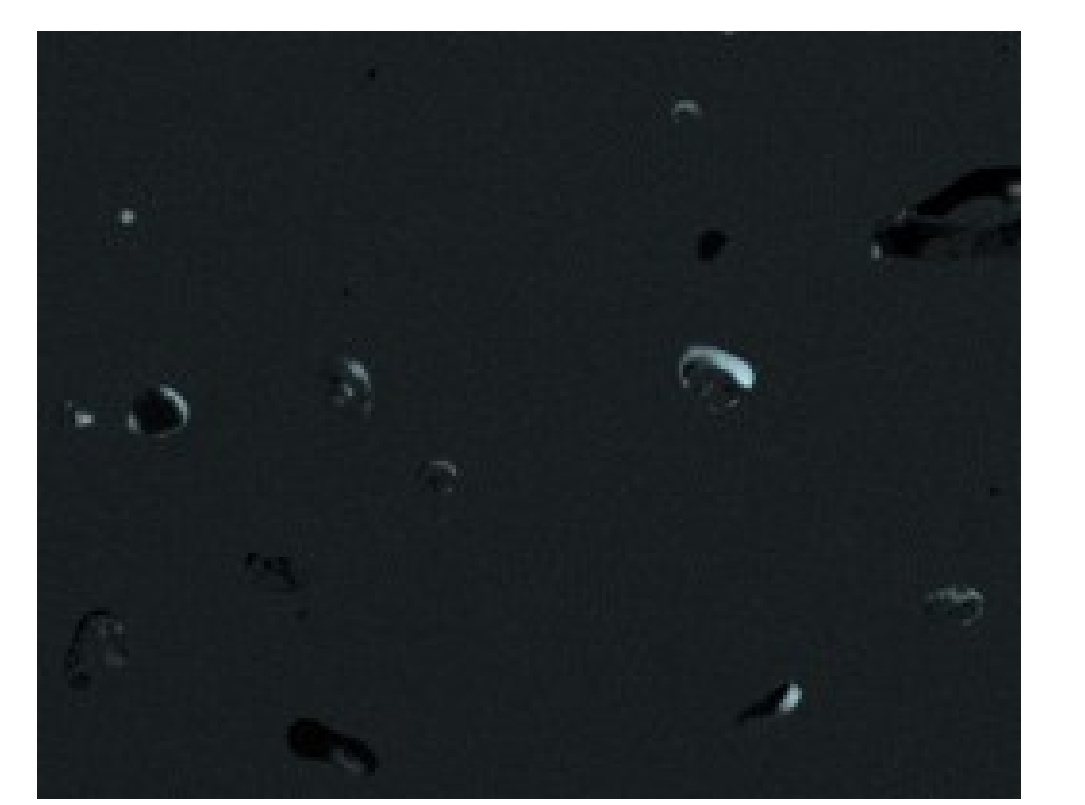

Ti K

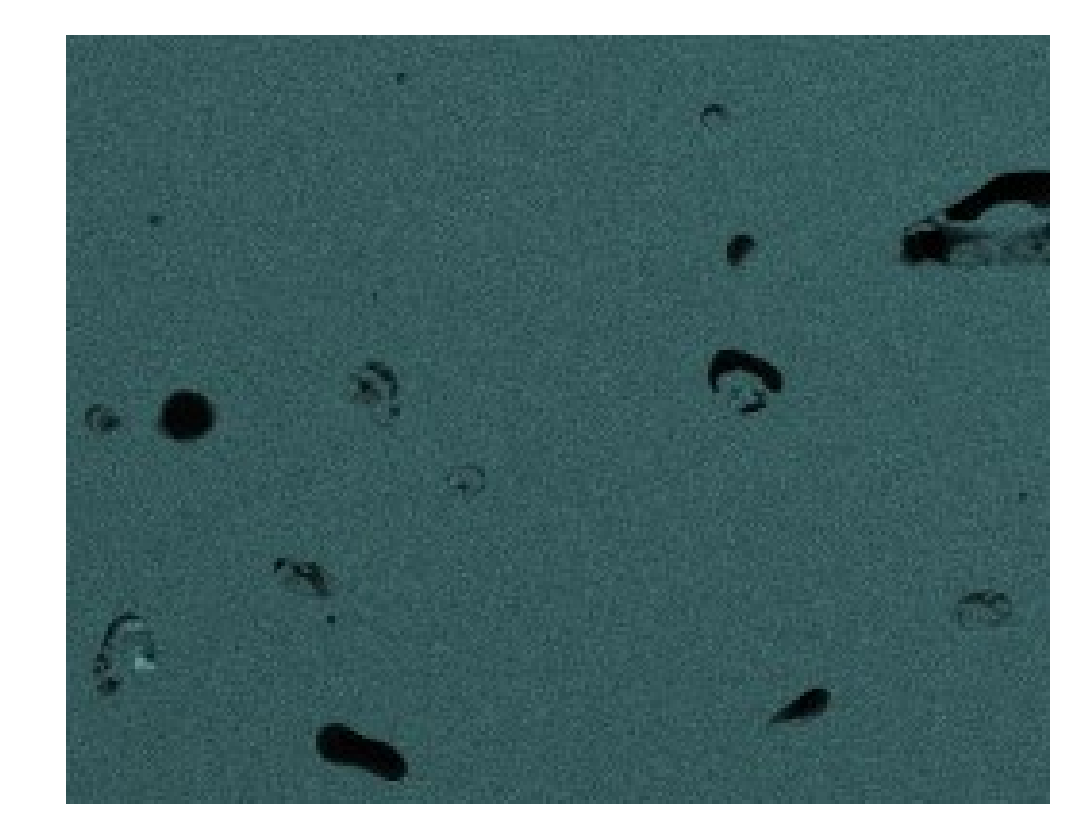

Zr L

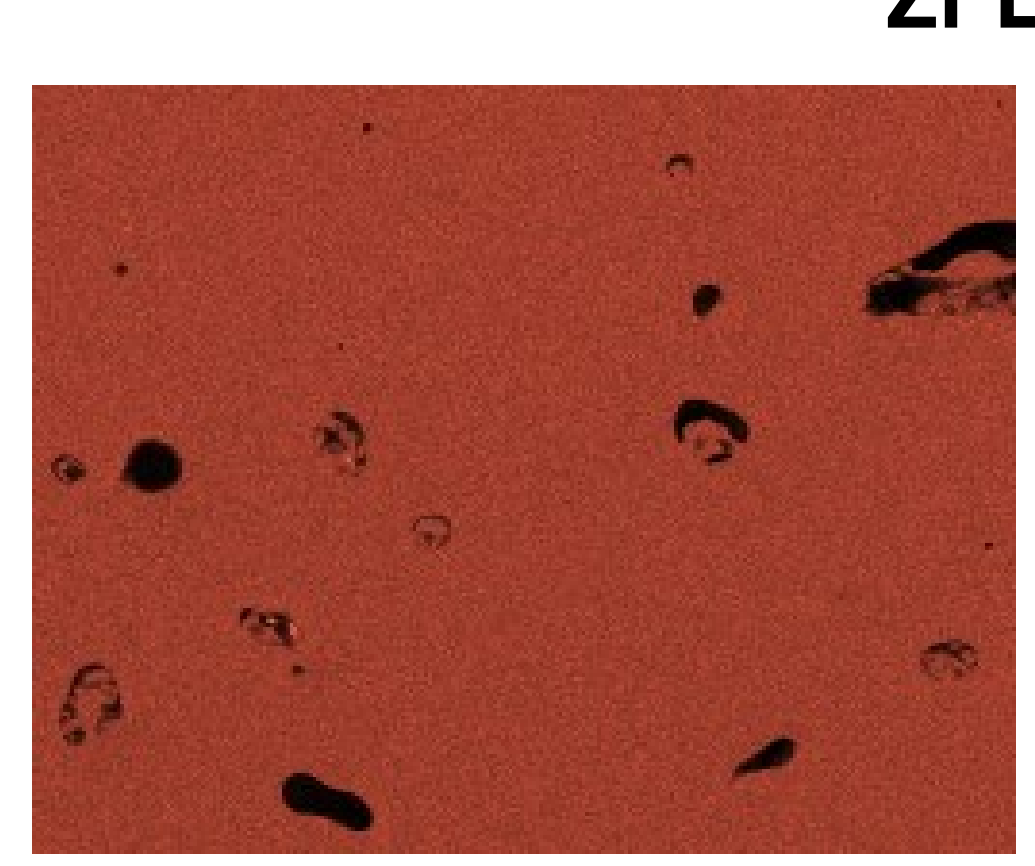

Nb L

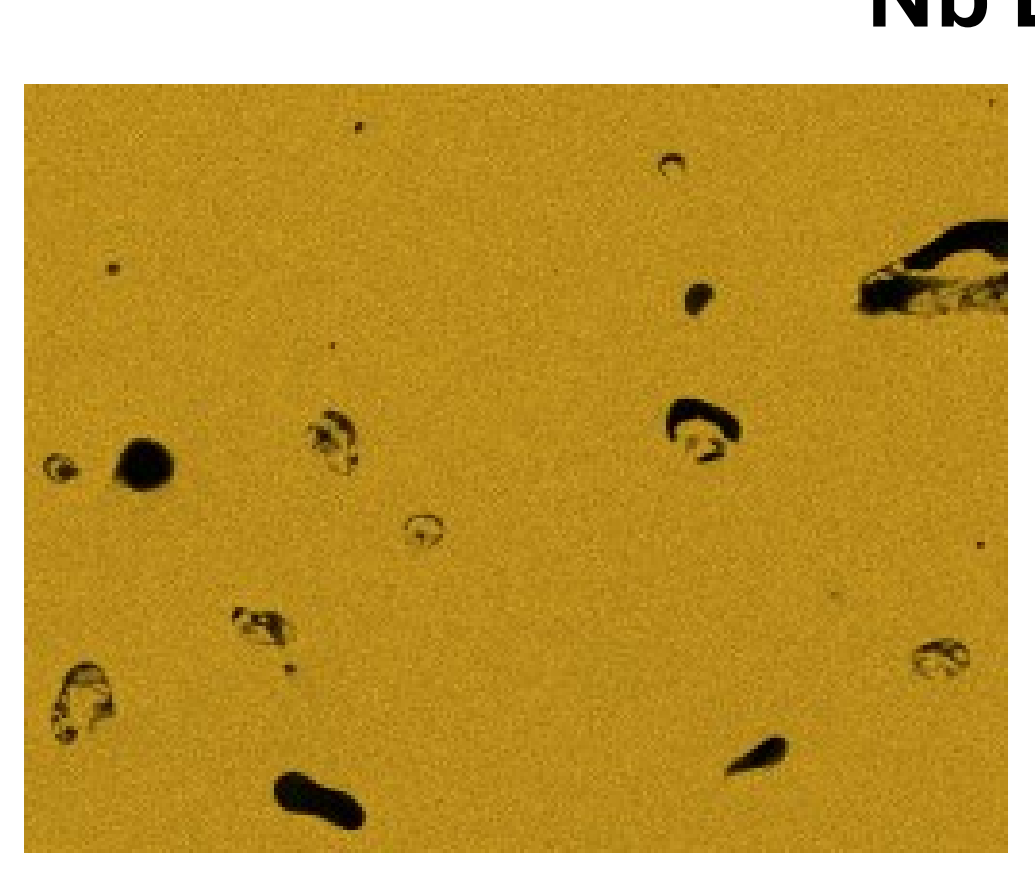

Ta L

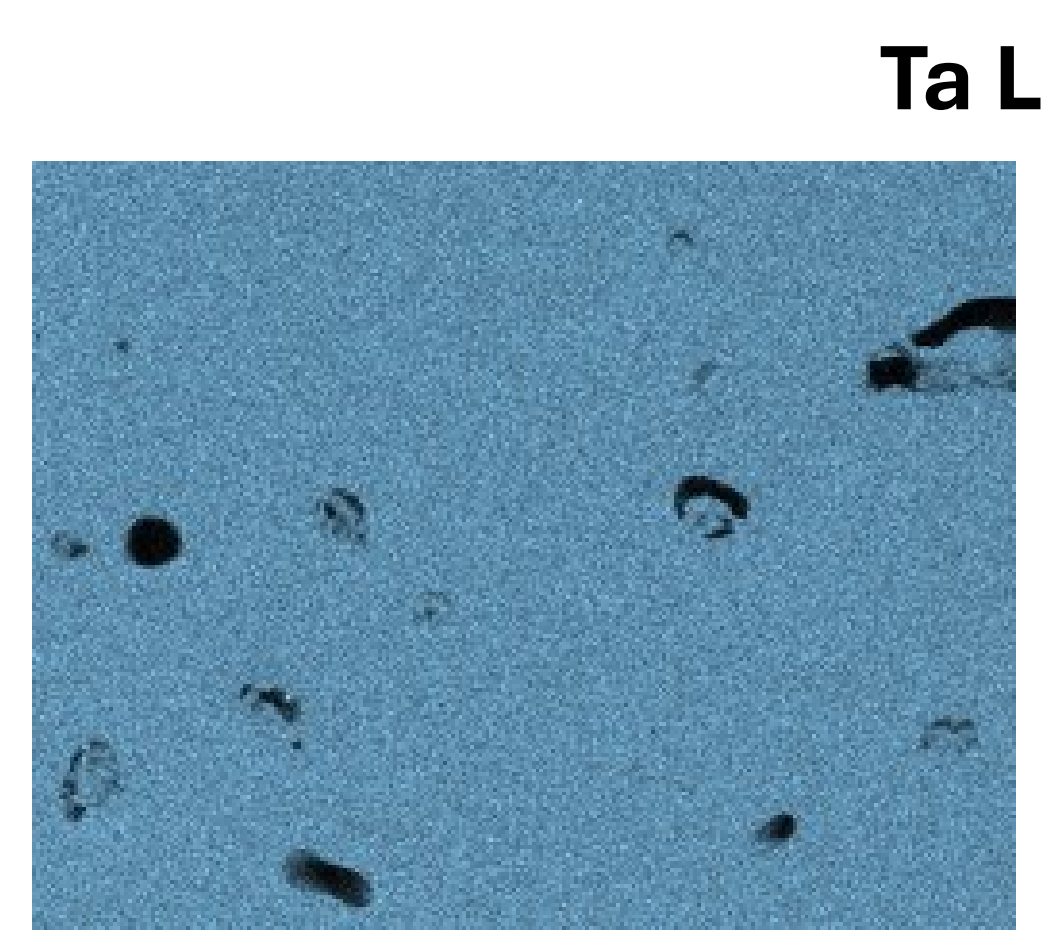

Nb K

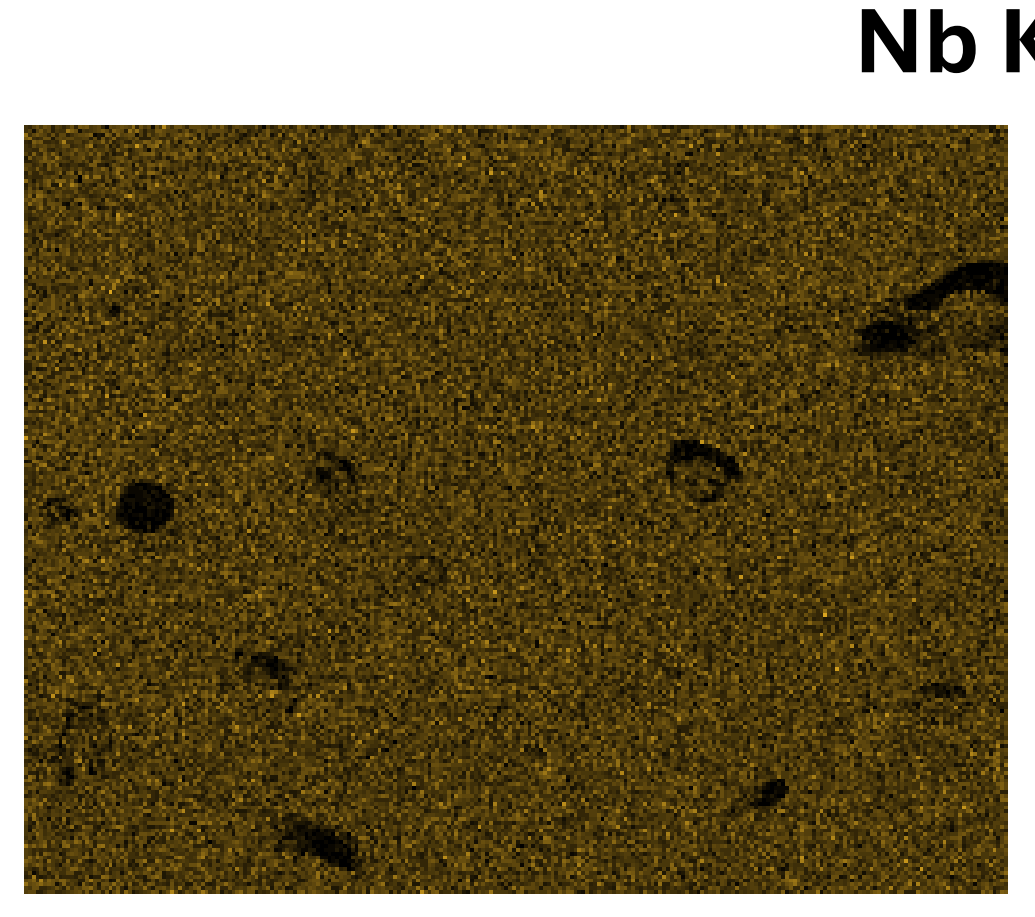

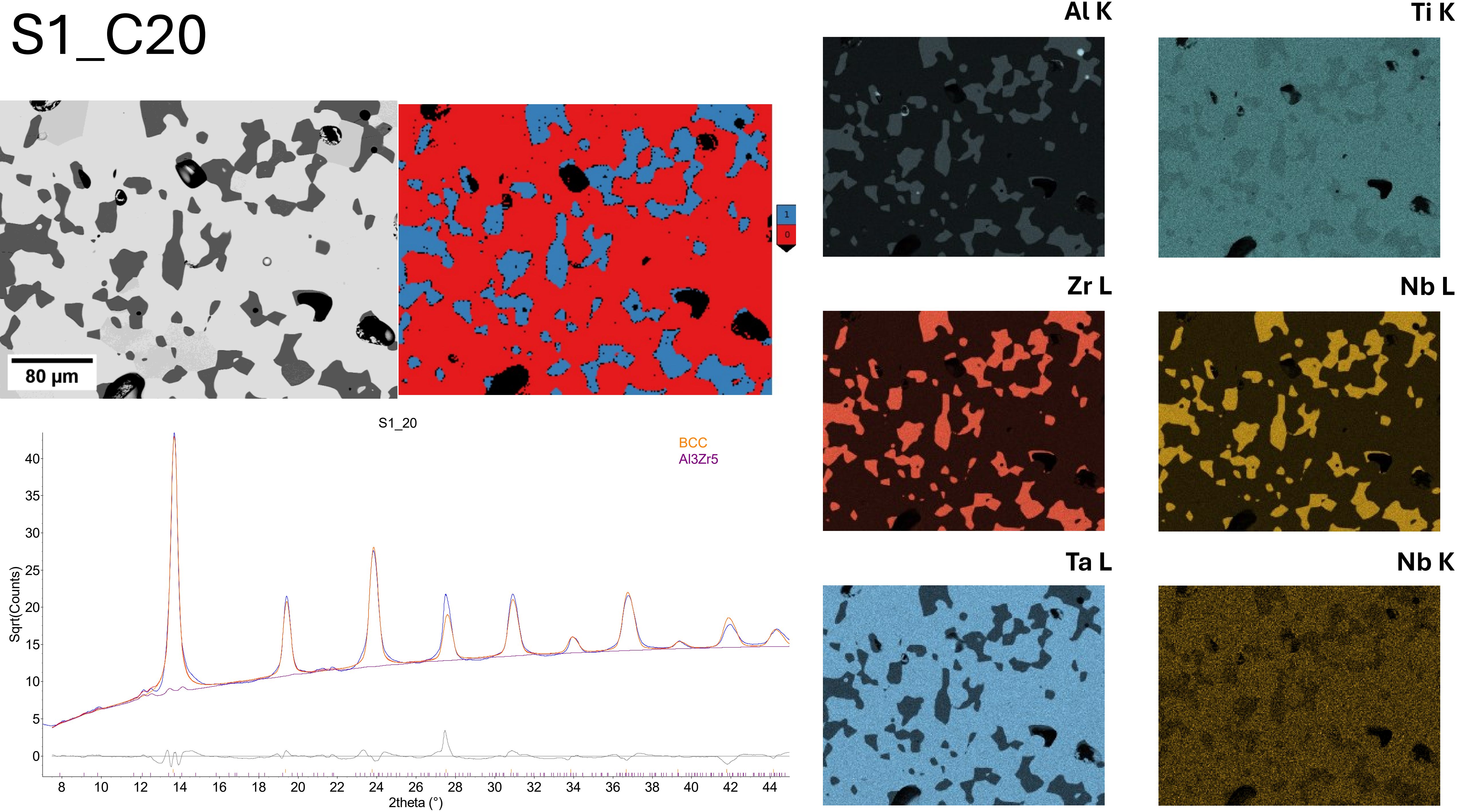

# S1_C21

100 µm

1
0

Al K
Ti K
Zr L
Nb L
Ta L
Nb K

S1_21
BCC
Al3Zr5
Sqrt(Counts)
2theta (°)

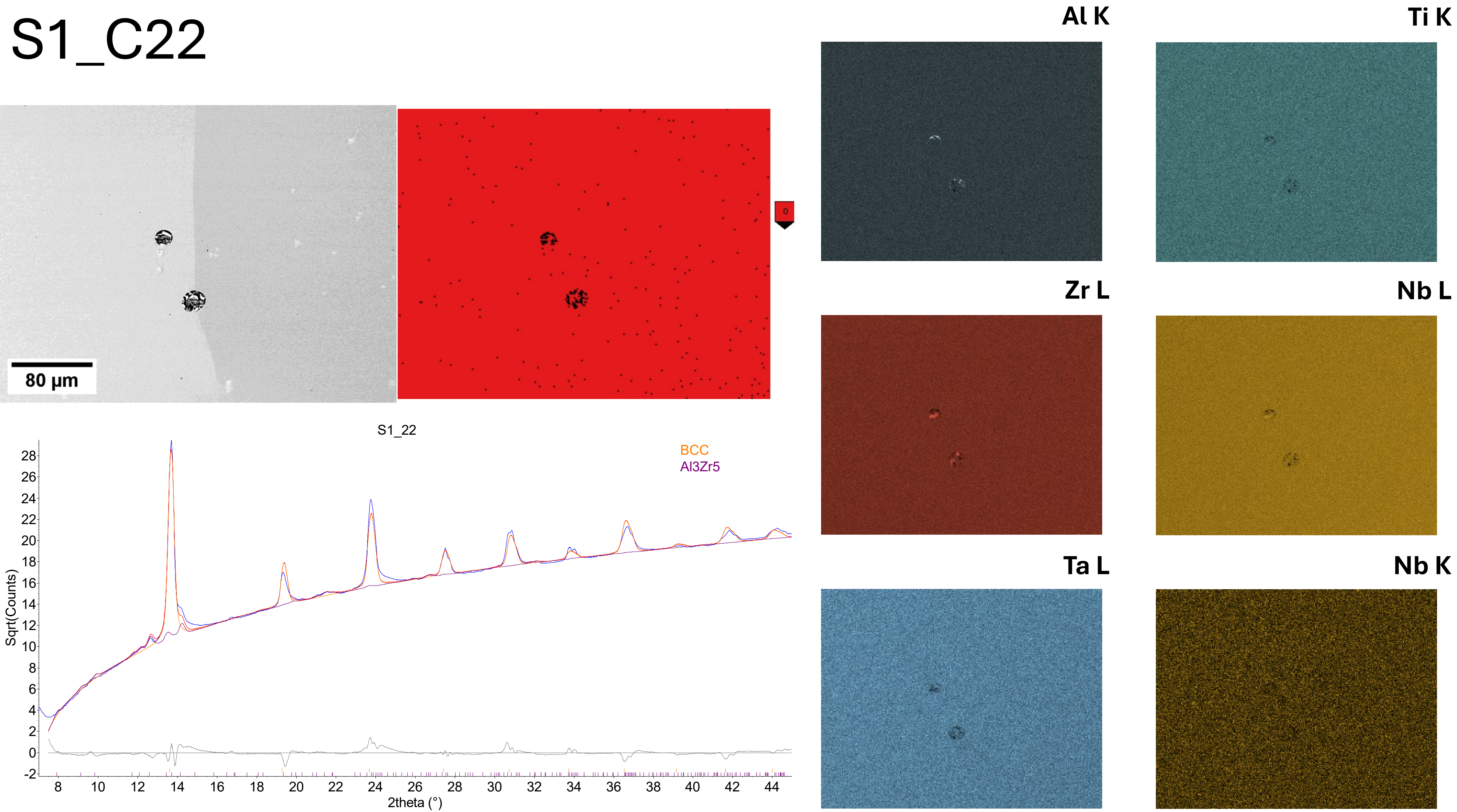

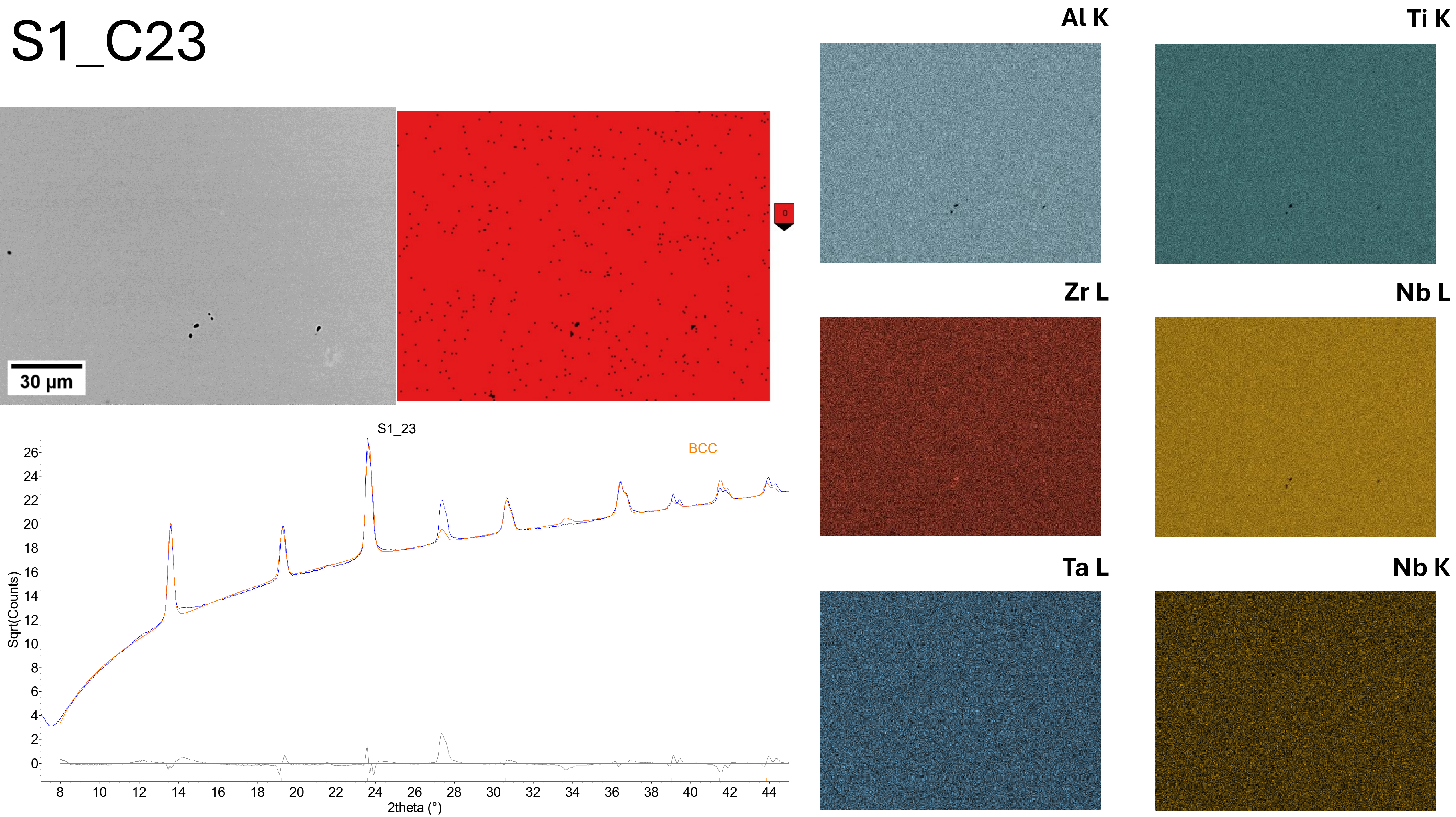

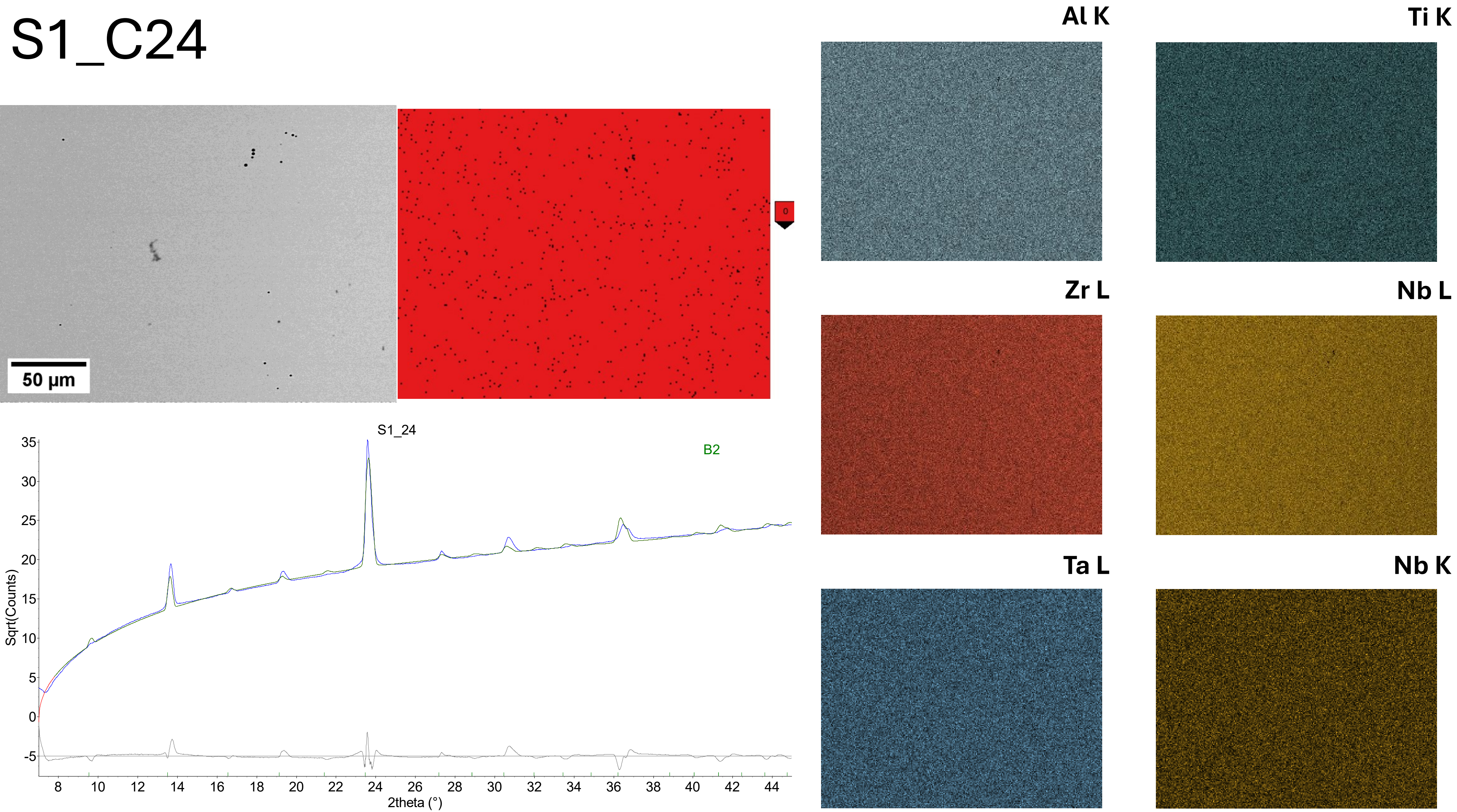

# S1_C30

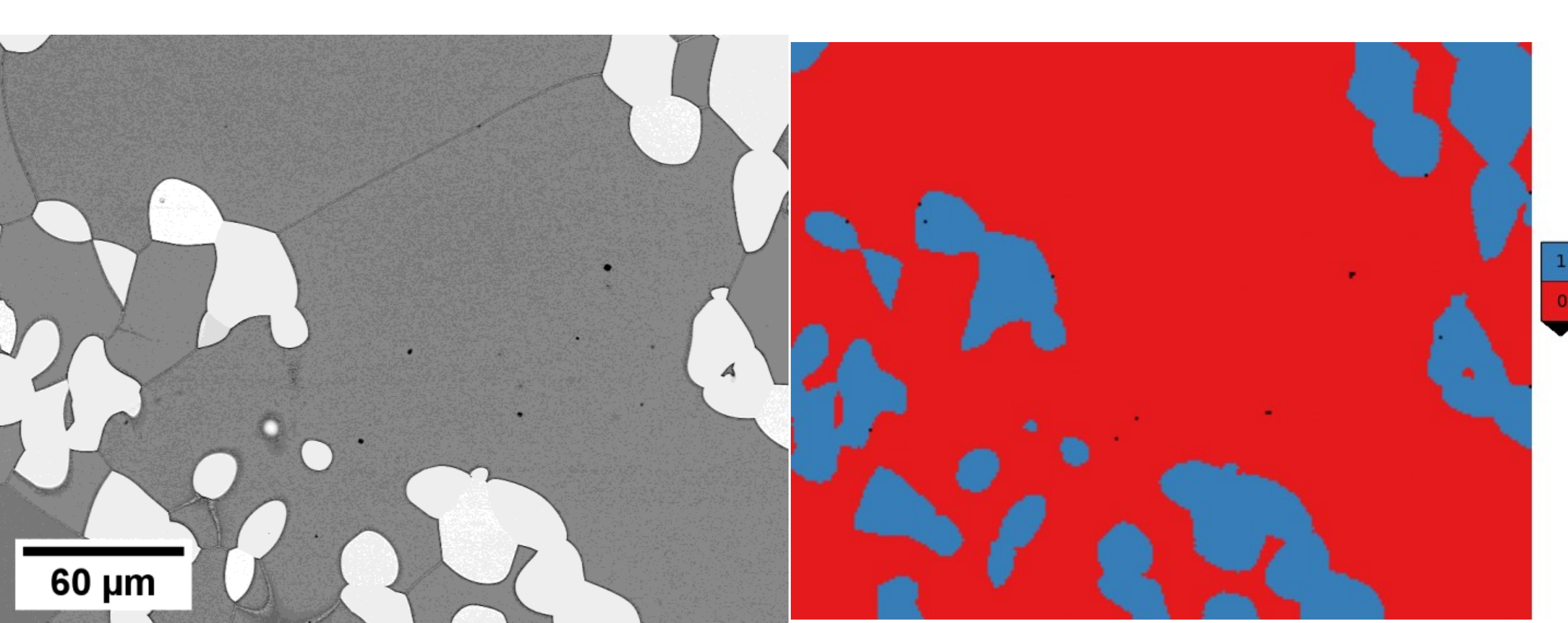

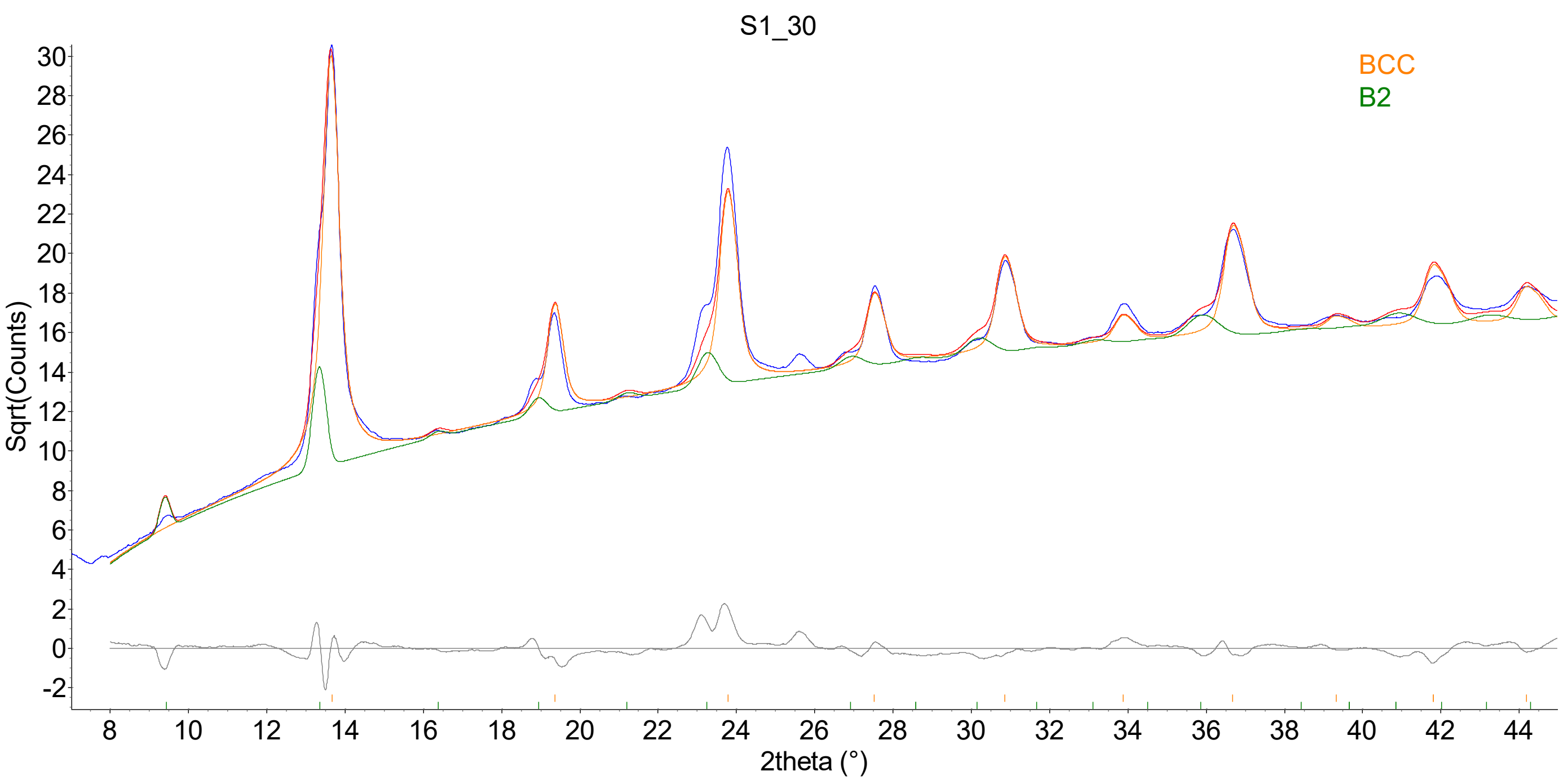

Al K

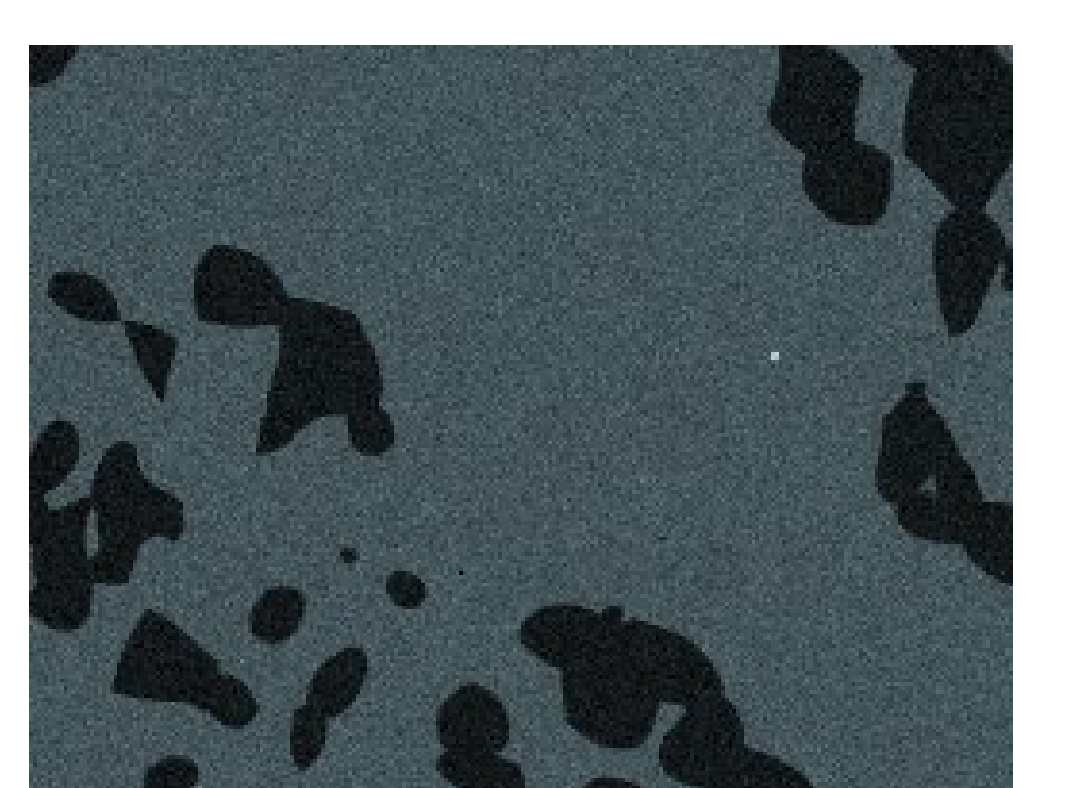

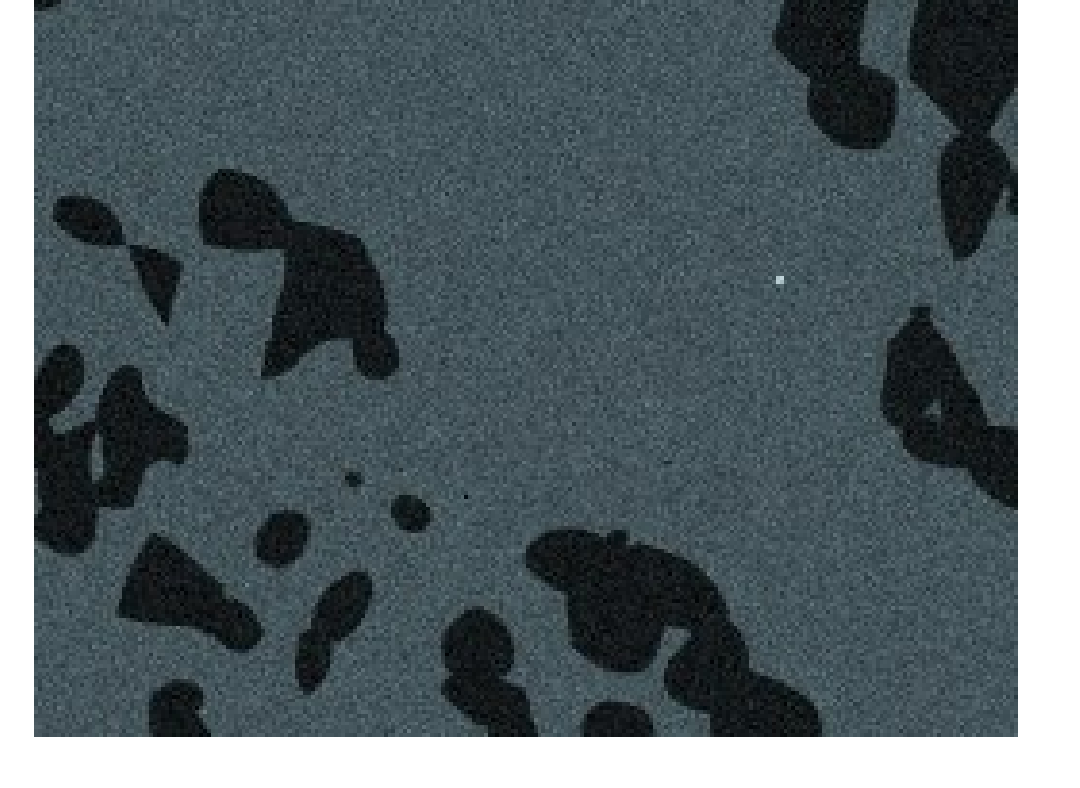

Ti K

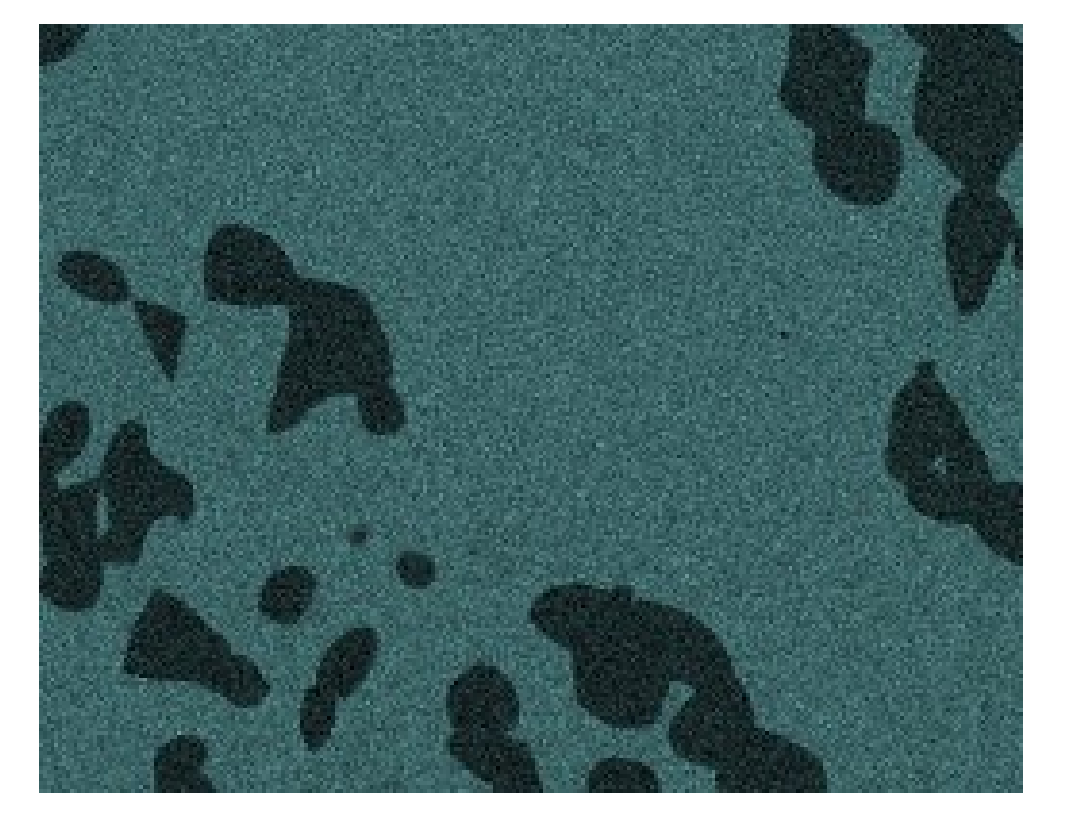

Zr L

Nb L

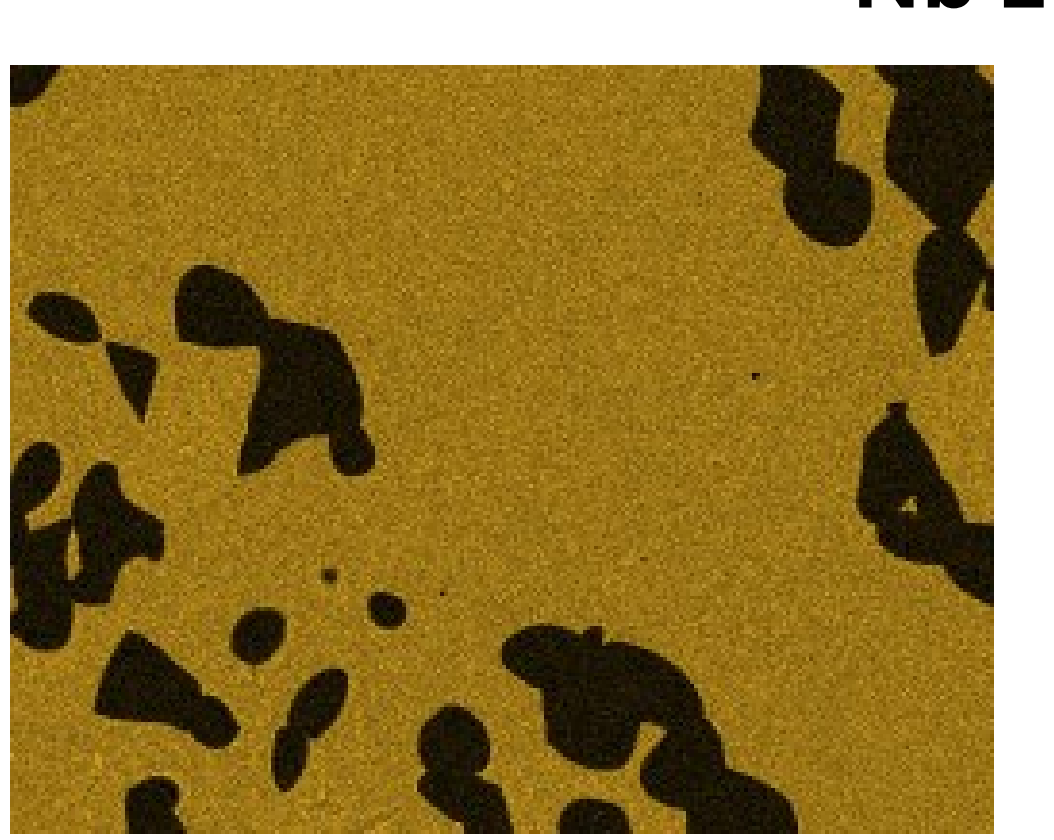

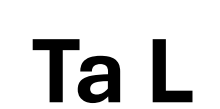

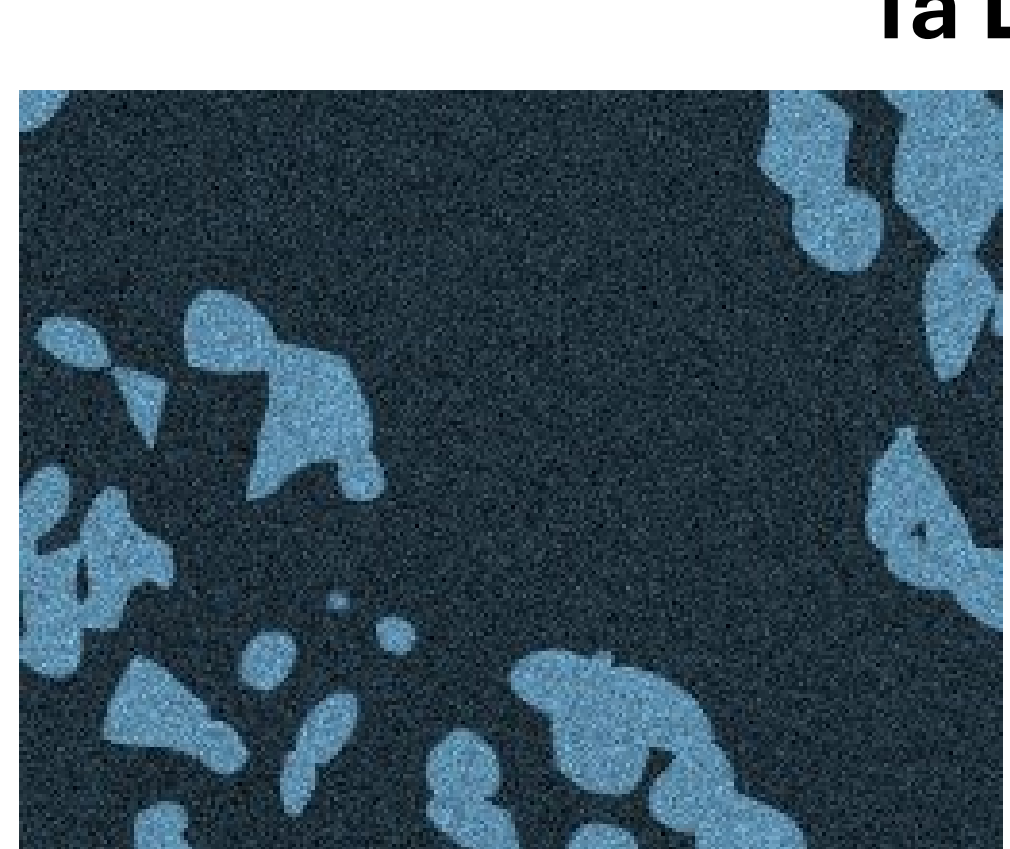

Nb K

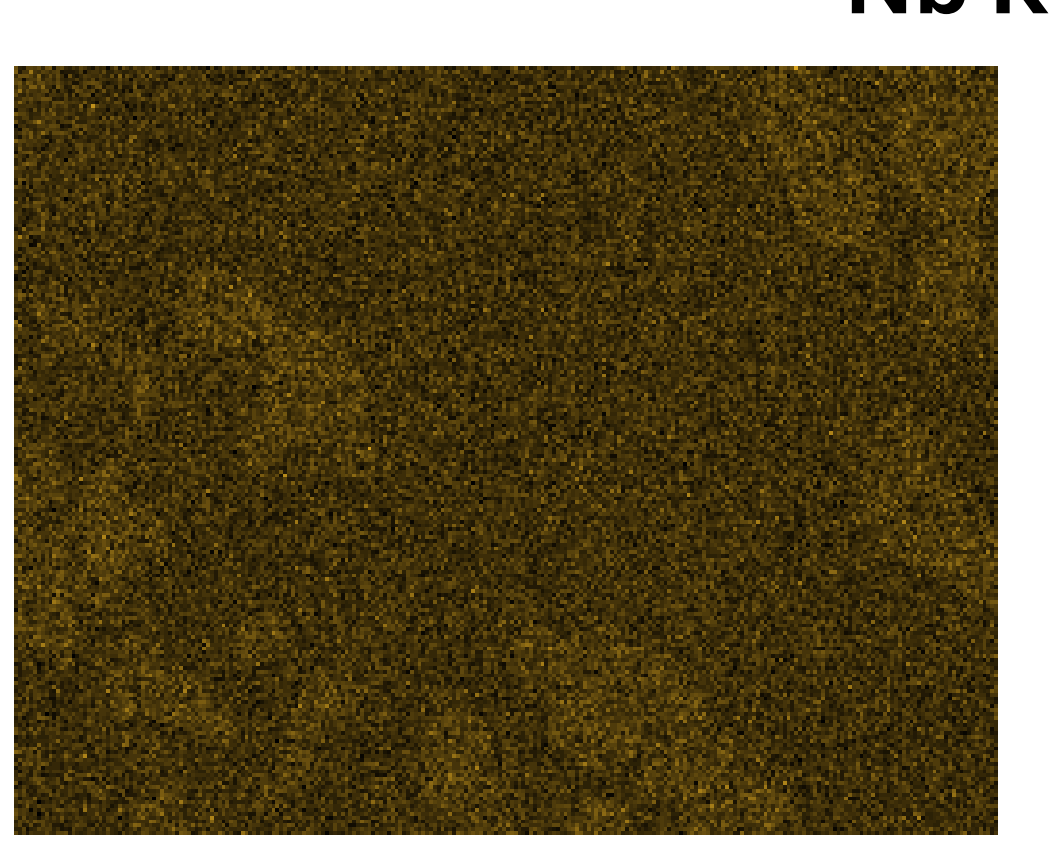

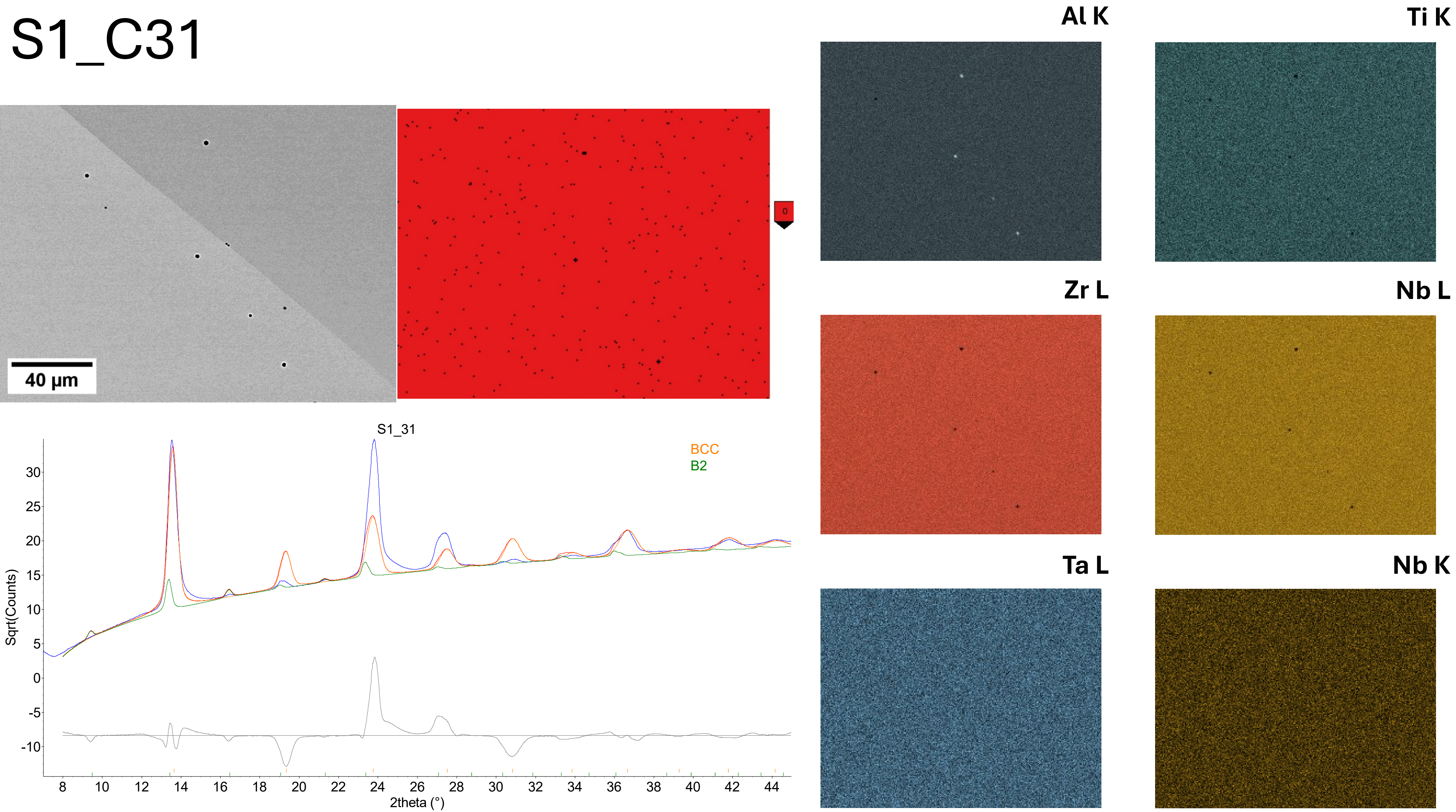

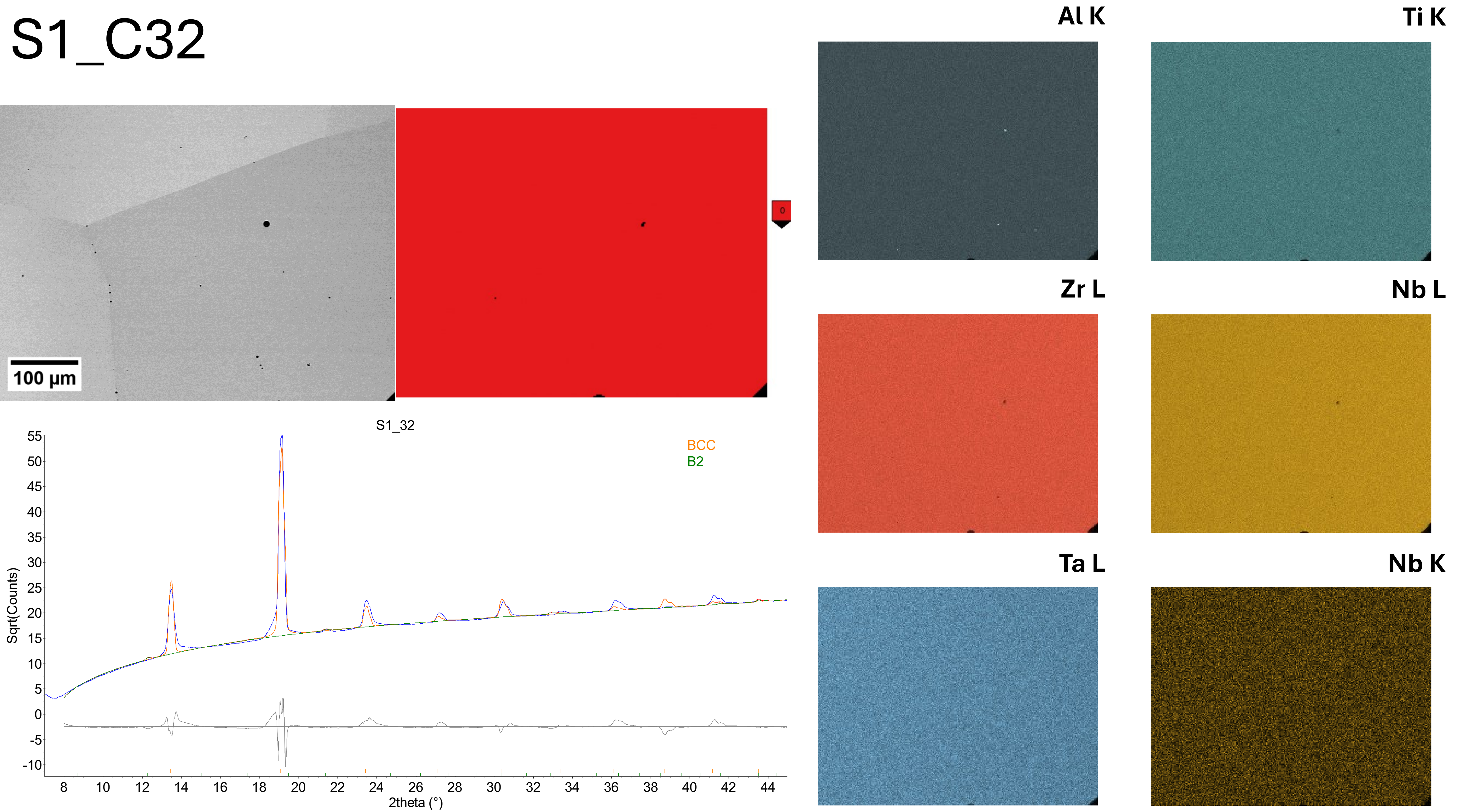

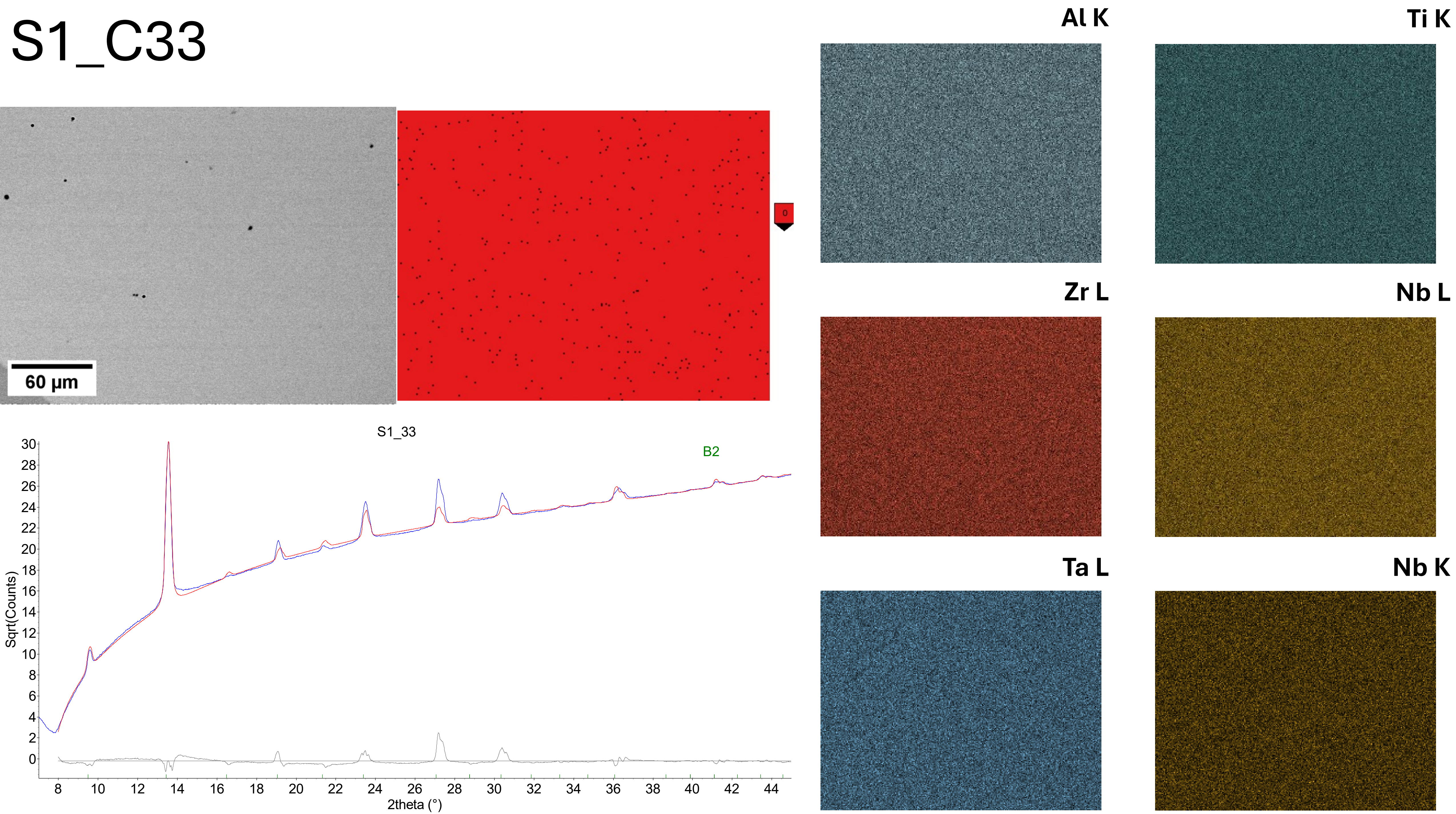

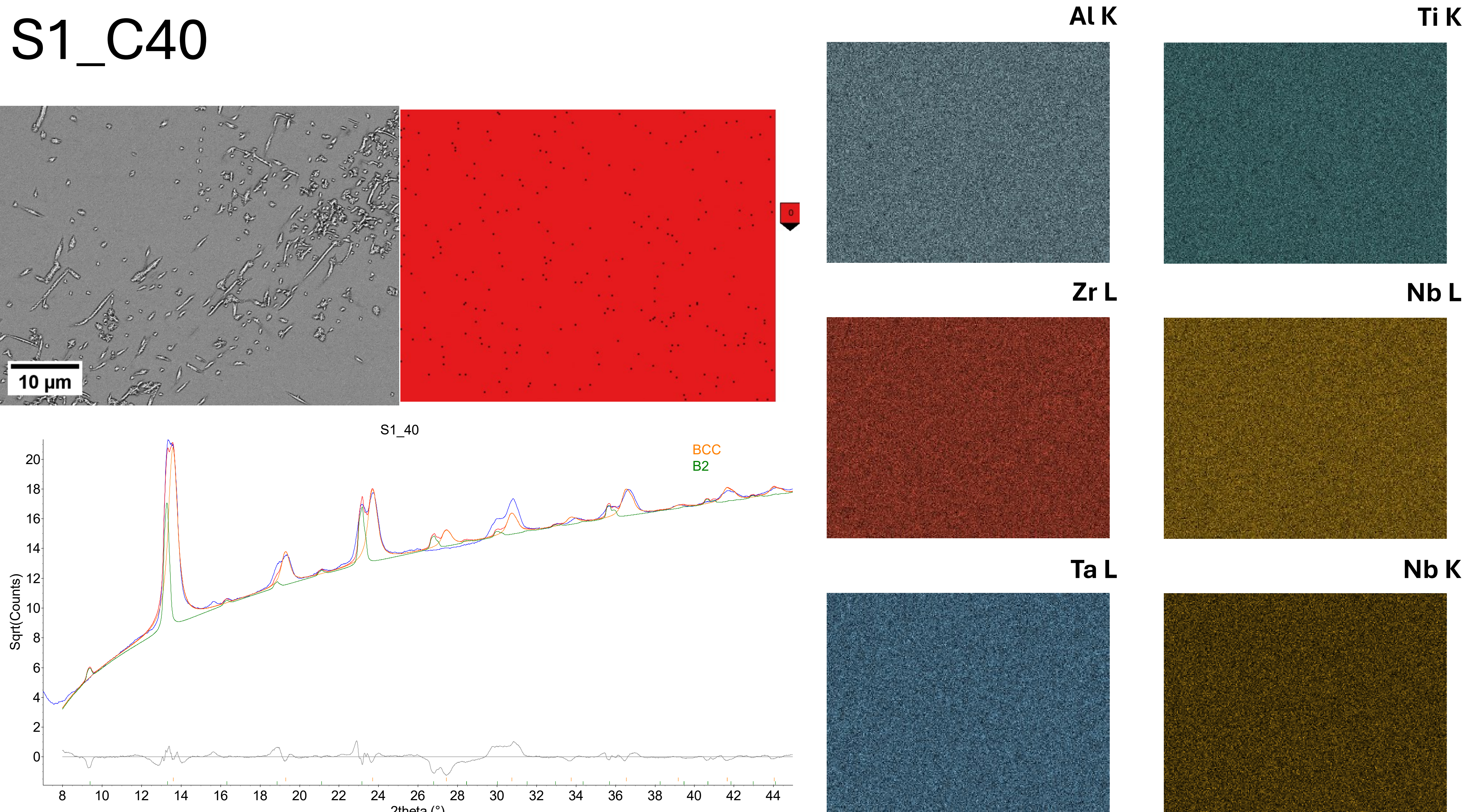

# S1_C41

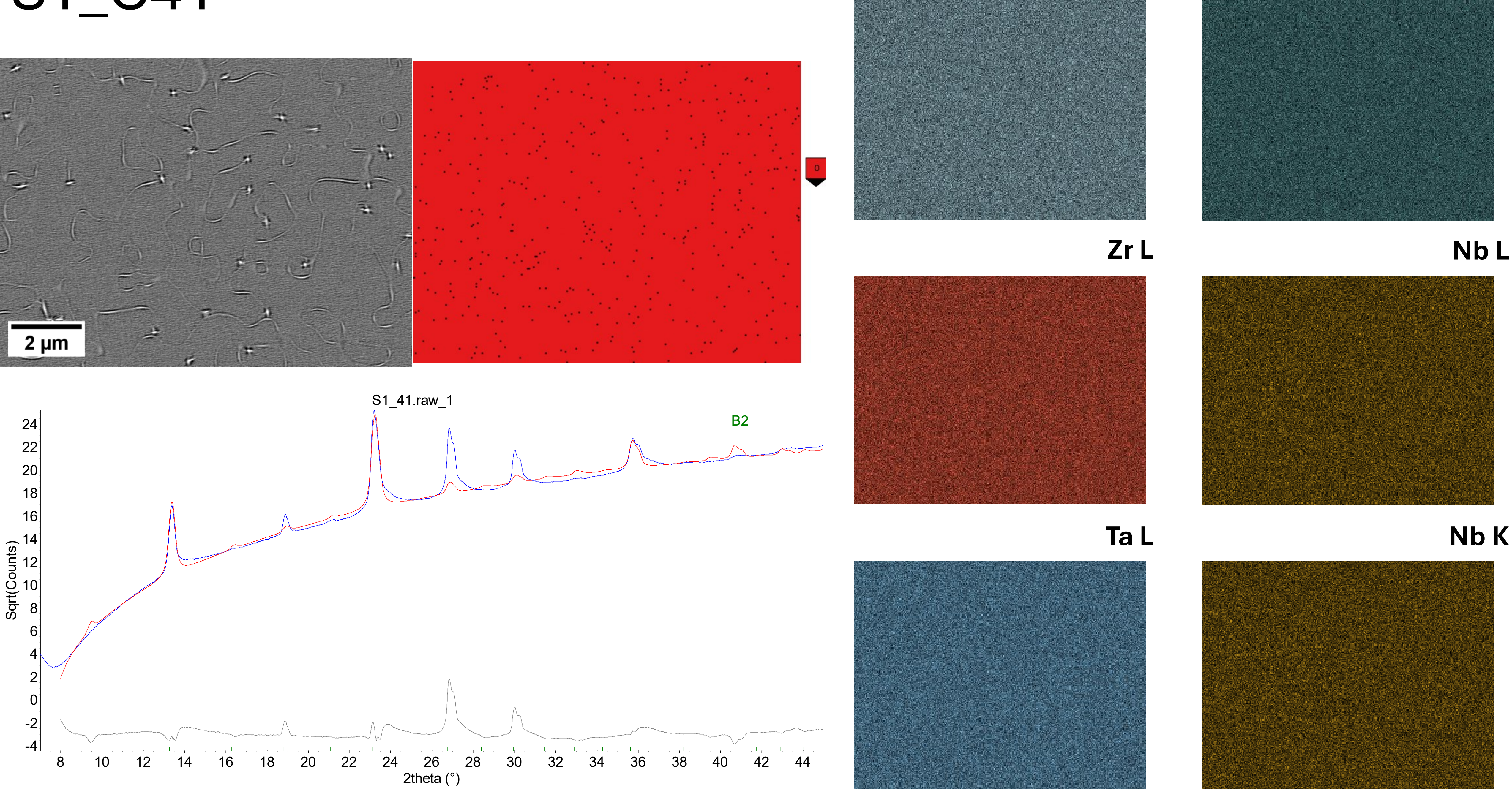

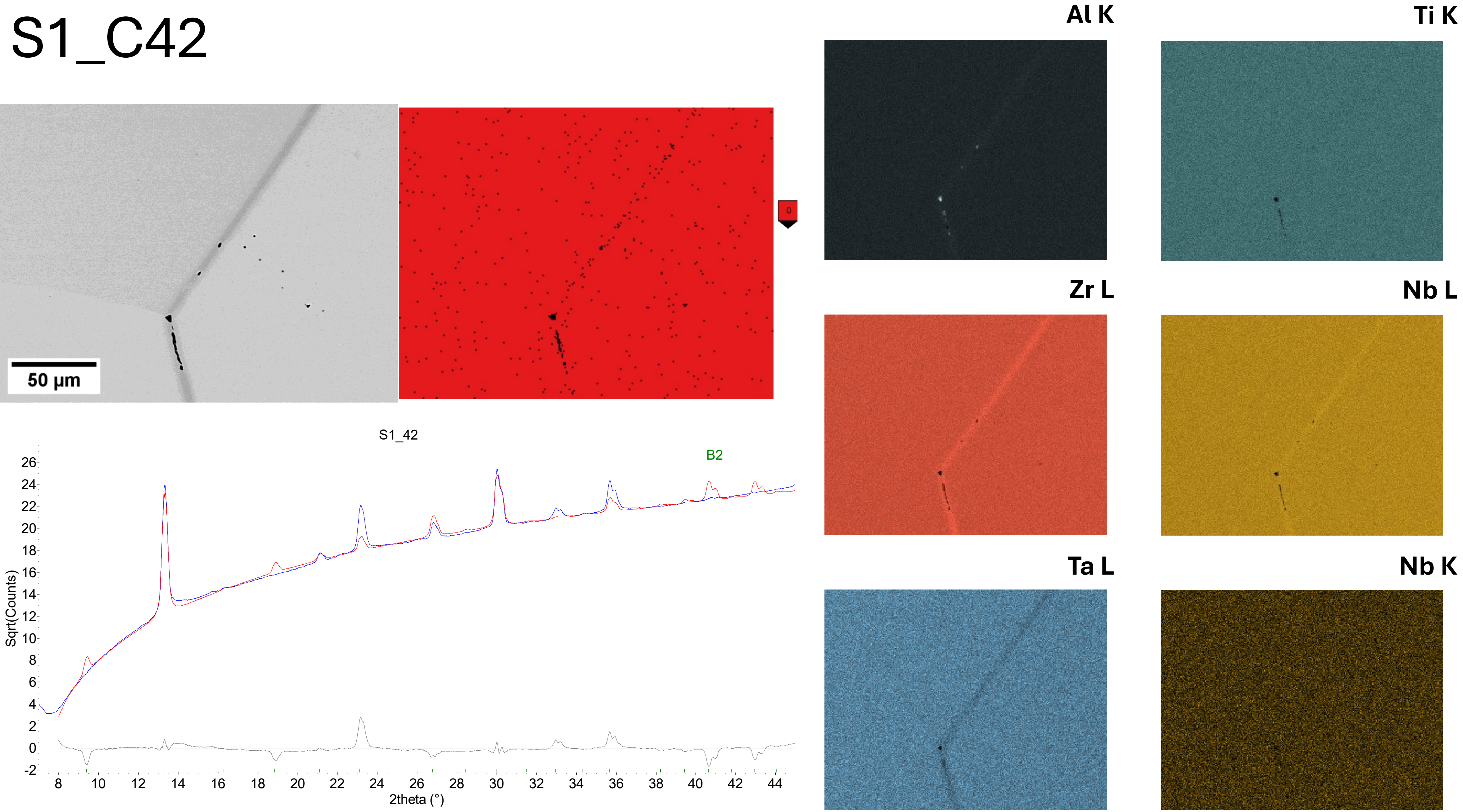

# Supplementary material B2: Phase identification: Sample S2

# **Al-Nb-Zr-Ta system**

Supplementary of paper:

## **Phase Equilibria of the Al-Ti-Nb-Zr-Ta System**

Jiří Kozlík [a*], František Lukáč [b], Mariano Casas-Luna [a], Jozef Veselý [a], Eliška Jača [a], Kateřina Ficková [a] , Stanislav Šašek [a], Kristína Bartha [a], Adam Strnad [a], Tomáš Chráska [b], Josef Stráský [a]

[a] *Charles University, Faculty of Mathematics and Physics, Department of Physics of Materials, Ke Karlovu 5, Prague 121 16, Czechia*

[b] *Institute of Plasma Physics of the Czech Academy of Sciences, U Slovanky 2525/1a, Prague 182 00, Czechia*

* Corresponding author: jiri.kozlik@matfyz.cuni.cz

# S2_C21

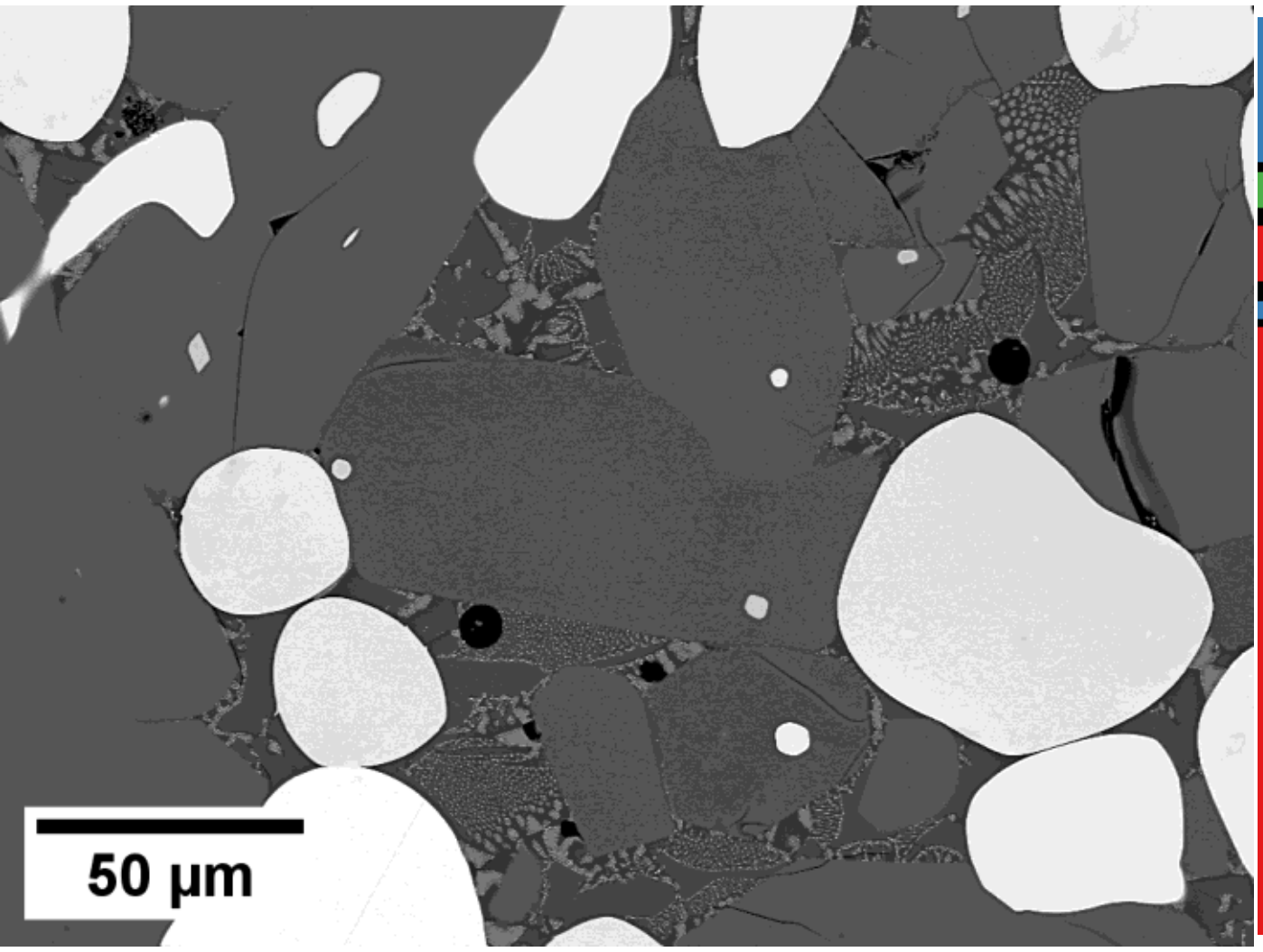

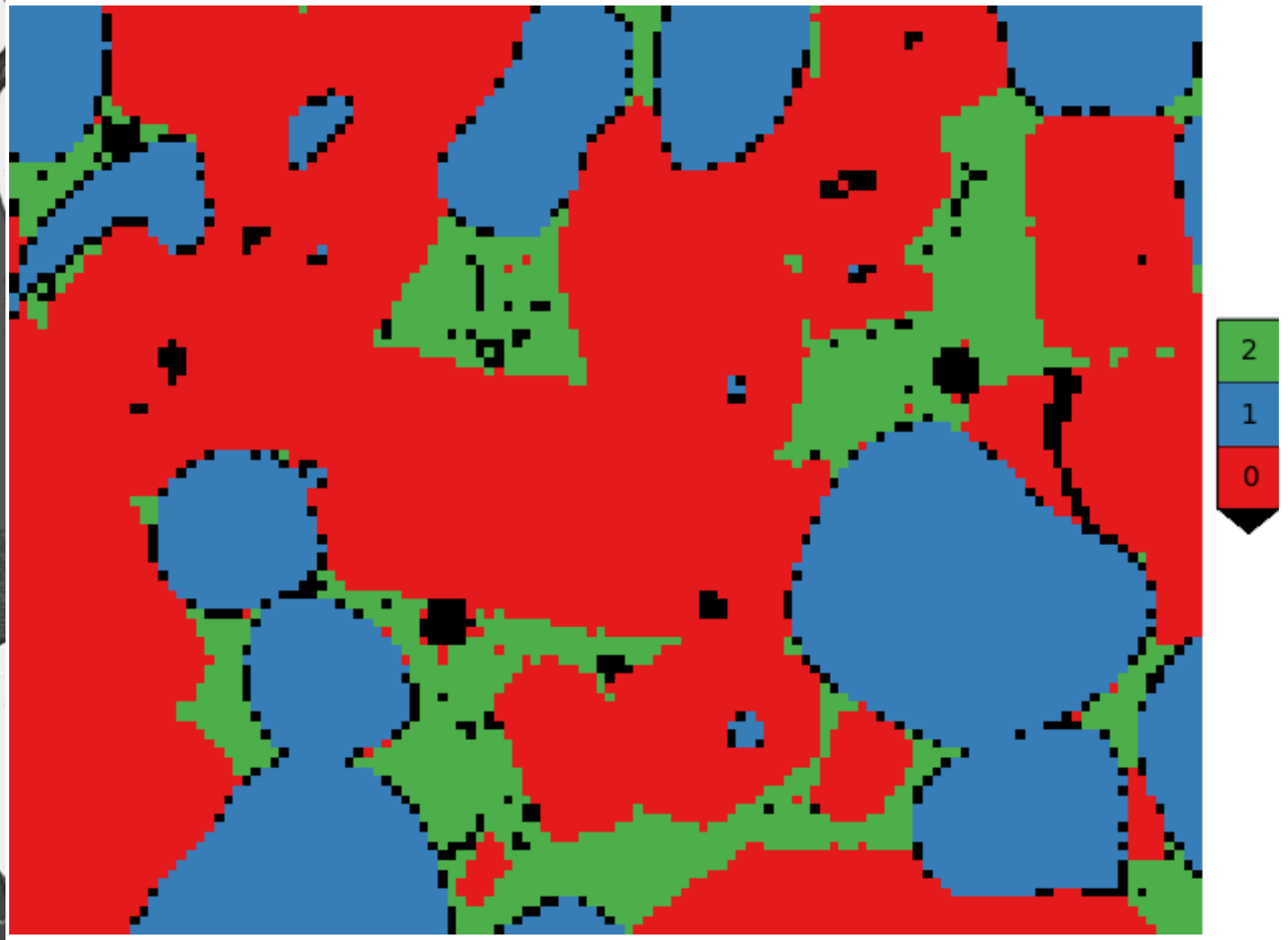

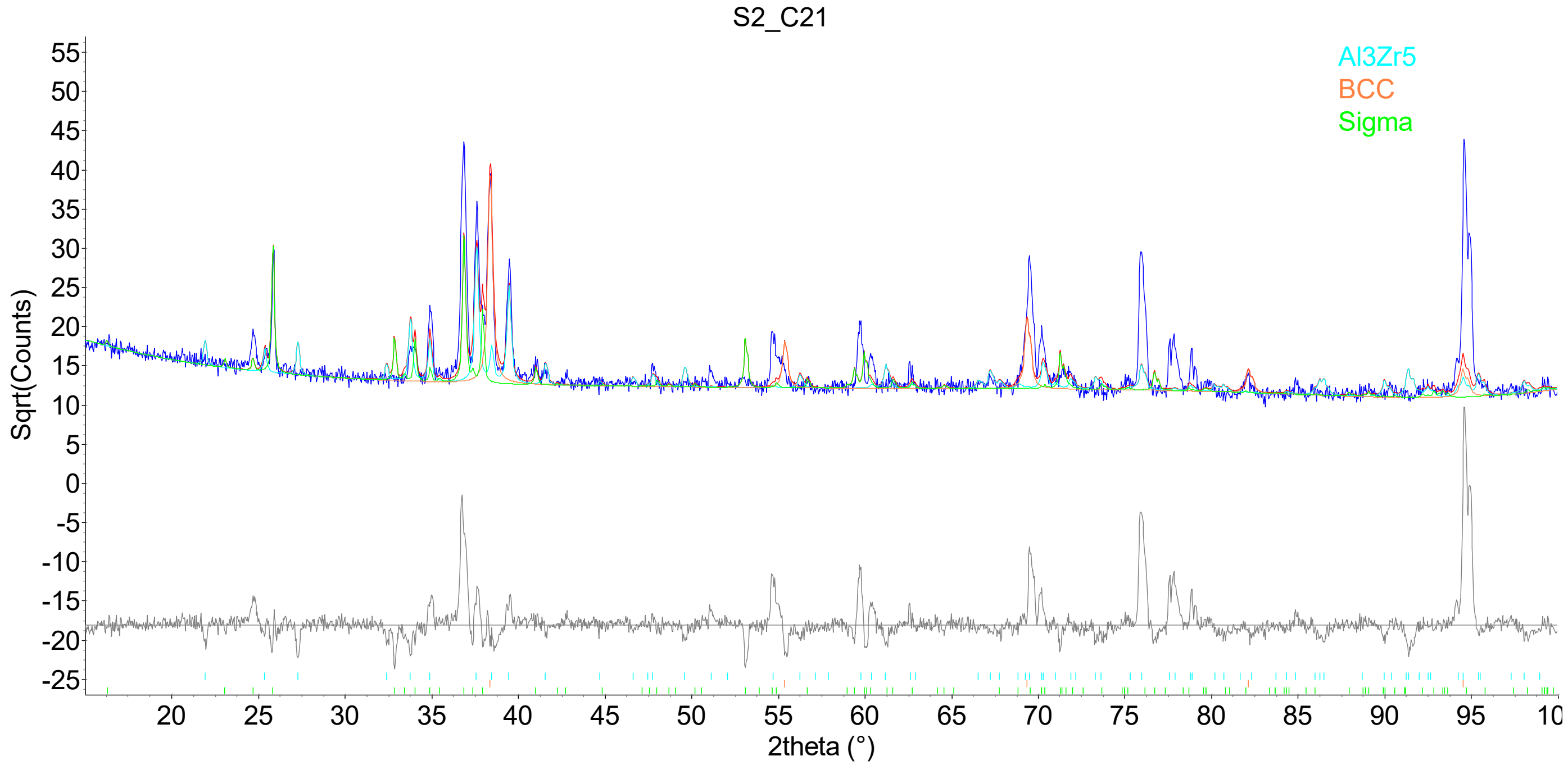

50 µm

2
1
0

S2_C21

Al3Zr5
BCC
Sigma

Sqrt(Counts)

2theta (°)

Al K

Zr L

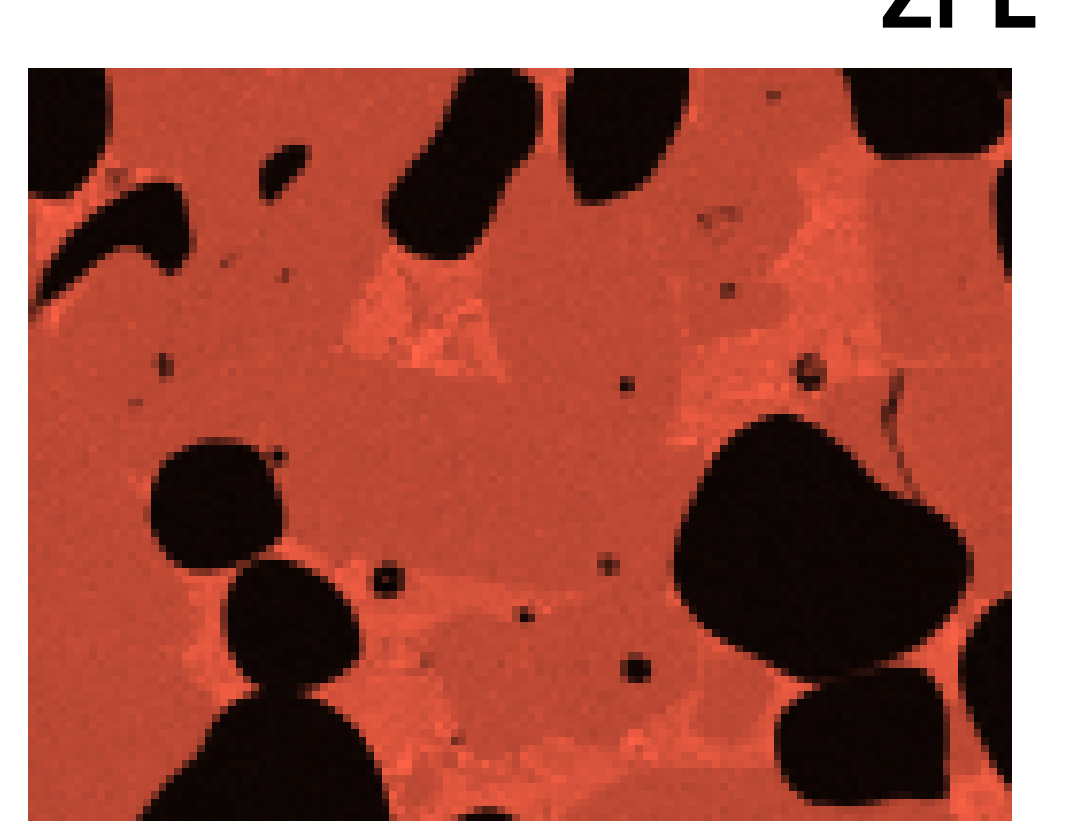

Nb L

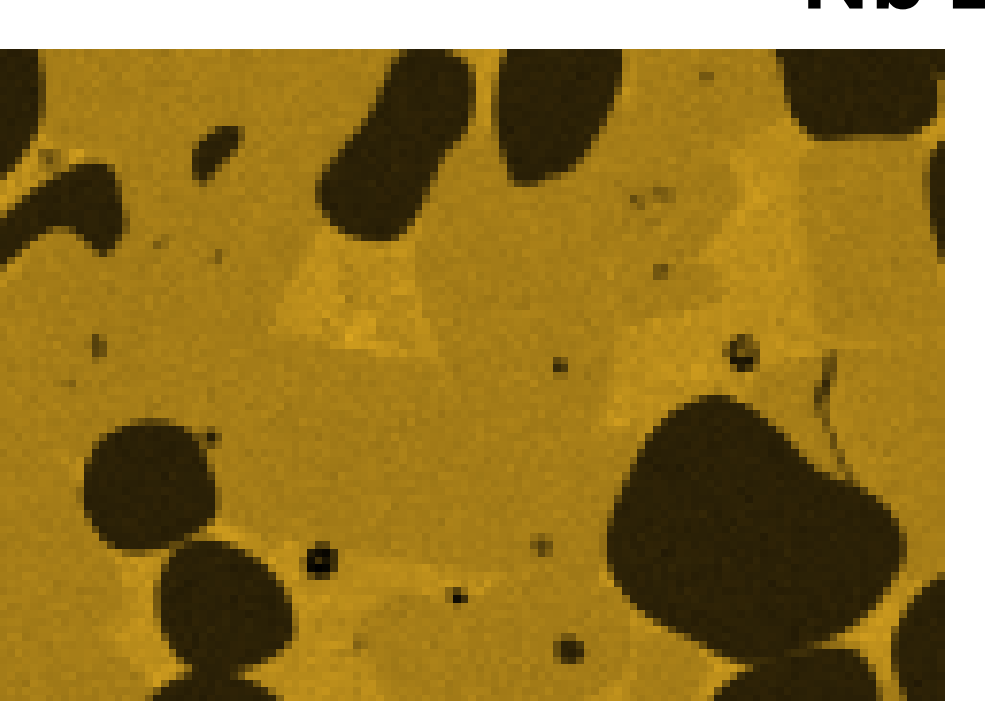

Ta L

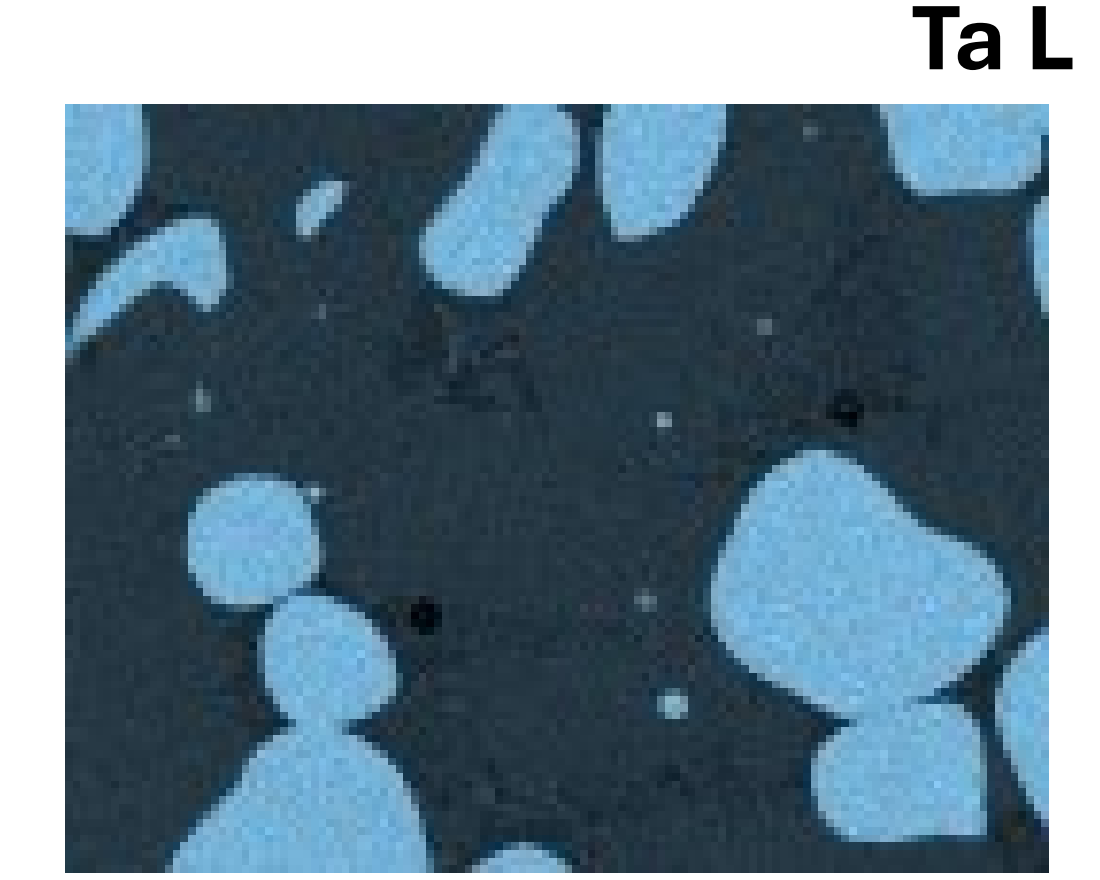

Nb K

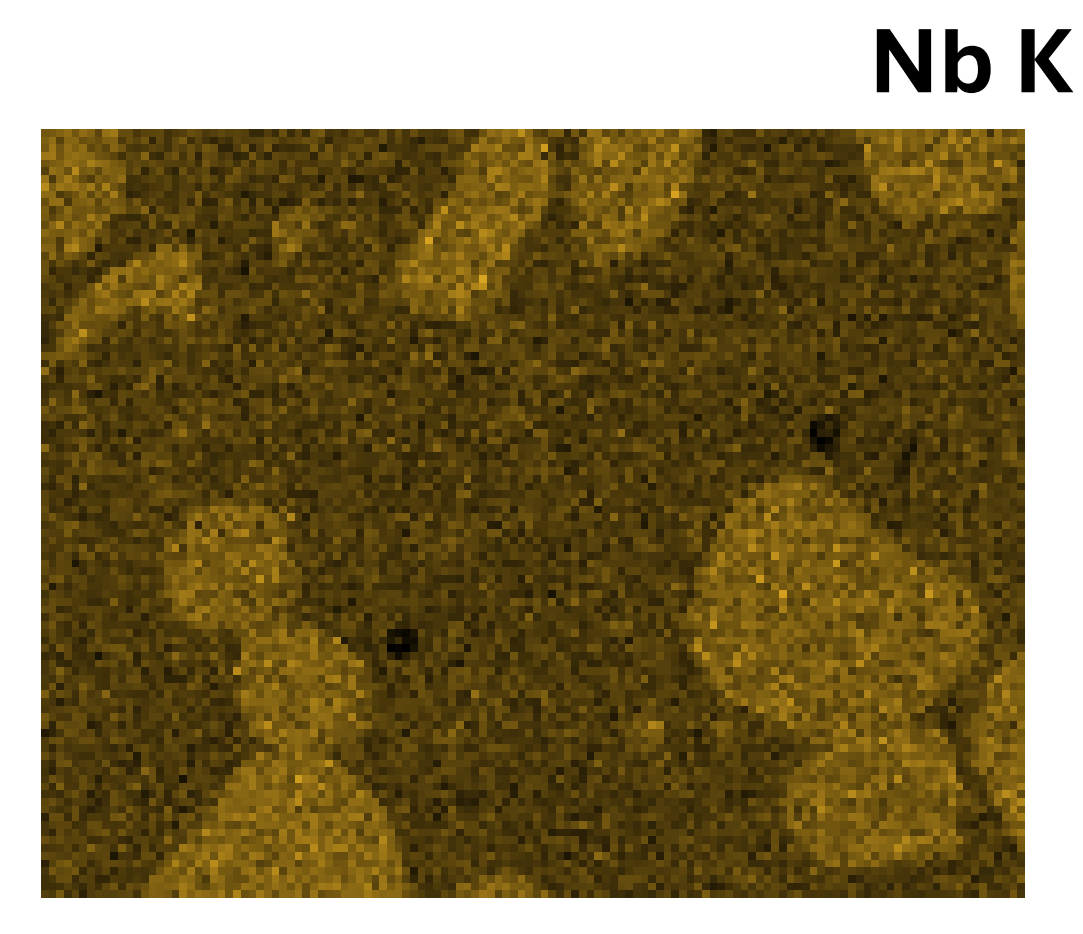

# S2_C22

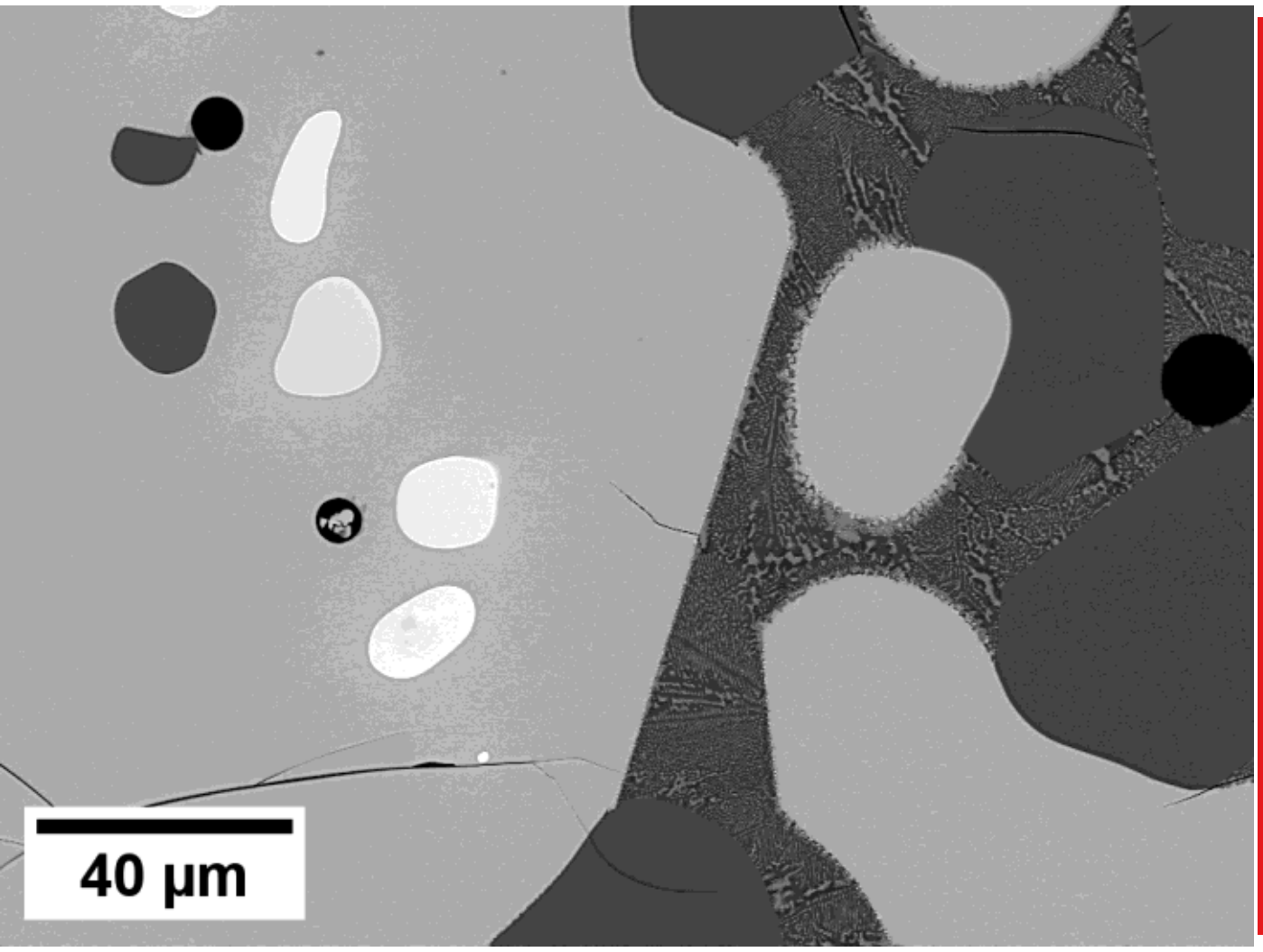

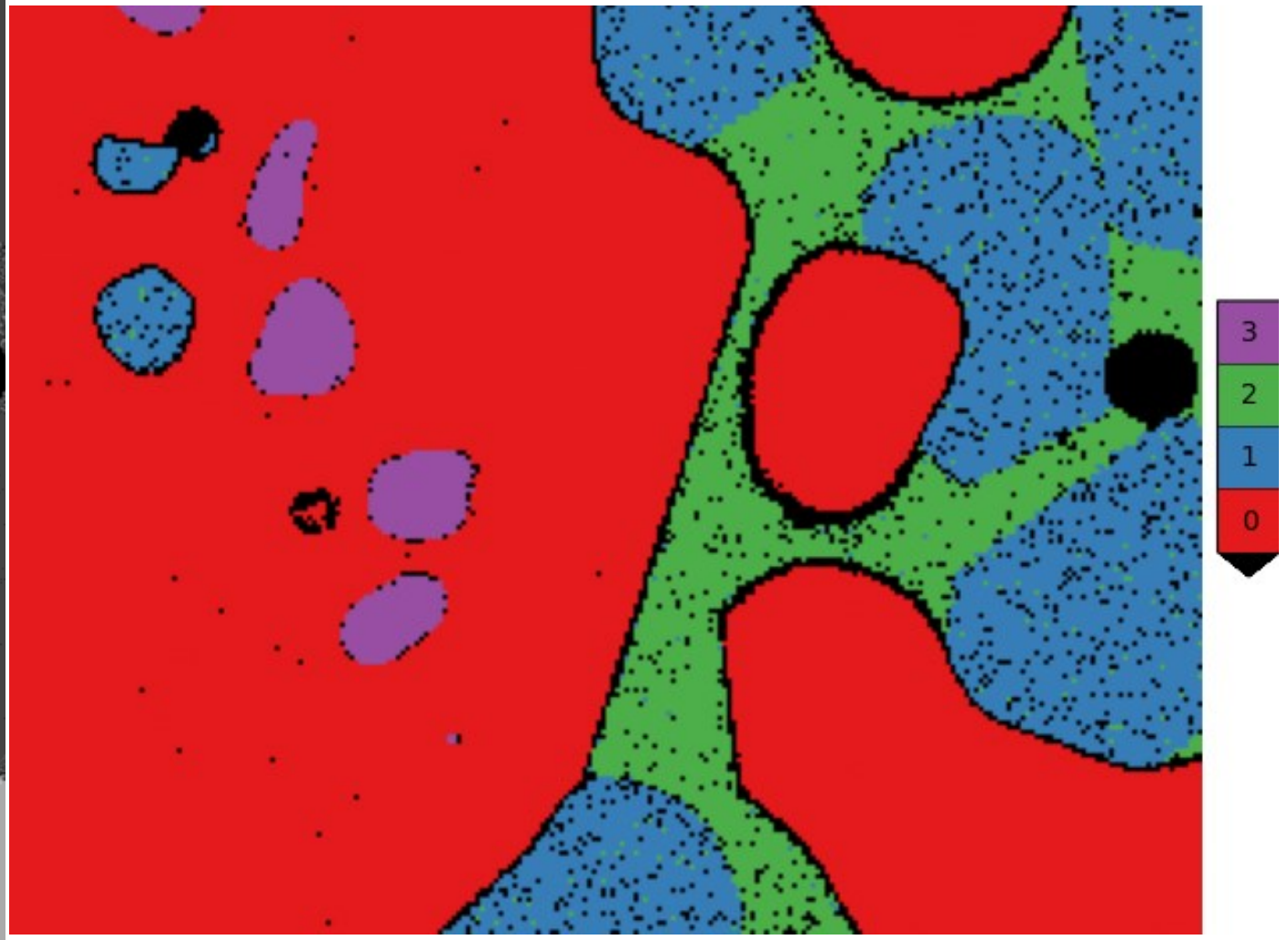

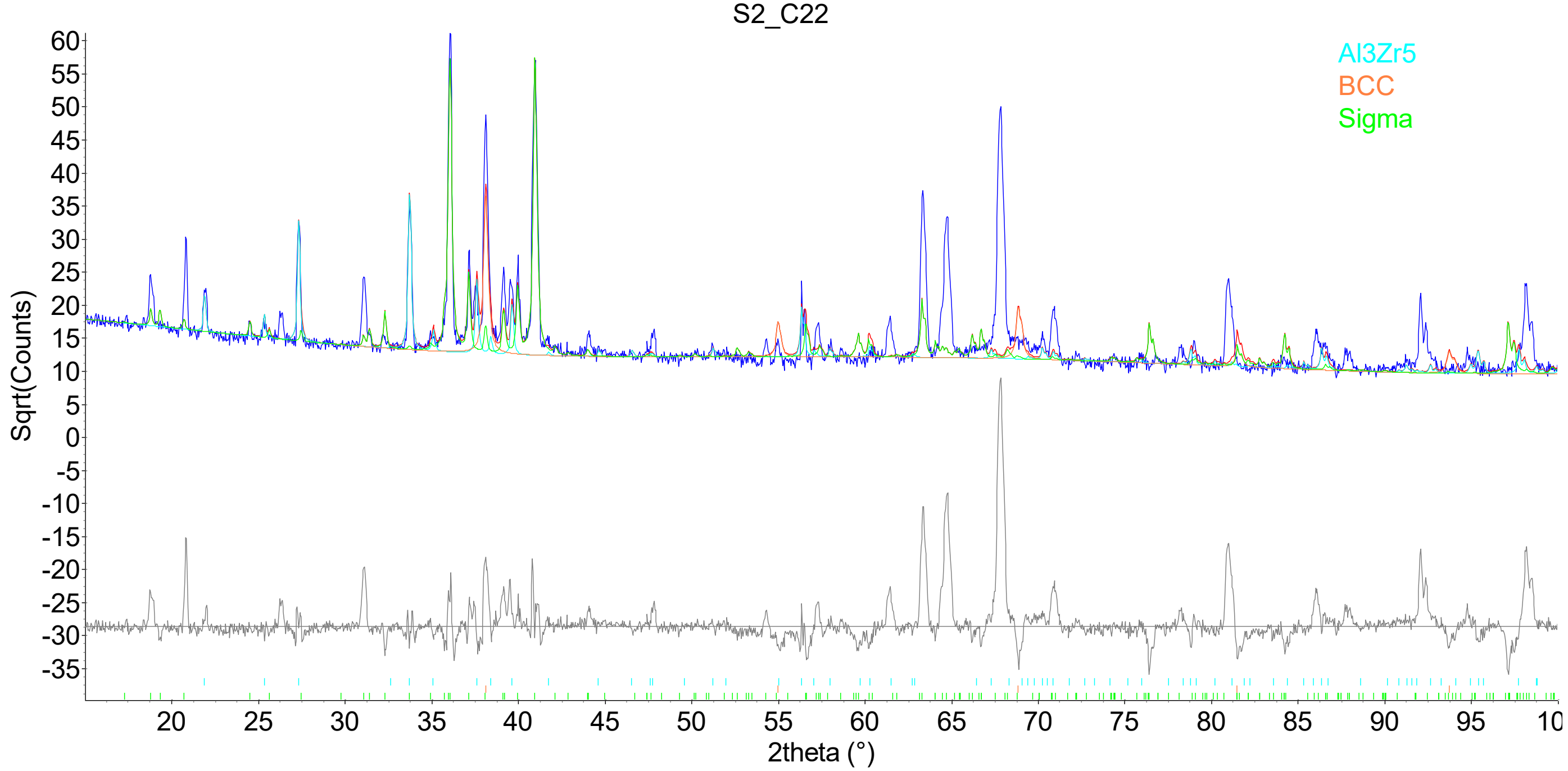

40 µm

3
2
1
0

S2_C22

Al3Zr5
BCC
Sigma

Sqrt(Counts)

2theta (°)

Al K

Zr L

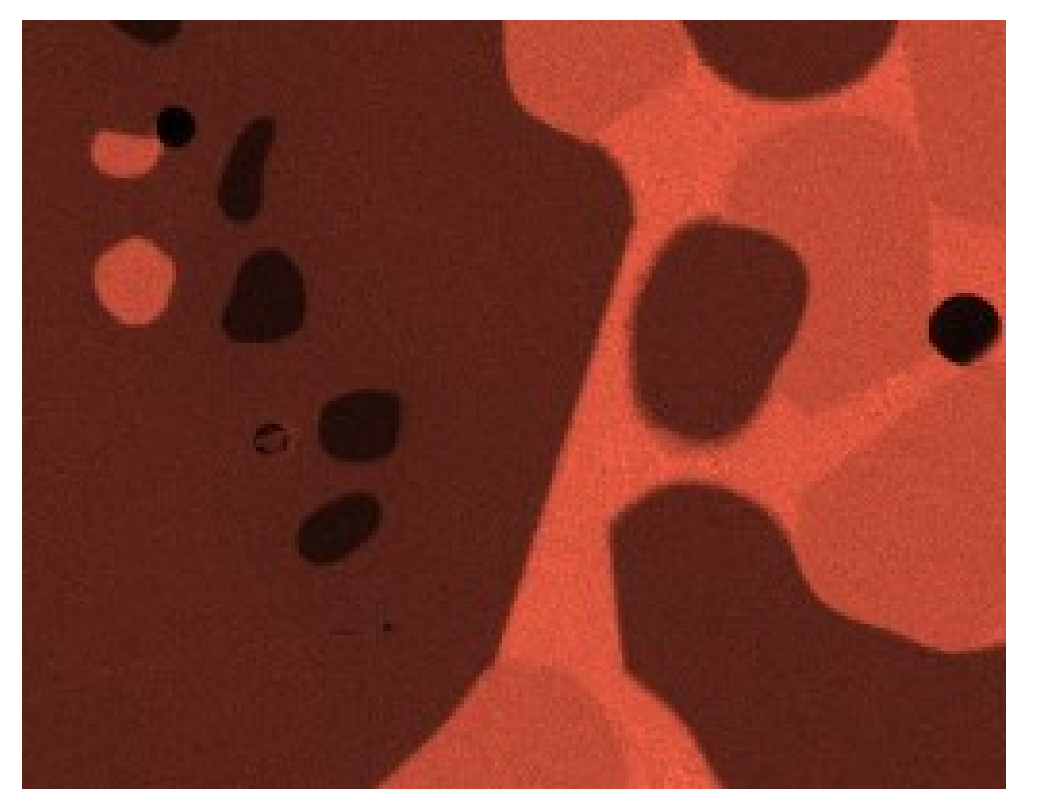

Nb L

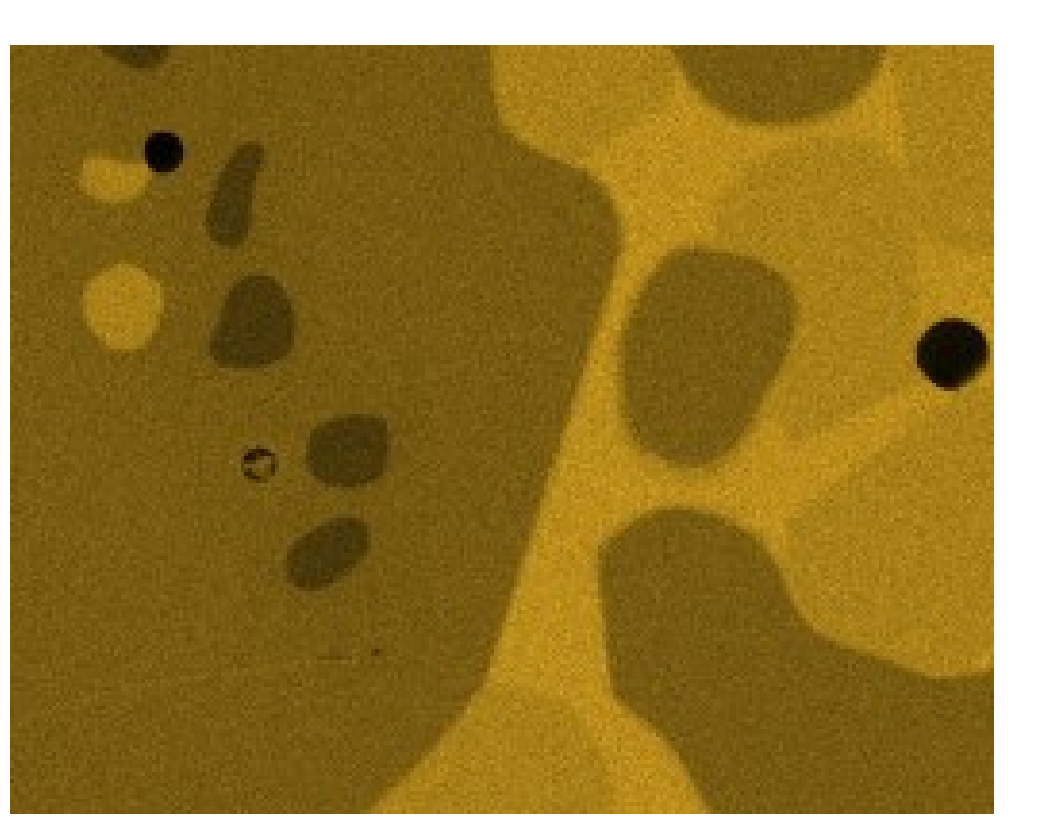

Ta L

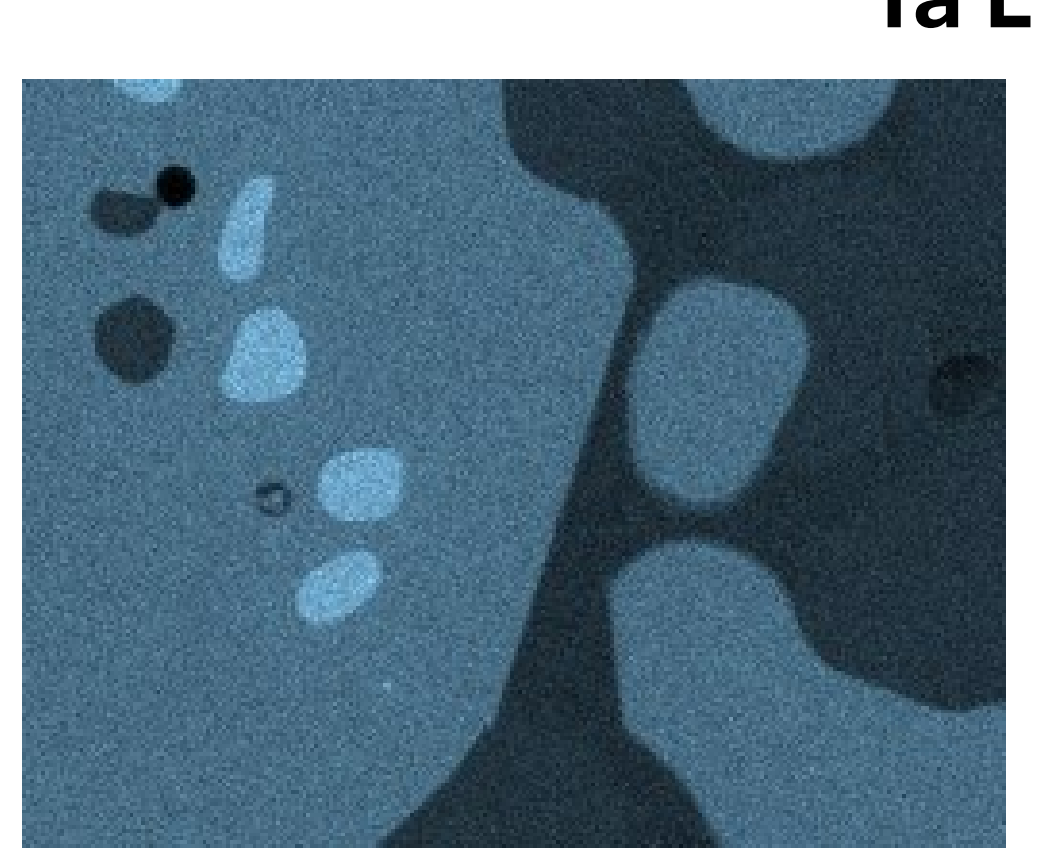

Nb K

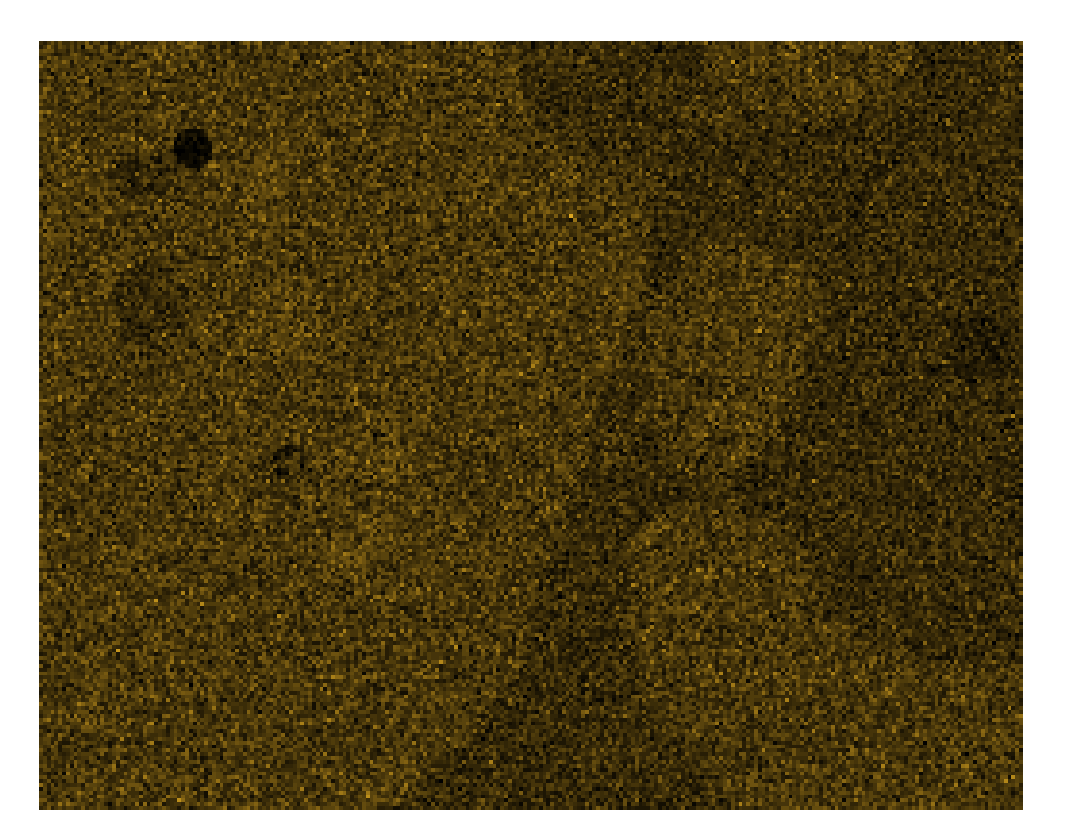

# S2_C23

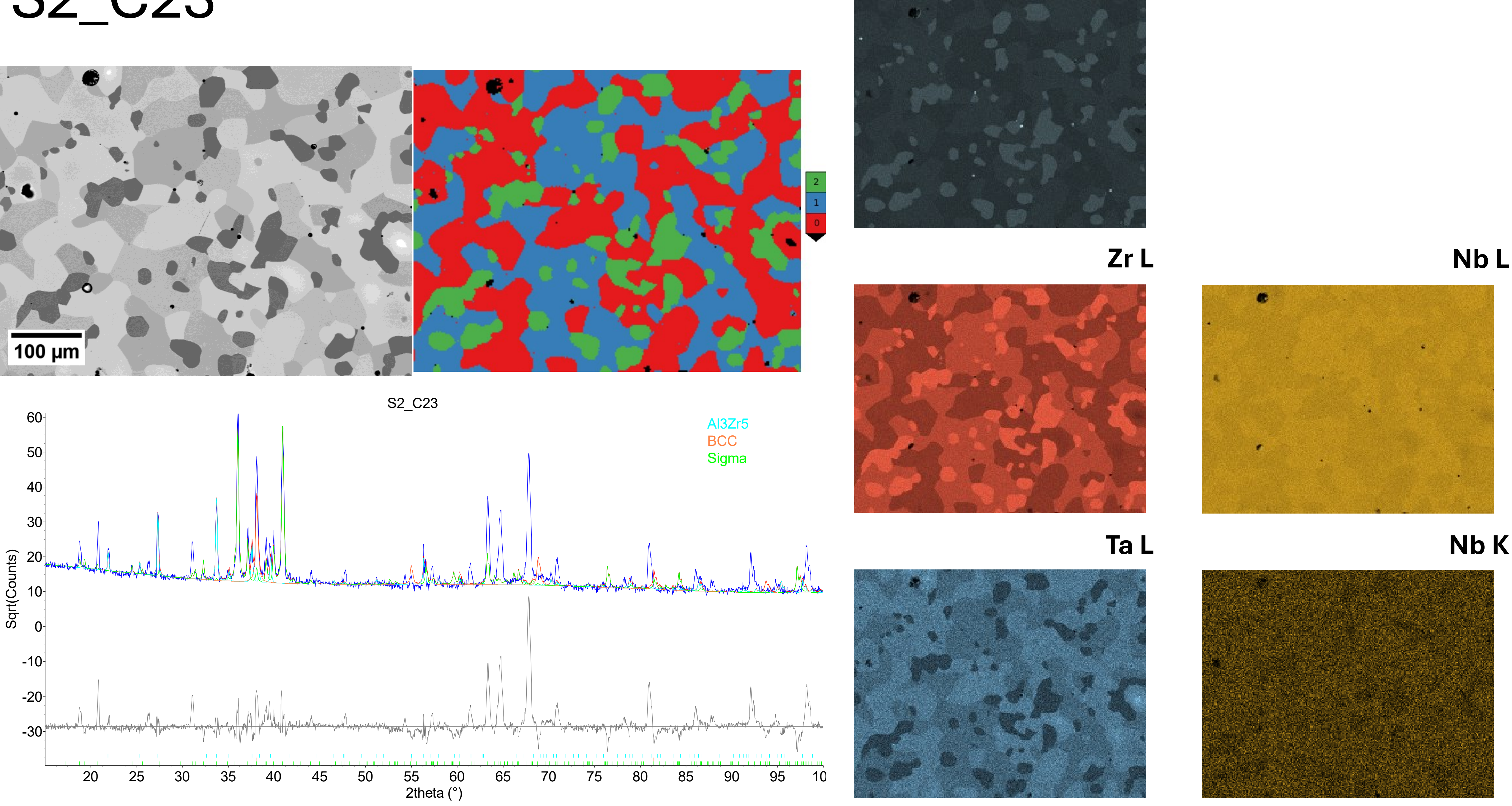

# S2_C24

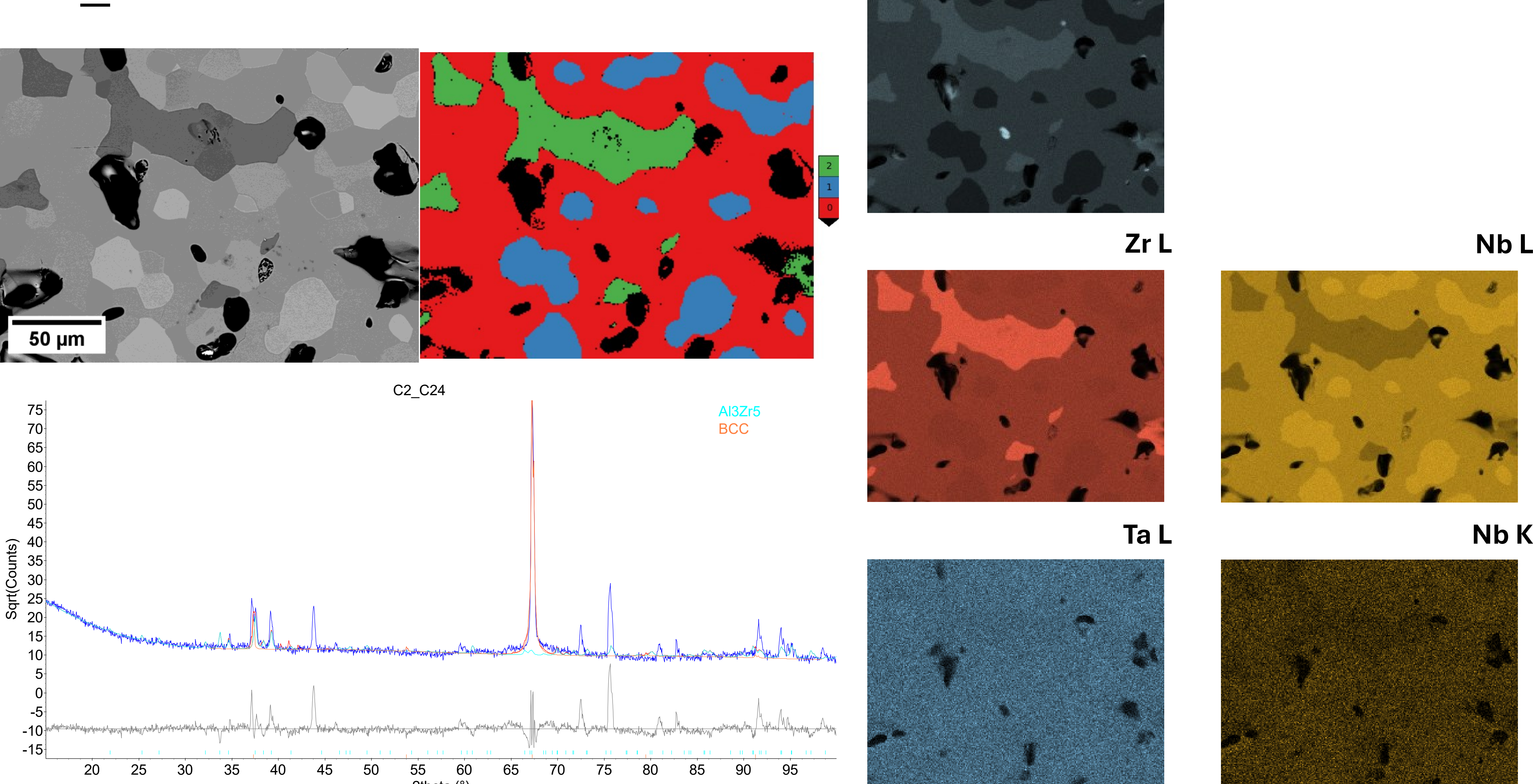

# S2_C31

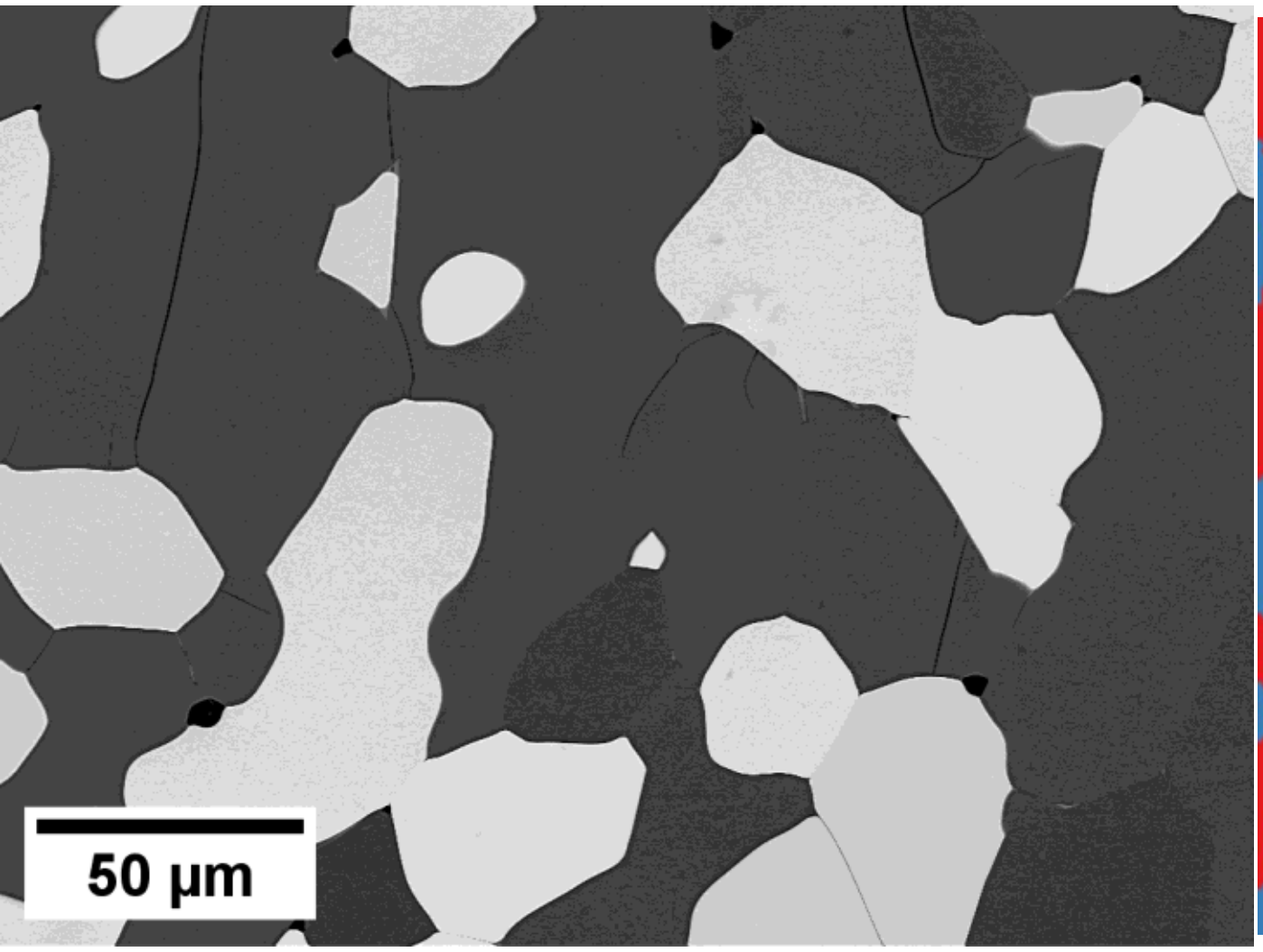

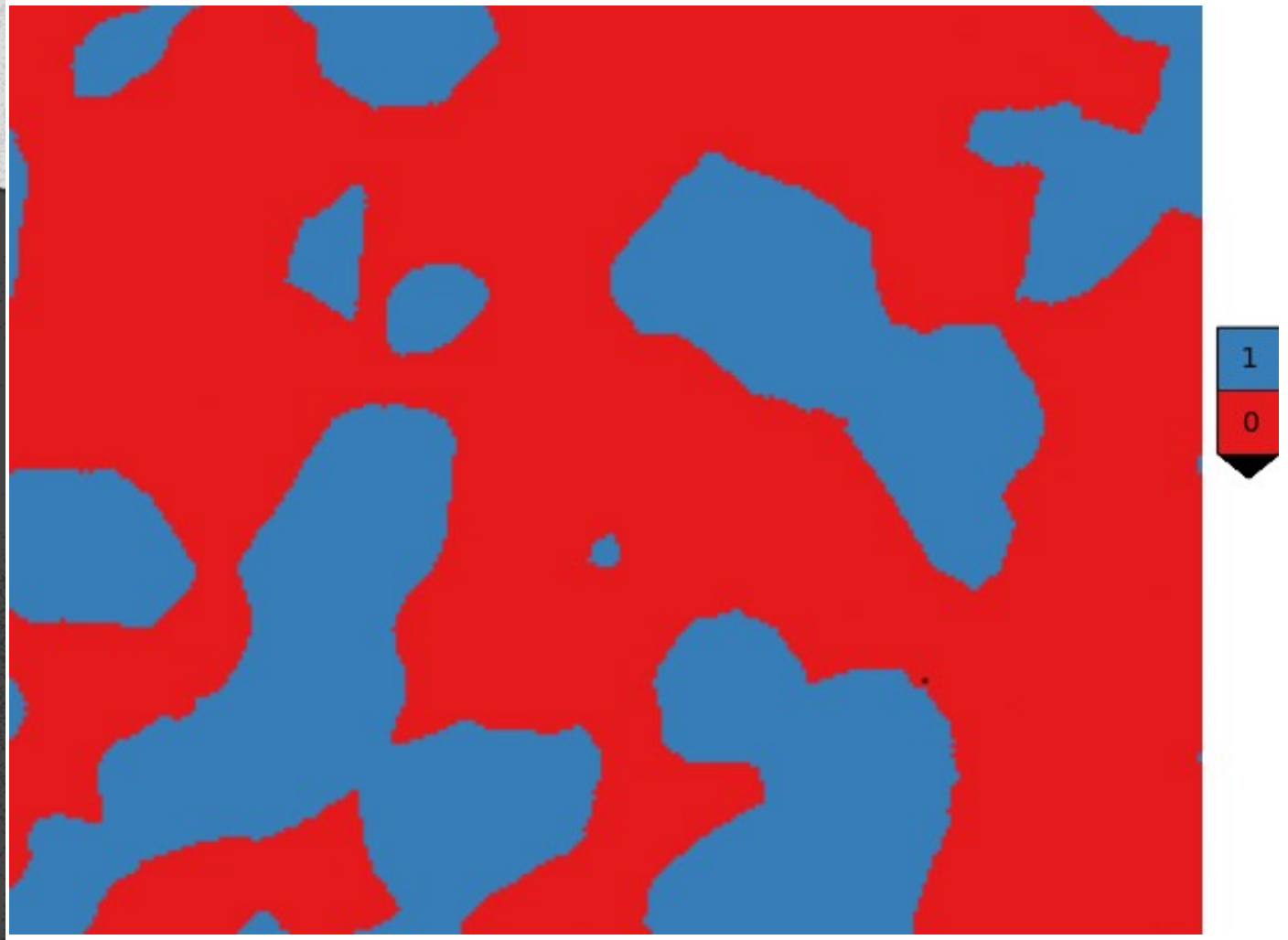

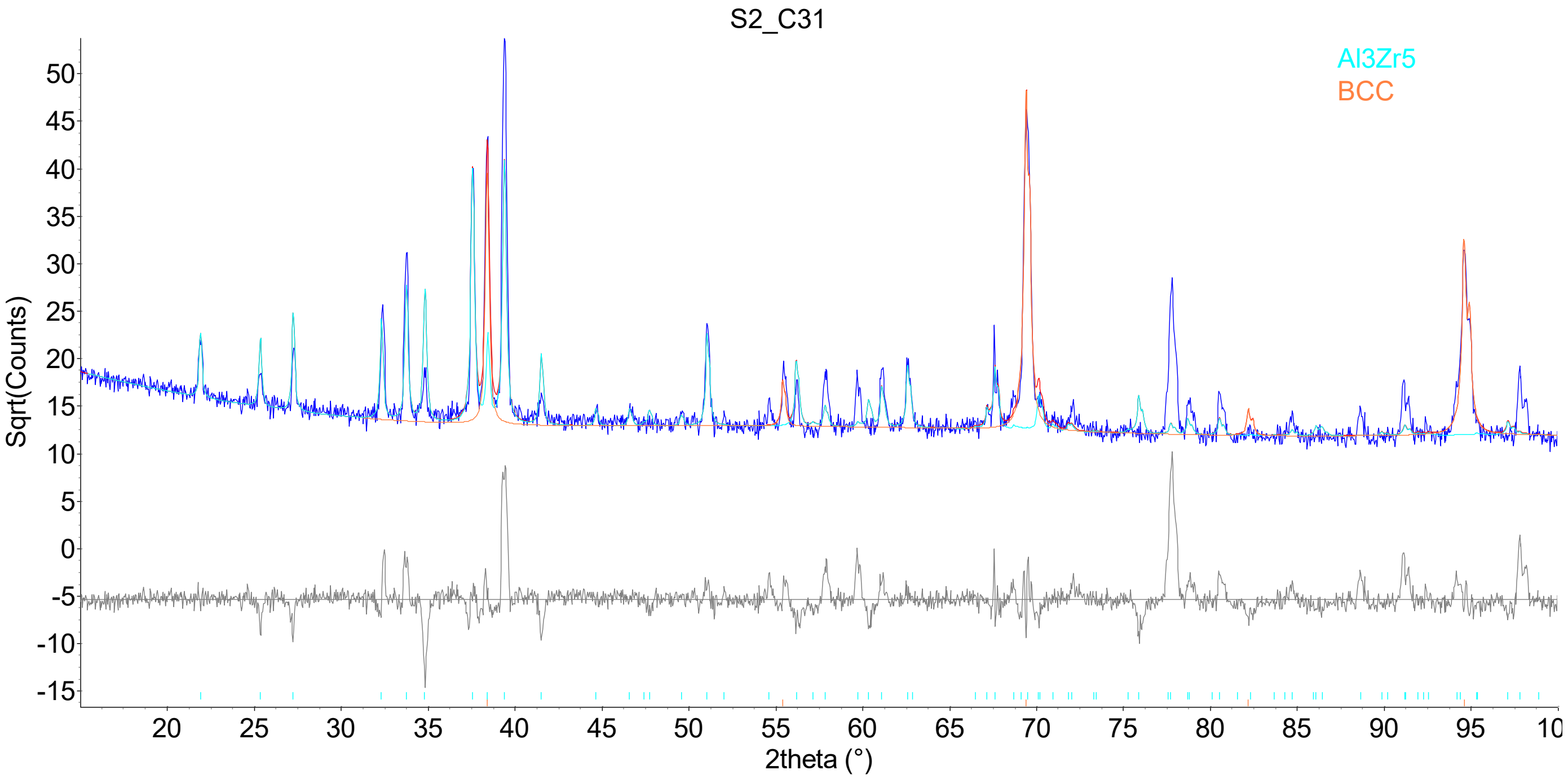

50 µm

1
0

S2_C31

Al3Zr5
BCC

Sqrt(Counts)

50
45
40
35
30
25
20
15
10
5
0
-5
-10
-15

20 25 30 35 40 45 50 55 60 65 70 75 80 85 90 95 10

2theta (°)

Al K

Zr L

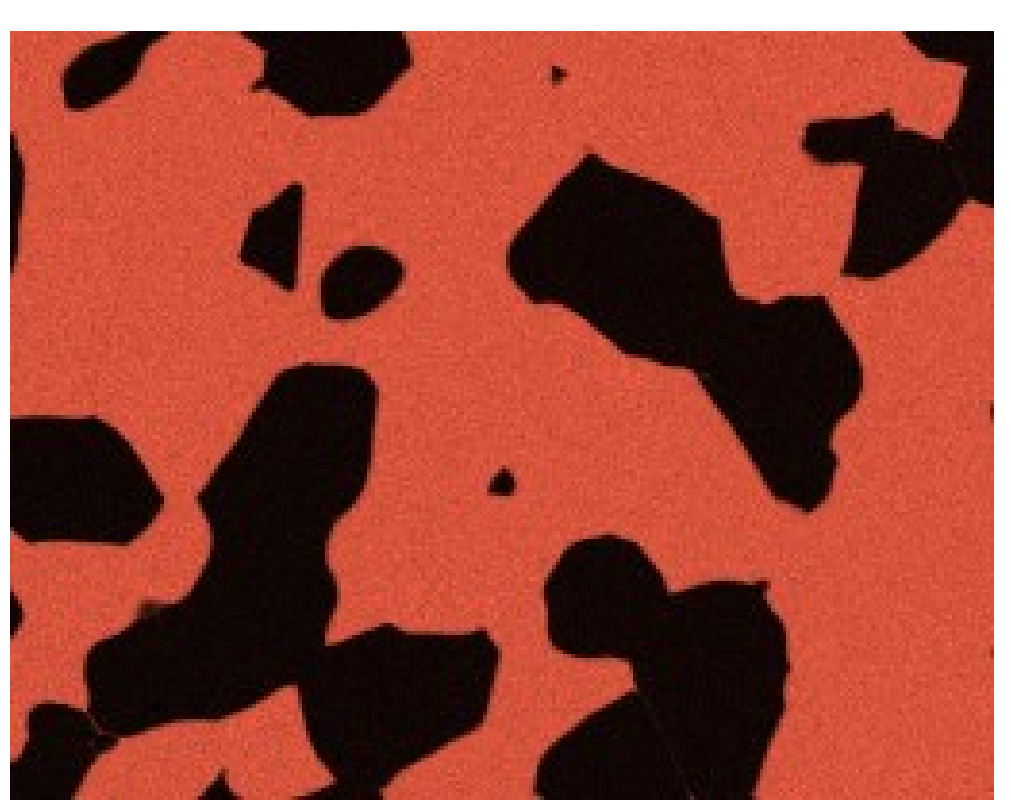

Nb L

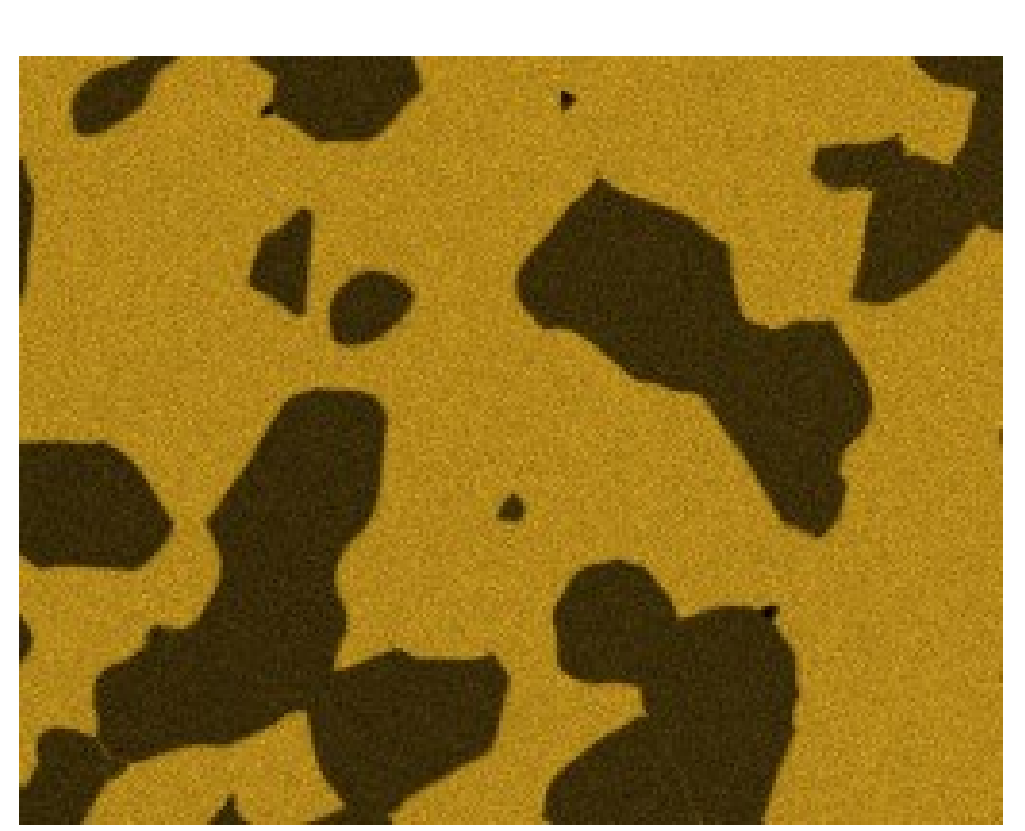

Ta L

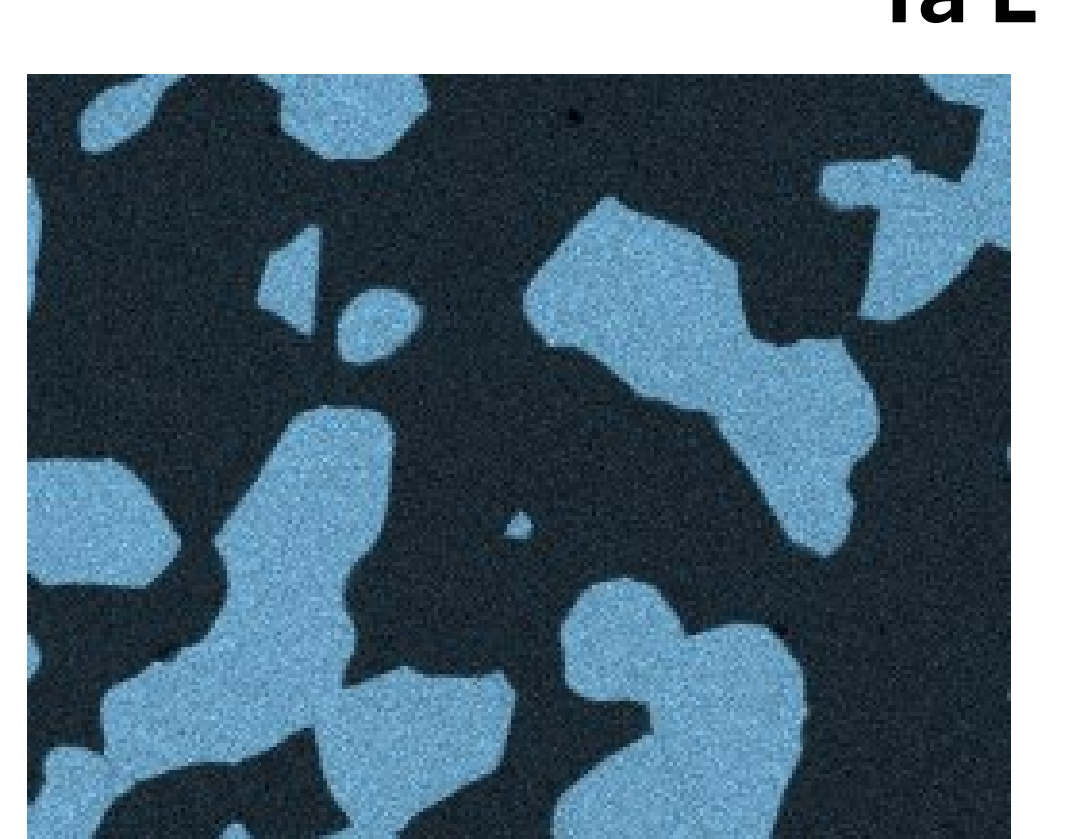

Nb K

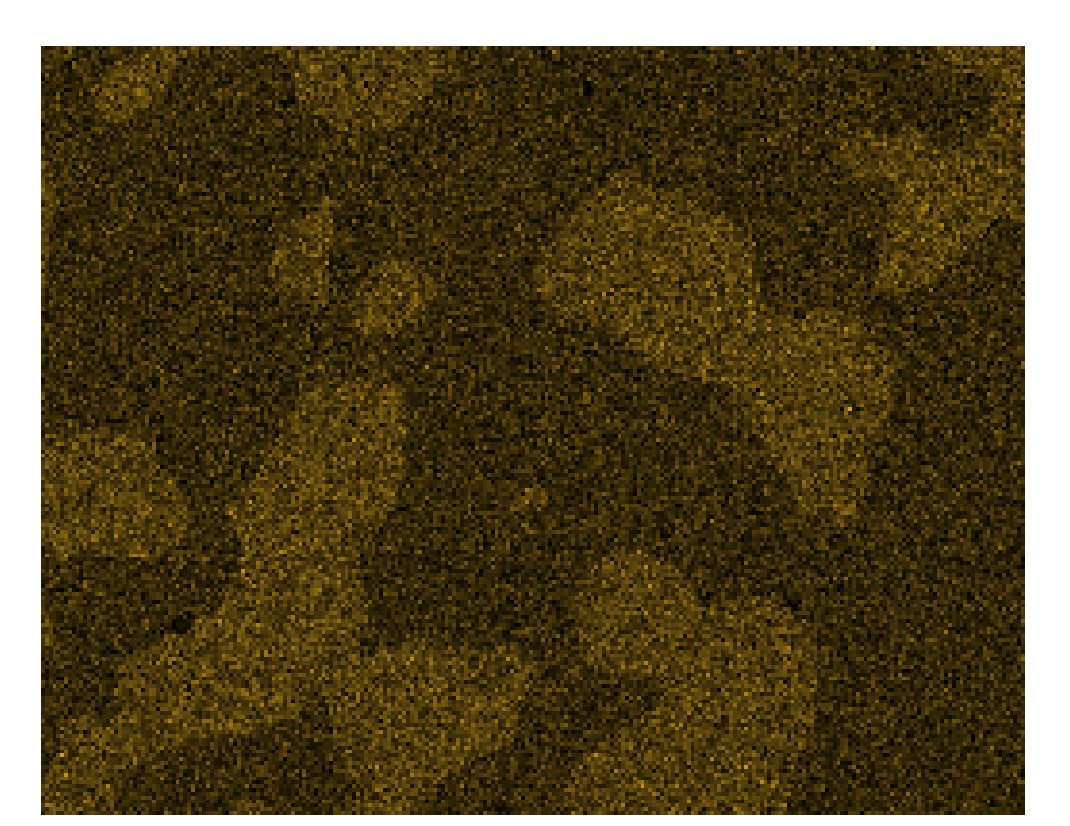

# S2_C32

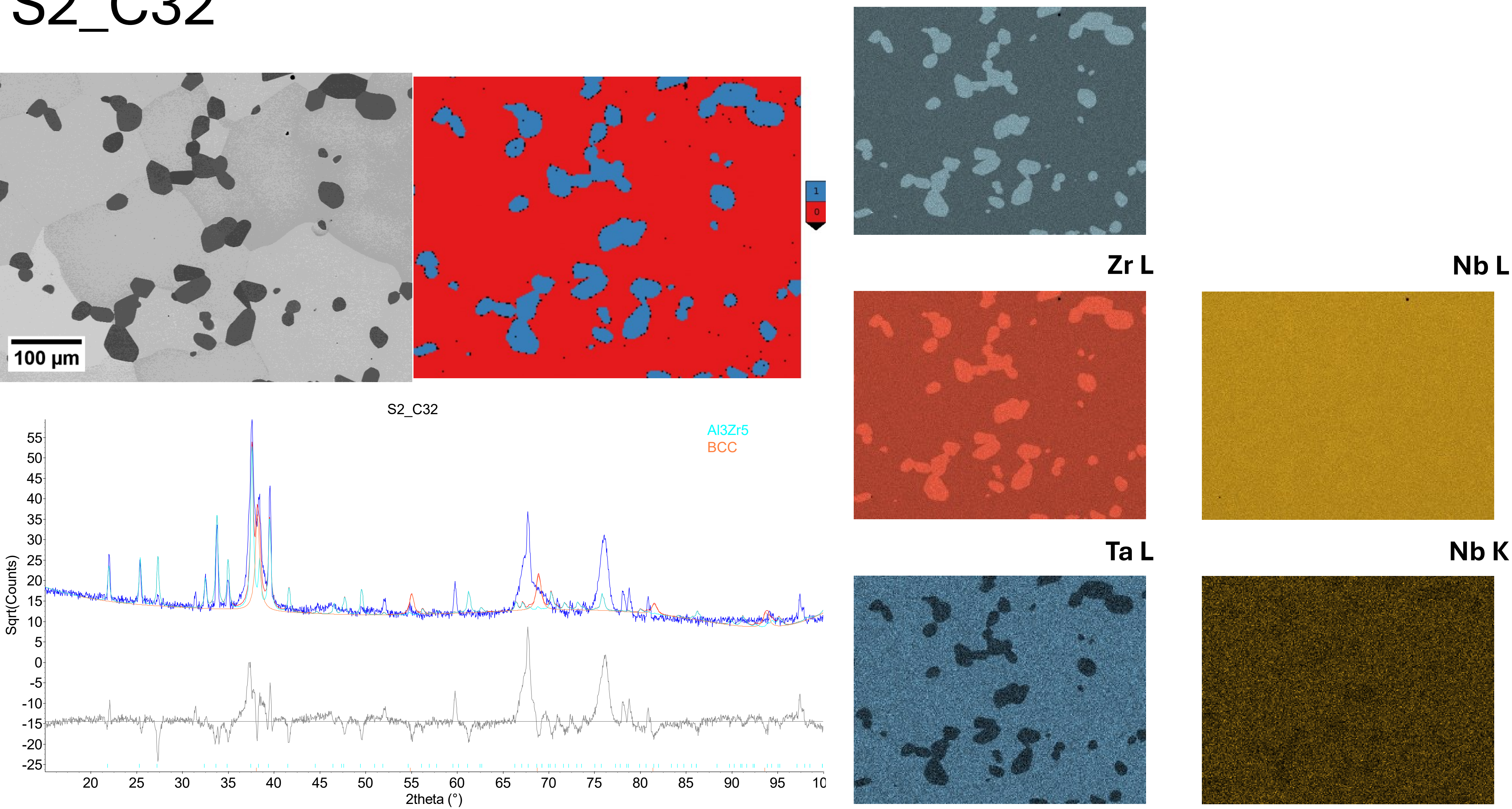

# S2_C33

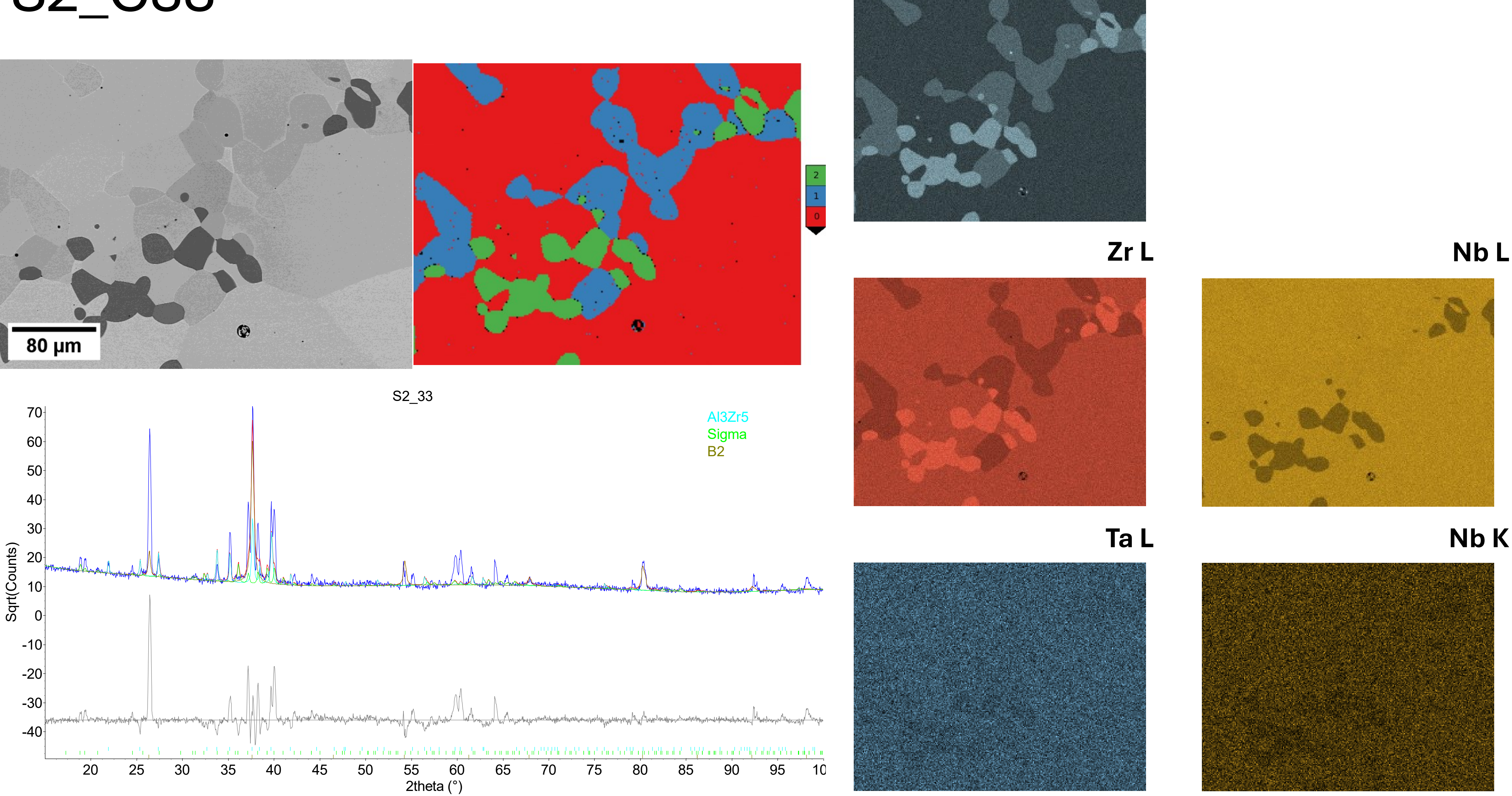

# Supplementary material B3:
# Phase identification: Sample S3
# **Ti-Nb-Zr-Ta system**

Supplementary of paper:

## **Phase Equilibria of the Al-Ti-Nb-Zr-Ta System**

Jiří Kozlík [a*], František Lukáč [b], Mariano Casas-Luna [a], Jozef Veselý [a], Eliška Jača [a], Kateřina Ficková [a] , Stanislav Šašek [a], Kristína Bartha [a], Adam Strnad [a], Tomáš Chráska [b], Josef Stráský [a]

[a] *Charles University, Faculty of Mathematics and Physics, Department of Physics of Materials, Ke Karlovu 5, Prague 121 16, Czechia*

[b] *Institute of Plasma Physics of the Czech Academy of Sciences, U Slovanky 2525/1a, Prague 182 00, Czechia*

* Corresponding author: jiri.kozlik@matfyz.cuni.cz

# S3_C11

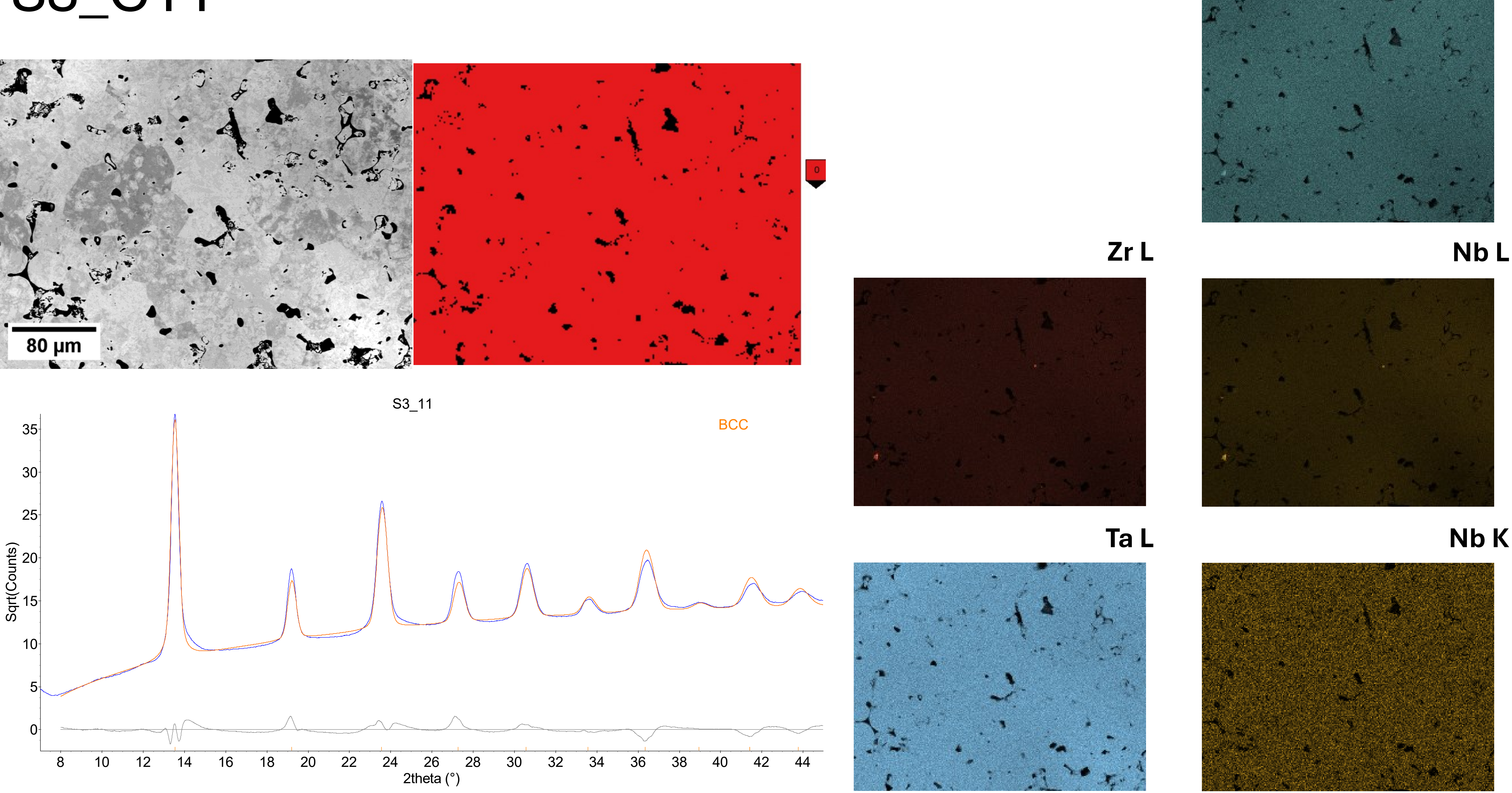

# S3_C12

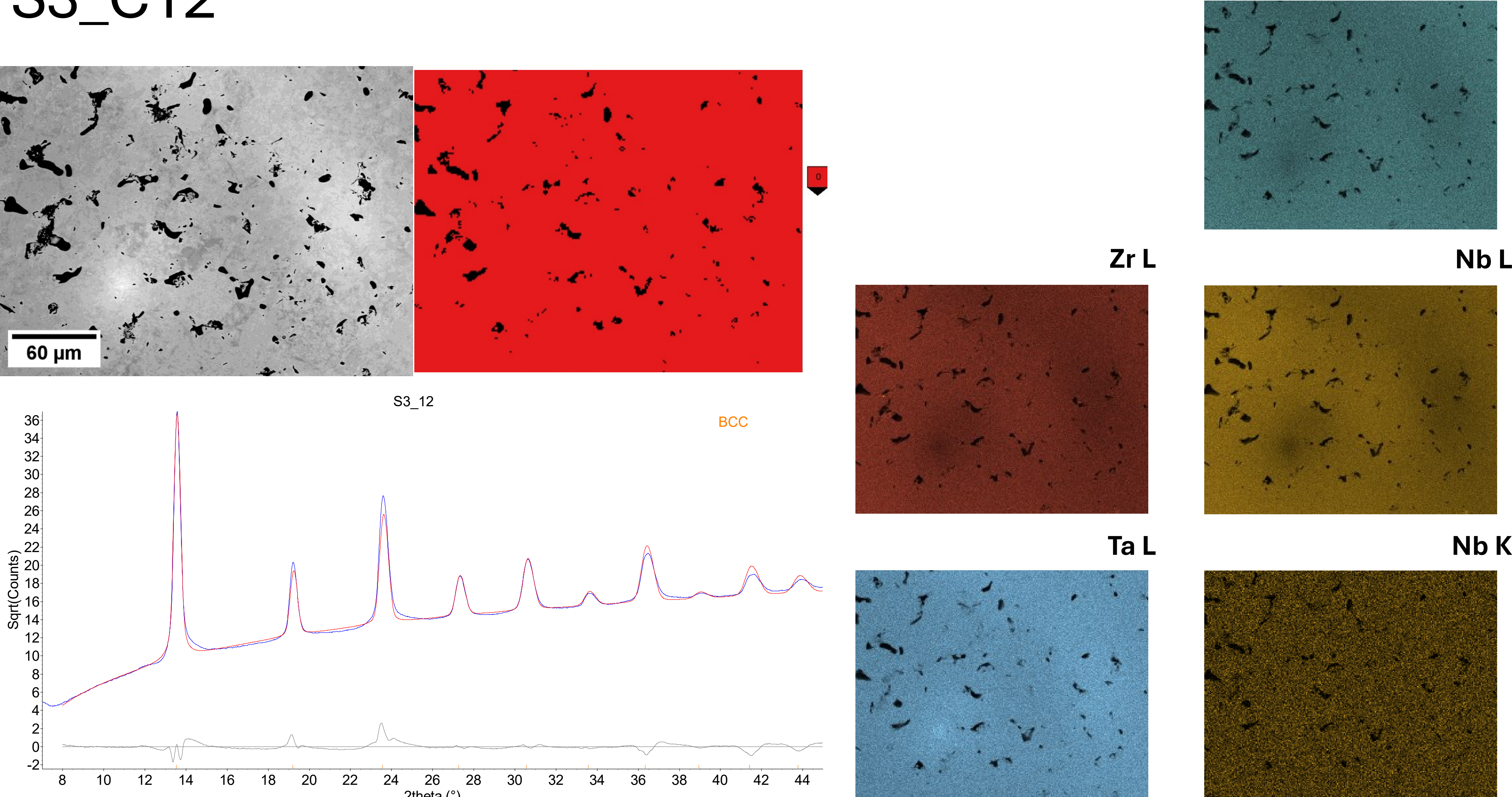

# S3_C13

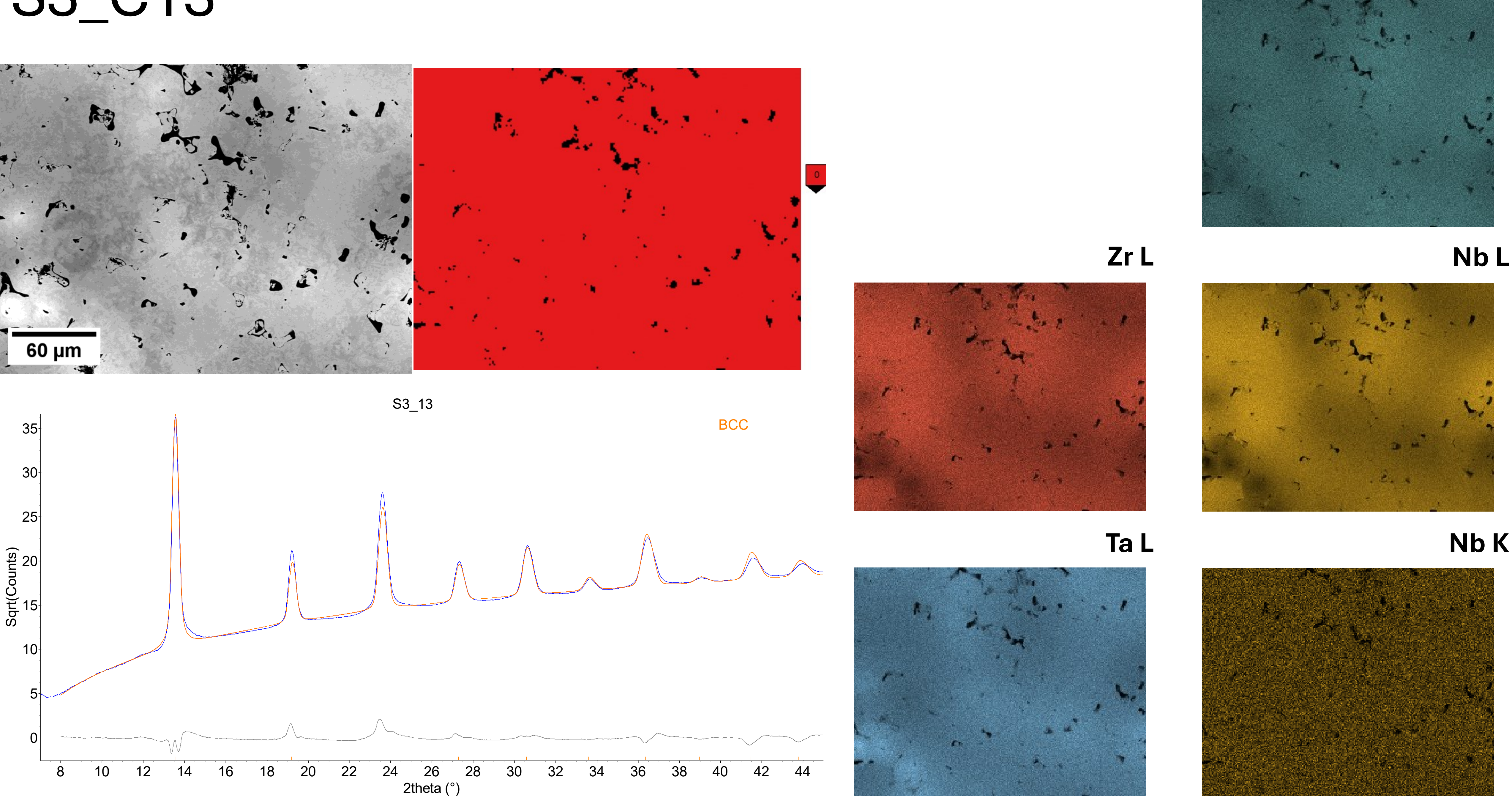

# S3_C14

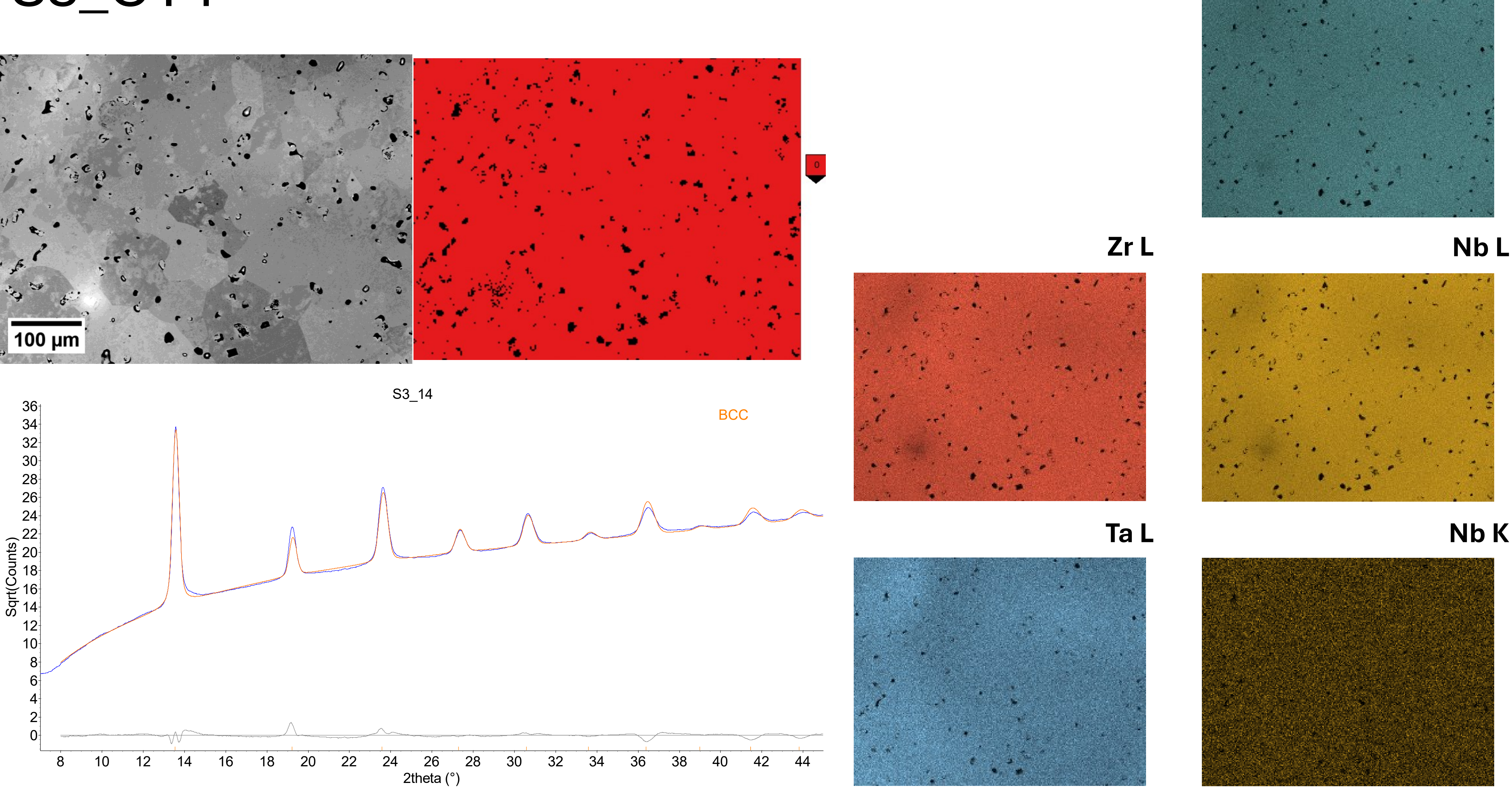

# S3_C20

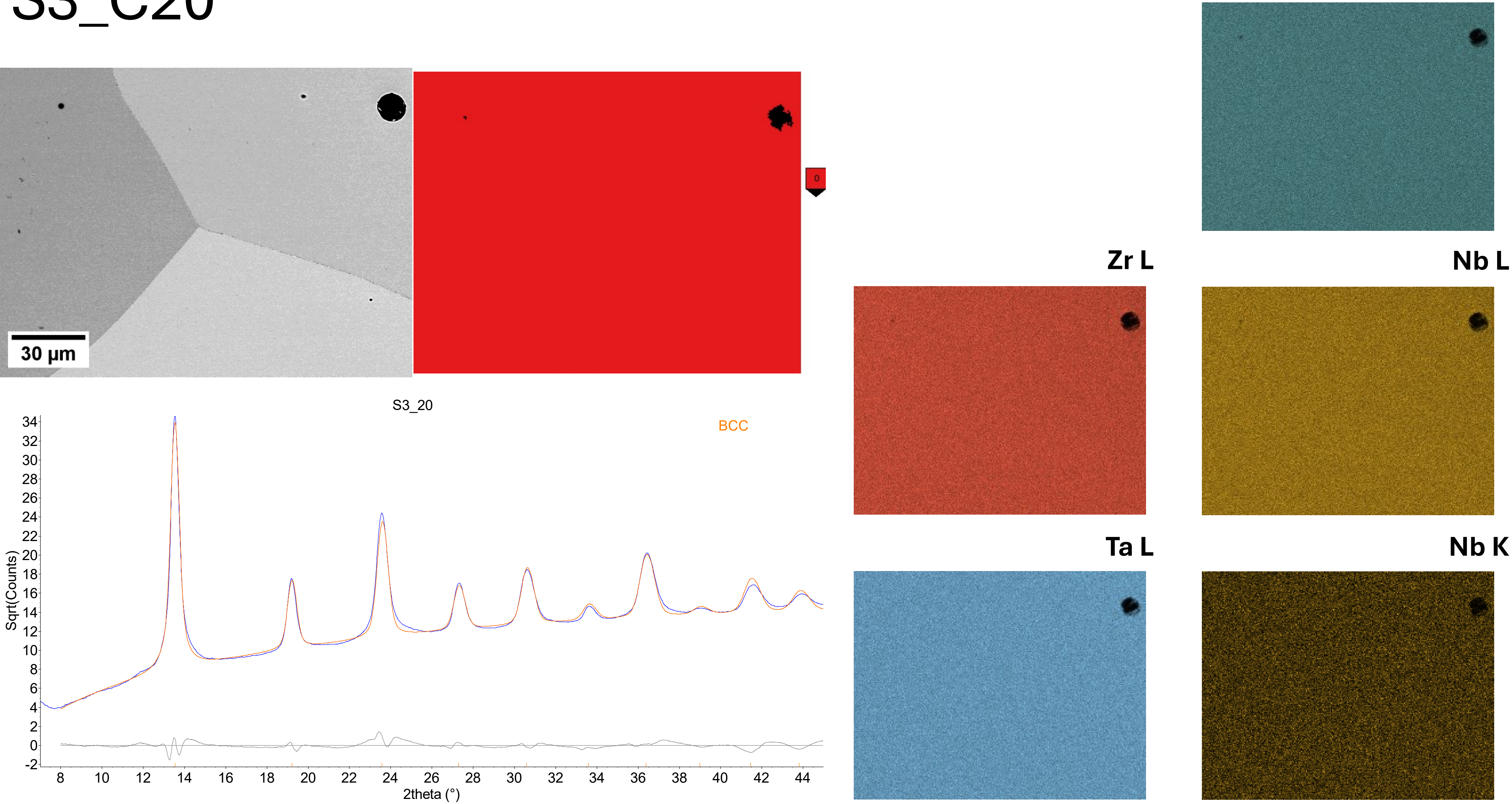

# S3_C21

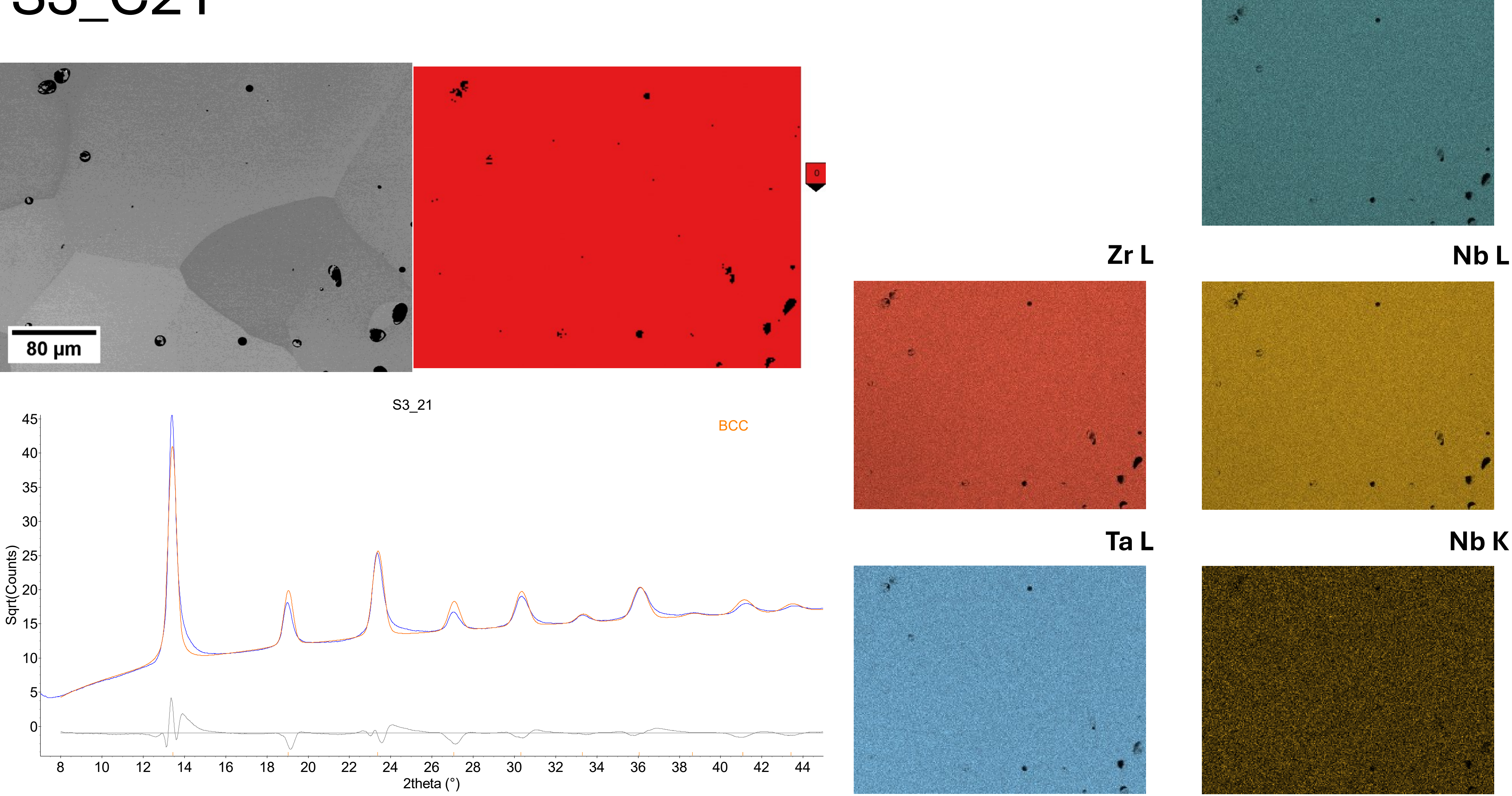

# S3_C22

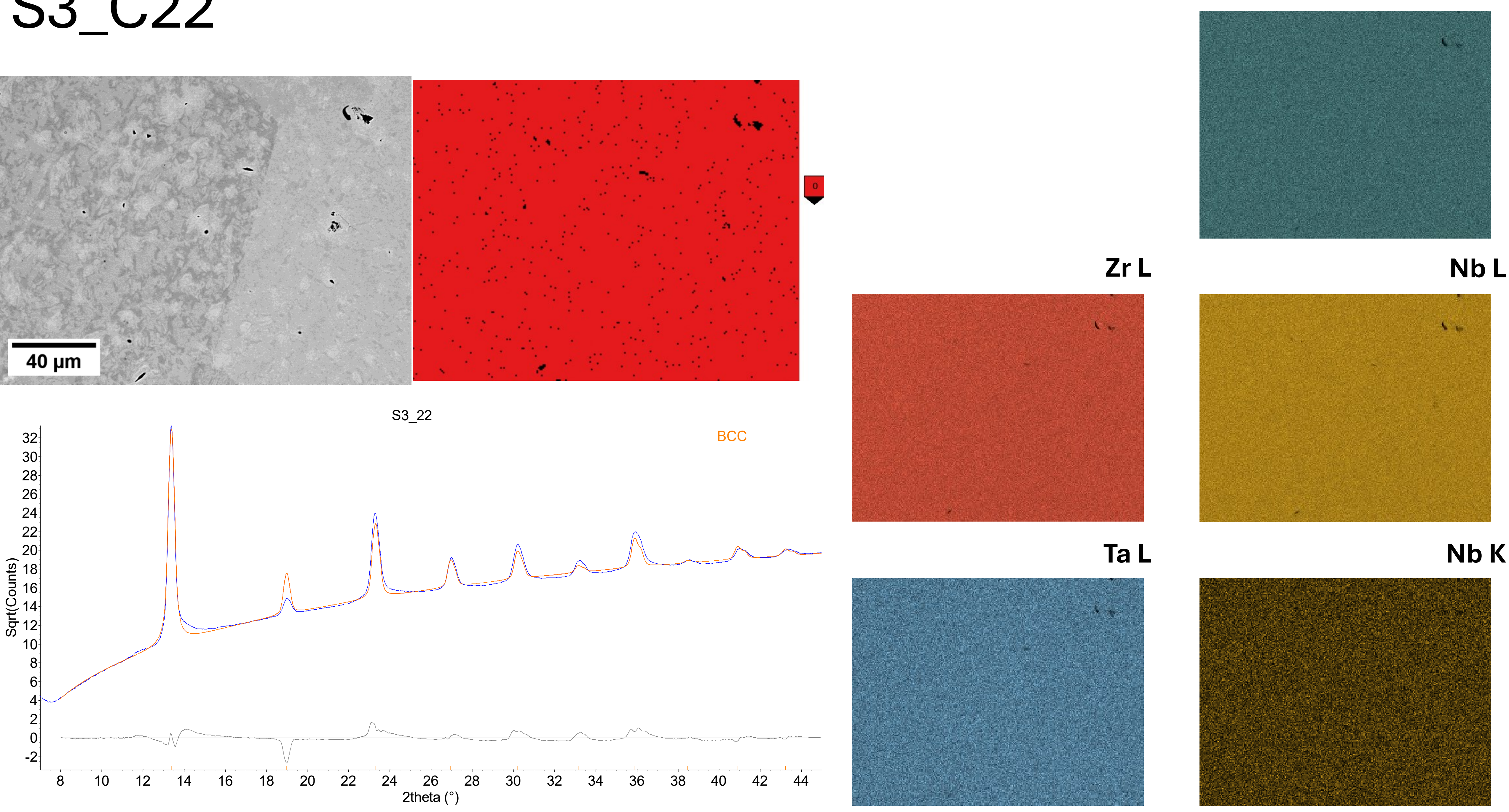

# S3_C23

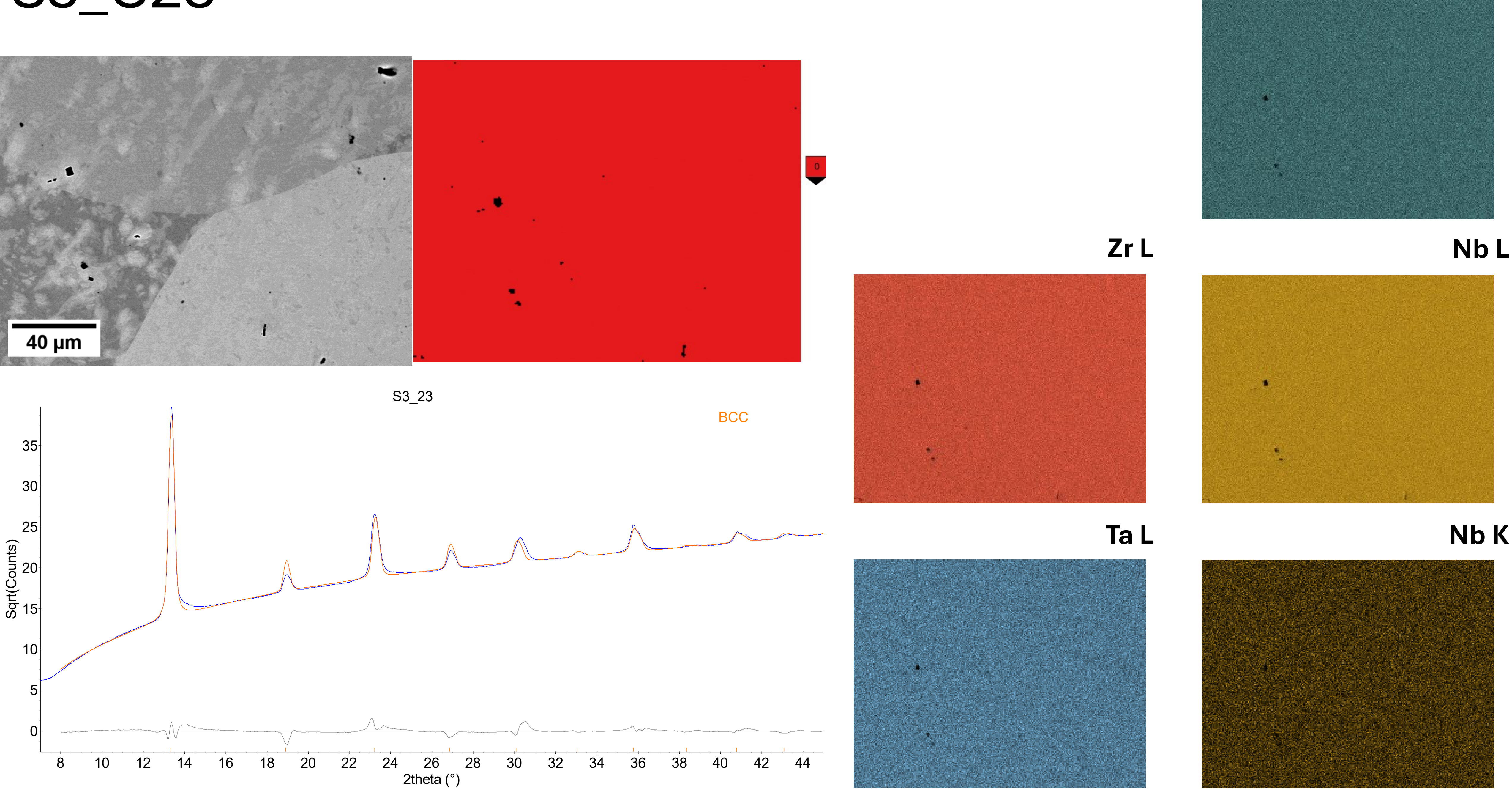

# S3_C24

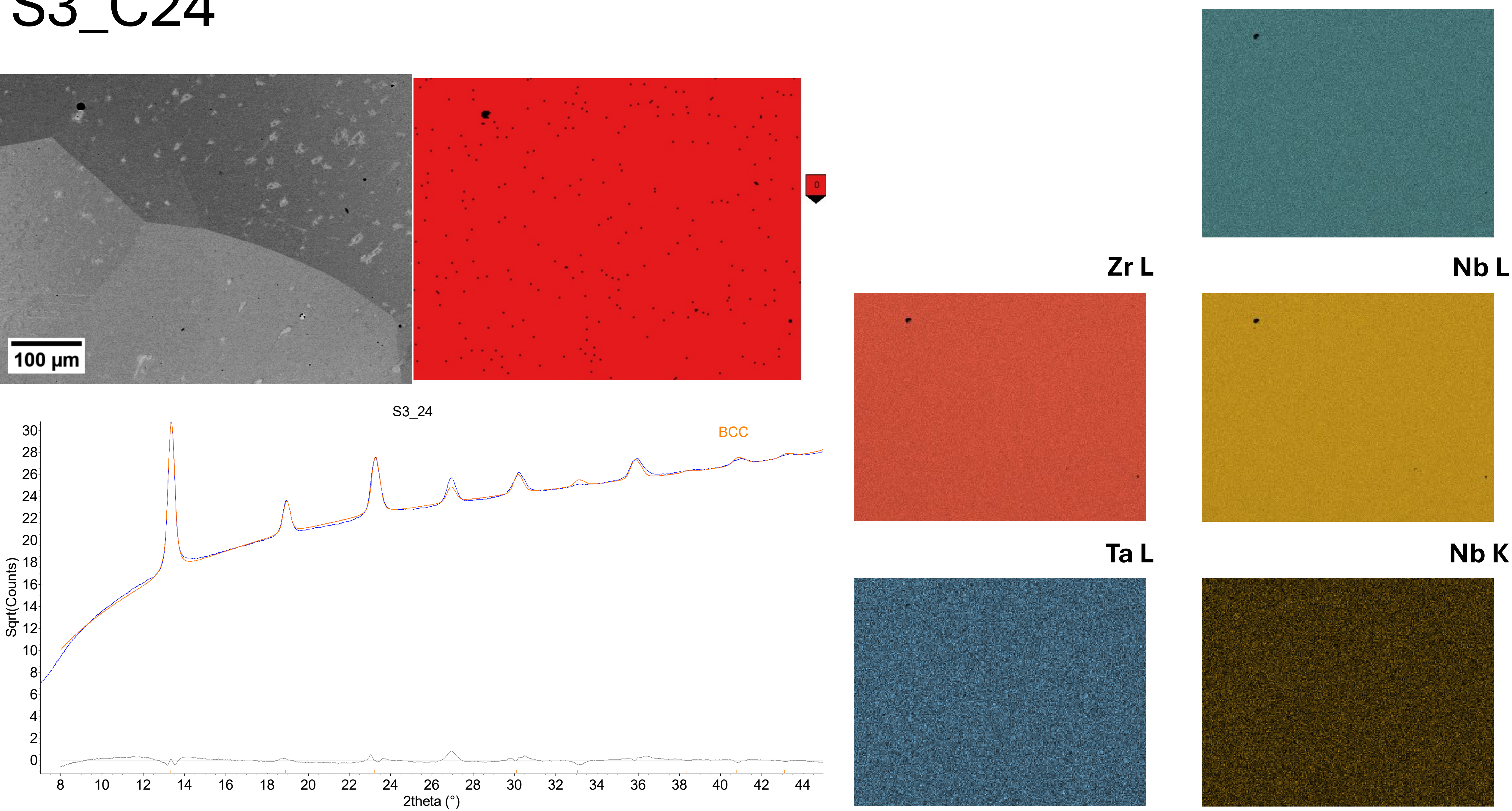

# S3_C30

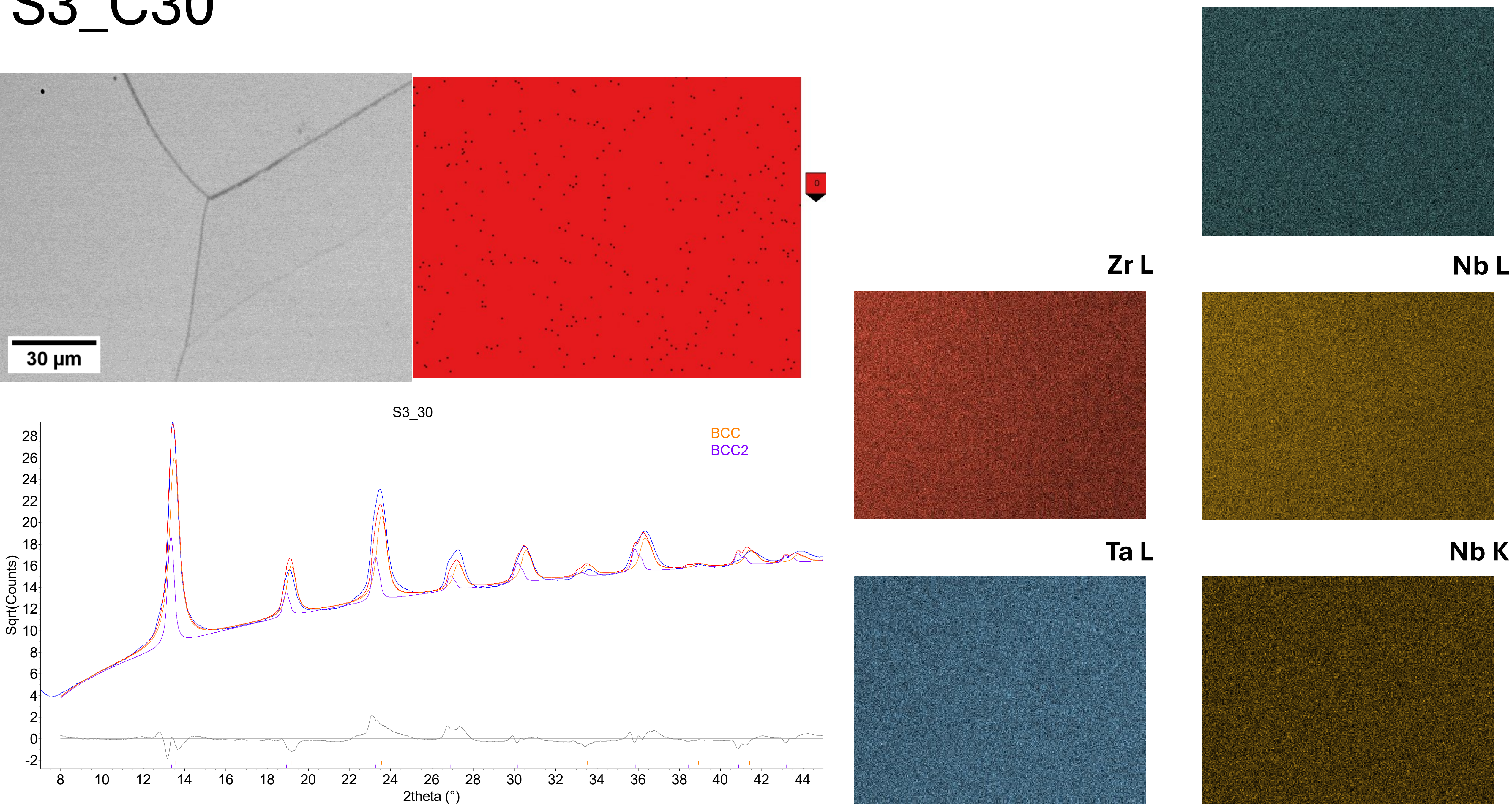

# S3_C31

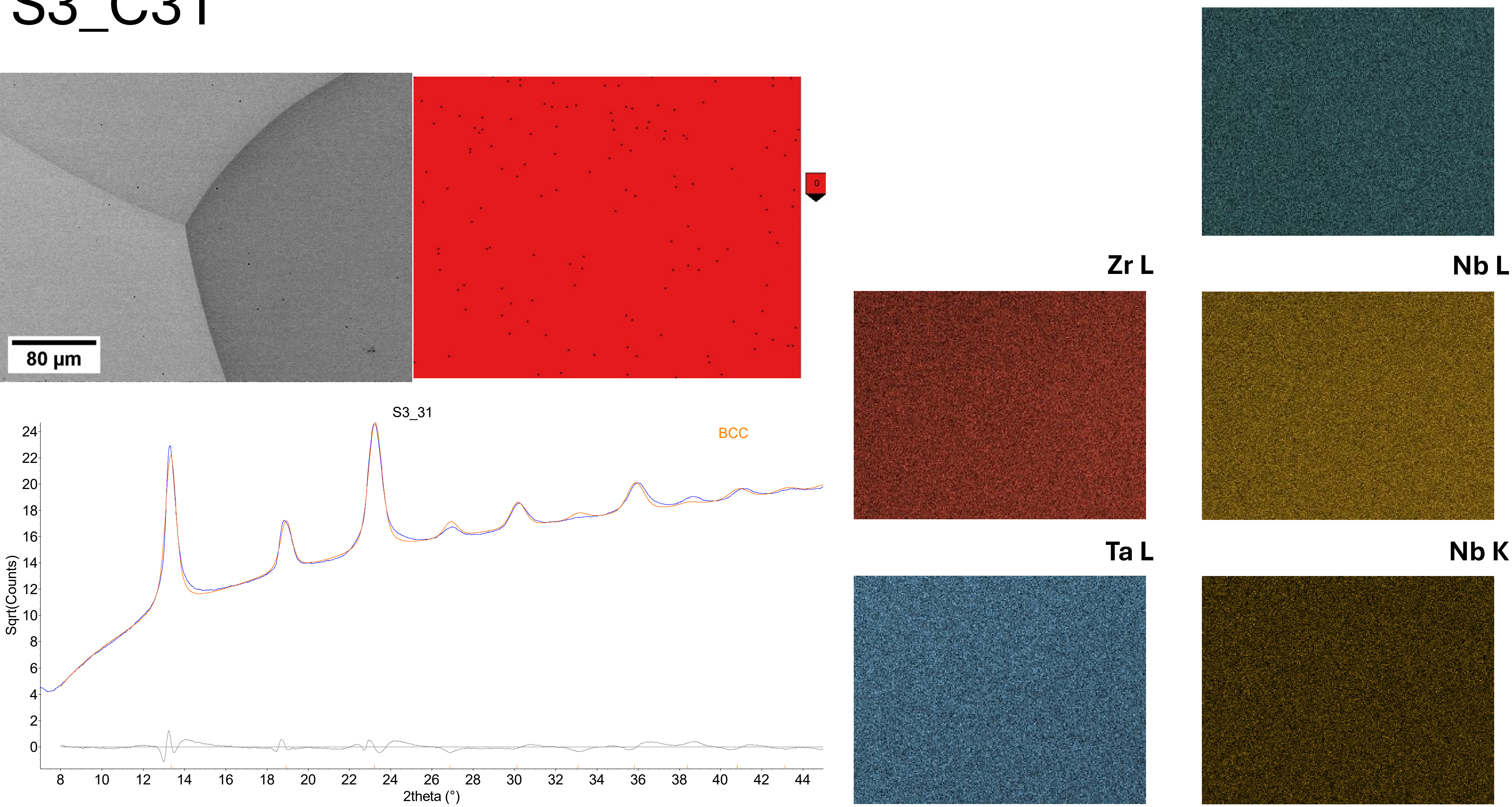

# S3_C32

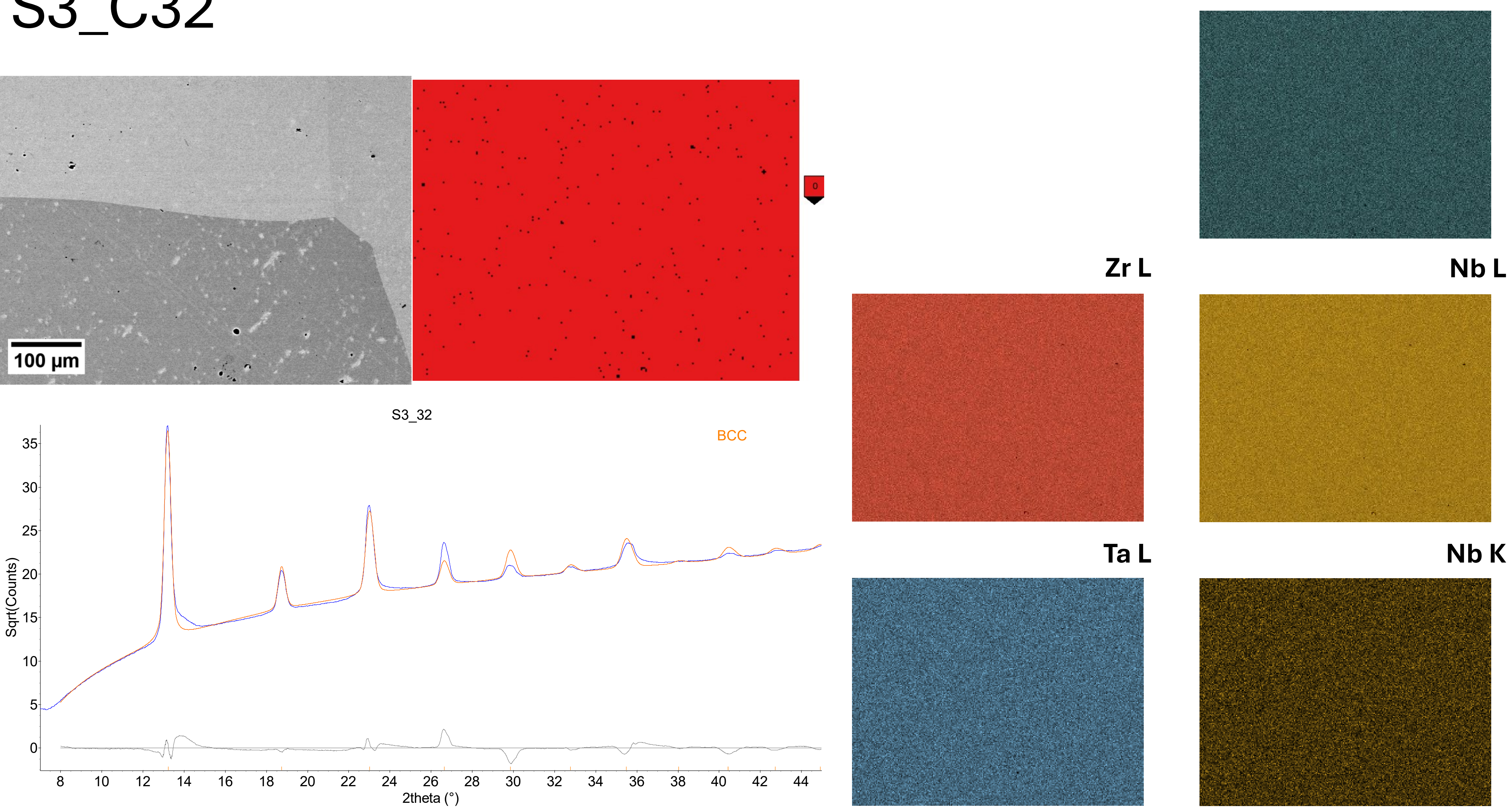

# S3_C33

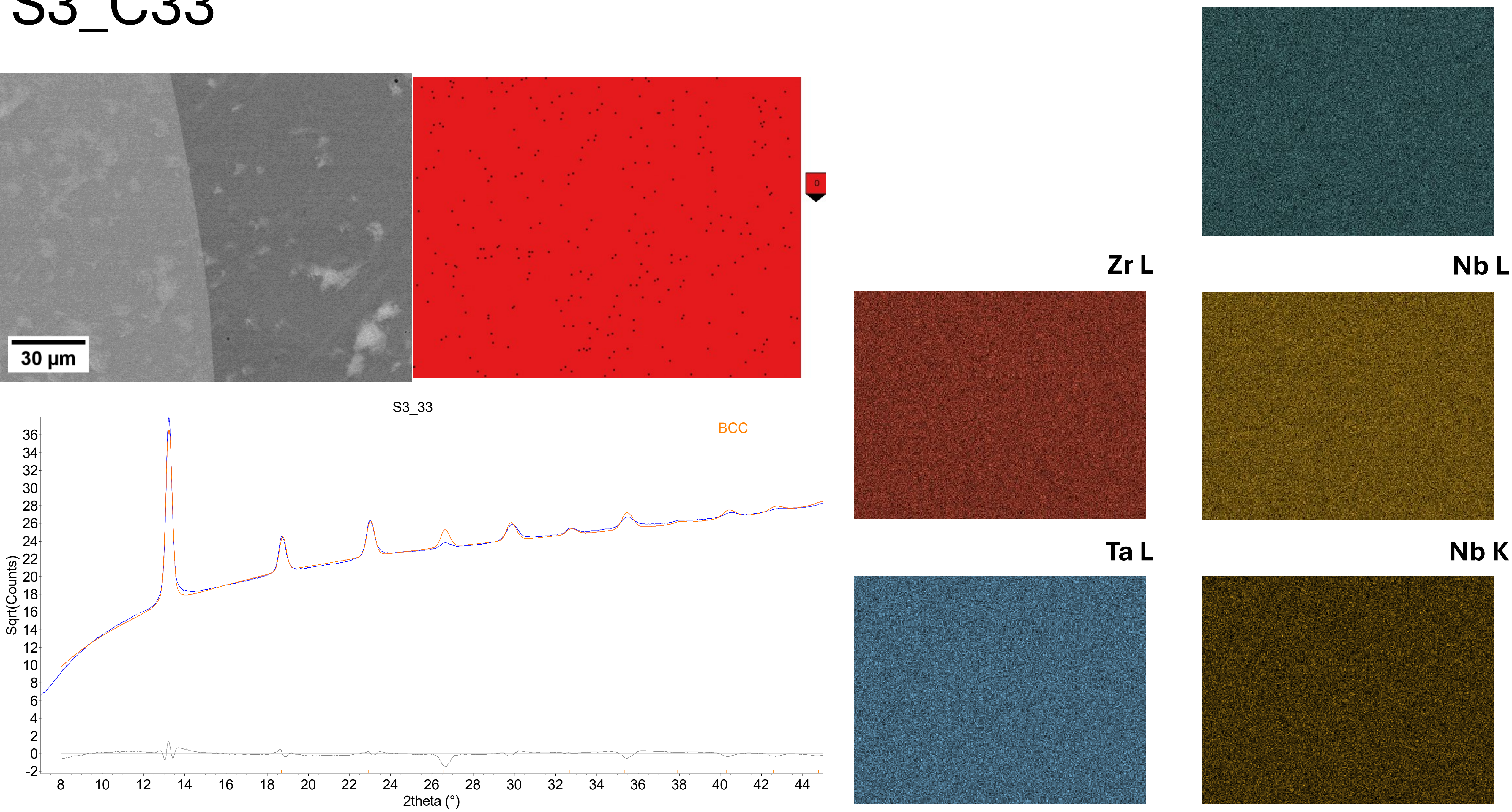

# S3_C40

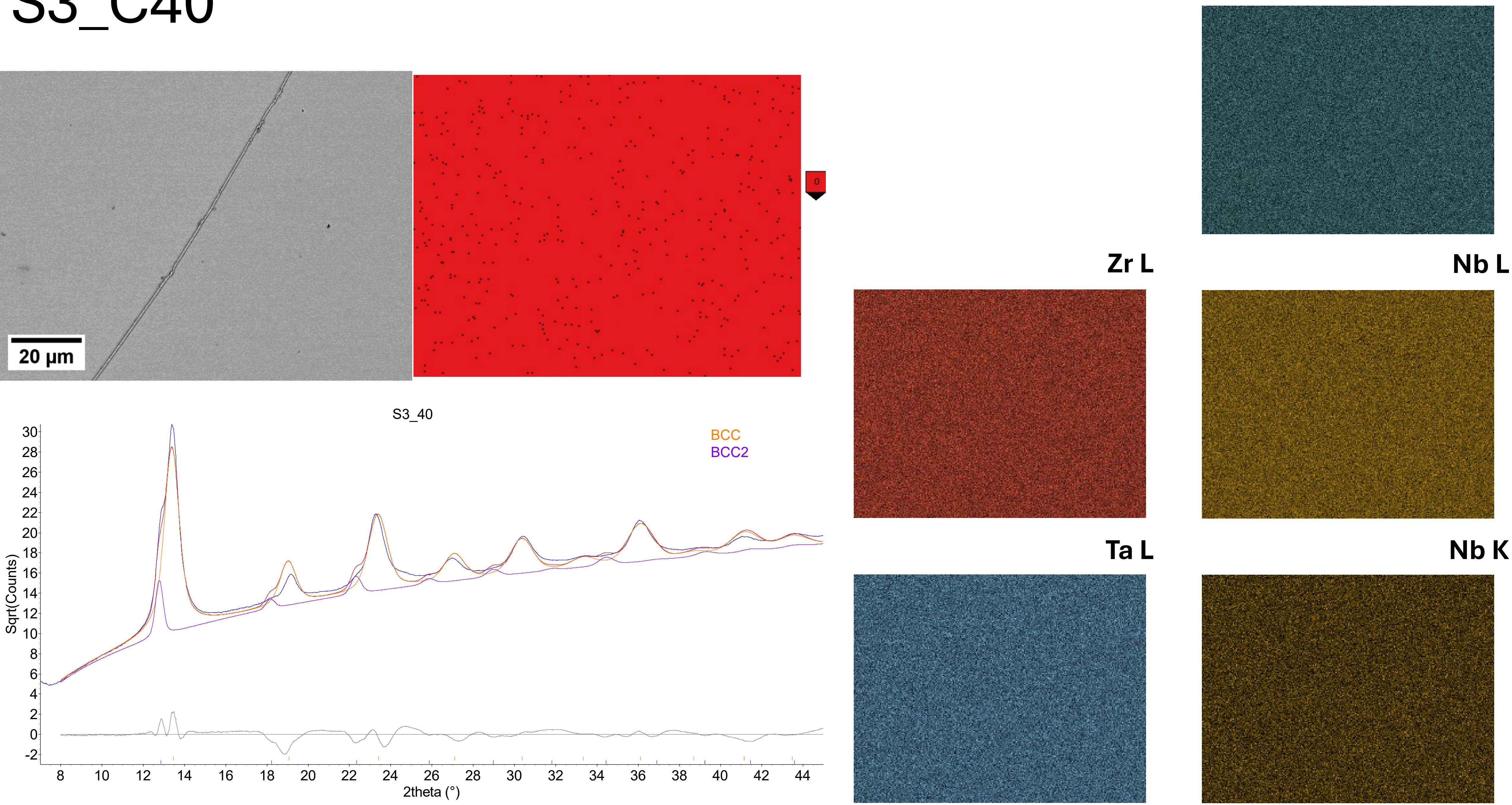

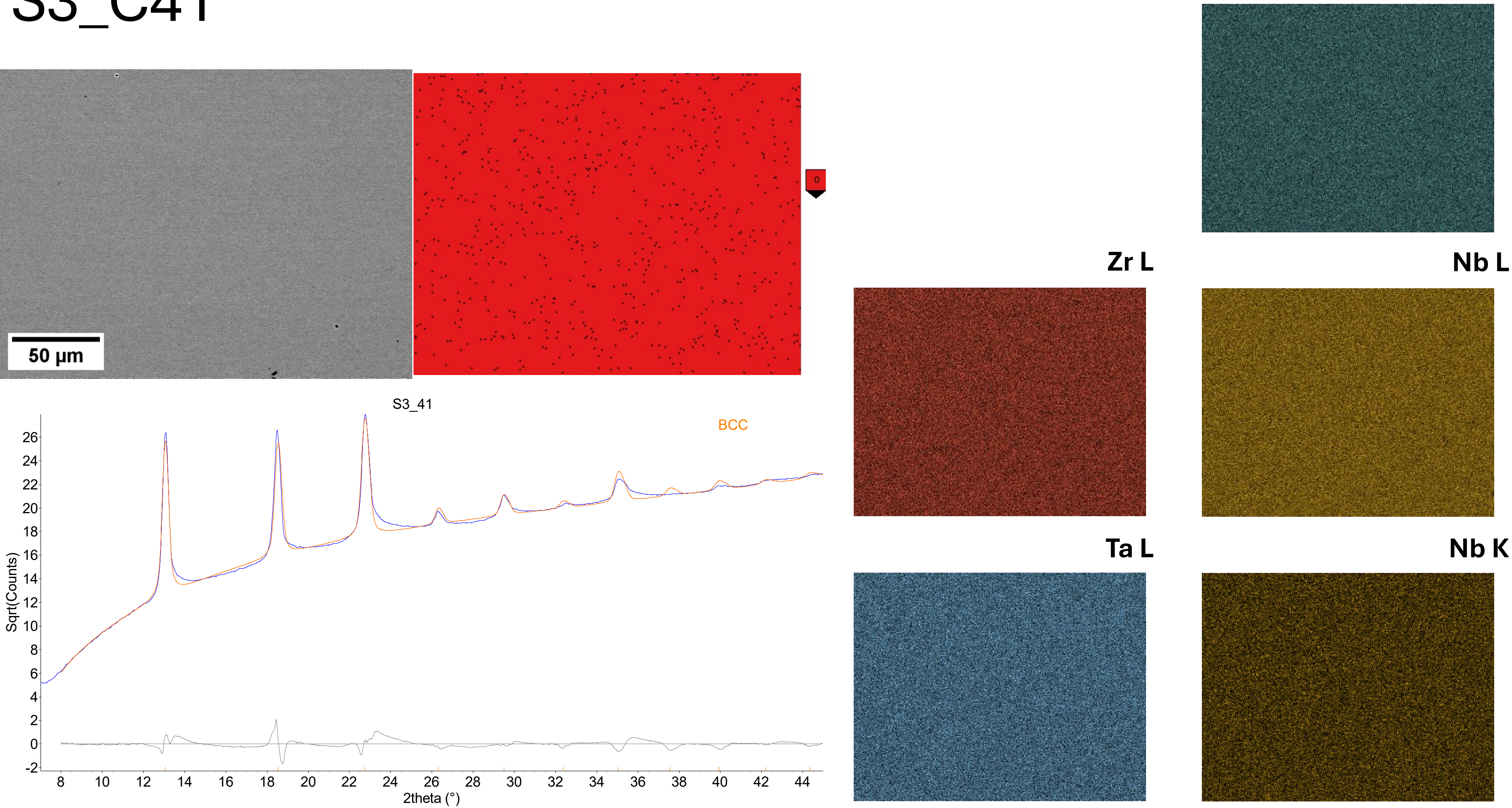

# S3_C42

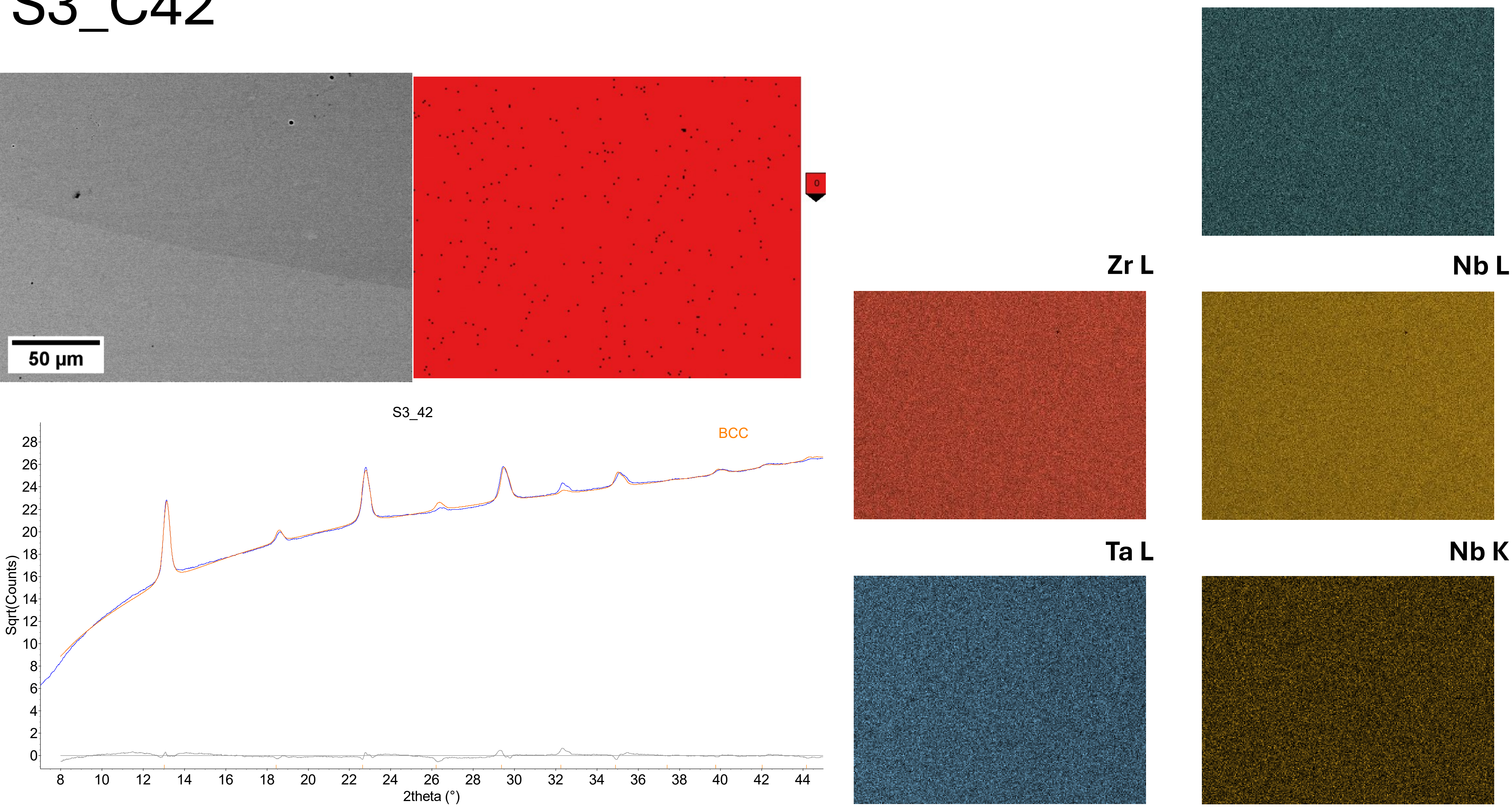

# Supplementary material C: Correlative EDS + EBSD analysis

compartments S2_C22, S2_C23, S2_C24 and S2_C33

Supplementary of paper:

## Phase Equilibria of the Al-Ti-Nb-Zr-Ta System

Jiří Kozlík [a*], František Lukáč [b], Mariano Casas-Luna [a], Jozef Veselý [a], Eliška Jača [a], Kateřina Ficková [a], Stanislav Šašek [a], Kristína Bartha [a], Adam Strnad [a], Tomáš Chráska [b], Josef Stráský [a]

[a] *Charles University, Faculty of Mathematics and Physics, Department of Physics of Materials, Ke Karlovu 5, Prague 121 16, Czechia*

[b] *Institute of Plasma Physics of the Czech Academy of Sciences, U Slovanky 2525/1a, Prague 182 00, Czechia*

* Corresponding author: jiri.kozlik@matfyz.cuni.cz

# S2_C22

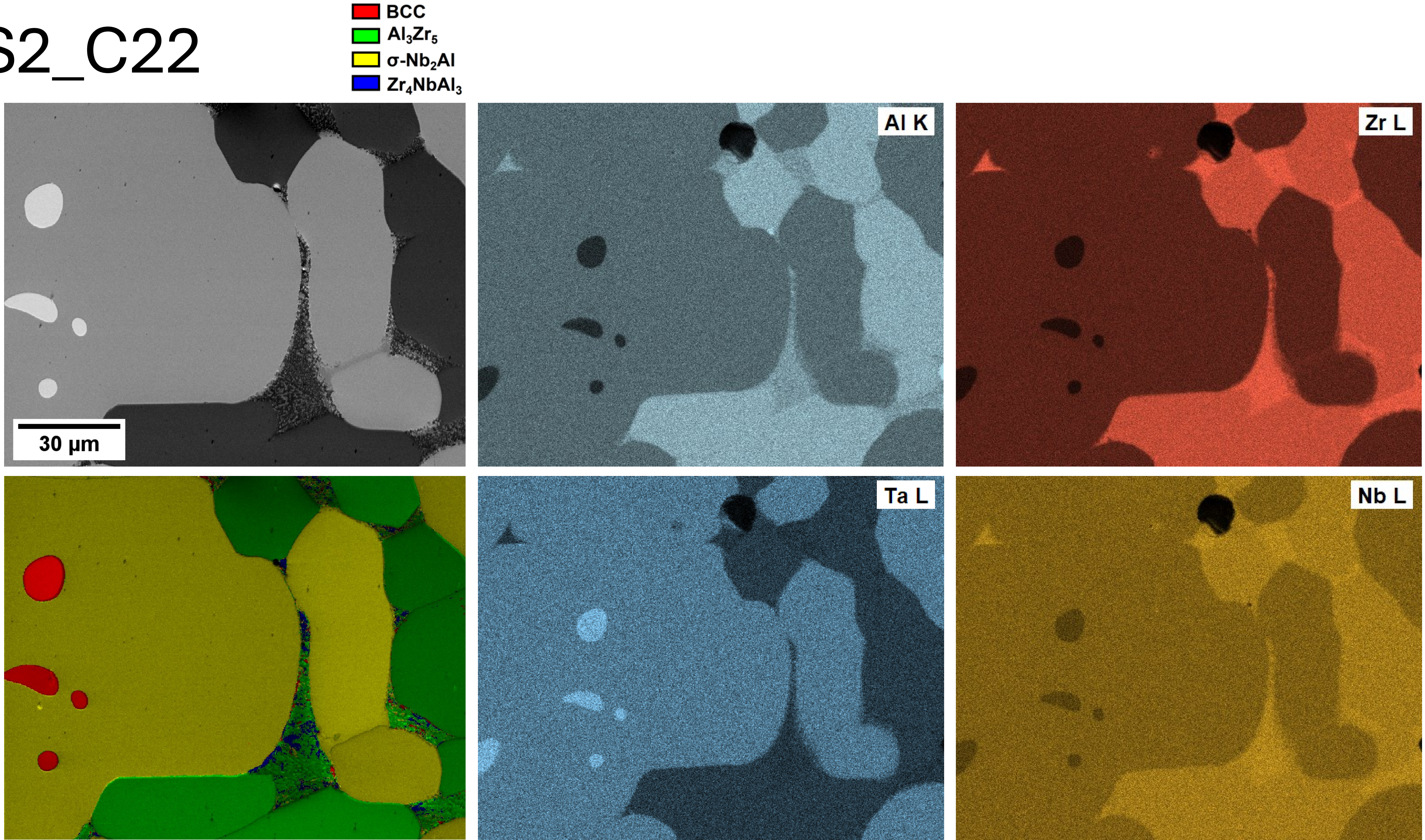

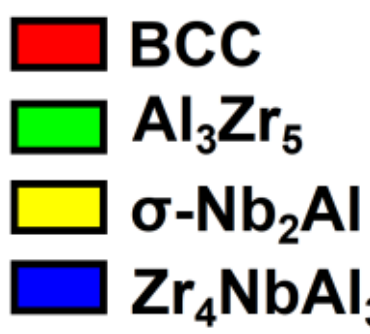

# S2_C23

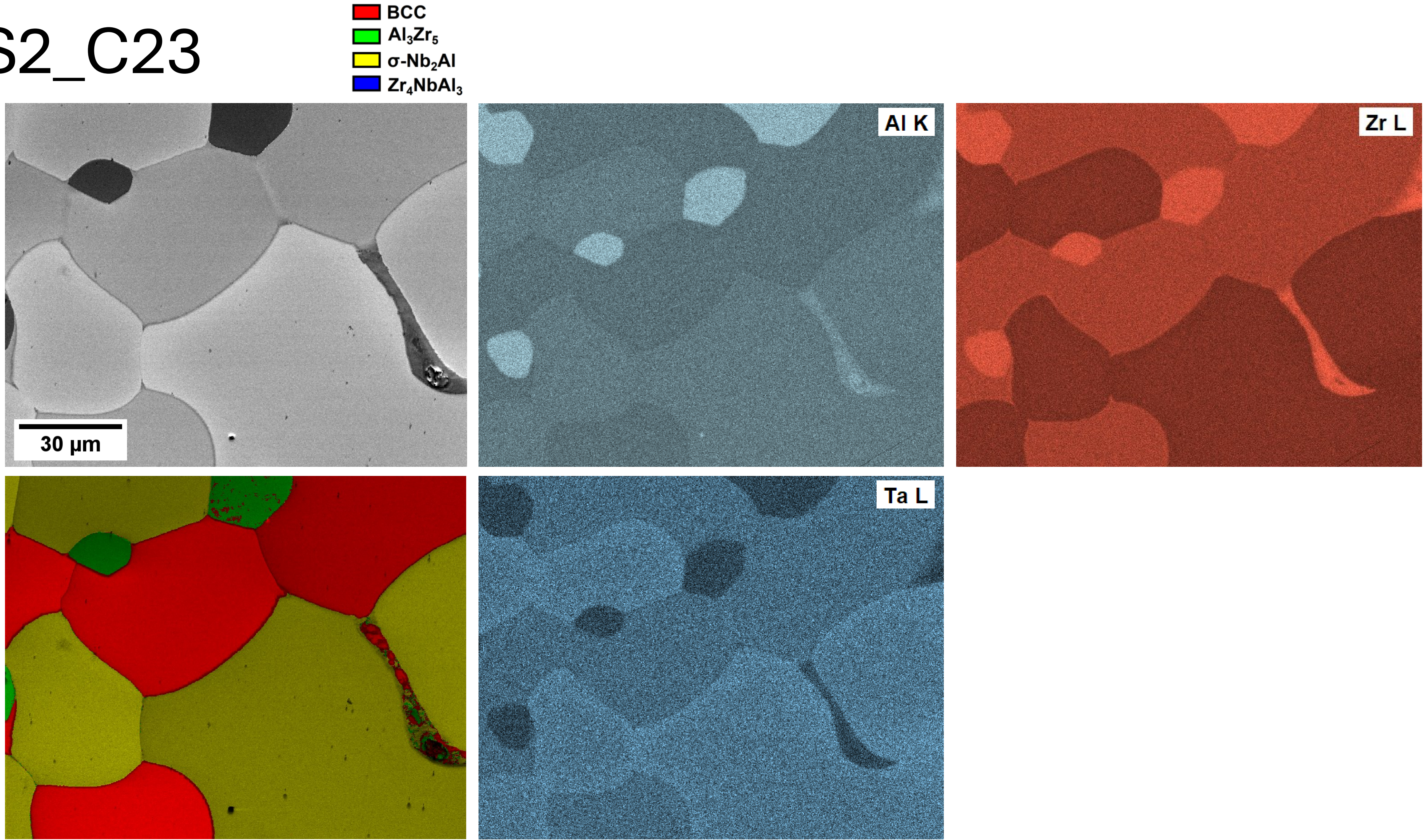

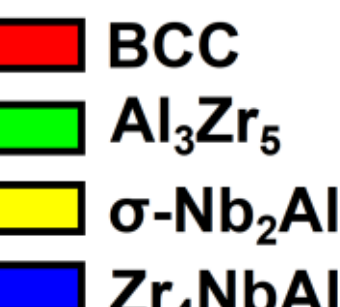

30 µm

Al K

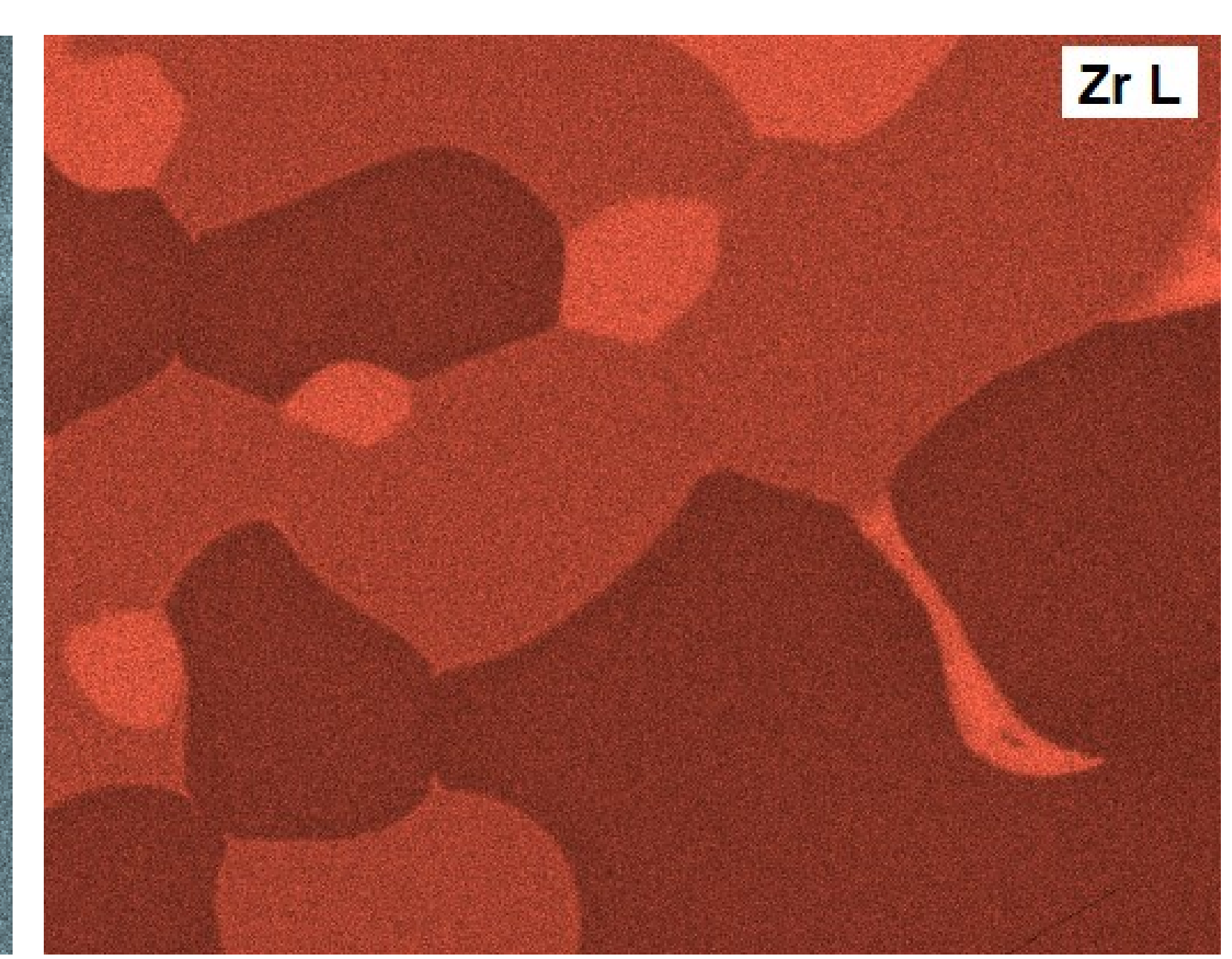

Ta L

# S2_C24

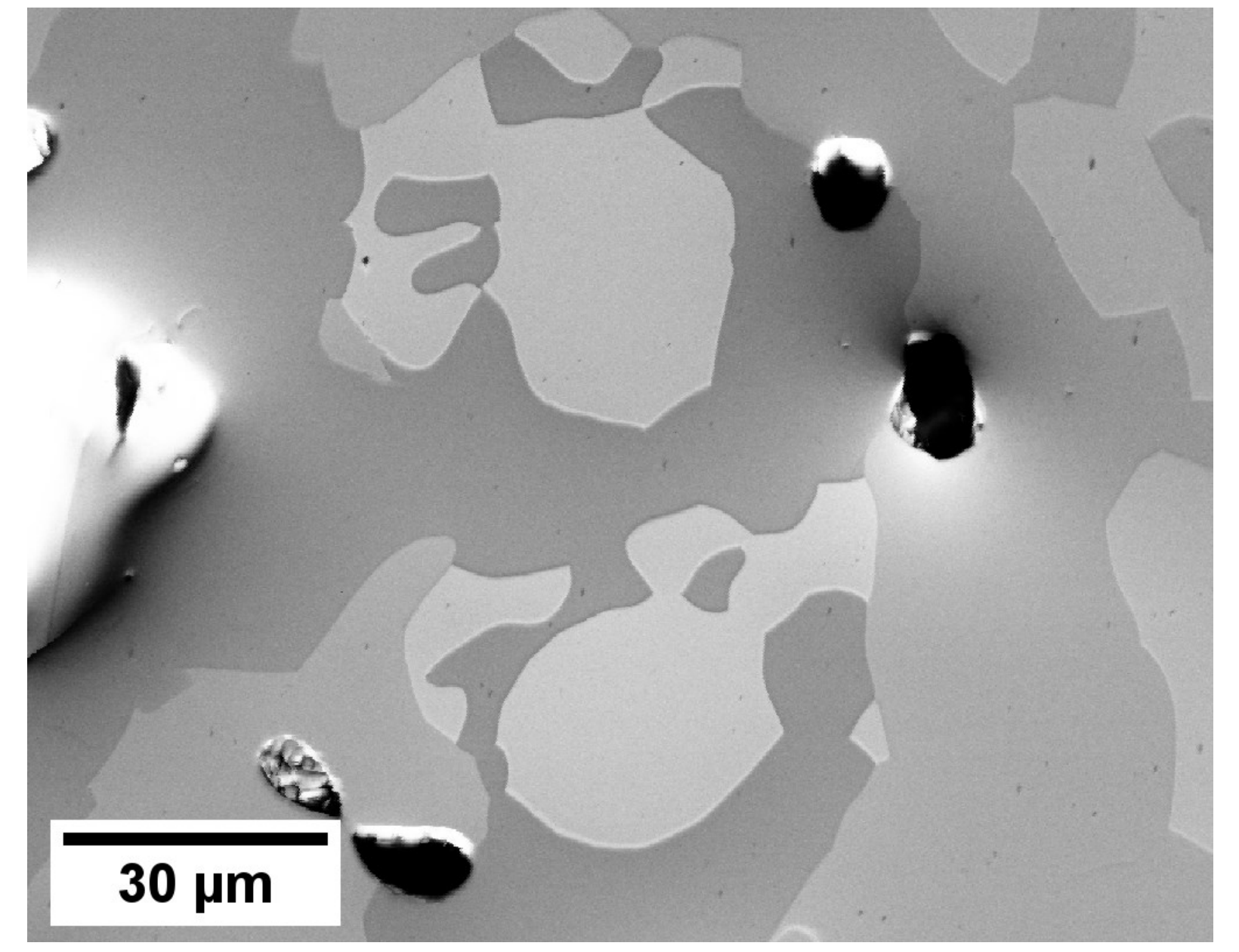

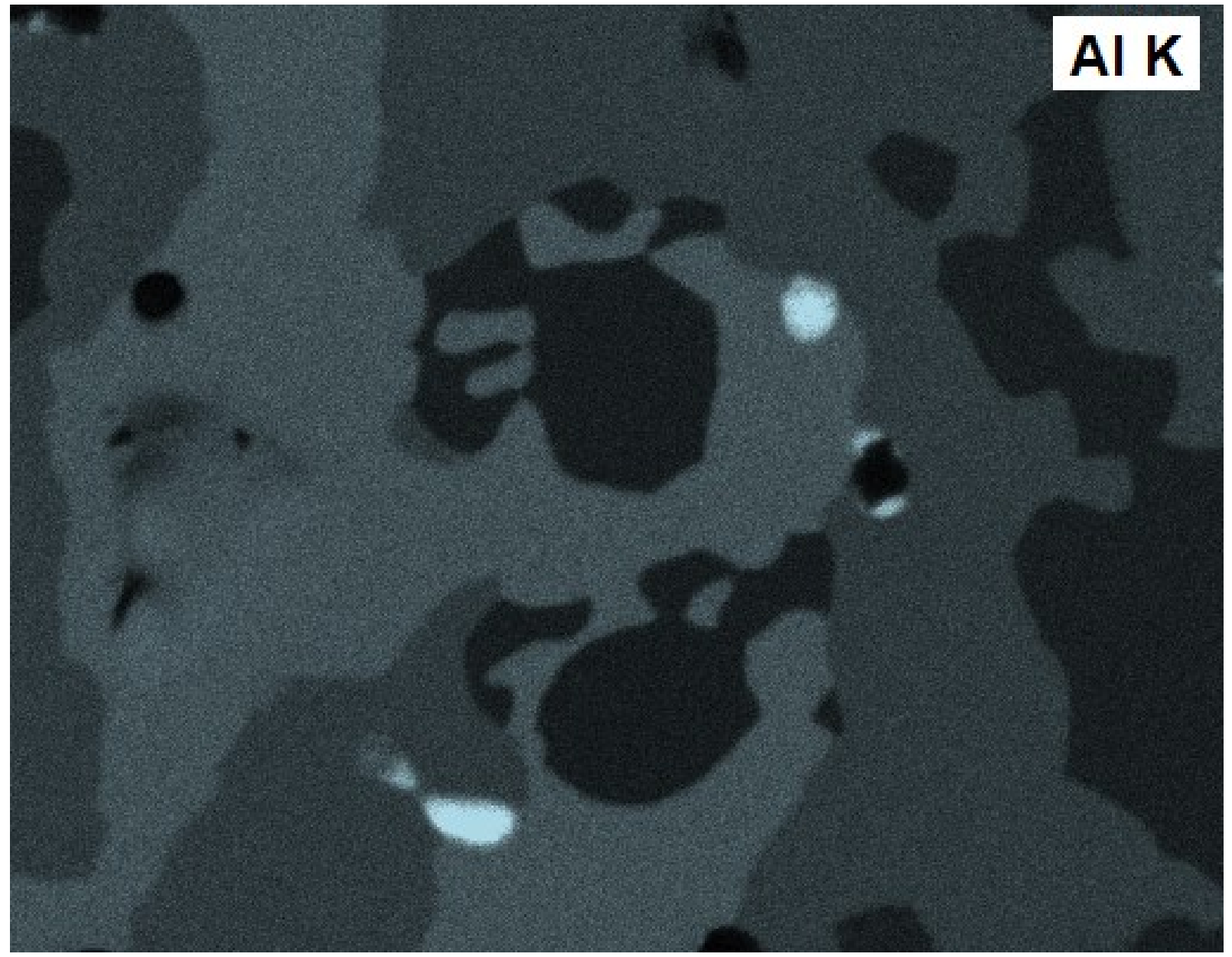

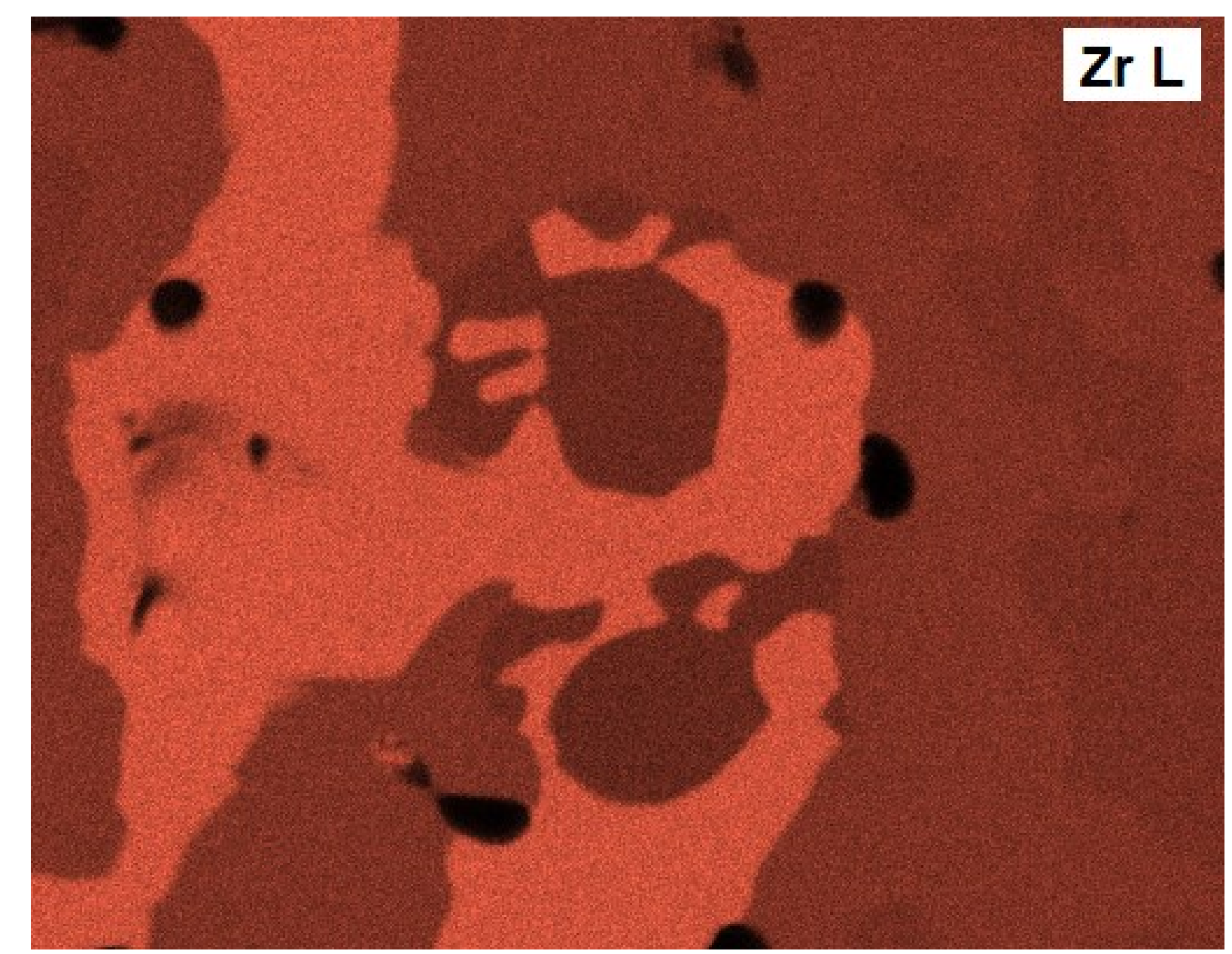

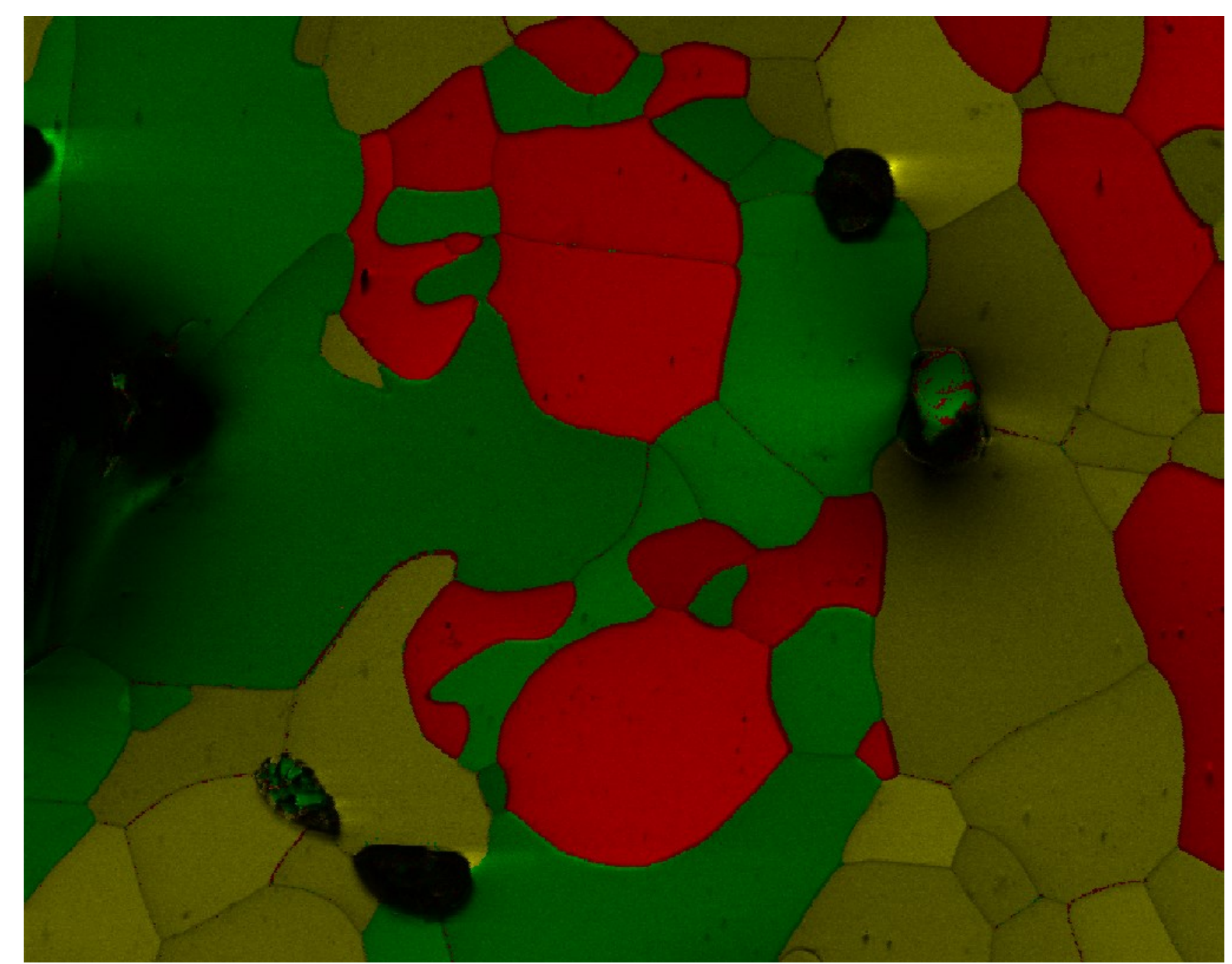

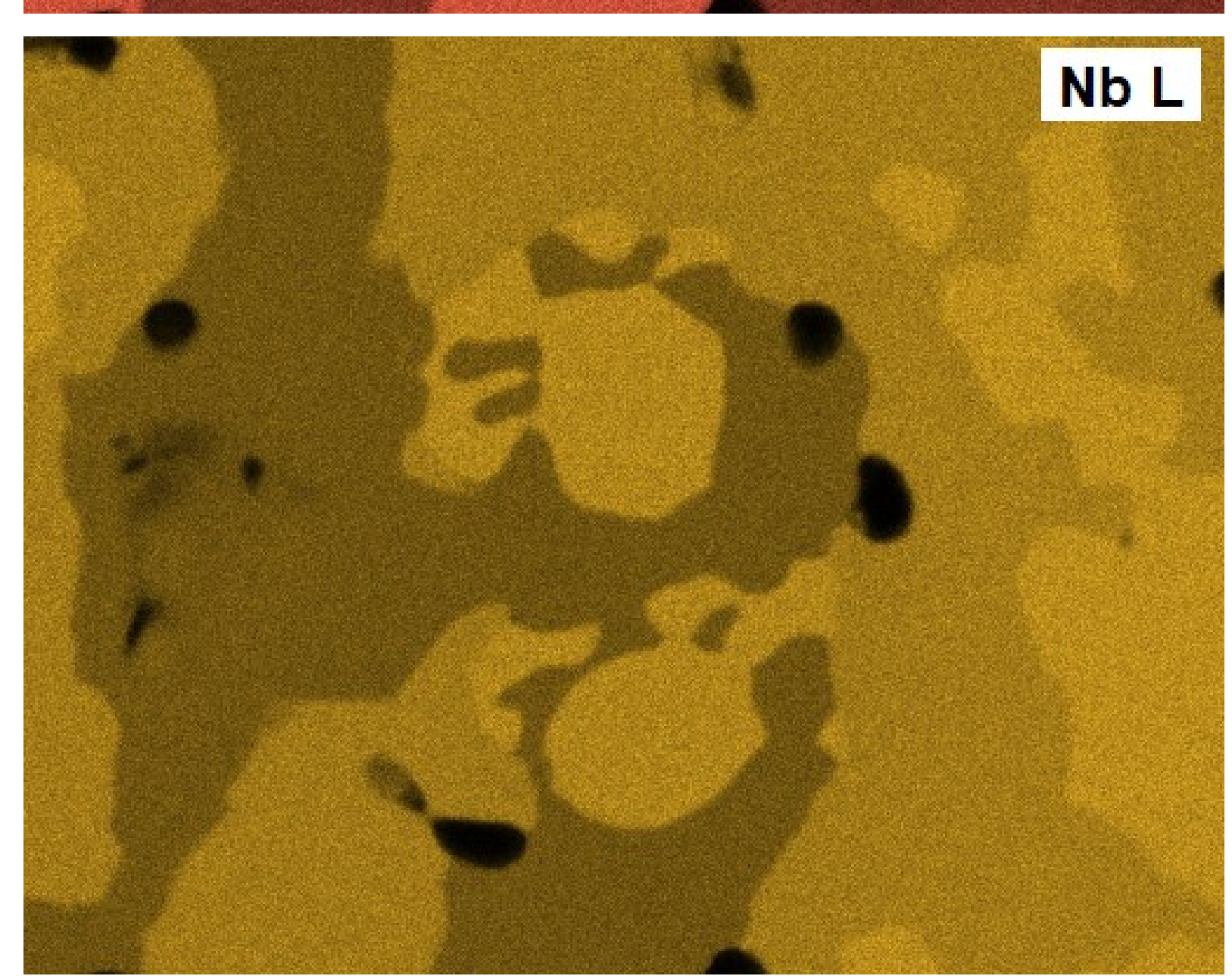

# S2_C33

- BCC
- $Al_3Zr_5$
- $\sigma$-$Nb_2Al$
- $Zr_4NbAl_3$

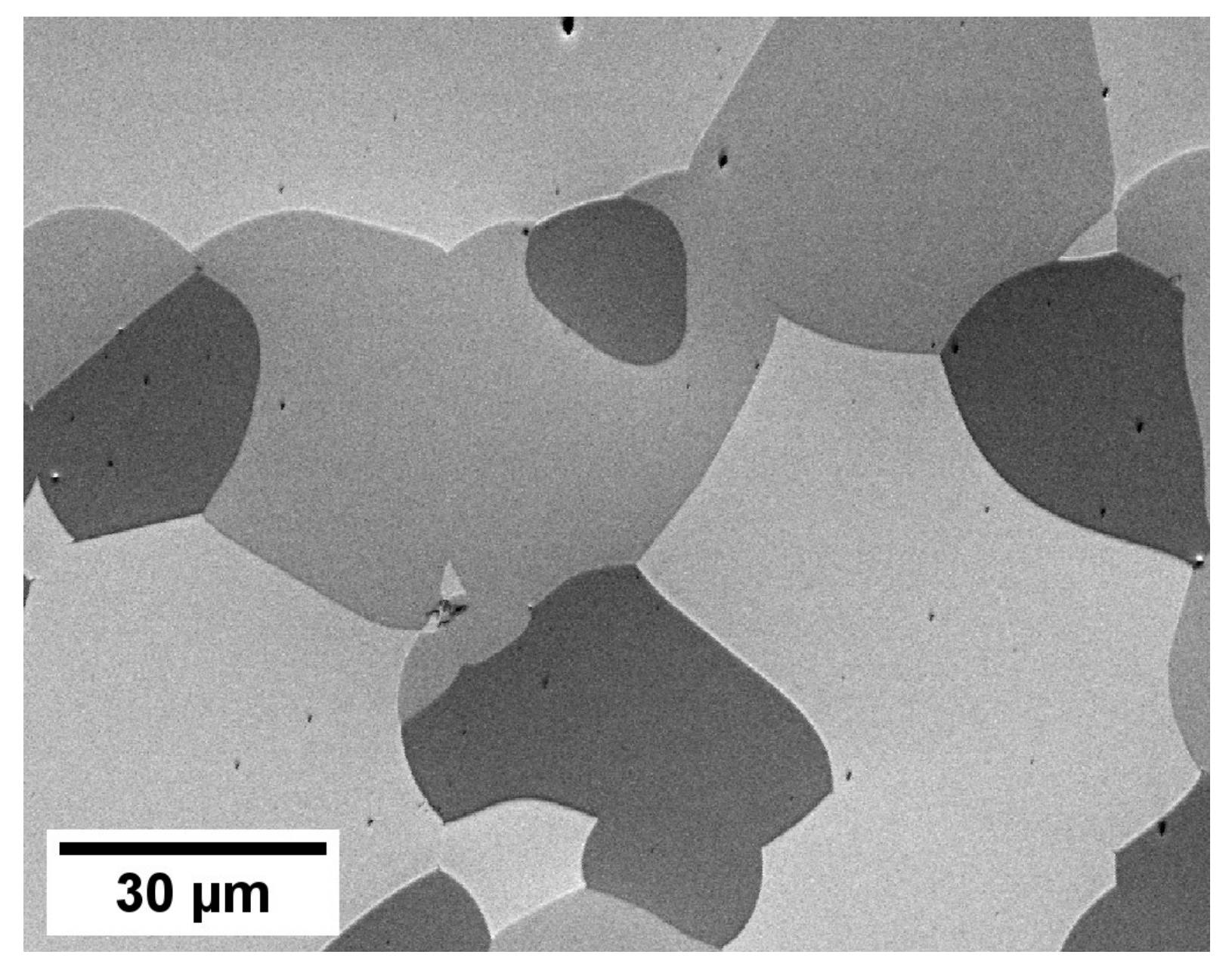

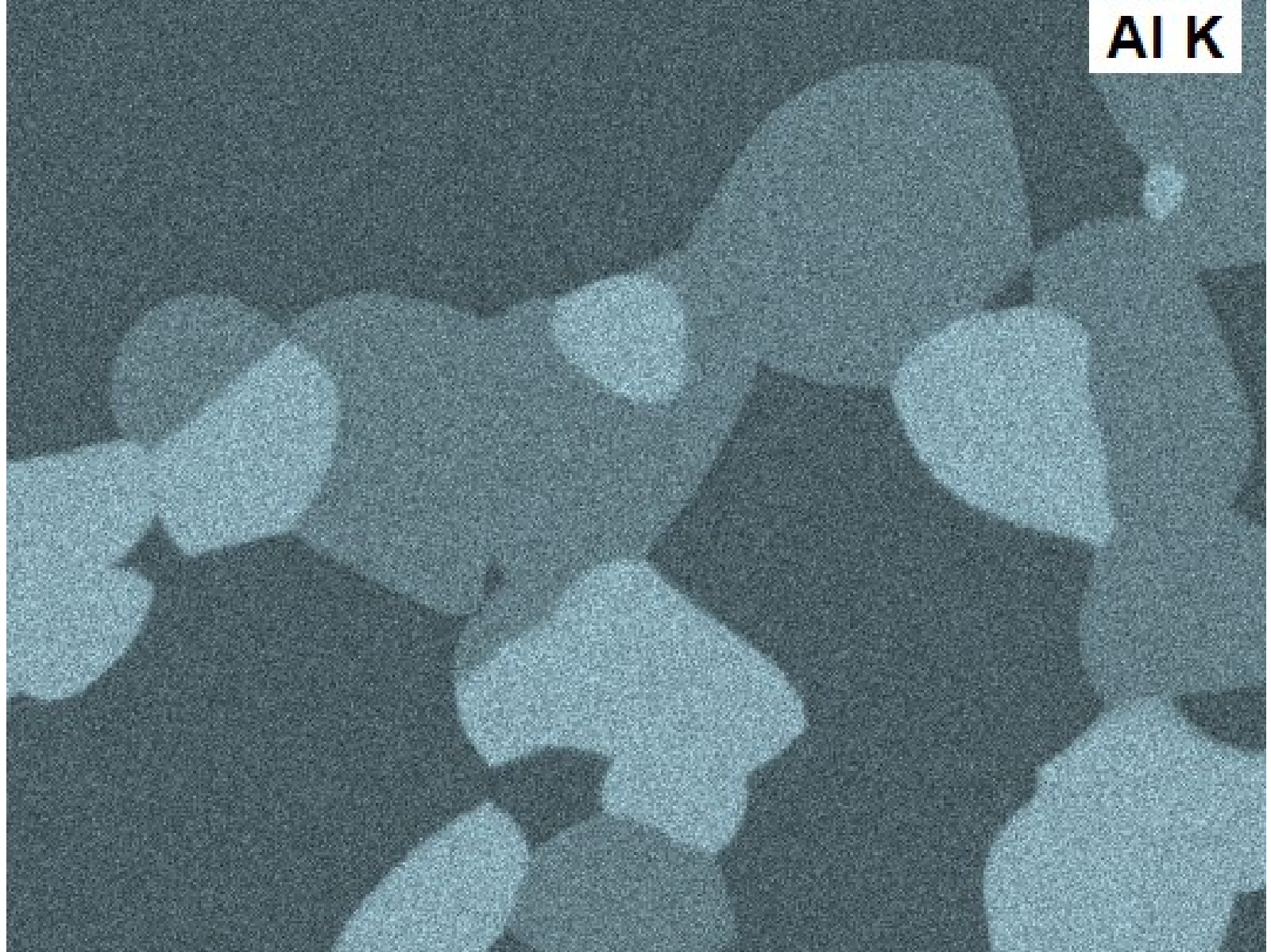

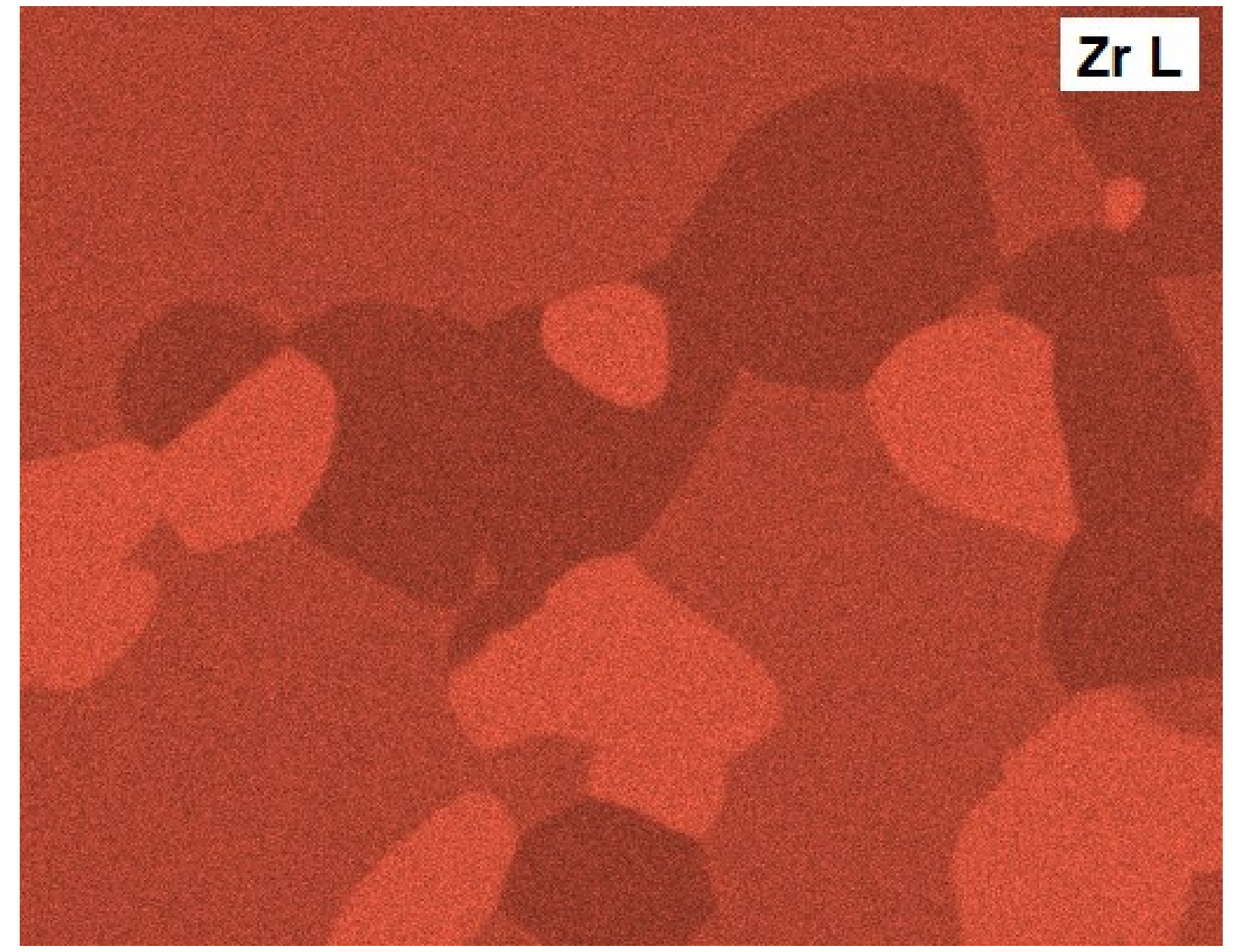

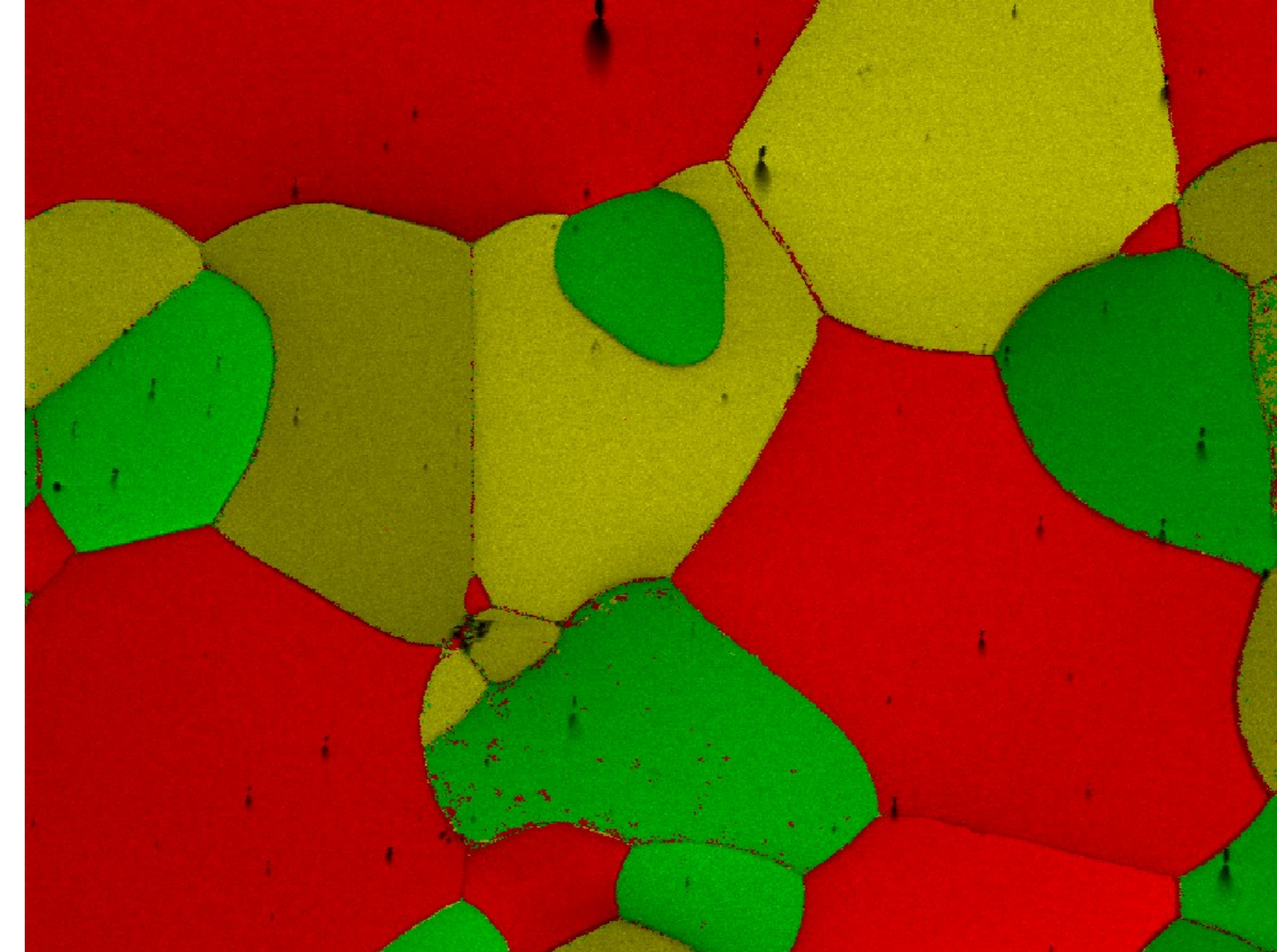

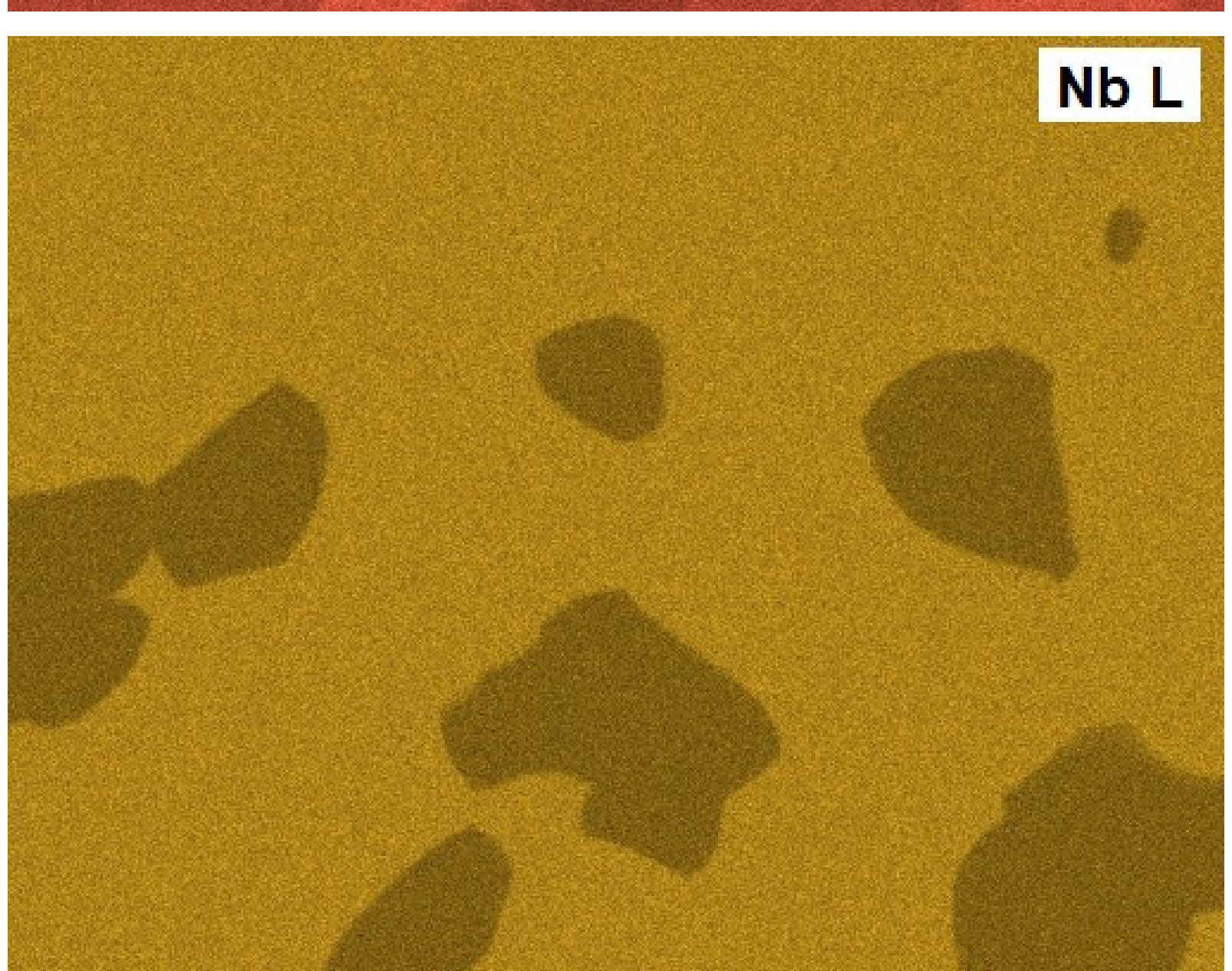